# The rules of long DNA-sequences and tetra-groups of oligonucleotides

Sergey V. Petoukhov

Head of Laboratory of Biomechanical System, Mechanical Engineering Research Institute of the Russian Academy of Sciences, Moscow

spetoukhov@gmail.com, http://petoukhov.com/

**Comment:** Some materials of this article were presented by the author in the keynote speeches at the following conferences: the International Belgrade Bioinformatics Conference 2018 (Belgrade, Serbia, 18-22 June 2018, http://belbi.bg.ac.rs/); the 2nd International Conference Artificial Intelligence, Medical Engineering, Education (Moscow, Russia, 1-3 October 2019); the 3rd International Conference on Computer Science, Engineering and Education Applications (Kiev, Ukraine, 21-22 January 2020). Also an author's presentation with elements of this article was done at the 6th International Conference in Code Biology (Friedrichsdorf, Germany, 3-7 June 2019, http://www.codebiology.org/conferences/Friedrichsdorf2019/).

**Abstract**. The article represents a new class of hidden symmetries in long sequences of oligonucleotides of single stranded DNA from their representative set. These symmetries are an addition to symmetries described by the second Chargaff's parity rule (%A ≅ %T and %G ≅ %C). These new symmetries and their rules concern the cooperative oligomer organization of long DNA sequences including complete sets of chromosomes of human and some model organisms. The rules on the equality of collective probabilities and the associated hyperbolic rules for the total sums of oligomers of length *n* were identified in long DNA sequences due to using the author's method of oligomer sums, which is also described in the article. These rules are considered as possible candidacies for the role of the universal rules of long DNA-sequences. A quantum-informational model of the described genetic symmetries is proposed on the basis of the known quantum-mechanic statement that quantum state of a multicomponent system is defined by the tensor product of quantum states of its subsystems. In this model, nitrogenous bases C, T, G, A of DNA are represented as computational basis states of 2-qubit quantum CTGA-systems. An important role of resonances, photons and photonic crystals in quantum-information genetics is noted.

## 1. Introduction.

Two Chargaff's parity rules are well known in genetics. They are important because they point to a kind of "grammar of biology" (these words were used by E.Chargaff in the title of his article [Chargaf, 1971]): a set of hidden rules that govern the structure of DNA. The first Chargaff's parity rule states that in any double-stranded DNA segment, the number of occurrences (or frequencies) of adenine A and thymine T are equal, and so are frequencies of cytosine C and guanine G [Chargaff, 1951, 1971]. The rule was an important clue that J.Watson and F.Crick used to develop their model of the double helix structure of DNA.

The second Chargaff's parity rule (CSPR) states that both %A ≅ %T and %G ≅ %C are approximately valid in single stranded DNA for long nucleotide sequences. Many works of different authors are devoted to confirmations and discussions of this second Chargaff's rule [Albrecht-Buehler, 2006, 2007; Baisnee, Hampson, Baldi, 2002; Bell, Forsdyke, 1999; Chargaff, 1971, 1975; Dong, Cuticchia, 2001; Forsdyke, 1995, 2002, 2006; Forsdyke, Bell, 2004; Mitchell, Bridge, 2006; Okamura, Wei, Scherer, 2007; Perez, 2013; Prabhu, 1993; Rapoport, Trifonov, 2012; Sueoka, 1995; Yamagishi, Herai, 2011]. Originally, CSPR is meant to be valid only to mononucleotide frequencies (that is quantities of monoplets) in single stranded DNA. "*But, it occurs that oligonucleotide frequencies follow a generalized Chargaff's second parity rule (GCSPR) where the frequency of an oligonucleotide is approximately equal to its complement reverse oligonucleotide frequency [Prahbu, 1993]. This is known in the literature as the Symmetry Principle*" [Yamagishi, Herai, 2011, p. 2]. The work [Prahbu, 1993] shows the implementation of the Symmetry Principle in long DNA-sequences for cases of complementary reverse n-plets with n = 2, 3, 4, 5 at least. (In literature, a few synonymes of the term "n-plets" are used: n-tuples, n-words or n-mers).

These parity rules, including generalized Chargaff's second parity rule for n-plets in long nucleotide sequences, concerns the equality of frequencies of two separate mononucleotides or two separate oligonucleotides, for example: the equality of frequencies of adenine and thymine; the equality of frequences of the doublet CA and its complement-reverse doublet TG; the equality of the triplets CAT and its complement-reverse triplet ATG, etc. By contrast to this, to study hidden symmetries in long sequences of oligonucleotides of single stranded DNA, we apply a comparative analysis of equalities not for probabilities of single oligonucleotides but for collective probabilities of sets of oligonucleotides, which form subgroups of so called tetra-groups. Below we explain the notion of these tetra-groups of oligonucleotides and represent new rules of probabilities in tetra-groups of long sequences of oligonucleotides in single stranded DNA of many organisms including complete sets of chromosomes of human and some model organisms.

## 2. Tetra-groups of oligonucleotides and collective probabilities of tetra-groups in long DNA-texts

Information in DNA strands is written by means of the tetra-group of nitrogenous bases: adenine A, cytosine C, guanine G and thymine T (in RNA the tetra-group of nitrogenous bases contains uracil U instead of thymine T). E.Chargaff has received both his parity rules by comparative analysis of frequencies of each of 4 members of this tetra-group of mononucleotides in DNA. He and his followers studied DNA sequences as sequences of mononucleotides, frequences of separate fragments of which were compared. In other words, Chargaff studied long DNA-texts in a form of texts of 1-letter words and calculated frequencies (or probabilities) of each of 4 members of the genetic tetra-group of 1-letter words A, T, C and G.

But it is obvious that each of long DNA-sequences (for example, the sequence CAGGTATCGAAT…) can be represented not only in the form of the text of 1-letter words (C-A-G-G-T-A-T-C-G-A-A-T-…) but also in the form of the text of 2-letter words (CA-GG-TA-TC-GA-AT-…) or in the form of the text of 3-letter words (CAG-GTA-TCG-AAT-…) or in the form of the text of n-letter words in a general case. We briefly call such representations "n-letter representations" of DNA-sequences. Any of such long DNA-texts of n-letter words can be considered as a collection of 4 subgroups of possible tetra-groups, each of which is defined by an attribute of one of 4 letters A, T, C, G at a certain position inside n-letter words of the DNA-text. As we know, till now nobody studied systematically frequencies and probabilities of subgroups of such tetra-groups in DNA-texts of n-letter words. The proposed set of representations of long DNA-sequences plays a key role in revealing hidden symmetries in these sequences. This article shows some results of author's study of hidden symmetries in long DNA-texts of n-letter words.

In contrast to Chargaff, in our approach, firstly, we analyze DNA sequences not as sequences of mononucleotides but as sequences of oligonucleotides of identical lengths: as sequences of doublets, or triplets, or 4-plets, or 5-plets, etc. Secondly, we compare not values of separate frequencies of individual oligonucleotides in long nucleotide sequences but values of sums of indvidual frequencies of all oligonucleotides, which belong to each of 4 subgroups of special tetra-groups of oligonucleotides of an identical length (such sum is called a collective frequency of the subgroup of the tetra-group). At the final stage of the analysis we compare collective probabilities (or percentage) of collective frequencies of different subgroups of such tetra-groups in the considered sequence. The mentioned tetra-groups are formed in each of considered cases by means of certain positional attributes of the letters A, T, C and G inside oligonucleotides of an identical length. Each of 4 subgroups of such tetra-group combines all oligonucleotides of the same length n (n-plets), which possess the identical letter at their certain position. To simplify the explanation, Fig. 1 shows the example of two tetra-groups, which are formed and studyed by us for the analysis of long sequences of doublets.

| Subgroups of tetra-groups | Composition of subgroups of doublets with identical letters at their first positions | Composition of subgroups of doublets with identical letters at their second position |
|---|---|---|
| A-subgroup | AA, AC, AG, AT | AA, CA, GA, TA |

| T-subgroup | TC, TA, TT, TG | CT, AT, TT, GT |
|---|---|---|
| C-subgroup | CC, CA, CT, CG | CC, AC, TC, GC |
| G-subgroup | GC, GA, GT, GG | CG, AG, TG, GG |

Fig. 1. Compositions of two tetra-groups of doublets with 4 doublets in each of their 4 subgroups.

In the first tetra-group in Fig. 1, the complete alphabet of 16 doublets is divided into 4 subsets with 4 doublets in each by the attribute of an identical letter on the first position in each of doublets. The complect of these 4 subsets is called the tetra-group of doublets on the basis of this attribute; each of the 4 subsets is called a subgroup of the tetra-group; each of 4 subgroup has its individual name with an indication of its characteristic letter (A-subgroup, T-subgroup, C-subgroup and G-subgroup). In the second tetra-group (Fig. 1, right) the complete set of 16 doublets is divided into 4 subsets with 4 doublets in each by the attribute of an identical letter on the second position in each of doublets. The name "tetra-group" is used since 4 subgroups exist here and 4 letters of DNA play a decisive role in the dismemberment of the set of m-plets on the regular subsets in question. It is obvious that corresponding subgroups of both tetra-groups of doublets in Fig. 1 (left and right columns) are interrelated on the basis of cyclic shifts of positions in doublets. For example, the set of doublets AA, AC, AG, AT in the A-subgroup of the first tetra-group (Fig. 1, left column) is transformed into the A-subgroup AA, CA, GA, TA of the second tetra-group (Fig. 1, right column) by the cyclic shift of positions in doublets. Similar mutual transformations of subgroups of corresponding tetra-groups on the basis of cyclic shifts are also valid in the case of alphabets of triplets, 4-plets, 5-plets, etc. Fig. 2 shows three tetra-groups of triplets, which are used for the analysis of long sequences of triplets.

| Subgroups of tetra-groups | Composition of subgroups of triplets with an identical letter at their 1st position | Composition of subgroups of triplets with an identical letter at their 2nd position | Composition of subgroups of triplets with an identical letter at their 3rd position |
|---|---|---|---|
| A-subgroup | AAA, AAC, AAG, AAT, ATA, ATC, ATG, ATT, ACA, ACC, ACG, ACT, AGA, AGC, AGG, AGT | AAA, CAA, GAA, TAA, AAT, CAT, GAT, TAT, AAC, CAC, GAC, TAC, AAG, CAG, GAG, TAG | AAA, CAA, GAA, TAA, CTA, ATA, TTA, GTA, CCA, CAA, CTA, CGA, CGA, AGA, TGA, GGA |
| T-subgroup | TAA, TAC, TAG, TAT, TTA, TTC, TTG, TTT, TCA, TCC, TCG, TCT, TGA, TGC, TGG, TGT | ATA, CTA, GTA, TTA, ATT, CTT, GTT, TTT, ATC, CTC, GTC, TTC, ATG, CTG, GTG, TTG | AAT, CAT, GAT, TAT, CTT, ATT, TTT, GTT, CCT, CAT, CTT, CGT, CGT, AGT, TGT, GGT |
| C- | CAA, CAC, CAG, CAT, CTA, CTC, CTG, CTT, CCA, CCC, CCG, CCT, | ACA, CCA, GCA, TCA, ACT, CCT, GCT, TCT, ACC, CCC, GCC, TCC, | AAC, CAC, GAC, TAC, CTC, ATC, TTC, GTC, CCC, CAC, CTC, CGC, |

| subgroup | CGA, CGC, CGG, CGT | ACG, CCG, GCG, TCG | CGC, AGC, TGC, GGC |
|---|---|---|---|
| G-subgroup | GAA, GAC, GAG, GAT, GTA, GTC, GTG, GTT, GCA, GCC, GCG, GCT, GGA, GGC, GGG, GGT | AGA, CGA, GGA, TGA, AGT, CGT, GGT, TGT, AGC, CGC, GGC, TGC, AGG, CGG, GGG, TGG | AAG, CAG, GAG, TAG, CTG, ATG, TTG, GTG, CCG, CAG, CTG, CGG, CGG, AGG, TGG, GGG |

Fig. 2. Compositions of three tetra-groups of triplets with 16 triplets in each of their 4 subgroups.

In a general case of a sequence of n-plets, the complete alphabet of $4^n$ n-plets is divided into 4 subgroups with $4^{n-1}$ n-plets in each by the attribute of an identical letter on the chosen position inside n-plets. In this case n tetra-groups of n-plets are formed:

- The tetra-group on the basis of the attribute of an identical letter on the 1st position of n-plets;
- The tetra-group on the basis of the attribute of an identical letter on the 2nd position of n-plets;
- .....
- The tetra-group on the basis of the attribute of an identical letter on the n-th position of n-plets.

We use the symbol $\Sigma_n$ (n = 1 ,2, 3, 4,…) to denote the total quantity of n-plets in the considered DNA-sequence of n-plets; for example, the expression $\Sigma_3$=100000 means that an analyzed sequence of triplets contains 100000 triplets. Let us define notions and symbols of collective frequences and collective probabilities of subgroups of tetra-groups in long sequences of n-plets, where each of n-plets has its individual frequency (or number of its occurrences): for example, doublets have their individual frequencies F(CC), F(CA), etc.

In a long sequence of n-plets, collective frequencies $F_n(A_k)$, $F_n(T_k)$, $F_n(C_k)$ and $F_n(G_k)$ of each of 4 subgroups of a tetra-group of n-plets are defined as the sum of all individual frequencies of n-plets belong to this subgroup (here the index n denotes the length of n-plets; the index k = 1, 2, 3, ..., n denotes the position of the letter in n-plets). For example, in the case of a sequence of triplets with the letters A, C, G and G on the second positions of triplets, these collective frequencies are denoted $F_3(A_2)$, $F_3(T_2)$, $F_3(C_2)$ and $F_3(G_2)$; correspondingly the expression $F_3(A_2)$= 50000 means that a considered sequence of triplets contains 50000 triplets with the letter A in their second positions. Fig. 3 shows appropriate definitions of collective frequencies $F_2(A_k)$, $F_2(T_k)$, $F_2(C_k)$ and $F_2(G_k)$ for the case of sequences of doublets (here k=1,2), which are analyzed from the standpoint of both tetra-groups from Fig. 1. Fig. 3 also shows – for the case of sequences of doublets – collective probabilities $P_n(A_k)$, $P_n(T_k)$, $P_n(C_k)$ and $P_n(G_k)$ of separate subgroups of tetra-groups; in general case these probalities are defined by expressions $P_n(A_k) = F_n(A_k)/\Sigma_n$, $P_n(T_k) = F_n(T_k)/\Sigma_n$, $P_n(C_k) = F_n(C_k)/\Sigma_n$ and $P_n(G_k) = F_n(G_k)/\Sigma_n$. Below we represent the tetra-group rules for these collective probabilities, which are sum of individual probabilities of separate n-plets. More precisely, in the case of a sequence of n-plets, the probability of each of 4 subgroups of a separate tetra-group is a sum of $4^{n-1}$ individual probabilities of such n-plets. For example, in the case of sequence of 5-plets, the probability $P_5(A_1)$ of the A-subgroup, which combines 5-plets with the letter A at their first position, is a sum of $4^4$=256 individual probabilities of 5-plets: $P_5(A_1)$ = P(AAAAA) + P(AAAAT) + P(AAAAC) +…. , etc.

| | |
|---|---|
| $F_2(A_1)$=F(AA)+F(AC)+F(AG)+F(AT)<br>$P_2(A_1) = F_2(A_1)/\Sigma_2$ | $F_2(A_2)$=F(AA)+F(CA)+F(GA)+F(TA)<br>$P_2(A_2) = F_2(A_2)/\Sigma_2$ |
| $F_2(T_1)$=F(TC)+F(TA)+F(TT)+F(TG)<br>$P_2(T_1) = F_2(T_1)/\Sigma_2$ | $F_2(T_2)$=F(CT)+F(AT)+F(TT)+F(GT)<br>$P_2(T_2) = F_2(T_2)/\Sigma_2$ |
| $F_2(C_1)$=F(CC)+F(CA)+F(CT)+F(CG)<br>$P_2(C_1) = F_2(C_1)/\Sigma_2$ | $F_2(C_2)$=F(CC)+F(AC)+F(TC)+F(GC)<br>$P_2(C_2) = F_2(C_2)/\Sigma_2$ |
| $F_2(G_1)$=F(GC)+F(GA)+F(GT)+F(GG)<br>$P_2(G_1) = F_2(G_1)/\Sigma_2$ | $F_2(G_2)$=F(CG)+F(AG)+F(TG)+F(GG)<br>$P_2(G_2) = F_2(G_2)/\Sigma_2$ |

Fig. 3. The definition of collective frequences $F_n(A_k)$, $F_n(T_k)$, $F_n(C_k)$, $F_n(G_k)$ and collective propabilities $P_n(A_k)$, $P_n(T_k)$, $P_n(C_k)$, $P_n(G_k)$ for long sequences of doublets. Left: the case of the tetra-group of doublets with an identical letter at their first position (Fig. 1). Right: the case of the tetra-group of doublets with an identical letter at their second position (Fig. 1). The symbols F(AA), F(AC), … denote individual frequencies of doublets.

Two subgroups of any tetra-group of n-plets with the complementary letters on the characteristic positions are conditionally called complementary subgroups of the appropriate tetra-group. For example, the A-subgroup and the T-subgroup are complementary subgroups in each of two tetra-groups in Fig. 1. Such complementary subgroups participate in one of the represented tetra-group rules of long sequences of n-plets in single stranded DNA.

## 3. The first, second and third rules of symmetries of collective probabilities of tetra-groups in long DNA-texts

It is generally accepted that long sequences contain more than 50 thousands or 100 thousands nucleotides [Albrecht-Buehler, 2006; Prahbu, 1993; Rapoport, Trifonov, 2012]. In this Section we show data of analysis of two sequences of Homo sapiens chromosomes, each of which has its length of exactly one million nucleotides (only two these sequences of such length are retrieved from Entrez Search Field of Genbank by the known range operator 1000000:1000001[SLEN], https://www.ncbi.nlm.nih.gov/Sitemap/samplerecord.html). Fig. 4 shows calculation data of the first of them from the standpoint of the proposed tetra-group approach: Homo sapiens chromosome 7 sequence, ENCODE region ENm012, accession NT_086368, version NT_086368.3, https://www.ncbi.nlm.nih.gov/nuccore/NT_086368.3. These data include collective frequencies $F_n(A_k)$, $F_n(T_k)$, $F_n(C_k)$, $F_n(G_k)$ and collective probabilities $P_n(A_k)$, $P_n(T_k)$, $P_n(C_k)$, $P_n(G_k)$ of subgroups of appropriate tetra-groups of doublets, triplets, 4-plets and 5-plets of the sequence. We use the data in Fig. 4 to formulate the suppositional general rules of symmetries (or rules of approximate equalities) of these collective probabilities $P_n(A_k)$, $P_n(T_k)$, $P_n(C_k)$, $P_n(G_k)$ in long texts of single stranded DNA. These rules can be briefly named "tetra-group rules" or "rules of tetra-group symmetries". Below the formulated tetra-group rules will be confirmed by similar analysis of a representative set of other long nucleotide sequences from the Genbank. By analogy with the

| NUCLEOTIDES | DOUBLETS | TRIPLETS | 4-PLETS | 5-PLETS |
|---|---|---|---|---|
| $\Sigma_1$ = 1000000 | $\Sigma_2$ = 500000 | $\Sigma_3$ = 333333 | $\Sigma_4$ = 250000 | $\Sigma_5$ = 200000 |
| $F_1(A_1)=307519$<br>$P_1(A_1)=0,3075$ | $F_2(A_1)=153652$<br>$P_2(A_1)=0,3073$ | $F_3(A_1)=102657$<br>$P_3(A_1)=0,3080$ | $F_4(A_1)=76990$<br>$P_4(A_1)=0,3080$ | $F_5(A_1)=61097$<br>$P_5(A_1)=0,3055$ |
| $F_1(T_1)=335023$<br>$P_1(T_1)=0,3350$ | $F_2(T_1)=167514$<br>$P_2(T_1)=0,3350$ | $F_3(T_1)=111609$<br>$P_3(T_1)=0,3348$ | $F_4(T_1)=83590$<br>$P_4(T_1)=0,3344$ | $F_5(T_1)=67004$<br>$P_5(T_1)=0,3350$ |
| $F_1(C_1)=176692$<br>$P_1(C_1)=0,1767$ | $F_2(C_1)=88158$<br>$P_2(C_1)=0,1763$ | $F_3(C_1)=58893$<br>$P_3(C_1)=0,1767$ | $F_4(C_1)=44181$<br>$P_4(C_1)=0,1767$ | $F_5(C_1)=35525$<br>$P_5(C_1)=0,1776$ |
| $F_1(G_1)=180766$<br>$P_1(G_1)=0,1808$ | $F_2(G_1)=90676$<br>$P_2(G_1)=0,1813$ | $F_3(G_1)=60174$<br>$P_3(G_1)=0,1805$ | $F_4(G_1)=45239$<br>$P_4(G_1)=0,1810$ | $F_5(G_1)=36374$<br>$P_5(G_1)=0,1819$ |
| | $F_2(A_2)=153867$<br>$P_2(A_2)=0,3077$ | $F_3(A_2)=102597$<br>$P_3(A_2)=0,3078$ | $F_4(A_2)=77218$<br>$P_4(A_2)=0,3089$ | $F_5(A_2)=61441$<br>$P_5(A_2)=0,3072$ |
| | $F_2(T_2)=167509$<br>$P_2(T_2)=0,3350$ | $F_3(T_2)=111658$<br>$P_3(T_2)=0,3350$ | $F_4(T_2)=83756$<br>$P_4(T_2)=0,3350$ | $F_5(T_2)=67042$<br>$P_5(T_2)=0,3352$ |
| | $F_2(C_2)=88534$<br>$P_2(C_2)=0,1771$ | $F_3(C_2)=58812$<br>$P_3(C_2)=0,1764$ | $F_4(C_2)=44168$<br>$P_4(C_2)=0,1767$ | $F_5(C_2)=35440$<br>$P_5(C_2)=0,1772$ |
| | $F_2(G_2)=90090$<br>$P_2(G_2)=0,1802$ | $F_3(G_2)=60266$<br>$P_3(G_2)=0,1808$ | $F_4(G_2)=44858$<br>$P_4(G_2)=0,1794$ | $F_5(G_2)=36077$<br>$P_5(G_2)=0,1804$ |
| | | $F_3(A_3)=102265$<br>$P_3(A_3)=0,3068$ | $F_4(A_3)=76662$<br>$P_4(A_3)=0,3066$ | $F_5(A_3)=61621$<br>$P_5(A_3)=0,3081$ |
| | | $F_3(T_3)=111755$<br>$P_3(T_3)=0,3353$ | $F_4(T_3)=83924$<br>$P_4(T_3)=0,3357$ | $F_5(T_3)=67118$<br>$P_5(T_3)=0,3356$ |
| | | $F_3(C_3)=58987$<br>$P_3(C_3)=0,1770$ | $F_4(C_3)=43977$<br>$P_4(C_3)=0,1759$ | $F_5(C_3)=35198$<br>$P_5(C_3)=0,1760$ |
| | | $F_3(G_3)=60326$<br>$P_3(G_3)=0,1810$ | $F_4(G_3)=45437$<br>$P_4(G_3)=0,1817$ | $F_5(G_3)=36063$<br>$P_5(G_3)=0,1803$ |
| | | | $F_4(A_4)=76649$<br>$P_4(A_4)=0,3066$ | $F_5(A_4)=61706$<br>$P_5(A_4)=0,3085$ |
| | | | $F_4(T_4)=83753$<br>$P_4(T_4)=0,3350$ | $F_5(T_4)=66970$<br>$P_5(T_4)=0,3348$ |
| | | | $F_4(C_4)=44366$<br>$P_4(C_4)=0,1775$ | $F_5(C_4)=35313$<br>$P_5(C_4)=0,1766$ |
| | | | $F_4(G_4)=45232$<br>$P_4(G_4)=0,1809$ | $F_5(G_4)=36011$<br>$P_5(G_4)=0,1801$ |
| | | | | $F_5(A_5)=61654$<br>$P_5(A_5)=0,3083$ |
| | | | | $F_5(T_5)=66889$<br>$P_5(T_5)=0,3344$ |
| | | | | $F_5(C_5)=35216$<br>$P_5(C_5)=0,1761$ |
| | | | | $F_5(G_5)=36241$<br>$P_5(G_5)=0,1812$ |

generalized Chargaff's second rule, in cases of these new rules it is assumed that the length “n” of considered n-plets is much smaller than the length of the studied sequence. At this stage, we study only long DNA-sequences of doublets, triplets, 4-plets and 5-plets.

Fig. 4. Collective frequencies $F_n(A_k)$, $F_n(T_k)$, $F_n(C_k)$ and $F_n(G_k)$ and also collective probabilities $P_n(A_k)$, $P_n(T_k)$, $P_n(C_k)$ and $P_n(G_k)$ ($n$ = 1, 2, 3, 4, 5 and $k \le n$) of subgroups of tetra-groups for sequences of n-plets, which have the same letter in their position k, in the case of the following sequence: Homo sapiens chromosome 7 sequence, 1000000 bp, encode region ENm012, accession NT_086368, version NT_086368.3, https://www.ncbi.nlm.nih.gov/nuccore/NT_086368.3. Collective probabilities $P_n(A_k)$, $P_n(T_k)$, $P_n(C_k)$ and $P_n(G_k)$ are marked by red for a visual comfort of their comparison each with other.

From data in Fig. 4 one can assume existence of the following tetra-group rules for long sequences of oligonucleotides (these rules are confirmed by similar analizes of a set of other long DNA-sequences represented below).

**The first tetra-group rule** (the rule of an approximate equality of the collective probabilities of n-plets with an identical letter in their fixed position k, regardless of the length n of the considered n-plets):

- In long sequences of n-plets of single stranded DNA, collective probabilites $P_n(X_k)$ ($X$ = A, T, C, G; $k \le n$; $n$ = 1, 2, 3, 4, 5,.. is not too large) of those subset of n-plets, which have the letter X in their position k, are approximately equal to the individual probability of the nucleotide X regardless on values n.

This first rule can be also formulated in another way:

- If a long sequence of single stranded DNA is represented in different forms of texts of n-letter words (n = 1, 2, 3, 4, 5.... is not too large), then - in these texts - probabilities of words with the letter X (X = A, T, C, G) in their position $k \le n$ are approximately equal to each other regardless on values n.

For example, Fig. 4 shows the following approximate symmetries:

- $P_1(A_1)=0,3075 \cong P_2(A_1)=0,3073 \cong P_3(A_1)=0,3080 \cong P_4(A_1)=0,3080 \cong P_5(A_1)=0,3055$;
- $P_1(T_1)=0,3350 \cong P_2(T_1)=0,3350 \cong P_3(T_1)=0,3348 \cong P_4(T_1)=0,3344 \cong P_5(T_1)=0,3350$;
- $P_1(C_1)=0,1767 \cong P_2(C_1)=0,1763 \cong P_3(C_1)=0,1767 \cong P_4(C_1)=0,1767 \cong P_5(C_1)=0,1776$;
- $P_1(G_1)=0,1808 \cong P_2(G_1)=0,1813 \cong P_3(G_1)=0,1805 \cong P_4(G_1)=0,1810 \cong P_5(G_1)=0,1819$.

For a more convenient vision of this rule and corresponding symmetrical relations, Fig. 5 reproduces separately data about collective frequencies $P_n(X_k)$ (where X = A, T, C, G) from Fig. 4. Each of tabular rows contains approximately identical values of collective probalities and – for emphasizing this fact – all its cells are marked by the same color.

| NUCLEOTIDES | DOUBLETS | TRIPLETS | 4-PLETS | 5-PLETS |
|---|---|---|---|---|
| $\Sigma_1$ = 1000000 | $\Sigma_2$ = 500000 | $\Sigma_3$ = 333333 | $\Sigma_4$ = 250000 | $\Sigma_5$ = 200000 |
| $P_1(A_1)=0,3075$ | $P_2(A_1)=0,3073$ | $P_3(A_1)=0,3080$ | $P_4(A_1)=0,3080$ | $P_5(A_1)=0,3055$ |
| $P_1(T_1)=0,3350$ | $P_2(T_1)=0,3350$ | $P_3(T_1)=0,3348$ | $P_4(T_1)=0,3344$ | $P_5(T_1)=0,3350$ |
| $P_1(C_1)=0,1767$ | $P_2(C_1)=0,1763$ | $P_3(C_1)=0,1767$ | $P_4(C_1)=0,1767$ | $P_5(C_1)=0,1776$ |
| $P_1(G_1)=0,1808$ | $P_2(G_1)=0,1813$ | $P_3(G_1)=0,1805$ | $P_4(G_1)=0,1810$ | $P_5(G_1)=0,1819$ |
| | $P_2(A_2)=0,3077$ | $P_3(A_2)=0,3078$ | $P_4(A_2)=0,3089$ | $P_5(A_2)=0,3072$ |
| | $P_2(T_2)=0,3350$ | $P_3(T_2)=0,3350$ | $P_4(T_2)=0,3350$ | $P_5(T_2)=0,3352$ |
| | $P_2(C_2)=0,1771$ | $P_3(C_2)=0,1764$ | $P_4(C_2)=0,1767$ | $P_5(C_2)=0,1772$ |
| | $P_2(G_2)=0,1802$ | $P_3(G_2)=0,1808$ | $P_4(G_2)=0,1794$ | $P_5(G_2)=0,1804$ |
| | | $P_3(A_3)=0,3068$ | $P_4(A_3)=0,3066$ | $P_5(A_3)=0,3081$ |
| | | $P_3(T_3)=0,3353$ | $P_4(T_3)=0,3357$ | $P_5(T_3)=0,3356$ |
| | | $P_3(C_3)=0,1770$ | $P_4(C_3)=0,1759$ | $P_5(C_3)=0,1760$ |
| | | $P_3(G_3)=0,1810$ | $P_4(G_3)=0,1817$ | $P_5(G_3)=0,1803$ |
| | | | $P_4(A_4)=0,3066$ | $P_5(A_4)=0,3085$ |
| | | | $P_4(T_4)=0,3350$ | $P_5(T_4)=0,3348$ |
| | | | $P_4(C_4)=0,1775$ | $P_5(C_4)=0,1766$ |
| | | | $P_4(G_4)=0,1809$ | $P_5(G_4)=0,1801$ |
| | | | | $P_5(A_5)=0,3083$ |
| | | | | $P_5(T_5)=0,3344$ |
| | | | | $P_5(C_5)=0,1761$ |
| | | | | $P_5(G_5)=0,1812$ |

Fig. 5. Collective probabilities $P_n(A_k)$, $P_n(T_k)$, $P_n(C_k)$ and $P_n(G_k)$ from Fig. 4.

The first tetra-group rule can be graphically illustrated by a particular example in Fig. 6 for the same long DNA-sequence (Fig. 4).

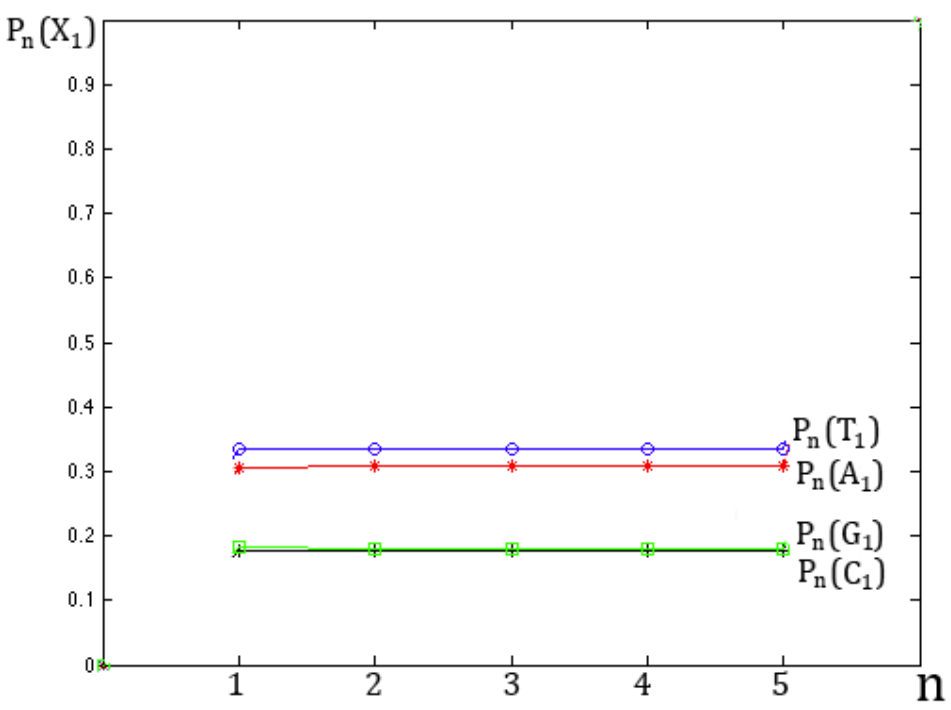

Fig. 6. The illustration of the dependence of collective probabilities $P_n(A_1)$, $P_n(T_1)$, $P_n(C_1)$ and $P_n(G_1)$ from the length n = 1, 2, 3, 4, 5 of n-plets in the case of Homo sapiens chromosome 7 sequence, 1000000 bp. Numerical data are taken from Fig. 4.

The first tetra-group rule and corresponding symmetrical relations can be briefly expressed by the following expression (1) for any of values n =1,2, 3, 4, ... under the condition of a fixed value of the index k:

$$P_1(X_1) \cong P_n(X_k)\ ,\ \ k \le n, \tag{1}$$

where X means any of letters A, T, C and G; n = 1, 2, 3, 4,... is not too large.

One should remind here that in the expression (1) various collective probabilities $P_n(X_k)$ are sum of individual probabilities of very different quantities of n-plets: the collective probability $P_2(A_1)$ is sum of individual probabilities of 4 doublets; the collective probability $P_3(A_1)$ is sum of individual probabilities of 16 triplets; the collective probability $P_4(A_1)$ is sum of individual probabilities of 64 tetraplets and the collective probability $P_5(A_1)$ is sum of individual probabilities of 256 pentaplets.

**The second tetra-group rule** (the rule of approximate equalities of collective probabilities of n-plets with an identical letter in their position k, regardless of the value of the positional index k in the considered n-plets):

- In long sequences of n-plets of single stranded DNA, collective probabilites $P_n(X_k)$ (X = A, T, C, G; $k \le n$; n = 1, 2, 3, 4, 5,.. is not too large) of those subset of n-plets, which have the letter X in their position k, are approximately equal to the individual probability of the nucleotide X regardless on values k.

This second rule can be also formulated in another way:

- If a long sequence of single stranded DNA is represented in different forms of texts of n-letter words (n = 1, 2, 3, 4, 5.... is not too large), then - in these texts - probabilities of words with the letter X (X = A, T, C, G) in their position $k \leq n$ are approximately equal to each other regardless on values k.

Fig. 5 facilitates a vision of this rule and corresponding symmetrical relations in the considered DNA-sequence: each of tabular columns contains approximately identical values of collective probalities and – for emphasizing this fact – all such cells are marked by the same color.

For example, one can see from data in Fig. 4 the following approximate symmetries:

- $P_5(A_1)$=0,3055 ≅ $P_5(A_2)$=0,3072 ≅ $P_5(A_3)$=0,3081 ≅ $P_5(A_4)$=0,3085 ≅ $P_5(A_5)$=0,3083;

- $P_5(T_1)$=0,3350 ≅ $P_5(T_2)$=0,3352 ≅ $P_5(T_3)$=0,3356 ≅ $P_5(T_4)$=0,3348 ≅ $P_5(T_5)$=0,3344;

- $P_5(C_1)$=0,1776 ≅ $P_5(C_2)$=0,1772 ≅ $P_5(C_3)$=0,1760 ≅ $P_5(C_4)$=0,1766 ≅ $P_5(C_5)$=0,1761;

- $P_5(G_1)$=0,1819 ≅ $P_5(G_2)$=0,1804 ≅ $P_5(G_3)$=0,1803 ≅ $P_5(G_4)$=0,1801 ≅ $P_5(G_5)$=0,1812.

The second tetra-group rule and corresponding approximate symmetries can be graphically illustrated by a particular example in Fig. 7 for the same long DNA-sequence (Fig. 4).

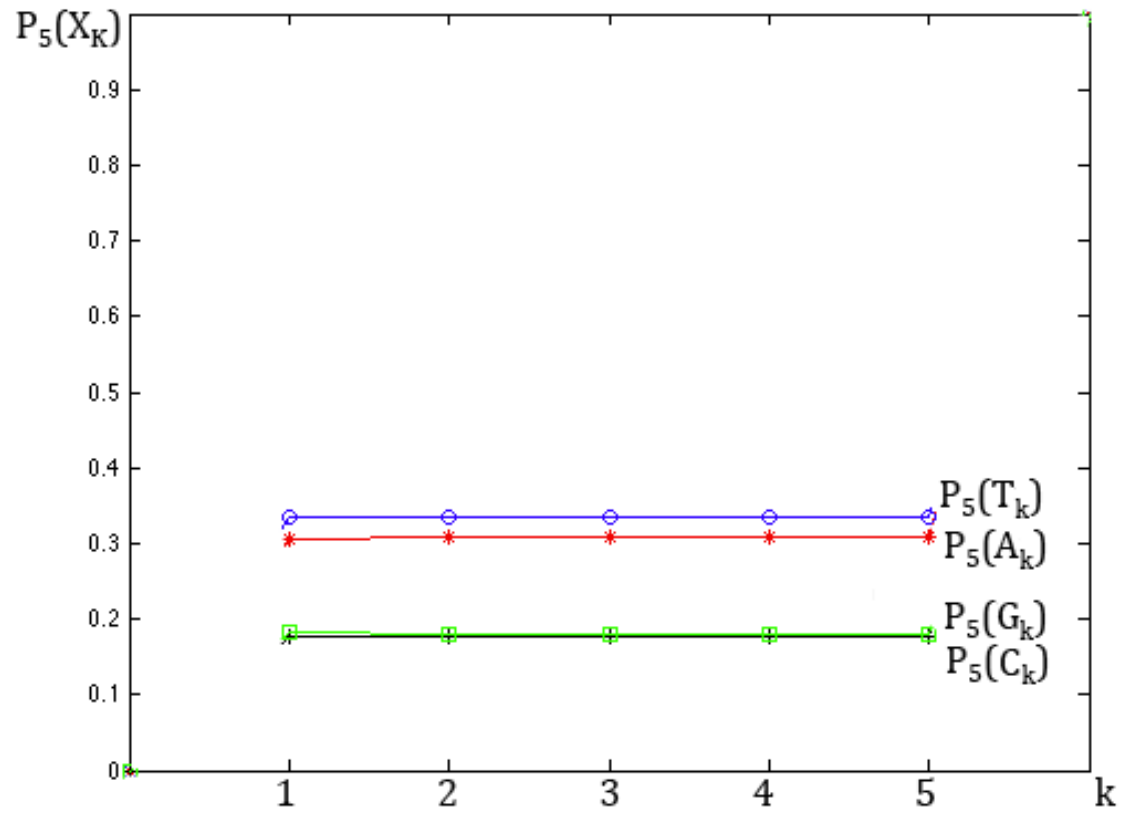

Fig. 7. The illustration of the dependence of collective probabilities $P_5(A_k)$, $P_5(T_k)$, $P_5(C_k)$ and $P_5(G_k)$ of appropriate subgroups of tetra-groups from the index k of a position of the attributive letter in 5-plets in the case of the sequence of 5-plets of Homo sapiens chromosome 7 sequence, 1000000 bp. Numerical data are taken from Fig. 4.

The second tetra-group rule and corresponding approximate symmetries can be briefly expressed by the following expression (2) for any of values k under the condition of a fixed value of a length n of n-plets:

$$P_n(X_1) \cong P_n(X_k), \quad k \leq n, \qquad (2)$$

where X means any of letters A, T, C and G; n = 1, 2, 3, 4,... is not too large.

One can see from expressions (1) and (2) that the first rule and the second rule can be jointly expressed in a brief way by the expression (3) without the mentioned conditions of a fixation of values n and k:

$$P_1(X_1) \cong P_n(X_k)\ , \quad k \le n, \qquad (3)$$

where X means any of letters A, T, C and G; n = 1, 2, 3, 4,… is not too large.

**The third tetra-group rule for those long DNA-sequences, which correspond to the second Chargaff's rule** (the rule of approximate equalities of collective probabilities of complementary subgroups of tetra-groups):

- in tetra-groups of long sequences of n-plets of those single-stranded DNA, that satisfy the second Chargaff's rule, collective probabilities $P_n(A_k)$ and $P_n(T_k)$ of the complementary A- and T-subgroups of tetra-groups are approximately equal to each other. The same is true for collective probabilities $P_n(C_k)$ and $P_n(G_k)$ of the complementary C- and G-subgroups of tetra-groups.

This third rule can be also formulated in another way:

- If a long sequence of those single-stranded DNA, that satisfy the second Chargaff's rule, is represented in different forms of texts of n-letter words (n = 1, 2, 3, 4, 5…. is not too large), then - in these texts - probabilities of words with the complementary letters A and T in their position k are approximately equal to each other. The same is true for probabilities of words with the complementary letters C and G in their position k.

This rule is expressed by expressions (4) for any of considered values of n and k, where k ≤ n and n = 1, 2, 3, 4, … is not too large:

$$P_n(A_k) \cong P_n(T_k) \text{ and } P_n(C_k) \cong P_n(G_k). \qquad (4)$$

For example, one can see from data in Fig. 4 for the case k=1 the following:

- $P_1(A_1)=0{,}3075 \cong P_1(T_1)=0{,}3350$; $P_2(A_1)=0{,}3073 \cong P_2(T_1)=0{,}3350$; $P_3(A_1)=0{,}3080 \cong P_3(T_1)=0{,}3348$; $P_4(A_1)=0{,}3080 \cong P_4(T_1)=0{,}3344$; $P_5(A_1)=0{,}3055 \cong P_5(T_1)=0{,}3350$;
- $P_1(C_1)=0{,}1767 \cong P_1(G_1)=0{,}1808$; $P_2(C_1)=0{,}1763 \cong P_2(G_1)=0{,}1813$; $P_3(C_1)=0{,}1767 \cong P_3(G_1)=0{,}1805$; $P_4(C_1)=0{,}1767 \cong P_4(G_1)=0{,}1810$; $P_5(C_1)=0{,}1776 \cong P_5(G_1)=0{,}1819$.

The similar situation is true for cases k = 2, 3, 4, 5 in Fig. 4.

We emphasize that approximately equal collective frequencies $F_n(A_k)$ and $F_n(T_k)$ of the complementary A- and T-subgroups of tetra-groups (as well as $F_n(C_k)$ and $F_n(G_k)$ of the complementary C- and G-subgroups), which are used in expressions (4), can differ significantly by values of individual frequences of n-plets in them. For example, for the sequence of doublets in Fig. 4, these collective frequencies are sum of the following individual frequencies of separate doublets:

- $F_2(A_1)$ = F(AA)+F(AC)+F(AG)+F(AT) = 52506+23434+31500+46212,

- $F_2(T_1) = F(TT)+F(TG)+F(TC)+F(TA) = 61483+35978+29290+40763$,
- $F_2(C_1) = F(CA)+F(CC)+F(CG)+F(CT) = 32721+32721+2800+33533$,
- $F_2(G_1) = F(GT)+F(GG)+F(GC)+F(GA) = 26281+19812+16706+27877$ (5)

One can see from the expression (5) that, for example, the individual frequency F(CG)=2800, which is used in the expression of the collective frequency $F_2(C_1)$ of the C-subgroup, differs by a factor of 6 from the individual frequency of the complementary doublet F(GC)=16706, which is used in the expression of the collective frequency $F_2(G_1)$ of the complementary G-subgroup. Correspondingly the individual probability $P(CG) = F(CG)/\Sigma_2 = 2800/500000 = 0{,}0056$ differs by a factor of 6 from the individual probability of the complementary doublet $P(GC) = F(GC)/\Sigma_2 = 16706/500000 = 0{,}0334$. This indicates that the described tetra-group rules can't be reduced to rules of individual n-plets, but they form a special class of rules of a collective organization in oligonucleotide sequences of single stranded DNA.

The third tetra-group rule can be considered as a generalization of the second Chargaff's parity rule, which states an approximate equality of individual frequences of complementary letters $F(A)\cong F(T)$ and $F(A)\cong F(T)$ (or probabilities $P(A)\cong P(T)$ and $P(A)\cong P(T)$) in long nucleotide sequences of single stranded DNA. In the case of the sequence in Fig. 4, the second Chargaff's parity rule is expressed by expressions $P_1(A_1)=0{,}3075 \cong P_1(T_1)=0{,}3350$ and $P_1(C_1)=0{,}1767 \cong P_1(G_1)=0{,}1808$. The level of accuracy of the second Chargaff's rule execution for this sequence is determined by the difference of probabilities: for the probabilities of complementary letters A and T in the mononucleotide sequence, this difference is equal to $P_1(T_1)-P_1(A_1) = 0{,}3350-0{,}3075 = 0{,}0275$, and for the probabilities of complementary letters C and G it is equal to $P_1(G_1)-P_1(C_1) = 0{,}1808-0{,}1767 = 0{,}0041$. For the analyzed sequence (Fig. 4), Fig. 8 shows that the same level of accuracy $P_1(T_1)-P_1(A_1)=0{,}0275$ is approximately executed for all differencies $P_n(T_k)-P_n(A_k)$, and that the same level of accuracy $P_1(G_1)-P_1(C_1)=0{,}0041$ is approximately executed for all differencies $P_n(G_k)-P_n(C_k)$. It testifies that the second Chargaff's parity rule can be considered as a particular case of the third tetra-group rule.

| **NUCLEOTIDES** | **DOUBLETS** | **TRIPLETS** | **4-PLETS** | **5-PLETS** |
|---|---|---|---|---|
| $P_1(T_1)-P_1(A_1)$ = 0,0275 | $P_2(T_1)-P_2(A_1)$ =0,0277 | $P_3(T_1)-P_3(A_1)$ = 0,0268 | $P_4(T_1)-P_4(A_1)$ = 0,0264 | $P_5(T_1)-P_5(A_1)$ = 0,0295 |
| $P_1(G_1)-P_1(C_1)$ = 0,0041 | $P_2(G_1)-P_2(C_1)$ = 0,005 | $P_3(G_1)-P_3(C_1)$ = 0,0038 | $P_4(G_1)-P_4(C_1)$ = 0,0043 | $P_5(G_1)-P_5(C_1)$ = 0,0043 |
| | $P_2(T_2)-P_2(A_2)$ = 0,0273 | $P_3(T_2)-P_3(A_2)$ = 0,0272 | $P_4(T_2)-P_4(A_2)$ = 0,0261 | $P_5(T_2)-P_5(A_2)$ = 0,028 |
| | $P_2(G_2)-P_2(C_2)$ = 0,0031 | $P_3(G_2)-P_3(C_2)$ = 0,0044 | $P_4(G_2)-P_4(C_2)$ = 0,0027 | $P_5(G_2)-P_5(C_2)$ = 0,0032 |
| | | $P_3(T_3)-P_3(A_3)$ = 0,0285 | $P_4(T_3)-P_4(A_3)$ = 0,0291 | $P_5(T_3)-P_5(A_3)$ = 0,0275 |
| | | $P_3(G_3)-P_3(C_3)$ = 0,004 | $P_4(G_3)-P_4(C_3)$ = 0,0058 | $P_5(G_3)-P_5(C_3)$ = 0,0043 |
| | | | $P_4(T_4)-P_4(A_4)$ = 0,0284 | $P_5(T_4)-P_5(A_4)$ = 0,0263 |
| | | | $P_4(G_4)-P_4(C_4)$ | $P_5(G_4)-P_5(C_4)$ |

| | | | = 0,0034 | = 0,0035 |
|---|---|---|---|---|
| | | | | $P_5(T_5)$-$P_5(A_5)$ = 0,0261 |
| | | | | $P_5(G_5)$-$P_5(C_5)$ = 0,0051 |

Fig. 8. Levels of accuracy between values of collective propabilities (shown in Fig. 4) of complemetary subgroups of tetra-groups of n-plets (n = 1, 2, 3, 4, 5) for the sequence Homo sapiens chromosome 7 sequence, encode region ENm012, accession NT_086368, version NT_086368.3, https://www.ncbi.nlm.nih.gov/nuccore/NT_086368.3. Numerical data are taken from Fig. 4.

The second Chargaff's parity rule is known in the literature as the Symmetry Principle. New kinds of approximate symmetries in long DNA-texts can be conditionally named as tetra-group symmetries. They expand scientific knowledge about Symmetry Principles in organization of long DNA-texts.

For long DNA-sequences of n-plets, the described tetra-group rules have a predictive power: knowing the collective probabilities of only two of 4 subgroups of one of the tetra-groups (for example, $P_1(A_1)$ and $P_1(C_1)$), one can predict approximate values of probabilities of other subgroups of this and other tetra-groups on the basis of the expressions (1-4).

Fig. 9 shows a joint representation of all three tetra-group rules described above: sectors of the same color contain approximately the same values of collective probabilities. Each of its rings corresponds to an appropriate length "n" of n-plets: the smallest ring corresponds to the case of doublets (in this case k = 1, 2); the next ring corresponds to the case of triplets (in this case k = 1, 2, 3), etc.

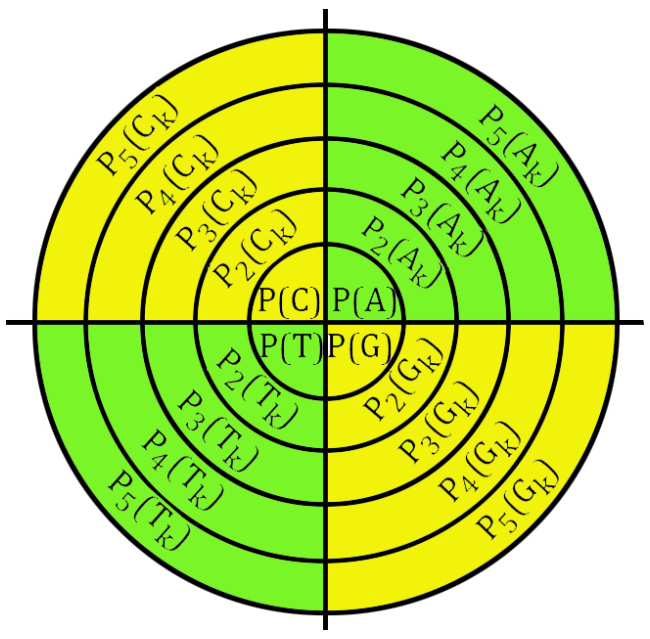

Fig. 9. The joint representation of the mentioned tetra-group rules. Sectors of the same color contain at each level approximately the same values of collective probabilities $P_n(A_k)$, $P_n(T_k)$, $P_n(C_k)$ and $P_n(G_k)$ in long DNA sequences.

Now let us turn to the second DNA-sequence with one million nucleotides, which was retrieved from Entrez Search Field of Genbank by the range operator 1000000:1000001[SLEN]: Homo sapiens chromosome 5 sequence, ENCODE region ENm002; accession NT_086358, version NT_086358.1, https://www.ncbi.nlm.nih.gov/nuccore/NT_086358.1. Fig. 10 shows our results of the analysis of this long sequence from the standpoint of tetra-groups of its oligonucleotides (by analogy with data in Fig. 4).

| **NUCLEOTIDES** | **DOUBLETS** | **TRIPLETS** | **4-PLETS** | **5-PLETS** |
|---|---|---|---|---|

| $\Sigma_1$ = 1000000 | $\Sigma_2$ =500000 | $\Sigma_3$ =333333 | $\Sigma_4$ = 250000 | $\Sigma_5$ = 200000 |
|---|---|---|---|---|
| $F_1(A_1)$=283765<br>$P_1(A_1)$=0,2838 | $F_2(A_1)$=142054<br>$P_2(A_1)$=0,2841 | $F_3(A_1)$=94805<br>$P_3(A_1)$=0,2844 | $F_4(A_1)$=70944<br>$P_4(A_1)$=0,2838 | $F_5(A_1)$=56865<br>$P_5(A_1)$=0,2843 |
| $F_1(T_1)$=270465<br>$P_1(T_1)$=0,2705 | $F_2(T_1)$=135082<br>$P_2(T_1)$=0,2702 | $F_3(T_1)$=90432<br>$P_3(T_1)$=0,2713 | $F_4(T_1)$=67590<br>$P_4(T_1)$=0,2704 | $F_5(T_1)$=53956<br>$P_5(T_1)$=0,2698 |
| $F_1(C_1)$= 226576<br>$P_1(C_1)$=0,2267 | $F_2(C_1)$=113153<br>$P_2(C_1)$=0,2263 | $F_3(C_1)$=75409<br>$P_3(C_1)$=0,2262 | $F_4(C_1)$=56607<br>$P_4(C_1)$=0,2264 | $F_5(C_1)$=45425<br>$P_5(C_1)$=0,227 |
| $F_1(G_1)$=219194<br>$P_1(G_1)$=0,2192 | $F_2(G_1)$=109711<br>$P_2(G_1)$=0,2194 | $F_3(G_1)$=72687<br>$P_3(G_1)$=0,2181 | $F_4(G_1)$=54859<br>$P_4(G_1)$=0,2194 | $F_5(G_1)$=43754<br>$P_5(G_1)$=0,2188 |
| | $F_2(A_2)$=141711<br>$P_2(A_2)$= 0,2834 | $F_3(A_2)$=94612<br>$P_3(A_2)$=0,2838 | $F_4(A_2)$=71111<br>$P_4(A_2)$=0,2844 | $F_5(A_2)$=56643<br>$P_5(A_2)$=0,2832 |
| | $F_2(T_2)$=135383<br>$P_2(T_2)$=0,2708 | $F_3(T_2)$=89797<br>$P_3(T_2)$=0,2694 | $F_4(T_2)$=67600<br>$P_4(T_2)$=0,2704 | $F_5(T_2)$=54278<br>$P_5(T_2)$=0,2714 |
| | $F_2(C_2)$=113423<br>$P_2(C_2)$=0,2268 | $F_3(C_2)$=75549<br>$P_3(C_2)$=0,2266 | $F_4(C_2)$=56616<br>$P_4(C_2)$=0,2265 | $F_5(C_2)$= 45153<br>$P_5(C_2)$=0,2258 |
| | $F_2(G_2)$=109483<br>$P_2(G_2)$=0,2190 | $F_3(G_2)$=73375<br>$P_3(G_2)$=0,2201 | $F_4(G_2)$=54673<br>$P_4(G_2)$=0,2187 | $F_5(G_2)$= 43926<br>$P_5(G_2)$=0,2196 |
| | | $F_3(A_3)$=94347<br>$P_3(A_3)$=0,2830 | $F_4(A_3)$=71110<br>$P_4(A_3)$=0,2844 | $F_5(A_3)$=56493<br>$P_5(A_3)$=0,2825 |
| | | $F_3(T_3)$=90236<br>$P_3(T_3)$=0,2707 | $F_4(T_3)$=67492<br>$P_4(T_3)$=0,2700 | $F_5(T_3)$=54239<br>$P_5(T_3)$=0,2712 |
| | | $F_3(C_3)$=75618<br>$P_3(C_3)$=0,2268 | $F_4(C_3)$=56546<br>$P_4(C_3)$=0,2262 | $F_5(C_3)$=45223<br>$P_5(C_3)$=0,2261 |
| | | $F_3(G_3)$=73132<br>$P_3(G_3)$=0,2194 | $F_4(G_3)$=54852<br>$P_4(G_3)$=0,2194 | $F_5(G_3)$=44045<br>$P_5(G_3)$=0,2202 |
| | | | $F_4(A_4)$=70600<br>$P_4(A_4)$=0,2824 | $F_5(A_4)$=56932<br>$P_5(A_4)$=0,28467 |
| | | | $F_4(T_4)$=67783<br>$P_4(T_4)$=0,2711 | $F_5(T_4)$=54087<br>$P_5(T_4)$=0,2704 |
| | | | $F_4(C_4)$=56807<br>$P_4(C_4)$=0,2272 | $F_5(C_4)$=45309<br>$P_5(C_4)$=0,2265 |
| | | | $F_4(G_4)$=54810<br>$P_4(G_4)$=0,2192 | $F_5(G_4)$=43672<br>$P_5(G_4)$=0,2184 |
| | | | | $F_5(A_5)$=56832<br>$P_5(A_5)$=0,2842 |
| | | | | $F_5(T_5)$=53905<br>$P_5(T_5)$=0,2695 |
| | | | | $F_5(C_5)$=45466<br>$P_5(C_5)$=0,2273 |
| | | | | $F_5(G_5)$=43797<br>$P_5(G_5)$=0,2190 |

Fig. 10. Collective frequencies $F_n(A_k)$, $F_n(T_k)$, $F_n(C_k)$ and $F_n(G_k)$ and also collective probabilities $P_n(A_k)$, $P_n(T_k)$, $P_n(C_k)$ and $P_n(G_k)$ ($n$ = 1, 2, 3, 4, 5 and $k \leq n$) of subgroups of tetra-groups for sequences of n-plets, which have the same letter in their position k, in the case of the following sequence: Homo sapiens chromosome 5 sequence, 1000000 bp, encode region ENm002; accession NT_086358, version NT_086358.1, https://www.ncbi.nlm.nih.gov/nuccore/NT_086358.1. Collective probabilities $P_n(A_k)$, $P_n(T_k)$, $P_n(C_k)$ and $P_n(G_k)$ are marked by red for a visual convenient of their comparison each with other.

One can see from data in Fig. 10 that they satisfy the three tetra-group rules by analogy with data in Fig. 4. Fig. 11 shows levels of accuracy between values of collective propabilities (shown in Fig. 10) of complemetary subgroups of tetra-groups of n-plets for this new long DNA-sequence. One can see that these levels of accuracy are approximately equal to the levels of accuracy in Fig. 8 for the previously considered sequence.

| **NUCLEOTIDES** | **DOUBLETS** | **TRIPLETS** | **4-PLETS** | **5-PLETS** |
|---|---|---|---|---|
| $P_1(A_1)-P_1(T_1)$ = 0,0133 | $P_2(A_1)-P_2(T_1)$ = 0,0139 | $P_3(A_1)-P_3(T_1)$ = 0,0131 | $P_4(A_1)-P_4(T_1)$= 0,0134 | $P_5(A_1)-P_5(T_1)$= 0,0145 |
| $P_1(C_1)-P_1(G_1)$ = 0,0075 | $P_2(C_1)-P_2(G_1)$ = 0,0069 | $P_3(C_1)-P_3(G_1)$ = 0,0081 | $P_4(C_1)-P_4(G_1)$= 0,007 | $P_5(C_1)-P_5(G_1)$= 0,0082 |
| | $P_2(A_2)$- $P_2(T_2$= 0,0126 | $P_3(A_2)-P_3(T_2)$= 0,0144 | $P_4(A_2)-P_4(T_2)$= 0,014 | $P_5(A_2)-P_5(T_2)$= 0,0118 |
| | $P_2(C_2)-P_2(G_2)$= 0,0078 | $P_3(C_2)-P_3(G_2)$= 0,0065 | $P_4(C_2)-P_4(G_2)$= 0,0078 | $P_5(C_2)-P_5(G_2)$= 0,0062 |
| | | $P_3(A_3)$- $P_3(T_3)$= 0,0123 | $P_4(A_3)$- $P_4(T_3)$= 0,0144 | $P_5(A_3)-P_5(T_3)$= 0,0113 |
| | | $P_3(C_3)-P_3(G_3)$ = 0,0074 | $P_4(C_3)$- $P_4(G_3)$ = 0,0068 | $P_5(C_3)-P_5(G_3)$= 0,0059 |
| | | | $P_4(A_4)-P_4(T_4)$= 0,0113 | $P_5(A_4)-P_5(T_4)$= 0,0143 |
| | | | $P_4(C_4)-P_4(G_4)$= 0,008 | $P_5(C_4)-P_5(G_4)$= 0,0081 |
| | | | | $P_5(A_5)-P_5(T_5)$= 0,0147 |
| | | | | $P_5(C_5)-P_5(G_5)$ = 0,0083 |

Fig. 11. Levels of accuracy between values of collective propabilities (shown in Fig. 6) of complemetary subgroups of tetra-groups of n-plets (n = 1, 2, 3, 4, 5) for the sequence Homo sapiens chromosome 5 sequence, 1000000 bp, encode region ENm002; accession NT_086358, version NT_086358.1, https://www.ncbi.nlm.nih.gov/nuccore/NT_086358.1.

And what about the implementation of tetra-group rules for the whole human genome, which contains about 3 billion nucleotides? Fig. 12 shows data, which are taken from [Perez, 2013, Table 3] about individual frequencies of all 64 triplets in the whole human genome.

| triplet | triplet frequency | triplet | triplet frequency | triplet | triplet frequency | triplet | triplet frequency |
|---|---|---|---|---|---|---|---|
| AAA | 36381293 | CAA | 17927956 | GAA | 18678084 | TAA | 19721149 |
| AAC | 13794251 | CAC | 14214421 | GAC | 8938833 | TAC | 10755607 |
| AAG | 18894716 | CAG | 19176935 | GAG | 15939419 | TAG | 12240281 |
| AAT | 23634011 | CAT | 17423117 | GAT | 12658530 | TAT | 19568343 |
| ACA | 19073189 | CCA | 17444649 | GCA | 13635427 | TCA | 18565027 |
| ACC | 11007307 | CCC | 12428986 | GCC | 11268094 | TCC | 14614789 |
| ACG | 2372235 | CCG | 2606672 | GCG | 2247440 | TCG | 2087242 |

| ACT | 15251455 | CCT | 16835177 | GCT | 13252828 | TCT | 20990387 |
|---|---|---|---|---|---|---|---|
| AGA | 20948987 | CGA | 2085226 | GGA | 14619310 | TGA | 18562015 |
| AGC | 13242724 | CGC | 2244432 | GGC | 11258126 | TGC | 13649076 |
| AGG | 16810797 | CGG | 2604253 | GGG | 12446600 | TGG | 17480496 |
| AGT | 15266057 | CGT | 2379612 | GGT | 11026602 | TGT | 19152113 |
| ATA | 19548709 | CTA | 12217331 | GTA | 10766854 | TTA | 19750578 |
| ATC | 12650299 | CTC | 15942742 | GTC | 8955434 | TTC | 18708048 |
| ATG | 17409063 | CTG | 19195946 | GTG | 14252868 | TTG | 18005020 |
| ATT | 23669701 | CTT | 18944797 | GTT | 13852086 | TTT | 36530115 |

Fig. 12. Frequencies of 64 triplets in the whole human genome in accordance with data from [Perez, 2013, Table 3].

In accordance with these data, the human genome contains 947.803.867 triplets and 2.843.411.601 (=947803867*3) nucleotides. From the data in Fig. 12, one can calculate propabilities $P_1(A)$, $P_1(T)$, $P_1(C)$ and $P_1(G)$ of members of the tetra-group of nucleotides A, T, C, G and also collective probabilities of subgroups of three tetra-groups of triplets: 1) $P_3(A_1)$, $P_3(T_1)$, $P_3(C_1)$, $P_3(G_1)$; 2) $P_3(A_2)$, $P_3(T_2)$, $P_3(C_2)$, $P_3(G_2)$; 3) $P_3(A_3)$, $P_3(T_3)$, $P_3(C_3)$ and $P_3(G_3)$. Results of such calculations are represented in Fig. 13.

| **NUCLEOTIDES** | **TRIPLETS** |
|---|---|
| $\Sigma_1$ = 2843411601 | $\Sigma_3$ = 947803867 |
| **$P_1(A)$ = 0,295359** | **$P_3(A_1)$ = 0,295372** |
| **$P_1(T)$ = 0,295847** | **$P_3(T_1)$ = 0,295821** |
| **$P_1(C)$ = 0,204341** | **$P_3(C_1)$ = 0,204338** |
| **$P_1(G)$ = 0,204453** | **$P_3(G_1)$ = 0,204469** |
| | **$P_3(A_2)$ = 0,295364** |
| | **$P_3(T_2)$ = 0,295841** |
| | **$P_3(C_2)$ = 0,204347** |
| | **$P_3(G_2)$ = 0,204448** |
| | **$P_3(A_3)$ = 0,295341** |
| | **$P_3(T_3)$ = 0,295879** |
| | **$P_3(C_3)$ = 0,204339** |
| | **$P_3(G_3)$ = 0,204441** |

Fig. 13. Probabilities of subgroups of tetra-groups of nucleotides and triplets in the case of the whole human genome. Initial data about frequencies of triplets are shown in Fig. 12 from [Perez, 2013, Table 3].

One can see in Fig. 13 the very high level of accuracy of compliance of the whole human genome with the tetra-group rules of these probabilities (up to the fourth decimal place in each of equalities $P_1(X) = P_3(X_1) = P_3(X_2) = P_3(X_3)$, where X = A, T, C or G):

$$P_1(A) = P_3(A_1) = P_3(A_2) = P_3(A_3) \approx 0{,}2953$$
$$P_1(T) = P_3(T_1) = P_3(T_2) = P_3(T_3) \approx 0{,}2958$$
$$P_1(C) = P_3(C_1) = P_3(C_2) = P_3(C_3) \approx 0{,}2043$$
$$P_1(G) = P_3(G_1) = P_3(G_2) = P_3(G_3) \approx 0{,}2044$$

Is it accidental that the highest species of living organisms - the human organism - has a genome with such high accuracy of realization of tetra-group symmetries (and of compliance with tetra-group rules)? In the course of biological evolution, does the accuracy of the realization of tetra-group symmetries in the genomes of biological species increase up to the highest accuracy in the human genome?

Values of tetra-group probabilities and levels of accuracy of the realization of tetra-group symmetries in the genomes of different organisms can be used as a new criteria for formal rankings different species of living organisms. Will these rankings of biological species correspond to the stages of evolutionary development of biological species or to some other biological characteristics of species? These and other questions - arising in connection with the described tetra-group symmetries in long DNA-texts - are open at this stage of our knowledge and subject to study.

The author also analyzed the implementation of tetra-group rules in a representative set of other long DNA-sequences including the complete set of 24 human chromosomes (see below). Many of these sequences were taken from their lists in articles of other authors in order to avoid suspicion of a special choise of DNA-texts. The results of this analysis are presented in Appendixes 1-7. They show that all considered sequences satisfy the three tetra-group rules and demonstrate the existence of symmetries of tetra-group probabilities in these long DNA-texts.

## 4. DNA-alphabets, genetic binary oppositions and the tensor product of matrices

It is desirable to create a mathematical model of origin of the described tetra-group symmetries of probabilities of oligonucleotides in long DNA-sequences. A great discovery of twentieth century physics was the probabilistic nature of physical phenomena at atomic scales, described in quantum mechanics. Molecules of heredity DNA and RNA belong to the microworld of quantum mechanics and they should obey to principles of quantum mechanics. In this Section the author proposes a possible models, which is connected with principles and formalisms of quantum mechanics and quantum informatics.

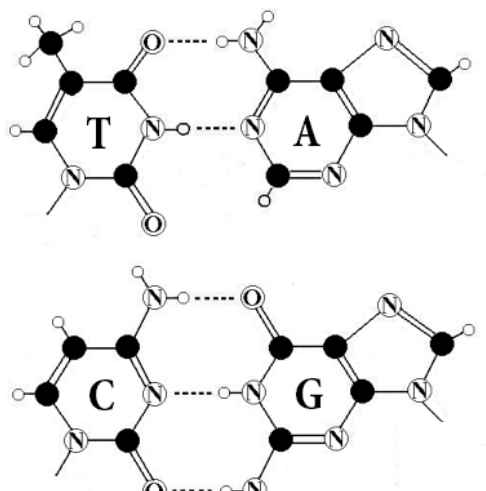

| № | Binary Symbols | C | A | G | T/U |
|---|---|---|---|---|---|
| 1 | $0_1$ — pyrimidines<br>$1_1$ — purines | $0_1$ | $1_1$ | $1_1$ | $0_1$ |
| 2 | $0_2$ — amino<br>$1_2$ — keto | $0_2$ | $0_2$ | $1_2$ | $1_2$ |
| 3 | $0_3$ — three hydrogen bonds;<br>$1_3$ — two hydrogen bonds | $0_3$ | $1_3$ | $0_3$ | $1_3$ |

Fig. 14. Left: the four nitrogenous bases of DNA: adenine A, guanine G, cytosine C, and thymine T. Right: three binary sub-alphabets of the genetic alphabet on the basis of three pairs of binary-oppositional traits or indicators.

Science does not know why the nucleotide alphabet of DNA has been created by nature from just four letters (A, T, C, G), and why just these very simple molecules were chosen for the DNA-alphabet (out of millions of possible molecules). But science knows [Fimmel, Danielli, Strüngmann, 2013; Petoukhov, 2008; Petoukhov, He, 2009; Stambuk, 1999] that these four molecules are interrelated by means of their symmetrical peculiarities into the united molecular ensemble with its three pairs of binary-oppositional traits or indicators (Fig. 14):

(1) Two letters are purines (A and G), and the other two are pyrimidines (C and T). From the standpoint of these binary-oppositional traits one can denote C = T = 0, A = G = 1. From the standpoint of these traits, any of the DNA-sequences are represented by a corresponding binary sequence. For example, GCATGAAGT is represented by 101011110;
(2) Two letters are amino-molecules (A and C) and the other two are keto-molecules (G and T). From the standpoint of these traits one can designate A = C = 0, G = T = 1. Correspondingly, the same sequence, GCATGAAGT, is represented by another binary sequence, 100110011;
(3) The pairs of complementary letters, A-T and C-G, are linked by 2 and 3 hydrogen bonds, respectively. From the standpoint of these binary traits, one can designate C = G = 0, A = T = 1. Correspondingly, the same sequence, GCATGAAGT, is read as 001101101.

Accordingly, each of the DNA-sequences of nucleotides is the carrier of three parallel messages on three different binary languages. At the same time, these three types of binary representations form a common logic set on the basis of logic operation of modulo-2 addition denoted by the symbol $\oplus$: modulo-2 addition of any two such binary representations of the DNA-sequence coincides with the third binary representation of the same DNA-sequence: for example, $101011110 \oplus 100110011 = 001101101$. One can be reminded here of the rules of the bitwise modulo-2 addition: $0 \oplus 0 = 0$; $0 \oplus 1 = 1$; $1 \oplus 0 = 1$; $1 \oplus 1 = 0$. (The logic operation of modulo-2 addition is actively used in quantum informatics, which will be considered below in a connection with DNA-texts).

Taking into account the phenomenologic fact that each of DNA-letters C, A, T and G is uniquely defined by any two kinds of mentioned binary-oppositional indicators (Fig. 14), these genetic letters can be represented as corresponding pairs of binary symbols, for example, from the standpoint of two first binary-oppositional indicators. It is convenient for us for further decription to show at the first position of each of letters its binary symbol from the second pair of binary-oppositional indicators (the indicator "amino or keto": C=A=0, T=G=1) and at the second positions of each of letters its binary symbol from the first pair of binary-oppostional indicators (the indicator "pyrimidine or purine": C=T=0, A=G=1). In this case the letter C is represented by the binary symbol $0_2 0_1$ (that is as 2-bit binary munber), A – by the symbol $0_2 1_1$, T – by the symbol $1_2 0_1$, G – by the symbol $1_2 1_1$. Using these representations of separate letters, each of 16 doublets is represented as concatenation of the binary symbols of its letters (that is as 4-bit binary number): for example, the doublet CC is represented as 4-bit binary number $0_2 0_1 0_2 0_1$, the doublet CA – as 4-bit binary number $0_2 0_1 0_2 1_1$, etc. By analogy, each of 64 triplets is represented as concatenation of the binary symbols of its letters (that is as 6-bit binary number): for example, the triplet CCC is represented as 6-bit binary number $0_2 0_1 0_2 0_1 0_2 0_1$, the triplet CCA – as 6-bit binary number $0_2 0_1 0_2 0_1 0_2 1_1$, etc. In general, each of n-plets is represented as concatenation of the binary symbols of its letters (below we will not show these indexes 2 and 1 of separate letters in binary representations of n-plets but will remember that each of positions corresponds to its own kind of indicators from the first or from the second set of indicators in Fig. 14).

It is convenient to represent DNA-alphabets of 4 nucleotides, 16 doublets, 64 triplets, ... $4^n$ n-plets in a form of appropriate square tables (Fig. 15), whose rows and columns are numerated by binary symbols in line with the following principle. Entries of each column are numerated by binary symbols in line with the first set of binary-oppositional indicators in Fig. 14 (for example, the triplet CAG and all other triplets in the same column are the combination "pyrimidine-purin-purin" and so this column is correspondingly numerated 011). By contrast, entries of each row are numerated by binary numbers in line with the second set of indicators (for example, the same triplet CAG and all other triplets in the same row are the combination "amino-amino-keto" and so this row is correspondingly numerated 001). In such tables (Fig. 15), each of 4 letters, 16 doublets, 64 triplets, ... takes automatically its own individual place and all components of the alphabets are arranged in a strict order.

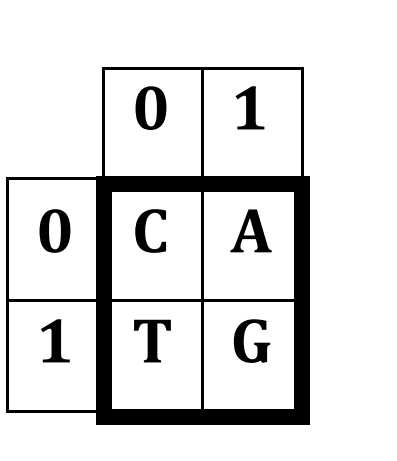

| | 0 | 1 |
|---|---|---|
| 0 | C | A |
| 1 | T | G |

| | 00 | 01 | 10 | 11 |
|---|---|---|---|---|
| 00 | CC | CA | AC | AA |
| 01 | CT | CG | AT | AG |
| 10 | TC | TA | GC | GA |
| 11 | TT | TG | GT | GG |

| | 000 | 001 | 010 | 011 | 100 | 101 | 110 | 111 |
|---|---|---|---|---|---|---|---|---|
| 000 | CCC | CCA | CAC | CAA | ACC | ACA | AAC | AAA |
| 001 | CCT | CCG | CAT | CAG | ACT | ACG | AAT | AAG |
| 010 | CTC | CTA | CGC | CGA | ATC | ATA | AGC | AGA |
| 011 | CTT | CTG | CGT | CGG | ATT | ATG | AGT | AGG |
| 100 | TCC | TCA | TAC | TAA | GCC | GCA | GAC | GAA |
| 101 | TCT | TCG | TAT | TAG | GCT | GCG | GAT | GAG |
| 110 | TTC | TTA | TGC | TGA | GTC | GTA | GGC | GGA |
| 111 | TTT | TTG | TGT | TGG | GTT | GTG | GGT | GGG |

Fig. 15. The square tables of DNA-alphabets of 4 nucleotides, 16 doublets and 64 triplets with a strict arrangement of all components. Each of tables is constructed in line with the principle of binary numeration of its column and rows (see explanations in text).

Here one can remind a historical fact that the same principle of constructing square tables with quite similar binary numerations of their columns and rows was used else in the Ancient Chinese book «I-Ching», which was written a few thousand years ago and which is the most ancient historical example of systematic usage of binary numbers. The famous table of 64 hexagrams in Fu-Xi's order exists there, which has many deep analogies with the genetic matrix of 64 triplets [Hu, Petoukhov, Petukhova, 2017; Petoukhov, 1999, 2008, 2017; Petoukhov, He, 2009]. The ancient Chinese claimed that this table is the universal archetype of nature. They knew nothing about the genetic code, but the genetic code is constructed in line with the "I-Ching".

For our article about tetra-group symmetries in long DNA-texts it is more important the following: these 3 separate genetic tables form the joint tensor (!) family of matrices since they are interrelated by the known operation of the tensor (or Kronecker) product of matrices. By definition, under tensor multiplication of two matrices, each of entries of the first matrix is multiplied with the whole second matrix [Bellman, 1960]. The second tensor power of the (2*2)-matrix [C, A; T, G] of 4 DNA-letters gives automatically the matrix of 16 doublets; the third tensor power of the matrix of the same matrix of 4 DNA-letters gives the matrix of 64 triplets with the same strict arrangement of entries (Fig. 16).

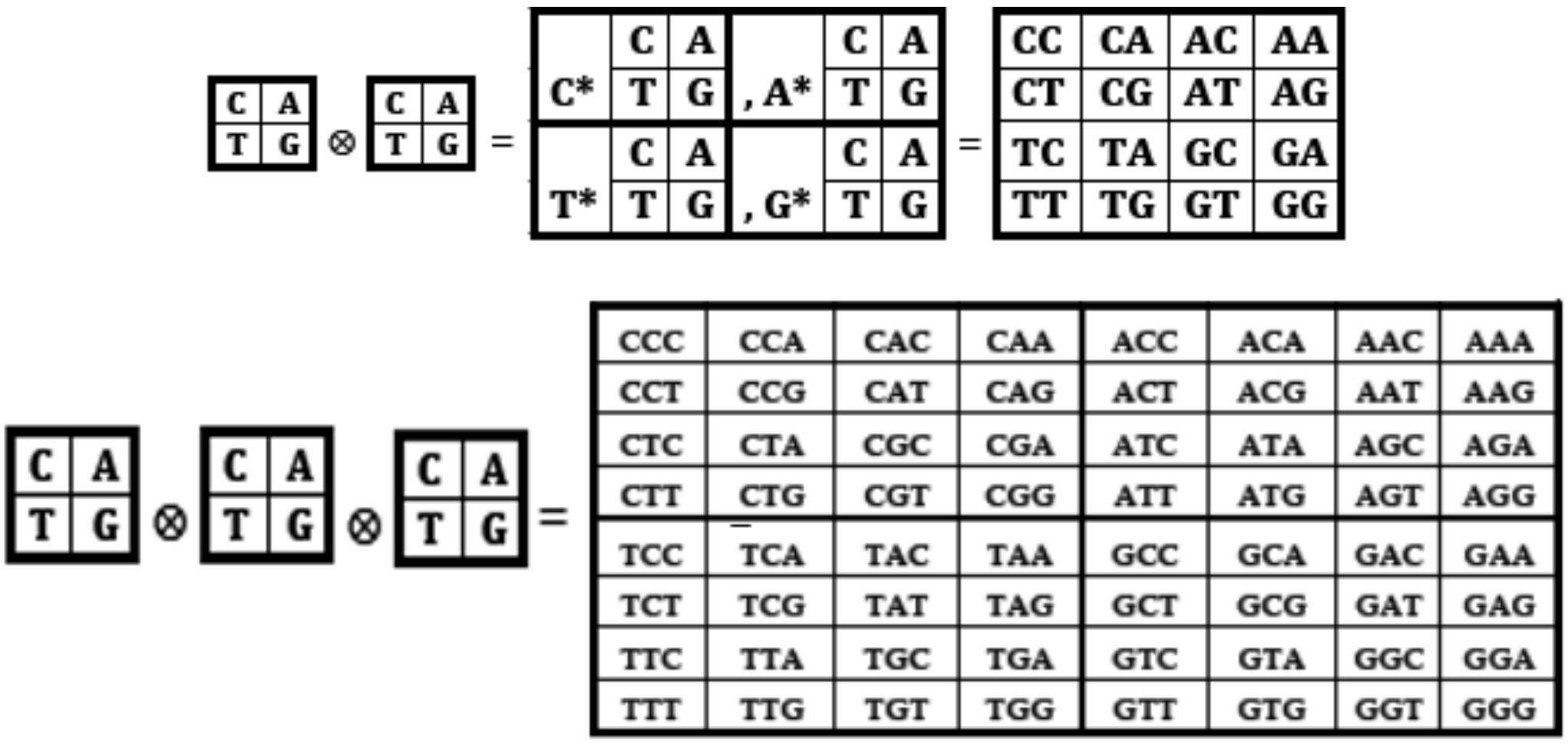

| CC | CA | AC | AA |
|---|---|---|---|
| CT | CG | AT | AG |
| TC | TA | GC | GA |
| TT | TG | GT | GG |

| CCC | CCA | CAC | CAA | ACC | ACA | AAC | AAA |
|---|---|---|---|---|---|---|---|
| CCT | CCG | CAT | CAG | ACT | ACG | AAT | AAG |
| CTC | CTA | CGC | CGA | ATC | ATA | AGC | AGA |
| CTT | CTG | CGT | CGG | ATT | ATG | AGT | AGG |
| TCC | TCA | TAC | TAA | GCC | GCA | GAC | GAA |
| TCT | TCG | TAT | TAG | GCT | GCG | GAT | GAG |
| TTC | TTA | TGC | TGA | GTC | GTA | GGC | GGA |
| TTT | TTG | TGT | TGG | GTT | GTG | GGT | GGG |

Fig. 16. The tensor family of genetic matrices [C, A; T, G]$^{(n)}$ (here tensor power n = 1, 2, 3) of DNA-alphabets of 4 nucleotides, 16 doublets and 64 triplets. The symbol ⊗ means the tensor product.

One can see from a comparison of Fig. 16 with Fig. 15 that the tensor powers of the genetic matrix of 4 letters [C, A; T, G] produce - in a new way - the same square tables of the DNA-alphabets with the same strict arrangements of alphabetic components without any mention about binary-oppositional traits of genetic letters C, A, T and G, which were used to construct the square tables in Fig. 15.

So, the structural organization of the system of DNA-alphabets is connected with the algebraic operation of the tensor product. It is important since the operation of the tensor product is well known in mathematics, physics and informatics, where it gives a way of putting vector spaces together to form larger vector spaces. The following quotation speaks about the meaning of the tensor product: «*This construction is crucial to understanding the quantum mechanics of multiparticle systems*» [Nielsen, Chuang, 2010, p. 71]. For us the most interesting is that the tensor product is one of basis instruments in quantum informatics. Notions and formalisms of quantum informatics are used in the next Section to simulate phenomena of the tetra-group symmetries in long DNA-texts and for creation of new quantum-informational approaches in mathematical genetics.

## 5. The quantum-information model to explain and predict symmetries of collective probabilities in tetra-groups of long DNA-texts

Are there systems in nature or in mathematical natural science, for which the rules of tetra-group symmetry of probabilities are fulfilled, similar to those observed in long DNA texts? Yes. Below we show that, in the field of quantum informatics, 2n-qubit systems in so called separable pure states exist, for which similar rules are fulfiled precisely (in contrast to their approximate implementation in long DNA-texts). Correspondingly the author proposes using these quantum states to model phenomena of tetra-group symmetries of probabilities in long DNA-texts. It should be mentioned that general thoughts and hypothesises about using principles of quantum informatics in living organisms are discussed in interesting publications of many authors [Igamberdiev, 1993, 2012; Matsuno, 1999, 2003, 2015; Matsuno, Paton, 2000; Simeonov, 2013; and others]. For example, the book [Penrose, 1996], concerning the topic of biological quantum computers, appeals to the fact that tubulin proteins exist in two different configurations, and they can switch between these configurations by analogy with triggers to provide bio-computer functions. In contrast to this "protein standpoint", our model approach testifies in favor that already the genetic level, which is the deepest level of living organisms, is connected with the principles of quantum computers.

DNA- and RNA-molecules belong to the microworld of quantum mechanics and perform the information function. By this reason their informational features should be studied on the basis of quantum informatics. Below the author proposes a possible model approach to DNA-texts from the standpoint of quantum informatics. This Section shows how notions and formalisms of quantum informatics can be introduced into the field of genetic informatics in connection with described tetra-group symmetries and rules of long DNA-texts. Here we use typical notions, denotations and formalisms of quantum informatics from the fundamental book [Nielsen, Chuang, 2010]. In particular we use the notion of quantum bits (or qubits) for model representations of elements of single stranded DNA. We also use ordinary Dirac notations for vectors and operations with them:

- $|\psi>$ means a vector, which is known also as a ket-vector;
- $<\psi|$ means a vector, which is dual to $|\psi>$ and known as a bra-vector;
- $<\varphi|\psi>$ means scalar (or inner) product between the vectors $<\varphi|$ and $|\psi>$;
- $|\varphi> \otimes |\psi>$ means the tensor product of $|\varphi>$ and $|\psi>$;
- $|\varphi>|\psi>$ is abbreviated notation for tensor product of $|\varphi>$ and $|\psi>$;
- $M^T$ – transpose of the M matrix;

- $<\varphi>|M|\psi>$ scalar product between $|\varphi>$ and $M|\psi>$.

In quantum informatics, such vector spaces H are considered, which are equipped with the scalar (or inner) product (so called Hilbert spaces). Let $H_1$ and $H_2$ be quantum mechanical state spaces, that is, finite dimensional Hilbert spaces with orthonormal basis states $|\alpha_i>$ and $|\beta_j>$, where i = 1, …, n and j = 1, …, m. By a postulate of quantum mechanics, the state space of the composite system is given by the tensor product $H_1 \otimes H_2$ with base states $\{|\alpha_i> \otimes |\beta_j>\}$, or in more compact notation $\{|\alpha_i\beta_j>\}$. "*The state space of a composite physical systems is the tensor product of the state spaces of the component physical systems. Moreover, if we have systems numbered 1 through n and system number i is prepared in the state $\rho_i$, then the joint state of the total system is $\rho_1 \otimes \rho_2 \otimes \ldots \rho_n$*" [Nielsen, Chuang, 2010, p. 102]. If a quantum state can be represented as a vector of a Hilbert space, such state is called a pure quantum state. If a pure state $|\psi> \in H_1 \otimes H_2$ can be written in the form $|\psi> = |\psi_1> \otimes |\psi_2>$, where $|\psi_i>$ is a pure state of the i-th subsystem, it is said to be separable. Otherwise it is called entangled.

As known, a quantum bit (or qubit) is a unit of quantim information. For two-level quantum systems used as qubits, the state $|0>$ is identified with the vector (1, 0), and similarly the state $|1>$ with the vector (0, 1). Two possible states for a qubit are the states $|0>$ and $|1>$, which correspond to the states of 0 and 1 for a classical bit. The difference between bits and qubits is that a qubit can be in a state other than $|0>$ or $|1>$. It is possible to form linear combinations of states, often called superpositions (6):

$$|\psi> = \alpha|0> + \beta|1>, \qquad (6)$$

The symbol $|\psi>$ means a state of a gubit. The numbers α and β can be complex numbers but in our case it is enough to think of them as real numbers. Put another way, the state of a qubit is a vector in a two-dimensional vector space. The standart notation for states in quantum mechanics is the Dirac notation "$|\ >$". The special states $|0>$ and $|1>$ are known as computational basis states, and form an orthonormal basis for this vector space [Nielsen, Chuang, 2010, p. 13]. As known, we cannot examine a qubit to determine its quantum state, that is, the values of $\alpha$ and $\beta$. Instead, quantum mechanics tells us that we can only acquire much more restricted information about the quantum state. When we measure a qubit we get either the result 0, with probability $|\alpha|^2$, or the result 1, with probability $|\beta|^2$. Naturally, $|\alpha|^2 + |\beta|^2 = 1$, since the probabilities must sum to one. Geometrically, we can interpret this as the condition that the qubit's state be normalized to length 1. Values $\alpha$ and $\beta$ are called amplitudes of probabilities. Thus, in general a qubit's state is a unit vector in a two-dimensional complex vector space. Let us emphasize again that when a qubit is measured, it only ever gives "0" or "1" as the measurement result – probabilistically.

In more general case, a system of "n" qubits is considered in quantum informatics. The computational basis states of this system are written in the form $|x_1x_2 \ldots x_n>$; a quantum state of such a system is specified by $2^n$ amplitudes [Nielsen, Chuang, 2010, p. 17]. In our model approach we interpret DNA-texts as quantum systems of many qubits.

In technical devices of quantum informatics, a qubit can be represented by many ways on the basis of different pairs of binary-oppositional indicators:

for example, by two electronic levels in an atom; by two kinds of polarization of a single photon (vertical polarization and horizontal polarization), etc.

In our model approach for genetic informatics, we represent qubits on the basis of different pairs of binary-oppositional indicators of adenine A, guanine G, cytosine C, and thymine T, which were shown above in Fig. 14. As we noted, each of these DNA bases can be uniquely defined by any two kinds of mentioned binary-oppositional indicators (Fig. 14). By analogy with the previous Section, to characterized each of the DNA-letters C, A, T, G we will use the first kind of indicators («pyrimidine or purine») and the second kind of indicators («three hydrogen bonds or two hydrogen bonds»). On the basis of each of these pairs of binary-oppositional indicators, a corresponding two-level quantum system can be formally introduced with a definition of its appropriate qubit (those qubits, which are introduced in genetic informatics on the basis of binary-oppositional indicators of genetic molecules, can be conditionally called «genetic qubits» or briefly «g-qubits»).

Let us introduce, firstly, the notion of a genetic qubit as a two-level quantum system on the basis of the indicators «pyrimidine or purine»: in this quantum system one level corresponds to the indicator «pyrimidine» and the second level – to the oppositional indicator «purine». In other words, such genetic qubit is represented by these oppositional indicators and the state of such qubit is a vector in its appropriate two-dimensional Hilbert space $H_1$. One can assume that the state $|0>$ corresponds to the state «pyrimidine», and the state $|1>$ - to the state «purine». By analogy with the expression (6), a state of such genetic qubit can be expressed by the expression (7), where $\alpha_0$ and $\alpha_1$ are amplitudes of probabilities of these computational basis states "pyrimidine" and "purine":

$$|\psi_1> = \alpha_0 |0> + \alpha_1 |1>, \qquad \alpha_0^2 + \alpha_1^2 = 1 \qquad (7)$$

In a particular case, a qubit (7) can represent a pure state of a quantum system in a form of a sequence, which consists of pyrimidines and purines.

Secondly, let us introduce the notion of another genetic qubit as a two-level quantum system on the basis of the indicators «three hydrogen bonds or two hydrogen bonds»: in this quantum system one level corresponds to the indicator «three hydrogen bonds»» and the second level – to the indicator «two hydrogen bonds». In other words, such genetic qubit is represented by these two indicators and the state of such qubit is a vector in its appropriate 2-dimensional Hilbert space $H_2$. One can assume that the state $|0>$ corresponds to the state «three hydrogen bonds», and the state $|1>$ - to the state «two hydrogen bonds». By analogy with the expression (6), a state of such genetic qubit can be expressed by the expression (8), where $\beta_0$ and $\beta_1$ are amplitudes of probabilities of these computational basis states:

$$|\psi_2> = \beta_0 |0> + \beta_1 |1>, \qquad \beta_0^2 + \beta_1^2 = 1 \qquad (8)$$

In a particular case, a qubit (8) can represent a pure state of a quantum system in a form of a sequence, which consists of elements with three and two hydrogen bonds.

So we have two different 2-dimensional Hilbert spaces $H_1$ and $H_2$, to which pure states of genetic qubits (7) and (8) belong correspondingly.

In our genetic case, the tensor product of the two-dimensional Hilbert space $H_1 \otimes H_2$ gives one four-dimensional Hilbert space with the following separable pure state of a quantum 2-qubit system:

$$|\psi_{12}> = |\psi_1> \otimes |\psi_2> = (\alpha_0|0> + \alpha_1|1>) \otimes (\beta_0|0> + \beta_1|1>) = \alpha_0\beta_0|00> + \alpha_0\beta_1|01> + \alpha_1\beta_0|10> + \alpha_1\beta_1|11> \quad (9)$$

Such 2-qubit system has four computational basis states denoted |00>, |01>, |10>, |11>. Amplitudes $\alpha_0\beta_0$, $\alpha_0\beta_1$, $\alpha_1\beta_0$ and $\alpha_1\beta_1$ of probabilities satisfy the normalization condition: $(\alpha_0\beta_0)^2 + (\alpha_0\beta_1)^2 + (\alpha_1\beta_0)^2 + (\alpha_1\beta_1)^2 = 1$. One can note that two kinds of indicators "pyrimidine-purine" and «three hydrogen bonds or two hydrogen bonds» for separate nitrogenous bases C, T, G, A define the following correspondence:

- Cytosine C corresponds to the computational basis state |00> of the 2-qubit system (9) since cytosine C is characterized by the indicator "pyrimidine", which is the computational basis state |0> in the first qubit (7), and also by the indicator "three hydrogen bonds", which is the computational basis state |0> in the second qubit (8). In the four-dimensional Hilbert space $H_1 \otimes H_2$, these computational basis states |0> and |0> of two genetic qubits (7) and (8) define the computational basis state |00> of the 2-qubit system (9);
- Thymine T corresponds to the computational basis state |01> of the 2-qubit system (9) since thymine T is characterized by the indicator "pyrimidine", which is the computational basis state |0> in the first qubit (7), and also by the indicator "two hydrogen bonds", which is the computational basis state |1> in the second qubit (8). In the four-dimensional Hilbert space $H_1 \otimes H_2$, these computational basis states |0> and |1> of two genetic qubits (7) and (8) define the computational basis state |01> of the 2-qubit system (9);
- Guanine G corresponds to the computational basis state |10> of the 2-qubit system (9) since guanine G is characterized by the indicator "purine", which is the computational basis state |1> in the first qubit (7), and also by the indicator "three hydrogen bonds", which is the computational basis state |0> in the second qubit (8). In the four-dimensional Hilbert space $H_1 \otimes H_2$, these computational basis states |1> and |0> of two genetic qubits (7) and (8) define the computational basis state |10> of the 2-qubit system (9).
- Adenine A corresponds to the computational basis state |11> of the 2-qubit system (9) since adenine A is characterized by the indicator "purine", which is the computational basis state |1> in the first qubit (7), and also by the indicator "two hydrogen bonds", which is the computational basis state |1> in the second qubit (8). In the four-dimensional Hilbert space $H_1 \otimes H_2$, these computational basis states |1> and |1> of two genetic qubits (7) and (8) define the computational basis state |11> of the 2-qubit system (9).

By this way, members of the tetra-group of nucleotides C, T, G and A get their representations as computational basis states of a 2-qubit system (9) in a four-dimensional Hilbert space $H_1 \otimes H_2$ with conditional denotations $|C>=|00>$, $|T>=|01>$, $|G>=|10>$ and $|A>=|11>$. (Below we will show that these computational basis states can be connected with pairs of photons of different frequencies, radiated by molecular elements of these dual indicators). Correspondingly the expression (9) can be rewritten in the following conditional form (10):

$$|\psi_{12}> = \alpha_0\beta_0\,|00> + \alpha_0\beta_1\,|01> + \alpha_1\beta_0\,|10> + \alpha_1\beta_1\,|11> = \\ = \alpha_0\beta_0\,|C> + \alpha_0\beta_1\,|T> + \alpha_1\beta_0\,|G> + \alpha_1\beta_1\,|A> \qquad (10)$$

We call the 2-qubit quantum system with its separable pure state (10) as a «monoplet CTGA-system». In a particular case, the state (10) of a 2-qubits monoplet CTGA-system can represent a separable pure state of a quantum system, which is a sequence of elements, where each of elements is pyrimidine or purin and simultaneously it has three or two hydrogen bonds by analogy with nucleotides. From (10), the probabilities of computational basis states of separate nucleotides C, T, G and A are equal to the following:

$$P_1(C_1) = (\alpha_0\beta_0)^2,\ P_1(T_1) = (\alpha_0\beta_1)^2,\ P_1(G_1) = (\alpha_1\beta_0)^2,\ P_1(A_1) = (\alpha_1\beta_1)^2 \qquad (11)$$

They should satisfy the normalization condition (12):

$$(\alpha_0\beta_0)^2 + (\alpha_0\beta_1)^2 + (\alpha_1\beta_0)^2 + (\alpha_1\beta_1)^2 = 1 \qquad (12)$$

To consider the case not single monoplets C, T, G, A but doublets CC, CT, CG, CA, TC, TT, TG, TA, GC, GT, GG, GA, AC, AT, AG, AA one should accordingly expand the 4-dimensional Hilbert space $H_1 \otimes H_2$ to the 16-dimensional Hilbert space $H_1 \otimes H_2 \otimes H_3 \otimes H_4$. Here the spaces $H_1$ and $H_2$ are related with the first letters of doublets and the spaces $H_3$ and $H_4$ are related with the second letters of doublets. These spaces $H_3$ and $H_4$ are defined in a close analogy with the spaces $H_1$ and $H_2$. Let us explain this.

By analogy with the expressions (6, 7), each of four letters C, T, G, A at the second position of doublets is interpreted firstly as a two-level quantum system on the basis of the oppositional indicators «pyrimidine or purine»: in this quantum system one level corresponds to the indicator «pyrimidine» and the second level – to the oppositional indicator «purine». In other words, a new genetic qubit arises on the basis of these oppositional indicators for the second letters inside 16 doublets; the state of such qubit is a vector in its appropriate two-dimensional Hilbert space $H_3$. One can assume that the state $|0>$ corresponds to the state «pyrimidine», and the state $|1>$ - to the state «purine». By analogy with the expression (7), a state of such genetic qubit can be denoted by the expression (13), where $\alpha_2$ and $\alpha_3$ are amplitudes of probabilities of these computational basis states "pyrimidine" and "purine" for the second letters in 16 doublets:

$$|\psi_3> = \alpha_2\,|0> + \alpha_3\,|1>, \qquad \alpha_2^2 + \alpha_3^2 = 1 \qquad (13)$$

Secondly, for the second letters of 16 doublets, the notion of another genetic qubit as a two-level quantum system on the basis of the oppositional indicators «three hydrogen bonds or two hydrogen bonds» is defined by analogy with the expression (8). In this quantum system, one level corresponds to the indicator «three hydrogen bonds»» and the second level – to the indicator «two hydrogen bonds». In other words, such genetic qubit is represented by these two indicators and the state of such qubit is a vector in its appropriate 2-dimensional Hilbert space $H_4$. One can assume that the state |0> corresponds to the state «three hydrogen bonds», and the state |1> - to the state «two hydrogen bonds». By analogy with the expression (8), a state of such genetic qubit can be expressed by the expression (14), where $\beta_2$ and $\beta_3$ are amplitudes of probabilities of these computational basis states:

$$|\psi_4\rangle = \beta_2\,|0\rangle + \beta_3\,|1\rangle, \qquad \beta_2^2 + \beta_3^2 = 1 \tag{14}$$

In such way, in the case of long DNA-texts of doublets, we have two additional 2-dimensional Hilbert spaces $H_3$ and $H_4$, to which pure states of genetic qubits (13) and (14) belong correspondingly. The tensor product of the two-dimensional Hilbert space $H_3 \otimes H_4$ gives one four-dimensional Hilbert space with the following separable pure state of a quantum 2-qubit system for the second letters inside 16 doublets:

$$\begin{aligned}|\psi_{34}\rangle = |\psi_3\rangle \otimes |\psi_4\rangle &= (\alpha_2\,|0\rangle + \alpha_3\,|1\rangle) \otimes (\beta_2\,|0\rangle + \beta_3\,|1\rangle) = \\ &\alpha_2\beta_2\,|00\rangle + \alpha_2\beta_3\,|01\rangle + \alpha_3\beta_2\,|10\rangle + \alpha_3\beta_3\,|11\rangle = \\ &\alpha_2\beta_2\,|C\rangle + \alpha_2\beta_3\,|T\rangle + \alpha_3\beta_2\,|G\rangle + \alpha_3\beta_3\,|A\rangle\end{aligned} \tag{15}$$

Such new 2-qubit system has four computational basis states denoted |00>, |01>, |10>, |11>. By analogy with the case of the Hilbert space space H1⊗H2 for the first letters of doublets with conditional denotations |C>=|00>, |T>=|01>, |G>=|10> and |A>=|11> (see the expression (10)), in the Hilbert space $H_3 \otimes H_4$ similar denotations are used for 2-qubit systems of the second letters of doublets: |C>=|00>, |T>=|01>, |G>=|10> and |A>=|11>. Amplitudes $\alpha_2\beta_2$, $\alpha_2\beta_3$, $\alpha_3\beta_2$ and $\alpha_3\beta_3$ of probabilities in (15) satisfy the normalization condition (16):

$$(\alpha_2\beta_2)^2 + (\alpha_2\beta_3)^2 + (\alpha_3\beta_2)^2 + (\alpha_3\beta_3)^2 = 1. \tag{16}$$

In the 16-dimensional Hilbert space $H_1 \otimes H_2 \otimes H_3 \otimes H_4$ for the case of 16 doublets, we have in our model approach the following 16 computational basis states with their appropriate amplitudes of probabilities in a long DNA-text:

$$\begin{aligned}&|\psi_{12}\rangle \otimes |\psi_{34}\rangle = \\ &\alpha_0\beta_0\alpha_2\beta_2|CC\rangle + \alpha_0\beta_0\alpha_2\beta_3|CT\rangle + \alpha_0\beta_0\alpha_3\beta_2|CG\rangle + \alpha_0\beta_0\alpha_3\beta_3|CA\rangle + \\ &\alpha_0\beta_1\alpha_2\beta_2|TC\rangle + \alpha_0\beta_1\alpha_2\beta_3|TT\rangle + \alpha_0\beta_1\alpha_3\beta_2|TG\rangle + \alpha_0\beta_1\alpha_3\beta_3|TA\rangle + \\ &\alpha_1\beta_0\alpha_2\beta_2|GC\rangle + \alpha_1\beta_0\alpha_2\beta_3|GT\rangle + \alpha_1\beta_0\alpha_3\beta_2|GG\rangle + \alpha_1\beta_0\alpha_3\beta_3|GA\rangle + \\ &\alpha_1\beta_1\alpha_2\beta_2|AC\rangle + \alpha_1\beta_1\alpha_2\beta_3|AT\rangle + \alpha_1\beta_1\alpha_3\beta_2|AG\rangle + \alpha_1\beta_1\alpha_3\beta_3|AA\rangle\end{aligned} \tag{17}$$

In the state (17) of a 4-qubit "doublet CTGA-system", 16 computational basis states are represented by 16 doublets: |0000>=|CC>, |0001>=|CT>, |0010>=|CG>, |0011>=|CA>, |0100>=|TC>, |0101>=|TT>, |0110>=|TG>,

|0111>=|TA>, |1000>=|GC>, |1001>=|GT>, |1010>=|GG>, |1011>=|GA>, |1100>=|AC>, |1101>=|AT>, |1110>=|AG>, |1111>=|AA>. In our model approach, these 16 computational basis states are interpreted as representations of appropriate 16 genetic doublets.

From the expression (17), we have the following collective probability $P_2(C_1)$ of all 4 doublets with the first letter C in them (they are collected in the first row of this expression):

$$P_2(C_1) = (\alpha_0\beta_0\alpha_2\beta_2)^2 + (\alpha_0\beta_0\alpha_2\beta_3)^2 + (\alpha_0\beta_0\alpha_3\beta_2)^2 + (\alpha_0\beta_0\alpha_3\beta_3)^2 =$$
$$(\alpha_0\beta_0)^2 *\{(\alpha_2\beta_2)^2 + (\alpha_2\beta_3)^2 + (\alpha_3\beta_2)^2 + (\alpha_3\beta_3)^2\} = (\alpha_0\beta_0)^2 = P_1(C) \qquad (18)$$

Here the sum in curly brackets is equal to1 according to the normalization condition (16). The expression (18) means that the collective probability $P_2(C_1)$ of 4 doublets CC, CT, CG and CA is equal to the individual probability $P_1(C)$ of the letter C in this quantum-informational model of collective and individual probabilities inside long DNA-texts. From the expression (17), similar calculations of collective probabilities $P_2(T_1)$, $P_2(G_1)$ and $P_2(A_1)$ of doublets with the first letters T, G, A in them give analogical results (19) of their equality to individual probabilities of letters T, G and A:

$$P_2(T_1) = P_1(T), \quad P_2(G_1) = P_1(G), \quad P_2(A_1) = P_1(A) \qquad (19)$$

These model results correspond to phenomenologic facts reflected in the the first rule of tetra-group symmetries in long DNA-texts (see Section 3 above).

The expression (17) allows a calculation of the collective probability $P_2(C_2)$ of all 4 doublets with the second letter C in them:

$$P_2(C_2) = (\alpha_0\beta_0\alpha_2\beta_2)^2 + (\alpha_0\beta_1\alpha_2\beta_2)^2 + (\alpha_1\beta_0\alpha_2\beta_2)^2 + (\alpha_1\beta_1\alpha_2\beta_2)^2 =$$
$$(\alpha_2\beta_2)^2 *\{(\alpha_0\beta_0)^2 + (\alpha_0\beta_1)^2 + (\alpha_1\beta_0)^2 + (\alpha_1\beta_1)^2\} = (\alpha_2\beta_2)^2 = P_1(C) \qquad (20)$$

Here the sum in curly brackets is equal to 1 according to the normalization condition (12). The expression (20) means that the collective probability $P_2(C_2)$ of all doublets with the second letter C (that is, CC, TC, GC, AC) is also equal to the individual probability $P_1(C)$ of the letter C. From the expression (17), similar calculations of collective probabilities $P_2(T_2)$, $P_2(G_2)$ and $P_2(A_2)$ of doublets with the second letters T, G, A in them give analogical results (21) of their equality to individual probabilities of letters T, G and A:

$$P_2(T_2) = P_1(T), \quad P_2(G_2) = P_1(G), \quad P_2(A_2) = P_1(A) \qquad (21)$$

These model results correspond to phenomenologic facts reflected in the the second rule of tetra-group symmetries in long DNA-texts (see Section 3 above).

To consider the case of 64 triplets one should accordingly expand the 16-dimensional Hilbert space $H_1 \otimes H_2 \otimes H_3 \otimes H_4$ to the 64-dimensional Hilbert space $H_1 \otimes H_2 \otimes H_3 \otimes H_4 \otimes H_5 \otimes H_6$. Here the spaces $H_1$ and $H_2$ are related with the first letters of doublets, the spaces $H_3$ and $H_4$ are related with the second letters of doublets and the additional spaces $H_5$ and $H_6$ are related with the third letters of triplets. These additional spaces $H_5$ and $H_6$ are defined in a close analogy with the spaces $H_1$, $H_2$, $H_3$ and $H_4$. Repeating for the spaces $H_5$ and $H_6$ those steps, which were made above to define the Hilbert spaces $H_3$ and $H_4$ for the second letters of doublets, we get for the third letter of triplets the following expressions:

$$|\psi_5> = \alpha_4\,|0> + \alpha_5\,|1>, \qquad \alpha_4^2 + \alpha_5^2 = 1 \tag{22}$$

$$|\psi_6> = \beta_4\,|0> + \beta_5\,|1>, \qquad \beta_4^2 + \beta_5^2 = 1 \tag{23}$$

$$\begin{aligned} |\psi_{56}> = |\psi_5> \otimes |\psi_6> &= (\alpha_4\,|0> + \alpha_5\,|1>) \otimes (\beta_4\,|0> + \beta_5\,|1>) = \\ &\alpha_4\beta_4\,|00> + \alpha_4\beta_5\,|01> + \alpha_5\beta_4\,|10> + \alpha_5\beta_5\,|11> = \\ &\alpha_4\beta_4\,|C> + \alpha_4\beta_5\,|T> + \alpha_5\beta_4\,|G> + \alpha_5\beta_5\,|A> \end{aligned} \tag{24}$$

$$(\alpha_4\beta_4)^2 + (\alpha_4\beta_5)^2 + (\alpha_5\beta_4)^2 + (\alpha_5\beta_5)^2 = 1 \tag{25}$$

Here we again use the denotations: $|C>=|00>$, $|T>=|01>$, $|G>=|10>$ and $|A>=|11>$ for the third letters of triplets in the 4-dimensional Hilbert space $H_5\otimes H_6$.

In the 64-dimensional Hilbert space $H_1\otimes H_2\otimes H_3\otimes H_4\otimes H_5\otimes H_6$ for the case of 64 triplets, we have the following 64 computational basis states with their appropriate amplitudes of probabilities:

$$\begin{aligned} &|\psi_{12}> \otimes |\psi_{34}> \otimes |\psi_{56}> = \\ &= \alpha_0\beta_0\alpha_2\beta_2\alpha_4\beta_4|CCC>+\alpha_0\beta_0\alpha_2\beta_2\alpha_4\beta_5|CCT>+\alpha_0\beta_0\alpha_2\beta_2\alpha_5\beta_4|CCG>+\alpha_0\beta_0\alpha_2\beta_2\alpha_5\beta_5|CCA> \\ &+ \alpha_0\beta_0\alpha_2\beta_3\alpha_4\beta_4|CTC>+\alpha_0\beta_0\alpha_2\beta_3\alpha_4\beta_5|CTT>+\alpha_0\beta_0\alpha_2\beta_3\alpha_5\beta_4|CTG>+\alpha_0\beta_0\alpha_2\beta_3\alpha_5\beta_5|CTA> \\ &+\alpha_0\beta_0\alpha_3\beta_2\alpha_4\beta_4|CGC>+\alpha_0\beta_0\alpha_3\beta_2\alpha_4\beta_5|CGT>+\alpha_0\beta_0\alpha_3\beta_2\alpha_5\beta_4|CGG>+\alpha_0\beta_0\alpha_3\beta_2\alpha_5\beta_5|CGA> \\ &+\alpha_0\beta_0\alpha_3\beta_3\alpha_4\beta_4|CAC>+\alpha_0\beta_0\alpha_3\beta_3\alpha_4\beta_5|CAT>+\alpha_0\beta_0\alpha_3\beta_3\alpha_5\beta_4|CAG>+\alpha_0\beta_0\alpha_3\beta_3\alpha_5\beta_5|CAA> \\ &+\alpha_0\beta_1\alpha_2\beta_2\alpha_4\beta_4|TCC>+\alpha_0\beta_1\alpha_2\beta_2\alpha_4\beta_5|TCT>+\alpha_0\beta_1\alpha_2\beta_2\alpha_5\beta_4|TCG>+\alpha_0\beta_1\alpha_2\beta_2\alpha_5\beta_5|TCA> \\ &+\alpha_0\beta_1\alpha_2\beta_3\alpha_4\beta_4|TTC>+\alpha_0\beta_1\alpha_2\beta_3\alpha_4\beta_5|TTT>+\alpha_0\beta_1\alpha_2\beta_3\alpha_5\beta_4|TTG>+\alpha_0\beta_1\alpha_2\beta_3\alpha_5\beta_5|TTA> \\ &+\alpha_0\beta_1\alpha_3\beta_2\alpha_4\beta_4|TGC>+\alpha_0\beta_1\alpha_3\beta_2\alpha_4\beta_5|TGT>+\alpha_0\beta_1\alpha_3\beta_2\alpha_5\beta_4|TGG>+\alpha_0\beta_1\alpha_3\beta_2\alpha_5\beta_5|TGA> \\ &+\alpha_0\beta_1\alpha_3\beta_3\alpha_4\beta_4|TAC>+\alpha_0\beta_1\alpha_3\beta_3\alpha_4\beta_5|TAT>+\alpha_0\beta_1\alpha_3\beta_3\alpha_5\beta_4|TAG>+\alpha_0\beta_1\alpha_3\beta_3\alpha_5\beta_5|TAA> \\ &+\alpha_1\beta_0\alpha_2\beta_2\alpha_4\beta_4|GCC>+\alpha_1\beta_0\alpha_2\beta_2\alpha_4\beta_5|GCT>+\alpha_1\beta_0\alpha_2\beta_2\alpha_5\beta_4|GCG>+\alpha_1\beta_0\alpha_2\beta_2\alpha_5\beta_5|GCA> \\ &+\alpha_1\beta_0\alpha_2\beta_3\alpha_4\beta_4|GTC>+\alpha_1\beta_0\alpha_2\beta_3\alpha_4\beta_5|GTT>+\alpha_1\beta_0\alpha_2\beta_3\alpha_5\beta_4|GTG>+\alpha_1\beta_0\alpha_2\beta_3\alpha_5\beta_5|GTA> \\ &+\alpha_1\beta_0\alpha_3\beta_2\alpha_4\beta_4|GGC>+\alpha_1\beta_0\alpha_3\beta_2\alpha_4\beta_5|GGT>+\alpha_1\beta_0\alpha_3\beta_2\alpha_5\beta_4|GGG>+\alpha_1\beta_0\alpha_3\beta_2\alpha_5\beta_5|GGA> \\ &+\alpha_1\beta_0\alpha_3\beta_3\alpha_4\beta_4|GAC>+\alpha_1\beta_0\alpha_3\beta_3\alpha_4\beta_5|GAT>+\alpha_1\beta_0\alpha_3\beta_3\alpha_5\beta_4|GAG>+\alpha_1\beta_0\alpha_3\beta_3\alpha_5\beta_5|GAA> \\ &+\alpha_1\beta_1\alpha_2\beta_2\alpha_4\beta_4|ACC>+\alpha_1\beta_1\alpha_2\beta_2\alpha_4\beta_5|ACT>+\alpha_1\beta_1\alpha_2\beta_2\alpha_5\beta_4|ACG>+\alpha_1\beta_1\alpha_2\beta_2\alpha_5\beta_5|ACA> \\ &+\alpha_1\beta_1\alpha_2\beta_3\alpha_4\beta_4|ATC>+\alpha_1\beta_1\alpha_2\beta_3\alpha_4\beta_5|ATT>+\alpha_1\beta_1\alpha_2\beta_3\alpha_5\beta_4|ATG>+\alpha_1\beta_1\alpha_2\beta_3\alpha_5\beta_5|ATA> \\ &+\alpha_1\beta_1\alpha_3\beta_2\alpha_4\beta_4|AGC>+\alpha_1\beta_1\alpha_3\beta_2\alpha_4\beta_5|AGT>+\alpha_1\beta_1\alpha_3\beta_2\alpha_5\beta_4|AGG>+\alpha_1\beta_1\alpha_3\beta_2\alpha_5\beta_5|AGA> \\ &+\alpha_1\beta_1\alpha_3\beta_3\alpha_4\beta_4|AAC>+\alpha_1\beta_1\alpha_3\beta_3\alpha_4\beta_5|AAT>+\alpha_1\beta_1\alpha_3\beta_3\alpha_5\beta_4|AAG>+\alpha_1\beta_1\alpha_3\beta_3\alpha_5\beta_5|AAA> \end{aligned} \tag{26}$$

In the state (26) of a 6-qubit "triplet CTGA-system", 64 computational basis states represent 64 triplets in our model approach:

|000000>=|CCC>, |000001>=|CCT>, |000010>=|CCG>, |000011>=|CCA>,
|000100>=|CTC>, |000101>=|CTT>, |000110>=|CTG>, |000111>=|CTA>,
|001000>=|CGC>, |001001>=|CGT>, |001010>=|CGG>, |001011>=|CGA>,
|001100>=|CAC>, |001101>=|CAT>, |001110>=|CAG>, |001111>=|CAA>,
|010000>=|TCC>, |010001>=|TCT>, |010010>=|TCG>, |010011>=|TCA>,
|010100>=|TTC>, |010101>=|TTT>, |010110>=|TTG>, |010111>=|TTA>,
|011000>=|TGC>, |011001>=|TGT>, |011010>=|TGG>, |011011>=|TGA>,
|011100>=|TAC>, |011101>=|TAT>, |011110>=|TAG>, |011111>=|TAA>,
|100000>=|GCC>, |100001>=|GCT>, |100010>=|GCG>, |100011>=|GCA>,
|100100>=|GTC>, |100101>=|GTT>, |100110>=|GTG>, |100111>=|GTA>,

|101000>=|GGC>, |101001>=|GGT>, |101010>=|GGG>, |101011>=|GGA>,
|101100>=|GAC>, |101101>=|GAT>, |101110>=|GAG>, |101111>=|GAA>,
|110000>=|ACC>, |110001>=|ACT>, |110010>=|ACG>, |110011>=|ACA>,
|110100>=|ATC>, |110101>=|ATT>, |110110>=|ATG>, |110111>=|ATA>,
|111000>=|AGC>, |111001>=|AGT>, |111010>=|AGG>, |111011>=|AGA>,
|111100>=|AAC>, |111101>=|AAT>, |111110>=|AAG>, |111111>=|AAA>. (27)

From the expression (26), we have the following collective probability $P_3(C_1)$ of all 16 triplets with the first letter C in them (they are collected in the first 4 rows of this expression):

$$
\begin{aligned}
P_3(C_1) = {} & (\alpha_0\beta_0\alpha_2\beta_2\alpha_4\beta_4)^2+(\alpha_0\beta_0\alpha_2\beta_2\alpha_4\beta_5)^2+(\alpha_0\beta_0\alpha_2\beta_2\alpha_5\beta_4)^2+(\alpha_0\beta_0\alpha_2\beta_2\alpha_5\beta_5)^2 + \\
& +(\alpha_0\beta_0\alpha_2\beta_3\alpha_4\beta_4)^2+(\alpha_0\beta_0\alpha_2\beta_3\alpha_4\beta_5)^2+(\alpha_0\beta_0\alpha_2\beta_3\alpha_5\beta_4)^2+(\alpha_0\beta_0\alpha_2\beta_3\alpha_5\beta_5)^2 \\
& +(\alpha_0\beta_0\alpha_3\beta_2\alpha_4\beta_4)^2+(\alpha_0\beta_0\alpha_3\beta_2\alpha_4\beta_5)^2+(\alpha_0\beta_0\alpha_3\beta_2\alpha_5\beta_4)^2+(\alpha_0\beta_0\alpha_3\beta_2\alpha_5\beta_5)^2 \\
& +(\alpha_0\beta_0\alpha_3\beta_3\alpha_4\beta_4)^2+(\alpha_0\beta_0\alpha_3\beta_3\alpha_4\beta_5)^2+(\alpha_0\beta_0\alpha_3\beta_3\alpha_5\beta_4)^2+(\alpha_0\beta_0\alpha_3\beta_3\alpha_5\beta_5)^2 = \\
& = (\alpha_0\beta_0\alpha_2\beta_2)^2 * \{(\alpha_4\beta_4)^2+(\alpha_4\beta_5)^2+(\alpha_5\beta_4)^2+(\alpha_5\beta_5)^2\} + \\
& + (\alpha_0\beta_0\alpha_2\beta_3)^2 * \{(\alpha_4\beta_4)^2+(\alpha_4\beta_5)^2+(\alpha_5\beta_4)^2+(\alpha_5\beta_5)^2\} + \\
& + (\alpha_0\beta_0\alpha_3\beta_2)^2 * \{(\alpha_4\beta_4)^2+(\alpha_4\beta_5)^2+(\alpha_5\beta_4)^2+(\alpha_5\beta_5)^2\} + \\
& + (\alpha_0\beta_0\alpha_3\beta_3)^2 * \{(\alpha_4\beta_4)^2+(\alpha_4\beta_5)^2+(\alpha_5\beta_4)^2+(\alpha_5\beta_5)^2\} = \\
& = (\alpha_0\beta_0\alpha_2\beta_2)^2 + (\alpha_0\beta_0\alpha_2\beta_3)^2 + (\alpha_0\beta_0\alpha_3\beta_2)^2 + (\alpha_0\beta_0\alpha_3\beta_3)^2 = \\
& = (\alpha_0\beta_0)^2 * \{(\alpha_2\beta_2)^2+(\alpha_2\beta_3)^2+(\alpha_3\beta_2)^2+(\alpha_3\beta_3)^2\} = (\alpha_0\beta_0)^2 = P_1(C). \qquad (28)
\end{aligned}
$$

In this calculations, the sums in curly brackets are equal to 1 according to the normalization conditions (16, 25).

This mathematical result (28) of the proposed model means that collective probability $P_3(C_1)$ of all 16 triplets with the first letter C is equal to the individual probability $P_1(C)$ of the letter C in the same long DNA-text.

From the expression (26), similar calculations of collective probabilities $P_3(T_1)$, $P_3(G_1)$ and $P_3(A_1)$ of triplets with the first letters T, G, A in them give analogical results (28) of their equality to individual probabilities of letters T, G and A:

$$P_3(T_1) = P_1(T), \quad P_3(G_1) = P_1(G), \quad P_3(A_1) = P_1(A) \qquad (29)$$

These model results (28, 29) correspond to phenomenologic facts reflected in the the first rule of tetra-group symmetries in long DNA-texts (see Section 3 above).

Similar calculations on the basis of the expression (26) give the following results:

- collective probabilities of all triplets with the second letters C, T, G, A in them are also equal to individual probabilities of letters C, T, G and A: $P_3(C_2) = P_1(C)$, $P_3(T_2) = P_1(T)$, $P_3(G_2) = P_1(G)$ and $P_3(A_2) = P_1(A)$;
- collective probabilities of all triplets with the third letters C, T, G, A in them are also equal to individual probabilities of letters C, T, G and A: $P_3(C_3) = P_1(C)$, $P_3(T_3) = P_1(T)$, $P_3(G_3) = P_1(G)$ and $P_3(A_3) = P_1(A)$.

These model results correspond to phenomenologic facts reflected in the second rule of tetra-group symmetries in long DNA-texts (see Section 3 above).

By analogy, in this quantum-informational model approach, one can consider cases of $4^n$ n-plets in appropriate $4^n$-dimensional Hilbert space (n = 4, 5, …) and get again and again confirmations of correspondence of this model to the described phenomenologic rules of approximate tetra-group symmetries of collective and individual probabilities in long DNA-texts. The model describes these symmetries of probabilities as exact symmetries in contrast to phenomenological data, where they exist as approximate symmetries.

An effective model should not only explain known phenomenologic data but also should predict unknown data to search them in natural systems. Let us show now that the proposed quantum-informational model approach not only explains the phenomenological facts, but also **has predictive power, allowing to open previously unknown properties of long DNA-texts**.

For this, consider the expression (26) of separable pure state of a quantum 6-qubit "triplet CTGA-system". The right part of the expression (26) contains 64 computational basis states |CCC>, |CCT>, …., |AAA>, which correpond to 64 genetic triplets. The set of 64 triplets contains the following 16 groups of triplets, each of which contains 4 triplets with one of 16 doublets on their identical positions. For example, the doublet CC defines one of $4^2$-groups of 4 triplets, each of which has this doublet in its beginning: CCC, CCT, CCG and CCA.

One should note that in previous Sections of this article we considered tetra-groups (or 4-groups), 4 subgroups of which were defined by monoplets: C-subgroups, T-subgroups, G-subgroups and A-subgroups (see Figs. 1, 2). Such 4-groups can be called as "4-groups of the first degree". But now we will pay our attention to the wider sets of the following $4^s$-groups (that is, tetra-groups of higher degrees s = 2, 3, 4, ...):

- $4^2$-groups, each of which contains 4 subgroups defined by one of 16 doublets (like the CC-subgroup of triplets CCC, CCT, CCG and CCA);
- $4^3$-groups, each of which contains 4 subgroups defined by one of 64 triplets (like the CCC-subgroup of tetraplets CCCC, CCCT, CCCG and CCCA);
- etc.

In the right part of the expression (26), each of 16 rows represents one of $4^2$-groups of triplets with one of 16 doublets in their first positions. Let us take the first row of (26) and calculate the collective probabilities $P_3(CC_1)$ of all 4 triplets CCC, CCT, CCG and CCG with the doublet CC in their beginning, using amplitudes of individual probabilities of the computational basis states |CCC>, |CCT>, |CCG> and |CCG>. The expression (30) shows the result of such calculation:

$$P_3(CC_1) =(\alpha_0\beta_0\alpha_2\beta_2\alpha_4\beta_4)^2 +(\alpha_0\beta_0\alpha_2\beta_2\alpha_4\beta_5)^2 +(\alpha_0\beta_0\alpha_2\beta_2\alpha_5\beta_4)^2 +(\alpha_0\beta_0\alpha_2\beta_2\alpha_5\beta_5)^2 =$$
$$= (\alpha_0\beta_0\alpha_2\beta_2)^2 * \{(\alpha_4\beta_4)^2 +(\alpha_4\beta_5)^2 +(\alpha_5\beta_4)^2 +(\alpha_5\beta_5)^2\} = (\alpha_0\beta_0\alpha_2\beta_2)^2 = P_2(CC) \qquad (30)$$

Here the sum in curly brackets is equal to 1 in accordance with the normalization condition (25). The model result (30) shows that the collective probability $P_3(CC_1)$ of the set of all 4 triplets CCC, CCT, CCG, CCA with the doublet CC in their beginning is equal to the individual probability $P_2(CC)$ of the doublet CC in the same long DNA-text.

Similar calculations of collective probabilities $P_3(CT_1)$, $P_3(CG_1)$, etc. of other 15 sets of 4 triplets in other 15 rows in (26) give analogical results (31) of their equality to individual probabilities of appropriate doublets:

$P_3(CT_1) = (\alpha_0\beta_0\alpha_2\beta_3)^2 = P_2(CT)$, $P_3(CG_1) = (\alpha_0\beta_0\alpha_3\beta_2)^2 = P_2(CG)$,
$P_3(CA_1) = (\alpha_0\beta_0\alpha_3\beta_3)^2 = P_2(CA)$, $P_3(TC_1) = (\alpha_0\beta_1\alpha_2\beta_2)^2 = P_2(TC)$,
$P_3(TT_1) = (\alpha_0\beta_1\alpha_2\beta_3)^2 = P_2(TT)$, $P_3(TG_1) = (\alpha_0\beta_1\alpha_3\beta_2)^2 = P_2(TG)$,

$$P_3(TA_1) = (\alpha_0\beta_1\alpha_3\beta_3)^2 = P_2(TA), \quad P_3(GC_1) = (\alpha_1\beta_0\alpha_2\beta_2)^2 = P_2(GC),$$
$$P_3(GT_1) = (\alpha_1\beta_0\alpha_2\beta_3)^2 = P_2(GT), \quad P_3(GG_1) = (\alpha_1\beta_0\alpha_3\beta_2)^2 = P_2(GG),$$
$$P_3(GA_1) = (\alpha_1\beta_0\alpha_3\beta_3)^2 = P_2(GA), \quad P_3(AC_1) = (\alpha_1\beta_1\alpha_2\beta_2)^2 = P_2(AC),$$
$$P_3(AT_1) = (\alpha_1\beta_1\alpha_2\beta_3)^2 = P_2(AT), \quad P_3(AG_1) = (\alpha_1\beta_1\alpha_3\beta_2)^2 = P_2(AG),$$
$$P_3(AA_1) = (\alpha_1\beta_1\alpha_3\beta_3)^2 = P_2(AA). \qquad (31)$$

From the expression (26), calculations of collective probabilities $P_3(CC_2)$ of all 4 triplets CCC, CCT, CCG and CCG with the doublet CC in their end, give the following:

$$P_3(CC_2) = (\alpha_0\beta_0\alpha_2\beta_2\alpha_4\beta_4)^2+(\alpha_0\beta_1\alpha_2\beta_2\alpha_4\beta_4)^2+(\alpha_1\beta_0\alpha_2\beta_2\alpha_4\beta_4)^2+(\alpha_1\beta_1\alpha_2\beta_2\alpha_4\beta_4)^2 =$$
$$= (\alpha_2\beta_2\alpha_4\beta_4)^2 *\{(\alpha_0\beta_0)^2+(\alpha_0\beta_1)^2+(\alpha_1\beta_0)^2+(\alpha_1\beta_1)^2\} = (\alpha_2\beta_2\alpha_4\beta_4)^2 = P_2(CC) \qquad (32)$$

Here the sum in curly brackets is equal to 1 in accordance with the normalization condition (12).

From the expression (26), similar calculations of collective probabilities $P_3(CT_2)$, $P_3(CG_2)$, etc. of other 15 sets of 4 triplets with identical doublets in their end give analogical results (33) of their equality to individual probabilities of appropriate doublets:

$$P_3(CT_2) = (\alpha_2\beta_2\alpha_4\beta_5)^2 = P_2(CT), \quad P_3(CG_2) = (\alpha_2\beta_2\alpha_5\beta_4)^2 = P_2(CG),$$
$$P_3(CA_2) = (\alpha_2\beta_2\alpha_5\beta_5)^2 = P_2(CA), \quad P_3(TC_2) = (\alpha_2\beta_3\alpha_4\beta_4)^2 = P_2(TC),$$
$$P_3(TT_2) = (\alpha_2\beta_3\alpha_4\beta_5)^2 = P_2(TT), \quad P_3(TG_2) = (\alpha_2\beta_3\alpha_5\beta_4)^2 = P_2(TG),$$
$$P_3(TA_2) = (\alpha_2\beta_3\alpha_5\beta_5)^2 = P_2(TA), \quad P_3(GC_2) = (\alpha_3\beta_2\alpha_4\beta_4)^2 = P_2(GC),$$
$$P_3(GT_2) = (\alpha_3\beta_2\alpha_4\beta_5)^2 = P_2(GT), \quad P_3(GG_2) = (\alpha_3\beta_2\alpha_5\beta_4)^2 = P_2(GG),$$
$$P_3(GA_2) = (\alpha_3\beta_2\alpha_5\beta_5)^2 = P_2(GA), \quad P_3(AC_2) = (\alpha_3\beta_3\alpha_4\beta_4)^2 = P_2(AC),$$
$$P_3(AT_2) = (\alpha_3\beta_3\alpha_4\beta_5)^2 = P_2(AT), \quad P_3(AG_2) = (\alpha_3\beta_3\alpha_5\beta_4)^2 = P_2(AG),$$
$$P_3(AA_2) = (\alpha_3\beta_3\alpha_5\beta_5)^2 = P_2(AA). \qquad (33)$$

One can consider these model results as predictions of a possible existence of additional rules of symmetries of collective probabilities of the following sets of n-plets in long DNA-texts in cases of the mentioned $4^s$-groups of oligonucleotides (s = 2, 3, 4, ...), for example:

- the CC-subgroup, CT-subgroup, CG-subgroup and CA-subroup of triplets;
- the CCC-subgroup, CCT-subgroup, CCG-subgroup, CCA-subgroup, CTC-subgroup, CTT-subgroup and other similar subgroups of 4-plets;
- the CCCC-subgroup, CCCT-subgroup, CCCG-subgroup, CCCA-subgroup and other similar subgroups of 5-plets;
- etc.

The Appendix 7 shows some initial confirmations of these theoretical predictions made on the basis of the described quantum-information model. Of course, these predictions should be checked for much more number of long DNA-texts including genomes and complete sets of chromosomes of different organisms. We hope to publish results of such checking some later.

## 6. About short DNA-texts

The last two Sections have shown a possibility and effectiveness of usage of quantum informatics to model properties of long DNA-texts by means of the formalisms of separable pure states of 2n-qubit quantum CTGA-systems. Now let us pay attention to a possible meaning for genetic systems other types of states of 2n-qubit quantum CTGA-systems. One can see from expression (10), which describes a separable pure state of 2-qubit system, that its amplitudes $\alpha_0\beta_0$, $\alpha_0\beta_1$, $\alpha_1\beta_0$, $\alpha_1\beta_1$ are closely interrelated with each other. Tetra-group symmetries in long DNA-texts say that these amplitudes in the expressions (10, 13) are interrelated in an appropriate manner. If one of these amplitudes is changed and the corresponding interrelation disappears, then the changed expression (10) will correspond to non-separable pure state.

In a general case, the tetra-group rules are not met for short DNA-texts (for example, for separate genes), and short DNA-texts cannot be represented in a form of separable quantum states $|\Psi> = |\Psi_1> \otimes |\Psi_2>$. In quantum informatics, a state of a composite system, which can't be written as a tensor product of states of its component systems, is called an entangled state [Nielsen, Chuang, 2010, p. 96; https://en.wikipedia.org/wiki/Separable_state ]. Correspondingly, genes and other relative short DNA-texts can be considered as quantum CTGA-systems in entangled states. In quantum informatics, entangled states play very important roles. The study and use of entangled states is one of the main problems of quantum computing: "*...entanglement is a key element in effects such as quantum teleportation, fast quantum algorithms, and quantum error-correstion. It is, in short, a resource of great utility in quantum computation and quantum information. … entangled states play a crucial role in quantum computation and quantum information*" [Nielsen, Chuang, 2010, p. XXIII and p. 96].

The quantum operator, which allows you to entangle two qubits $|x>$ and $|y>$, is called the operator CNOT and is given by the expression (36):

$$P_{12}|x, y> = |x, x \oplus y> \qquad (36)$$

where $x \oplus y$ is the logic operation of modulo-2 addition. Above in the Section 3 we shown that the DNA-alphabet of nucleotides C, T, G, A is connected with the logic operation of modulo-2 addition: any DNA-text is the carrier of three parallel messages on three different binary languages, and these three types of binary representations of the DNA-text form a common logic set on the basis of logic operation of modulo-2 addition. This connection of DNA-texts with the logic operation of modulo-2 addition can be used to represented fragments of long DNA-texts as entangled states of quantum CTGA-systems for study hidden requliarities and meanings of genetic messages. Beside this, the connection of DNA-structures with the logic operation of modulo-2 addition testifies additionally in favor of adequacy of the quantum-information approach to analyze DNA-texts. One can add that a possible important meaning of entangled states for functioning biological ensembles of protein molecules was discussed in [Matsuno, Paton, 2000].

One should note here also that, as known, Hadamard matrices («Hadamard gates») play essential role in quantum computers; they are used in in quantum mechanics in the form of unitary operators, etc. But structures of genetic alphabets are naturally connected with Hadamard matrices [Petoukhov, 2008, 2010, 2011, 2016, 2017a; Petoukhov, He, 2009]. It additionally testifies in favor of quantum-informational basises of genetic informatics.

In quantum mechanics, matrix operators play a significant role. One of them is the density operator (or the density matrix) that describes quantum systems in pure and mixed states. All the postulates of quantum mechanics can be reformulated in terms of the density operator language [Nielsen, Chuang, 2010, p. 99]. Quantum informatics possesses a rich set of other useful notions and mathematical formalisms including quantum search algorithms, quantum algorithms for encoding information with error-corrections, the quantum Fourier transform, the Schmidt decomposition, quantum logic, trace distances, stabilizer codes, von Neumann entropy, quantum circuits, etc. In our opinion, apllications of these notions and formalisms in genetic informatics will be very useful for progress of theoretical biology, medicine, systems of artificial intelligence and other fields including the quantum computer science itself. In particular, it will lead to development of quantum-informational genetics as a perspective scientific direction.

## 7. Resonances, photons and quantum-information genetics

From the point of view of quantum mechanics, the interaction of molecules is based on the emission and absorption of photons with the participation of resonance correspondences.

In modern molecular biology, when considering the interaction of molecules, the idea of the stereochemical correspondence of molecules is usually used, that is, the idea of the correspondence of the spatial configurations of molecules (the model of the "lock-and-key" correspondence). In our quantum-informational approach, quite another idea is in the centre of attention: the idea of exchanges of information among energetic states of interrelated molecules by means of photons on principles of resonances. In our approach, nucleotides and their combinations are represented by not their spatial configurations but, first of all, by their energetic peculiarities, which provide opportunities to exchange intermolecular information due to the emission and absorption of photons of certain energies (wavelengths) with the participation of resonance correspondences.

This Section is devoted to additional explanations of this quantum-informational approach, which is connected with the fundamental notion "resonance" and with photons as carriers of intermolecular information and also as the force carriers for electromagnetic force. Below we represent arguments why one can think about a connection of tetra-group symmetries in long DNA-texts with phenomena of resonances.

Quantum mechanics is closely connected with phenomena of resonances. This science has begun from 1900 year with the pioneer work of M.Planck (1936), who has analyzed a great set of resonant oscillators inside the special cavity to receive his famous law of electromagnetic radiation emitted by a black body in thermal equilibrium. One can say that Planck has represented the matter as a set of vibrating oscillators and set the task to study the equilibrium, which was established in the result of the exchange of energy between the oscillators and radiation. This work has introduced into science the Planck's constant as the quantum of action, central in quantum mechanics. It was originally the proportionality constant between the minimal increment of energy of a hypothetical electrically charged oscillator in the mentioned cavity and the frequency of its associated electromagnetic wave.

Later, after more than 50 years of successful development of quantum mechanics, E. Schrodinger emphasised the basic meaning of resonances: "*The one thing which one has to accept and which is the inalienable consequence of the wave-equation as it is used in every problem, under the most various forms, is this: that the interaction between two microscopic physical systems is controlled by a peculiar law of resonance*» [Schrodinger, 1952, p. 115]. In considering an exact balance in nature between bundles of energy, lost by one system and gained by another, he noted: «*I maintain that it can in all cases be understood as a resonance phenomenon*» (ibid, p. 114). He wrote in his resonance concept of quantum interactions that chemical reactions, including photochemical reactions, can be explained on the base of resonances. One of examples considered in his article was a production of water molecules $H_2O$ from a suitable mixture of hydrogen gas $H_2$ and oxygen gas $O_2$ under action of ultraviolet light. In this example, "*wave-mechanically the gaseous mixture is represented by a vibration of the combined system, and, by the way, not by one proper vibration since there is anyhow the vast variety of translational and rotational modes, and, of course, the electronic modes. The gaseous compound, $H_2O$, is represented by an entirely different vibration of the same system*" (ibid, p. 118).

His book [Schrodinger, 1944] declared that the chromosome is an aperiodic crystal since its atoms are connected each with other by forces of the same nature that atoms in crystals. This standpoint attracts an attention to a very important role of vibrations and resonances in crystals. The presence of the interaction of atoms in the crystal lattice together with the resonance phenomenon leads to the fact that oscillatory motions of lattice elements are combined in a collective oscillation process in a form of a wave propagating in the crystal. In the course of the normal vibrations, all the atoms in the crystal lattice oscillate about their equilibrium positions by harmonic law with the same frequency. As it is known, in a quantum description of small oscillations of a crystal, it is possible to interpreted normal fluctuations of the crystal as special quasiparticles, which are quanta of the field of elastic vibrations of the crystal and which are called phonons. The theory of phonons is one of the bases of physics of crystalls. One can hope that a similar resonance approach can usefully serve in genetics.

The notion "resonance" was introduced into quantum mechanics by W. Heisenberg in 1926 year in connection with analyzes of multi-body systems. He emphasized that in quantum mechanics the phenomenon of resonances has much more general character than in classical physics. In classic theory, two periodic oscillating systems come into their own resonance only in the case when a frequency of a separate sub-system doesn't depend on energy of the system and when this frequency is approximately equal in both sub-systems. In quantum mechanics, two atomic systems come into their resonance only in the case when a frequency of absorption of one system coincides with a frequency of emitting another system, or vice versa [Heisenberg, 1926, §2]. Quantized electromagnetic field is represented as a set of oscillators.

L.Pauling used ideas of resonances in quantum mechanical systems in his theory of resonance in structural chemistry. His book [Pauling, 1940] about this theory is the most quoted among scientific books of the XX century. The theory was developed to explain the formation of hybrid bonds in molecules. The actual molecule, as Pauling proposed, is a sort of hybrid, a structure that resonates between the two alternative extremes; and whenever there is a resonance between the two forms, the structure is stabilized. His theory uses the fundamental principle of a minimal energy because – in resonant combining of parts into a single unit – each of members of the ensemble requires less energy for performing own work than when working individually. Of course, this fundamental principle can be used in many other cases of resonances in different systems as the physical base. The principle of energetic minimum in resonance processes has some correlations with the principle of relaxation in morphogenetic processes proposed in [Igamberdiev, 2012].

Concerning problems of bioenergetics, McClare (1974) put forward the hypothesis about long-lived vibrational excitations in proteins, which could play an important role in protein functions. The results of researches on this hypothesis was summarized in the work [Turin, 2009]: "*Colin McClare was the first scientist to envision the potential importance of long-lived vibrational excitations in proteins. He did so at a time when there was no experimental evidence for their existence, no plausible mechanism to make them happen, and no interest in the possibility among his peers… Thirty years later (after his death in 1977 – S.P.), we can see that McClare had got one fundamental thing right: vibrational energy can be stored and transmitted in proteins*». This hypothesis of McClare is associated with the idea of Bauer (1935) that living systems work in expense of non-equilibrium, and the external energy is used not directly to perform work but to support the stable non-equilibrium state; most of this energy is transformed into the kinetic energy.

For a theme of a general importance of resonances in living matter, interesting materials are represented in works [Ji, 2012, 2015, 2017]. Their author has postulated analogy between enzymic catalysis and blackbody radiation, which was modelled by Planck due to his idea about huge number of resonances. This postulation was made by Ji on the base, first of all, of his observations that some important biological phenomena have their graphical representations analogical to ones of blackbody radiations. This author has proposed a generalization of the Planck equation for modeling different biological phenomena, having long-tailed histograms, from the general standpoint: protein folding, single-molecule enzymology, whole-cell transcriptomics, T-cell receptor variable region diversity, brain neurophysiology, RNA levels in budding yeast, human breast cancer tissues, codon profile in the human genome, etc. By analogy with the principle of quantization of energy in quantum mechanics, Ji postulates a quantization of free energy levels in enzymes. He also proposes an original theory of molecular machines with using Franck-Condon mechanisms concerning vibronic transitions, which are simultaneous changes in electronic and vibrational energy levels of a molecule.

His thoughts about musical-wave harmony in organization of living matter are correlated with musical aspects of our concept of systemic-resonance genetics, where an organism is considered as a very complex and developing music synthesizer and some kinds of so called «genetic music» are developed and represented in musical concerts of Moscow P.I.Tchaikovsky Conservatory [Hu, Petoukhov, 2017; Petoukhov, 2015b].

The articles [Petoukhov, 2015a, 2016; Petoukhov, Petukhova, 2017b] describe our concept about the important role of resonances in genetic structures. This concept is based on impressive analogies of some genetic structures and phenomena, including Mendelian laws, with eigenvalues and eigenvectors of tensor families of matrix operators representing oscillatory systems with many degrees of freedom (these operators can be interpreted as quantum operators). In these works we described data to the idea about existence of biological resonant computers inside any organism, which are based on binary-oppositional resonances in genetic systems.

Let us repeate that the interaction of molecules is based on the emission and absorption of photons (particles of light) with the participation of resonance correspondences. Therefore, special attention should be paid to the important role of photons in genetic informatics.

As known, a photon is a type of elementary particle, the quantum of electromagnetic field including electromagnetic radiation such as light, and the force carrier for electromagnetic field (in particle physics, force carriers or messenger particles or intermediate particles are particles that give rise to forces between other particles (https://en.wikipedia.org/wiki/Force_carrier). The photon has zero rest mass and always moves at the speed of light within a vacuum. A photon has two possible polarization states. Like all elementary particles, photons are currently best explained by quantum mechanics and exhibit wave-particle duality, exhibiting properties of both waves and particles. The photon's wave and quanta qualities are two observable aspects of a single phenomenon, and cannot be described by any mechanical model; a representation of this dual property of light, which assumes certain points on the wavefront to be the seat of the energy, is not possible. The quanta in a light wave cannot be spatially localized (https://en.wikipedia.org/wiki/Photon).

Photon energy is the energy carried by a single photon. The amount of energy is directly proportional to the photon's electromagnetic frequency and inversely proportional to the wavelength. Photon energy is solely a function of the photon's wavelength. Other factors, such as the intensity of the radiation, do not affect photon energy. In other words, two photons of light with the same «color» and therefore, same frequency, will have the same photon energy. The equation for photon energy E is the following: $E = hc/\lambda = hf$ , where «h» is the Planck constant, «c» is the speed of light in vacuum, «λ» is the photon's wavelength and «f» is the photon's wavefrequency.

Photons, which are radiated by different molecular elements, can differ by their frequencies. In systems of genetic molecules, where each of nitrogenous bases C, T, G, A is a carrier of its individual pair of binary-oppositional molecular indicators (Fig. 14), each of these two molecular indicators can define a radiation of photons with their individual frequencies. Jointly the two molecular indicators can radiate a beam of photons with two kinds of frequencies, which defines its appropriate computational basis state of a 2-qubit system in the expression (10) with the conditional denotations |C>=|00>, |T>=|01>, |G>=|10> and |A>=|11>. The existence of DNA-texts is accompanied by a rich set of appropriate beams of multicolor photons to provide a cooperative informational functioning of ensembles of genetic elements. From this point of view, nitrogenous bases A, C, G, T/U and A and their combinations in DNA and RNA are resonance determinant of frequencies of biophotonic ensembles within living bodies (or briefly, "biophotons determinant"). The reading and transmission of genetic information from DNA and RNA molecules occurs by means of a set of resonance frequencies of biological photons. DNA encodes quantum states of biophoton beams. The photons language is a serious candidacy for the role of a basic language of molecular-genetic information. Figuratively speaking, from this point of view, life in its information aspects is woven from the light.

Photons are actively studied in modern science as elements of quantum computers and devices of quantum cryptography. In models of quantum computers, conventional light polarizers are used to create pure and mixed states of n-qubit systems of light beams. The idea of ensembles of «multicolor» photons for a creation of n-qubit states, which was noted by us above in the connection with DNA-texts, was independently used in the recent engineering work of Canadian scientists [Caspani et al., 2016]. This work has revealed a new perspective way to create quantum computers. For increasing dimensionality of the photon quantum state, its authors used the ability to generate multiple photon pairs on a frequency comb, correpsponding to resonances in specifically designed microcavities. The posibility of fully exploiting the polarization degree of freedom, even for integrated devices, exists for further achievements.

One should note that, as known, living bodies posses inherited opportunities to manage photonic beams using physical principles of photonic crystalls with their properties of photon gratings, etc. Many biological phenomena of structural coloration and of animal reflectors are built on this, including a beautiful coloring of butterfly wings, peacock feathers, etc. (see details and lists of references in https://en.wikipedia.org/wiki/Photonic_crystal, https://en.wikipedia.org/wiki/Animal_reflectors, https://en.wikipedia.org/wiki/Structural_coloration). A photonic crystal is a periodic optical nanostructure that affects the motion of photons. Photonic crystals contain regularly repeating regions of high and low dielectric constant. Photons (behaving as waves) either propagate through this structure or not, depending on their wavelength. This gives rise to distinct optical phenomena, such as inhibition of spontaneous emission, high-reflecting omni-directional mirrors, and low-loss-waveguiding. The periodicity of the photonic crystal structure must be around half the wavelength of the electromagnetic waves to be diffracted.

We believe that spatial characteristics of ensembles of genetic and other biological molecules, that form complex diffraction structures, play the managing role of photonic crystals in the problem of controlling biophoton beams that are generated and absorbed by these molecules. In particular, the spatial configuration of genetic molecules as biophotonic crystals is an important factor in controlling the processes of transmission of genetic information from DNA and RNA molecules with using biophoton beams generated by them. Not without reason, Schrödinger called chromosomes aperiodic crystals [Schrodinger, 1944].

In our opinion, the inherited processes of the morphogenesis of living bodies are also determined to a large extent by bio-photon beams, the course of which is not accidental, but is strictly organized by a system of spatial characteristics of ensembles of genetic and other biological molecules as biophotonic crystalls. In the course of ontogeny, on the basis of electromagnetic (photonic) interactions, new molecular materials are involved into a naturally growing biological body, which leads to the appropriate growth of the managing system of biophotonic crystals and to the growth of numbers of photon beams. Of course, quantum-mechanic laws of resonances in molecular photonic interactions play a basic role. On this basis, we develop our concept of the "morpho-resonance field" to model morphogenetic phenomena [Petoukhov, 2015c,d, 2016].

Classical electrodynamics describes a photon as an electromagnetic wave with its circular right or left polarization. To the theme "life and photons", one can add many interesting connections of these polarization properties of photons with inherited properties of living bodies, for example, the following:

- One of the biggest mysteries of nature is the asymmetry of biological molecules, accompanied by a preferred direction of the rotation - to the left or to the right - of the polarization plane of light by these molecules (this was discovered by Louis Pasteur). For example, all biological amino acids (except the simplest glycine, which is symmetric), from which the proteins of all living organisms are composed, exist only in one of two possible asymmetric forms - in the left form. Amino acids in this form rotate the plane of polarization of light to the left. Our body doesn't use amino acids with the opposite right form, rotating the plane of polarization of light to the right. Biological catalysts - enzymes -, being built asymmetrically, act only on one optical antipode, without touching the another. The same asymmetry with respect to the right and left is inherent not only in amino acids, but also in the nucleotides that form DNA and RNA. The reason for this is the asymmetry of the components of the sugar, which is part of the nucleotides and which provides the optical activity of DNA and RNA molecules: they rotate the plane of polarization of light to the right.

- Millions of species of living organisms (insects, mollusks, arthropods, etc.) are endowed with inherited ability to see in polarized light (a human organism does not possess this ability).

At the end of the Section, one can say that our quantum-informational approach for modeling and analysis of the genetic system provides new materials for understanding the role of photons in the organization and functioning of the genetic system.

## 8. The rules of symmetries of collective probabilities in tetra-groups in the complete set of human chromosomes. The fourth rule of symmetries of tetra-group probabilities (for complete sets of chromosomes)

This Section is devoted to results of author's study of tetra-group symmetries in the complete set of human chromosomes. Fig. A2/1-A2/24 in the Appendix 2 show resulting tables of tetra-group symmetries of all 24 human chromosomes, initial data about which were taken from the GenBank. Let us note some important aspects of the obtained results.

Human organisms contain 22 autosomes and 2 sex chromosomes X and Y. These chromosomes contain long DNA molecules, the lengths of texts in which lie in the range from 50 to 250 million letters approximately. Autosomes are numbered from 1 to 22.

First of all, our results demonstrate very high accuracy of implementation of all the rules of tetra-group symmetries for each of human chromosomes. Fig. 18 shows the average values of probabilities $P_n(A_k)$, $P_n(T_k)$, $P_n(C_k)$ and $P_n(G_k)$, expressed in percentage, for each of the 24 chromosomes ($n = 1, 2, 3, 4, 5$; $k \leq n$) where the numerical values of probabilities in percentage are rounded to the third decimal place. Fluctuations of the probabilities around these averages are also shown. One should note that these fluctuations are very small.

| CHROMOSOME | AVERAGE VALUE OF $P_n(A_k)$ AND FLUCTUATIONS (%) | AVERAGE VALUE OF $P_n(T_k)$ AND FLUCTUATIONS (%) | AVERAGE VALUE OF $P_n(C_k)$ AND FLUCTUATIONS (%) | AVERAGE VALUE OF $P_n(G_k)$ AND FLUCTUATIONS (%) |
|---|---|---|---|---|
| 1 | 29,100±0,005 | 29,176±0,006 | 20,850±0,006 | 20,874±0,005 |
| 2 | 29,845±0,005 | 29,927±0,006 | 20,087±0,006 | 20,142±0,006 |
| 3 | 30,131±0,006 | 30,204±0,005 | 19,805±0,006 | 19,861±0,006 |
| 4 | 30,862±0,005 | 30,895±0,006 | 19,097±0,006 | 19,147±0,006 |
| 5 | 30,176±0,006 | 30,317±0,006 | 19,712±0,006 | 19,794±0,006 |
| 6 | 30,210±0,006 | 30,205±0,006 | 19,794±0,006 | 19,791±0,006 |
| 7 | 29,602±0,006 | 29,701±0,006 | 20,330±0,006 | 20,368±0,006 |
| 8 | 29,939±0,007 | 29,898±0,007 | 20,081±0,007 | 20,081±0,006 |
| 9 | 29,280±0,007 | 29,263±0,007 | 20,736±0,007 | 20,721±0,007 |
| 10 | 29,172±0,007 | 29,286±0,007 | 20,741±0,007 | 20,801±0,007 |
| 11 | 29,202±0,007 | 29,258±0,007 | 20,741±0,007 | 20,799±0,007 |
| 12 | 29,571±0,007 | 29,663±0,007 | 20,349±0,007 | 20,417±0,007 |
| 13 | 30,688±0,007 | 30,788±0,006 | 19,259±0,007 | 19,265±0,007 |
| 14 | 29,452±0,006 | 29,697±0,006 | 20,397±0,007 | 20,453±0,007 |
| 15 | 28,956±0,007 | 29,009±0,006 | 20,974±0,006 | 21,061±0,007 |
| 16 | 27,575±0,007 | 27,840±0,007 | 22,214±0,007 | 22,370±0,007 |
| 17 | 27,303±0,006 | 27,382±0,006 | 22,581±0,007 | 22,734±0,007 |
| 18 | 30,106±0,006 | 30,136±0,007 | 19,866±0,006 | 19,893±0,007 |
| 19 | 25,911±0,006 | 26,151±0,006 | 23,878±0,007 | 24,060±0,007 |
| 20 | 27,780±0,006 | 28,097±0,006 | 22,009±0,007 | 22,113±0,007 |
| 21 | 29,644±0,006 | 29,463±0,007 | 20,465±0,006 | 20,428±0,006 |
| 22 | 26,512±0,007 | 26,483±0,007 | 23,393±0,006 | 23,611±0,007 |
| X | 30,185±0,005 | 30,290±0,005 | 19,706±0,005 | 19,819±0,006 |
| Y | 29,855±0,005 | 30,120±0,005 | 20,011±0,006 | 20,015±0,005 |

Fig. 18. The table of average values of probabilities $P_n(A_k)$, $P_n(T_k)$, $P_n(C_k)$ and $P_n(G_k)$ in percentage for each of 24 human chromosomes (n = 1, 2, 3, 4, 5; k ≤ n). Fluctuations of the probabilities are also shown.

Here let us explain how these average values and fluctuations in relation to them are calculated, for example, in the case of the average value of $P_n(A_k)$ for the human chromosome № 1. In the Appendix 2, each of tables of tetra-group symmetries (Fig. A2/1-A2/24) contains 15 cells with $P_n(A_k)$. When each of these probabilities $P_n(A_k)$ is calculated up to the seventh decimal place for this DNA-text, the following 15 values arise: $P_1(A_1)$=0,2910013, $P_2(A_1)$=0,2910218, $P_2(A_2)$=0,2909808, $P_3(A_1)$=0,2910489, $P_3(A_2)$=0,2909651, $P_3(A_3)$=0,2909899, $P_4(A_1)$=0,2910234, $P_4(A_2)$=0,2909995, $P_4(A_3)$=0,2910202, $P_4(A_4)$=0,2909621, $P_5(A_1)$=0,2909899, $P_5(A_2)$= 0,2909896, $P_5(A_3)$= 0,2910202, $P_5(A_4)$= 0,2909677, $P_5(A_5)$= 0,2910397. The sum of these 15 numbers divided by 15 gives their average value 0,2910013. The probability $P_3(A_1)$=0,2910489 has the strongest deviation from this average value 0,29100134. This deviation is equal to 0,0000476 ≈ 0,00005. Thus, in percentage denotation, we get for the probabilities $P_n(A_k)$ their average value 29,100% and the deviation (fluctuation) 0,005% from it. The corresponding expression (29,100±0,005)% is shown in the table in Fig. 18 for the chromosome № 1. By analogy, values in all other cells of the table in Fig. 18 are calculated.

One can see from the table in Fig. 18 that average values of $P_n(A_k)$ and $P_n(T_k)$ are almost equal; the average values of $P_n(C_k)$ and $P_n(G_k)$ are also almost equal. Fig. 19 shows graphically average values of these probabilities.

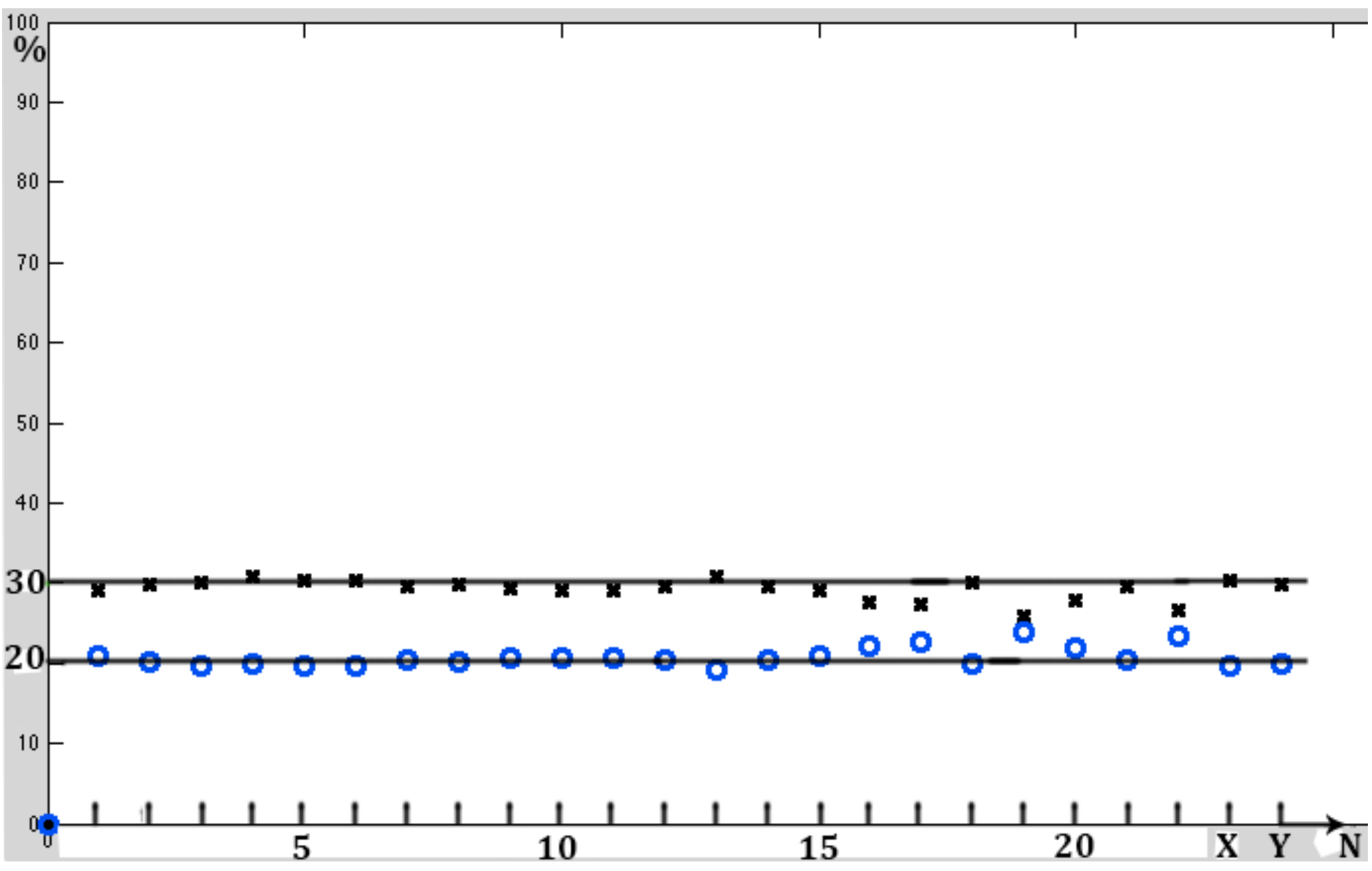

Fig. 19. The graphical representation of the average values of the probabilities $P_n(A_k)$, $P_n(T_k)$, $P_n(C_k)$ and $P_n(G_k)$ from Fig. 18 for all 24 human chromosomes. The abscissa axis contains numberings N of chromosomes, and the ordinate axis contains average values of these probabilities in percent. The symbol "o" corresponds the average values of $P_n(C_k) \approx P_n(G_k)$, and the symbol "x"

corresponds the average values of $P_n(A_k) \approx P_n(T_k)$. Direct lines correspond values 20% and 30%.

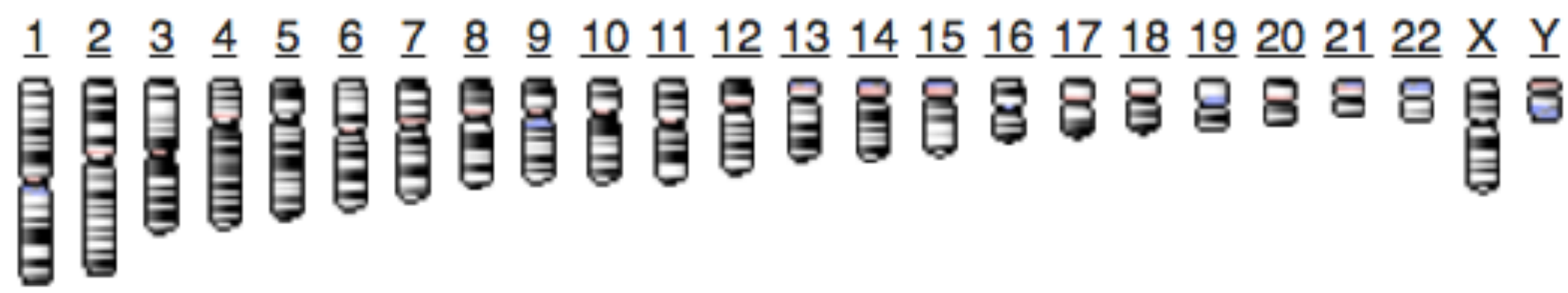

Fig. 20. Human chromosomes (https://www.ncbi.nlm.nih.gov/genome/51 )

These 24 human chromosomes differ greatly in their molecular dimensions, their sequences of letters, kinds and quantities of genes in them, cytogenetic bands (which shows biochemical specifity of different parts of chromosomes), etc. (Fig. 20). Taking into account these great differences among human chromosomes, it was very unexpected for the author to reveal that the 24 human chromosomes are similar each to other from the point of view of tetra-group symmetries of their DNA-texts. It means that all these chromosomes are not completely individual objects but they are closely interrelated each other in relation to tetra-group symmetries of their DNA-texts. As can be seen from the tables of tetra-group symmetries in Fig. 18, 19 and in the Appendix 2 (Fig. A2/1-A2/24), the maximum similarity in average values of collective probabilities of subgroups of the tetra-groups exists for DNA-texts of chromosomes №№ 1-18, 20, 21, X and Y (that is for 22 chromosomes from 24 chromosomes). Only two chromosomes № 19 and №22 have their DNA-texts with noticeable deviations of these probabilities in their average values in 15-20% in comparison with the average values of the probabilities in DNA-texts of other 22 chromosomes.

It seems that the average values of the probabilities of $P_n(A_k) \approx P_n(T_k)$ are concentrated around the value of 30% and the average values of the probabilities of $P_n(C_k) \approx P_n(G_k)$ are concentrated around the value of 20%. In theory of musical harmony, the ratio 30/20 = 3/2 is called “quint” (or "fifth").

It is obvious that such collective text phenomena in the complete set of human chromosomes cannot be explained on the basis of ideas of stereochemical interrelations among genetic molecules by the principle “lock-and-key”. Here another direction of thoughts is needed. We think that quantum information approach should be used to explain such collective informational phenomena.

In the next Section we represent results of our study of tetra-group symmetries in complete sets of chromosomes of some model organisms, which are traditionally used in the study of genetics, development and disease. All these results show that the represented tetra-group rules are implemented not only for separate long DNA-texts but also for studied complete sets of chromosomes of eukaryotes. These initial results allow putting forward the hypothesis about the validity of the **fourth rule of tetra-group symmetries, which concerns complete sets of chromosomes of different organisms**:

- In the complete set of chromosomes of each of eukaryot organisms, characteristics of the tetra-group symmetries of separate chromosomes are approximately equal to each other for all chromosomes.

Further researches are needed to check a degree of universality of this rule.

## 9. Tetra-group rules in complete sets of chromosomes of model organisms: *Caenorhabditis elegans*, *Drosophila melanogaster*, *Arabidopsis thaliana*, *Mus musculus*

This Section is devoted to consideration of perculiarities of tetra-group symmetries in complete sets of chromosomes of a few model organisms, which are used long ago in the study of genetics, development and disease. Tables of tetra-group symmetries of these sets of chromosomes are represented in Appendixes 3-5. Data of these tables confirm the implementation of the tetra-group rules for chromosomes of these organisms and they show also that – for each of the organisms - characteristics of tetra-group symmetries of each of chromosomes are approximately equal for all chromosomes.

We begin with a nematode *Caenorhabditis elegans*. It is a free-living soil nematode with 959 somatic cells in its body. *Caenorhabditis elegans* was the first multicellular eukaryotic organism whose genome was completely sequenced. Its nuclear genome is approximately 100 Mb, distributed among six chromosomes (see tables in the Appendix 3). Fig. 21 and 22 shows the table and the corresponding graphical represenation of average values of probabilities $P_n(A_k)$, $P_n(T_k)$, $P_n(C_k)$ and $P_n(G_k)$ (all initial data were taken from the CenBank (https://www.ncbi.nlm.nih.gov/genome?term=caenorhabditis%20elegans). These average values were calculated from the tables in Fig. A3/1-A3/6 by analogy with the described calculation of average values of the probabilities in cases of human chromosomes (Fig. 18).

| CHROMOSOME | AVERAGE VALUE OF $P_n(A_k)$ (%) | AVERAGE VALUE OF $P_n(T_k)$ (%) | AVERAGE VALUE OF $P_n(C_k)$ (%) | AVERAGE VALUE OF $P_n(G_k)$ (%) |
|---|---|---|---|---|
| 1 | 32,088 | 32,167 | 17,886 | 17,863 |
| 2 | 31,927 | 31,871 | 18,123 | 18,078 |
| 3 | 32,246 | 32,093 | 17,769 | 17,892 |
| 4 | 32,647 | 32,760 | 17,347 | 17,247 |
| 5 | 32,263 | 32,309 | 17,741 | 17,689 |
| X | 32,435 | 32,361 | 17,607 | 17,596 |

Fig. 21. The table of average values of probabilities $P_n(A_k)$, $P_n(T_k)$, $P_n(C_k)$ and $P_n(G_k)$ in percentage for each of 6 chromosomes of a nematode *Caenorhabditis elegans* ($n = 1, 2, 3, 4, 5; k \leq n$).

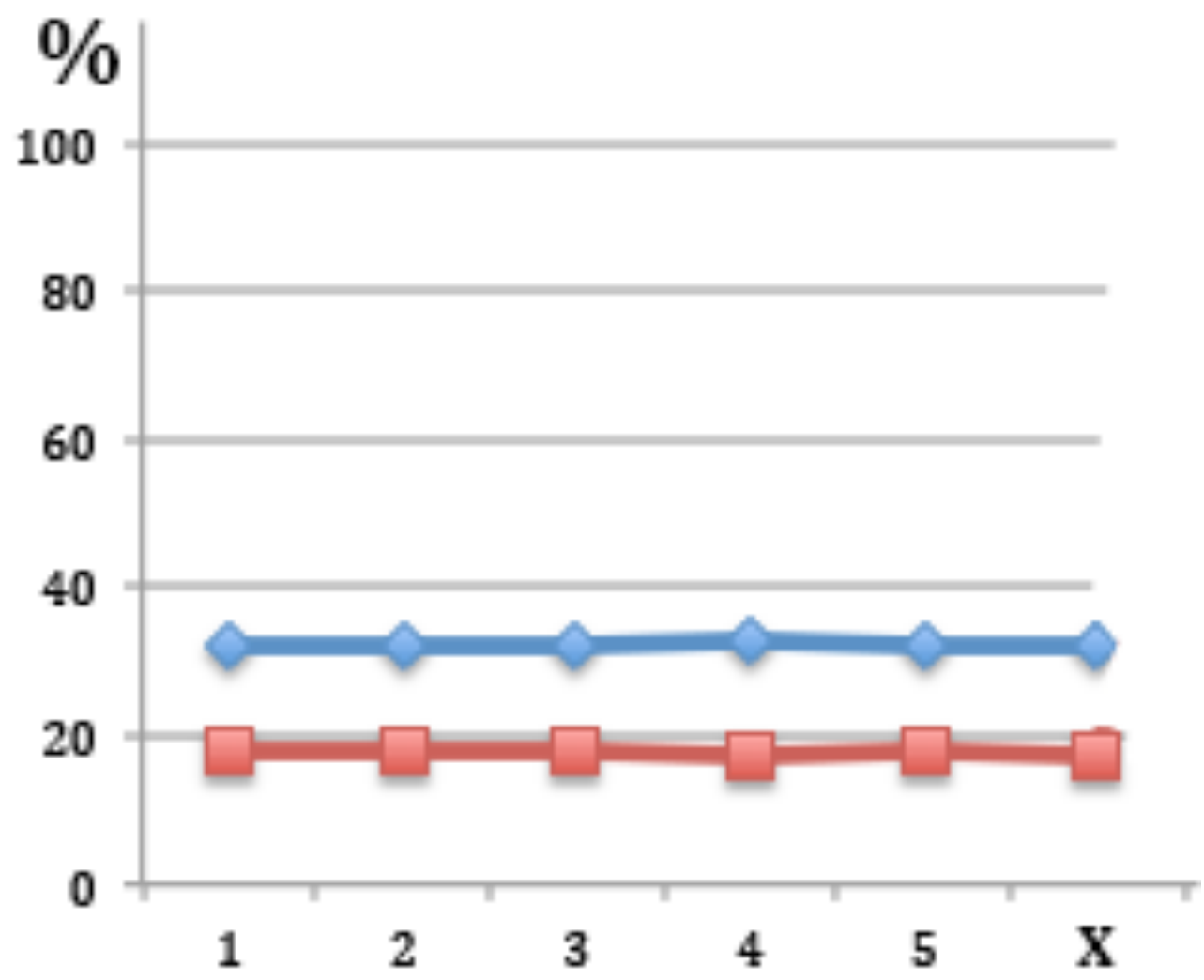

Fig. 22. The graphic representation of average values of probabilities $P_n(A_k)$, $P_n(T_k)$, $P_n(C_k)$ and $P_n(G_k)$ in percentage for each of 6 chromosomes of a nematode *Caenorhabditis elegans* ($n = 1, 2, 3, 4, 5$; $k \leq n$). The abscissa axis contains numberings of chromosomes, and the ordinate axis contains average values of these probabilities in percent. The symbol of a blue diamond corresponds the average values of $P_n(A_k) \approx P_n(T_k)$, and the symbol of a red square corresponds the average values of $P_n(C_k) \approx P_n(G_k)$.

Now let us turn to the second model organism - *Drosophila melanogaster*, which is studied in biology labs for over eighty years. Tables of tetra-group symmetries of its set of 7 chromosomes are represented in the Appendix 4. Fig. 23 and 24 shows the table and the corresponding graphical represenation of average values of probabilities $P_n(A_k)$, $P_n(T_k)$, $P_n(C_k)$ and $P_n(G_k)$ for all chromosomes of *Drosophila melanogaster* (all initial data about these chromosomes were taken from the CenBank - https://www.ncbi.nlm.nih.gov/genome/?term=drosophila+melanogaster). These average values were calculated from the tables in Fig. A4/1-A4/7 by analogy with the described calculation of average values of the probabilities in cases of human chromosomes (Fig. 18).

| CHROMOSOME | AVERAGE VALUE OF $P_n(A_k)$ (%) | AVERAGE VALUE OF $P_n(T_k)$ (%) | AVERAGE VALUE OF $P_n(C_k)$ (%) | AVERAGE VALUE OF $P_n(G_k)$ (%) |
|---|---|---|---|---|
| X | 28,679 | 28,858 | 21,195 | 21,267 |
| 2L | 29,145 | 29,074 | 20,889 | 20,892 |
| 2R | 28,769 | 28,621 | 21,342 | 21,269 |
| 3L | 29,092 | 29,287 | 20,813 | 20,807 |
| 3R | 28,719 | 28,693 | 21,320 | 21,269 |
| 4 | 31,945 | 32,805 | 17,471 | 17,779 |
| Y | 30,992 | 29,583 | 20,023 | 19,403 |

Fig. 23. The table of average values of probabilities $P_n(A_k)$, $P_n(T_k)$, $P_n(C_k)$ and $P_n(G_k)$ in percentage for each of 7 chromosomes of *Drosophila melanogaster* (n = 1, 2, 3, 4, 5; k ≤ n).

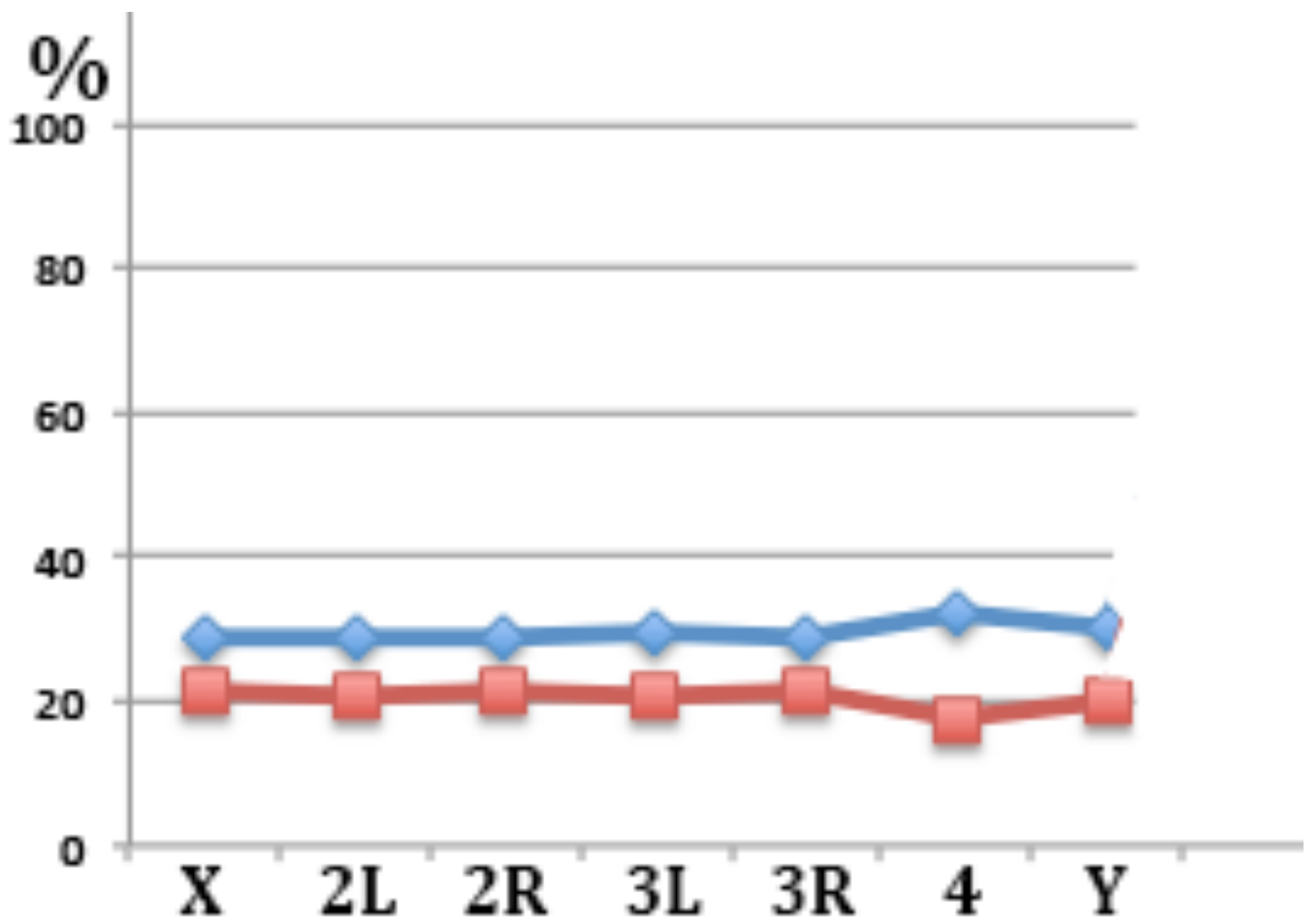

Fig. 24. The graphic representation of average values of probabilities $P_n(A_k)$ , $P_n(T_k)$, $P_n(C_k)$ and $P_n(G_k)$ in percentage for each of 7 chromosomes of *Drosophila melanogaster*. The abscissa axis contains numberings of chromosomes, and the ordinate axis contains average values of these probabilities in percent. The symbol of a blue diamond corresponds the average values of $P_n(A_k) \approx P_n(T_k)$, and the symbol of a red square corresponds the average values of $P_n(C_k) \approx P_n(G_k)$.

The third model organism is a plant *Arabidopsis thaliana*. This small flowering plant is used for over fifty years to study plant mutations and for classical genetic analysis. It became the first plant genome to be fully sequenced. Tables of tetra-group symmetries of its set of 5 chromosomes are represented in the Appendix 5. Fig. 25 and 26 shows the table and the corresponding graphical represenation of average values of probabilities $P_n(A_k)$, $P_n(T_k)$, $P_n(C_k)$ and $P_n(G_k)$ for all chromosomes of *Arabidopsis thaliana* (initial data about the chromosomes were taken from the CenBank - https://www.ncbi.nlm.nih.gov/genome/4, the column RefSeq). These average values were calculated from the tables in Fig. A5/1-A5/5 by analogy with the described calculation of average values of the probabilities in cases of human chromosomes (Fig. 18).

| CHROMOSOME | AVERAGE VALUE OF $P_n(A_k)$ (%) | AVERAGE VALUE OF $P_n(T_k)$ (%) | AVERAGE VALUE OF $P_n(C_k)$ (%) | AVERAGE VALUE OF $P_n(G_k)$ (%) |
|---|---|---|---|---|
| 1 | 32,083 | 32,043 | 17,960 | 17,913 |
| 2 | 32,066 | 32,070 | 17,989 | 17,876 |
| 3 | 31,913 | 31,756 | 18,156 | 18,175 |
| 4 | 31,969 | 31,827 | 18,144 | 18,060 |
| 5 | 31,976 | 32,088 | 17,920 | 18,018 |

Fig. 25. The table of average values of probabilities $P_n(A_k)$, $P_n(T_k)$, $P_n(C_k)$ and $P_n(G_k)$ in percentage for each of 5 chromosomes of *Arabidopsis thaliana* (n = 1, 2, 3, 4, 5; k ≤ n).

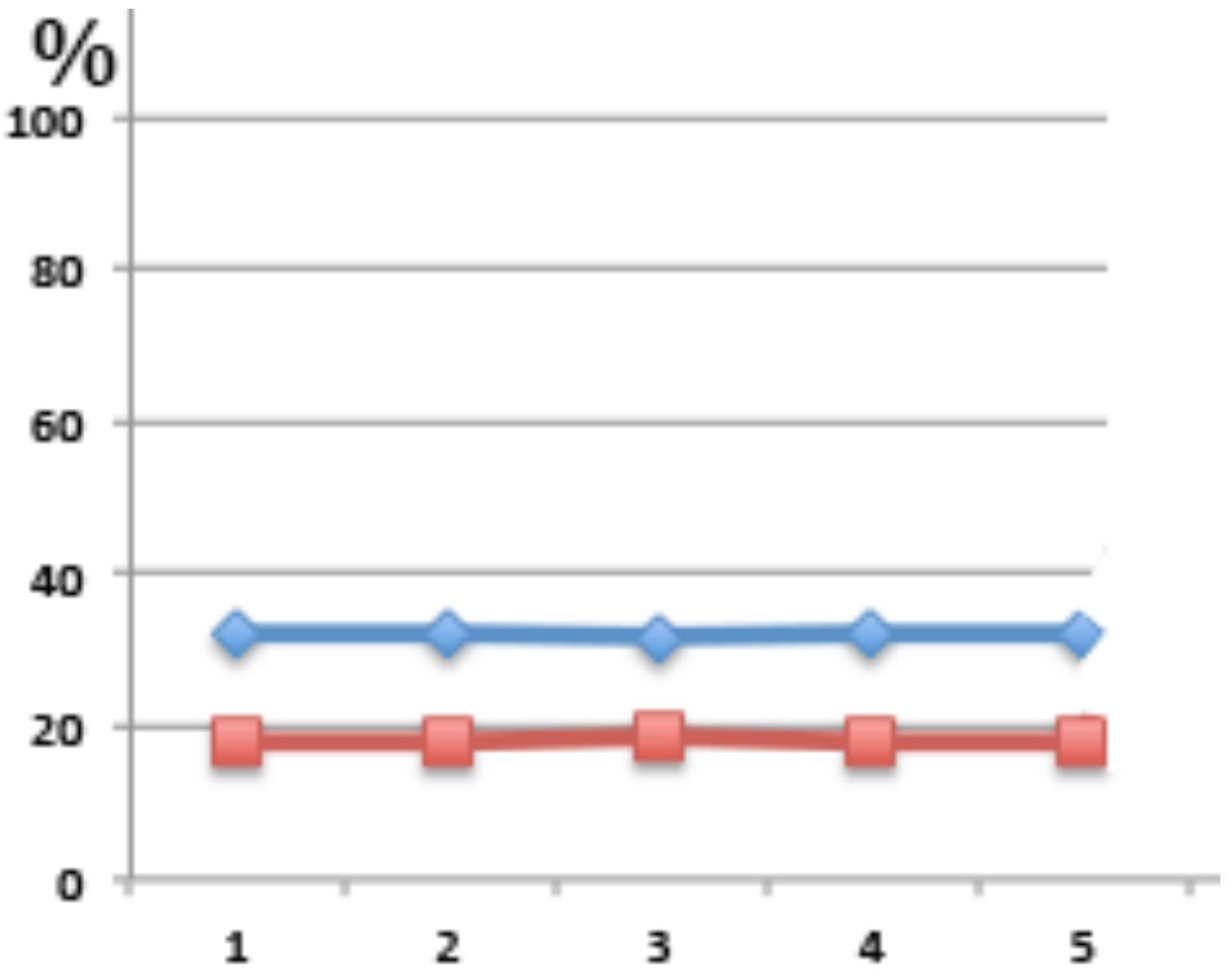

Fig. 26. The graphic representation of average values of probabilities $P_n(A_k)$ , $P_n(T_k)$, $P_n(C_k)$ and $P_n(G_k)$ in percentage for each of 7 chromosomes of *Arabidopsis thaliana*. The abscissa axis contains numberings of chromosomes, and the ordinate axis contains average values of these probabilities in percent. The symbol of a blue diamond corresponds the average values of $P_n(A_k) \approx P_n(T_k)$, and the symbol of a red square corresponds the average values of $P_n(C_k) \approx P_n(G_k)$.

One more analysed organism is the laboratory mouse, which is a major model organism for basic mammalian biology, human disease, and genome evolution. Data about the fulfillment of the described rules of tetra-group symmetries in the complete set of chromosomes of Mus musculus (house mouse) are represented in the Appendix 6, where Fig. A6/22 shows the table of fluctuations of collective probabilities for DNA-texts of all chromosomes and Fig. A6/23 shows shows the graphical representation of these collective probabilities for all chromosomes of Mus musculus.

It seems interesting in the future to study the characteristics of tetra-group symmetries for karyotypes of a variety of different organisms; these characteristics (approximate constants of karyotypes?) can be useful for comparative analysis of organisms, for example, in problems of biological evolution.

## 10. Fractal genetic nets and the fifth and sixth rules of symmetries of collective probabilities of tetra-groups. On a fractal grammar of long DNA-texts.

The tetra-group symmetries are connected with the fractal-like principle: in n-letter representations of a long DNA-text, the individual probabilities of tetra-group subgroups of n-plets of a lower order are repeated in the collective probabilities of the corresponding subgroups of n-plets of higher orders.

Another connection of long DNA-texts with fractals was described in the article [Petoukhov, Svirin, 2012] where the notion of "fractal genetic nets" (FGN) has been introduced. Each FGN of texts can contain different fractal genetic trees (FGT). In that article we have shown hidden regularities in long DNA-texts in a connection with Chargaff's thoughts about a "grammar of biology": our results testified in favor of existence of new symmetry principles in long nucleotide sequences in an addition to the known symmetry principle on the basis of the generalized Chargaff's second parity rule. Some new symmetry principles dealing with FGT and FGN have been formulated there.

Below we represent our results that show the implementation of the tetra-group symmetries in convoluted long DNA-texts at different levels of different kinds of fractal genetic nets (or trees). But initially let us remind about FGN.

In line with our article [Petoukhov, Svirin, 2012], FGT of various types are constructed by the method of sequential positional convolutions of a long DNA-text into a set of ever-shorter texts. Fig. 27 explains a construction of FGT of various types by an example of the FGT for a long DNA-text, which is represented as a sequence $S_0$ of 3-letter words (the sequence of triplets). In each triplet, 0, 1 and 2 numbers its three positions correspondingly. At the first level of the text convolution, an initial long sequence $S_0$ of triplets is transformed by means of a positional convolution into three new sequences of nucleotides $S_{1/0}$, $S_{1/1}$ and $S_{1/2}$, each of which is 3 times shorter in comparison with the initial sequence $S_0$ (in this notation of sequences, numerator of the index shows the level of the convolution, and the denominator - the position of the triplets, which is used for the convolution): the sequence $S_{1/0}$ includes one by one all the nucleotides that are in the initial position "0" of triplets of the original sequence $S_0$ ; the sequence $S_{1/1}$ includes one by one all the nucleotides that are in the middle position "1" of triplets of the original sequence $S_0$ ; the sequence $S_{1/2}$ includes one by one all the nucleotides that are in the last position "2" of triplets of the original sequence $S_0$ . At the final stage of the first level of the positional convolution, each of the sequences of nucleotides $S_{1/0}$ , $S_{1/1}$ , $S_{1/2}$ is represented as a sequence of triplets, where three positions inside each of triplets are numbered again by 0, 1 and 2. To construct the second level of the convolution, each of the sequences $S_{1/0}$ , $S_{1/1}$ , $S_{1/2}$ is transformed by means of the same positional convolution into three new sequences: $S_{1/0}$ is convolved into $S_{2/00}$ , $S_{2/01}$ , $S_{2/02}$; $S_{1/1}$ – into $S_{2/10}$, $S_{2/11}$, $S_{2/12}$; $S_{1/2}$ – into $S_{2/20}$, $S_{2/21}$, $S_{2/22}$ . Similarly, the third level and subsequent levels of the convolution are constructed to form a multi-level tree of sequences of triplets. This tree is called "the fractal genetic tree for the triplet convolution" or briefly "FGT-3". Texts at lower levels of any FGT can be figuratively called "daughter texts" of the initial long DNA-text $S_0$.

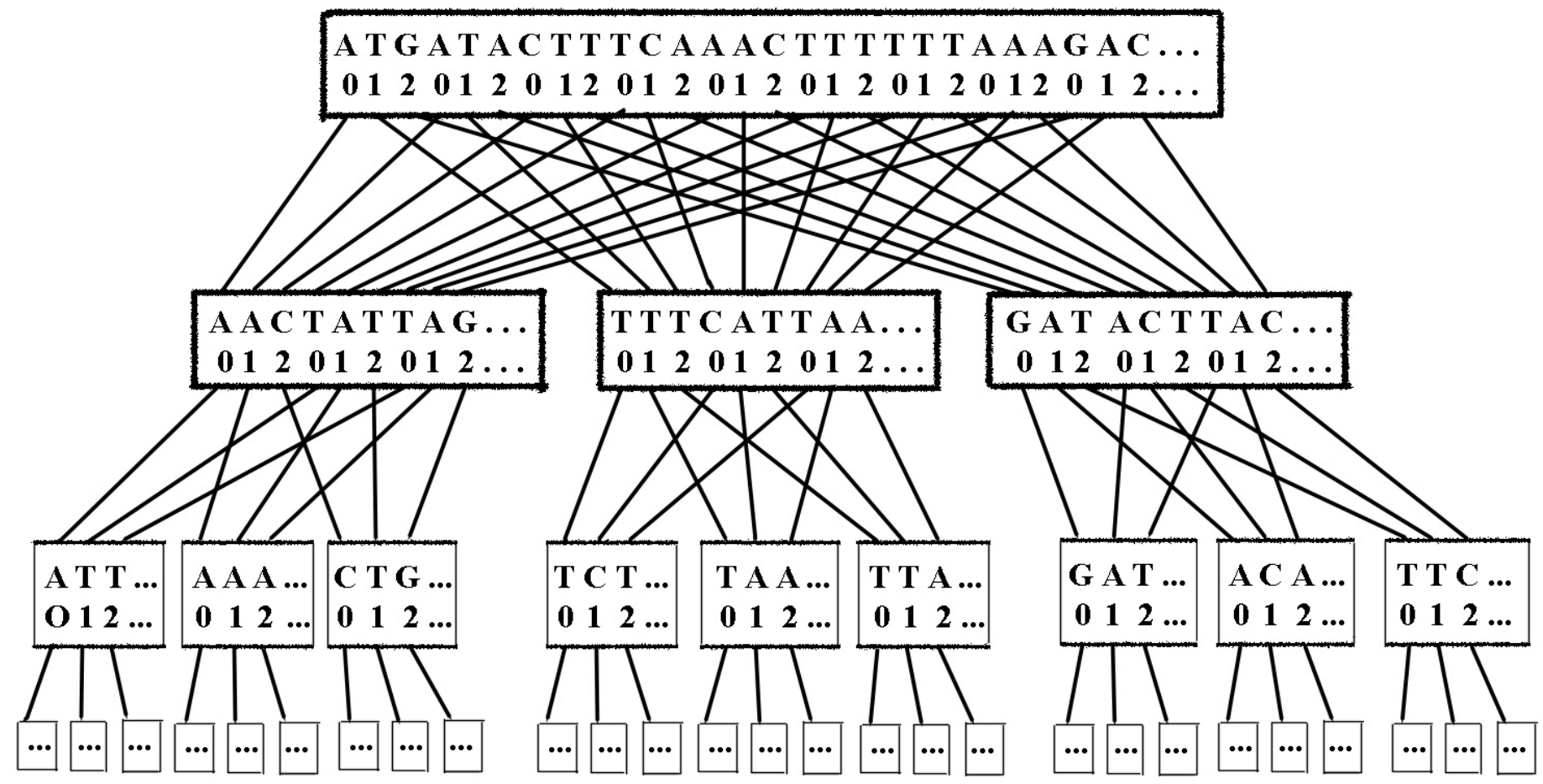

Fig. 27. The scheme of a fractal genetic tree (FGN-3) of a long DNA-text, which is represented as a sequence of triplets (from [Petoukhov, Svirin, 2012]).

This FGT possesses a fractal-like character if the enumeration of positions is only taken into account: each of long sequences of this FGT can be taken as an initial sequence to form a similar genetic tree on its basis (Fig. 4). In general case, the FGT can be built not only for DNA-texts of triplets, but also for DNA-texts of other n-plets (n = 2, 4, 5, ...) by means of the repeated positional convolution of each of texts from the previous level into "n" daughter texts of the next level of the convolution. This way one can build FGT-2, FGT-4, FGT-5, etc. for n = 2, 3, 4, 5,… correspondingly. A set of these FGT-2, FGT-3, FGT-4, FGT-5, … forms a net of separate trees of the initial and daughter texts; FGN is a set of such separate trees.

It is obvious that in the general case of FGT-n (where n = 2, 3, 4, 5,…), when we move to its next convolution level, the number of texts increases by a factor of n-fold and their lenghts are shortened by a factor of n. A FGT-n has at its first level of convolution n daughter texts; at its second level of convolution – $n^2$ daughter texts; and at p-th level – $n^p$ daughter texts. The total quantity N of texts in a FGT-n, which has p levels of convolution (p = 1, 2, 3, 4, …, ), is equal to the sum of daugther texts at all p levels:

$$N = 1+n+n^2+n^3+\ldots+n^p \qquad (37)$$

Each of daughter texts at different levels of a FGT-n is very individual and differs from other texts in this FGT-n in the general case.

For a long DNA-text of any biological organism, one can study implementation of the described rules of tetra-group symmetries in long texts at different levels of the convolution in cases of FGT-2, FGT-3, FGT-4, etc. Our own results of initial study of fractal genetic trees FGT-2, FGT-3, FGT-4 for enough long DNA-texts of different organisms show the implementation of these tetra-group rules in all daughter texts at initial levels of the FGT-2, FGT-3 and FGT-4. Moreover, for each of tested long DNA-tests, values of the probabilities $P_n(A_k)$, $P_n(T_k)$, $P_n(C_k)$ and $P_n(G_k)$ are approximately repeated in all daughter texts at different initial levels of these tested cases of FGT-n. In other words, for each daughter text at each level of these fractal genetic trees, values of each of these kinds of probabilities $P_n(A_k)$, $P_n(T_k)$, $P_n(C_k)$ and $P_n(G_k)$ (where n = 1, 2, 3, 4, 5;

$k \leq n$) lie in a narrow numerical interval (it is called an interval of fluctuations of the probability values at an appropriate level of a FGT-n). Fig. 28 and 29 illustrate this phenomenologic fact in a compressive form of tables showing these fluctuation intervals of probabilities $P_n(A_k)$, $P_n(T_k)$, $P_n(C_k)$ and $P_n(G_k)$ for daughter texts at each of two first levels of the FGT-2 and FGT-3 in cases of human sex chromosomes X and Y (their DNA-texts contain 156040895 and 57227415 letters correspondingly). Let us remind that - in line with the expression (37) – the quantity of different texts at the considered levels (from the level 0 till the level 2) is equal to $1+2+2^2 = 7$ in the case of FGT-2 and $1+3+3^2 = 13$ in the case of FGT-3.

| | Level 0 | Level 1/0 | Level 1/1 | Level 2/00 | Level 2/01 | Level 2/10 | Level 2/11 |
|---|---|---|---|---|---|---|---|
| $P_n(A_k)\in$ | 0.3017÷ 0.3019 | 0.3017÷ 0.3019 | 0.3016÷ 0.3020 | 0.3016÷ 0.3017 | 0.3017÷ 0.3019 | 0.3015÷ 0.3015 | 0.3015÷ 0.3019 |
| $P_n(T_k)\in$ | 0.3028÷ 0.3029 | 0.3028÷ 0.3028 | 0.3027÷ 0.3027 | 0.3025÷ 0.3029 | 0.3028÷ 0.3029 | 0.3027÷ 0.3031 | 0.3027÷ 0.3027 |
| $P_n(C_k)\in$ | 0.197÷ 0.1971 | 0.1969÷ 0.1971 | 0.1969÷ 0.197 | 0.1967÷ 0.1971 | 0.1969÷ 0.1971 | 0.1968÷ 0.197 | 0.1971÷ 0.1971 |
| $P_n(G_k)\in$ | 0.1981÷ 0.1982 | 0.1981÷ 0.1982 | 0.1981÷ 0.1982 | 0.1981÷ 0.1982 | 0.1981÷ 0.1982 | 0.198÷ 0.1984 | 0.1981÷ 0.1983 |

| | Level 0 | Level 1/0 | Level 1/1 | Level 1/2 | Level 2/00 | Level 2/01 |
|---|---|---|---|---|---|---|
| $P_n(A_k)\in$ | 0.3017÷ 0.3019 | 0.3016÷ 0.3020 | 0.3017÷ 0.3017 | 0.3017÷ 0.3019 | 0.3015÷ 0.3018 | 0.3016÷ 0.3019 |
| $P_n(T_k)\in$ | 0.3028÷ 0.3029 | 0.3027÷ 0.3029 | 0.3028÷ 0.3029 | 0.3027÷ 0.3029 | 0.3024÷ 0.3029 | 0.3025÷ 0.3028 |
| $P_n(C_k)\in$ | 0.1970÷ 0.1971 | 0.1970÷ 0.1970 | 0.1970÷ 0.1974 | 0.1967÷ 0.1971 | 0.1969÷ 0.1971 | 0.1967÷ 0.1968 |
| $P_n(G_k)\in$ | 0.1981÷ 0.1982 | 0.1981÷ 0.1981 | 0.1980÷ 0.1980 | 0.1981÷ 0.1982 | 0.198÷ 0.1981 | 0.1979÷ 0.1985 |

| | Level 2/02 | Level 2/10 | Level 2/11 | Level 2/12 | Level 2/20 | Level 2/21 | Level 2/22 |
|---|---|---|---|---|---|---|---|
| $P_n(A_k)\in$ | 0.3015÷ 0.3019 | 0.3018÷ 0.3019 | 0.3015÷ 0.3018 | 0.3014÷ 0.3014 | 0.3016÷ 0.3019 | 0.3016÷ 0.3017 | 0.3015÷ 0.3018 |
| $P_n(T_k)\in$ | 0.3025÷ 0.3026 | 0.3025÷ 0.3025 | 0.3026÷ 0.3030 | 0.3024÷ 0.3030 | 0.3025÷ 0.3029 | 0.3025÷ 0.3030 | 0.3027÷ 0.3030 |
| $P_n(C_k)\in$ | 0.1966÷ 0.1973 | 0.1967÷ 0.1969 | 0.1968÷ 0.1969 | 0.1969÷ 0.1974 | 0.1968÷ 0.1971 | 0.1966÷ 0.1968 | 0.1967÷ 0.1971 |
| $P_n(G_k)\in$ | 0.1980÷ 0.1981 | 0.1977÷ 0.1987 | 0.1979÷ 0.1983 | 0.1978÷ 0.1982 | 0.1980÷ 0.1981 | 0.1980÷ 0.1985 | 0.1978÷ 0.1981 |

Fig. 28. Tables of fluctuation intervals of probabilities $P_n(A_k)$, $P_n(T_k)$, $P_n(C_k)$ and $P_n(G_k)$ for the set of all texts at each of levels of convolutions in the FGT-2 (upper table) and FGT-3 (bottom tables) in the case of the human sex chromosome X (NCBI Reference Sequence: NC_000023.11). Here n = 1, 2, 3, 4, 5; $k \leq n$.

|  | Level 0 | Level 1/0 | Level 1/1 | Level 2/00 | Level 2/01 | Level 2/10 | Level 2/11 |
|---|---|---|---|---|---|---|---|
| $P_n(A_k)\in$ | 0.2983÷0.2987 | 0.2982÷0.2982 | 0.2979÷0.2987 | 0.2980÷0.2989 | 0.2981÷0.2981 | 0.2981÷0.2987 | 0.2976÷0.2992 |
| $P_n(T_k)\in$ | 0.3009÷0.3012 | 0.3008÷0.3021 | 0.3007÷0.3014 | 0.3009÷0.3013 | 0.3005÷0.3020 | 0.3005÷0.3010 | 0.3005÷0.3009 |
| $P_n(C_k)\in$ | 0.1998÷0.1998 | 0.1997÷0.1997 | 0.1995÷0.1995 | 0.1996÷0.1999 | 0.1997÷0.1997 | 0.1996÷0.2001 | 0.1995÷0.1995 |
| $P_n(G_k)\in$ | 0.1998÷0.2003 | 0.1994÷0.2001 | 0.1999÷0.2004 | 0.1995÷0.1998 | 0.1994÷0.2002 | 0.1995÷0.2002 | 0.1996÷0.2003 |

|  | Level 0 | Level 1/0 | Level 1/1 | Level 1/2 | Level 2/00 | Level 2/01 |
|---|---|---|---|---|---|---|
| $P_n(A_k)\in$ | 0.2983÷0.2987 | 0.2981÷0.2988 | 0.2984÷0.2989 | 0.298÷0.2982 | 0.2981÷0.2988 | 0.2976÷0.2984 |
| $P_n(T_k)\in$ | 0.3009÷0.3012 | 0.3008÷0.3012 | 0.3007÷0.3013 | 0.3008÷0.3012 | 0.300÷0.3013 | 0.3010÷0.3019 |
| $P_n(C_k)\in$ | 0.1998÷0.1998 | 0.1995÷0.2004 | 0.1999÷0.2002 | 0.1999÷0.1999 | 0.199÷0.1999 | 0.1994÷0.1994 |
| $P_n(G_k)\in$ | 0.1998÷0.2003 | 0.1996÷0.1996 | 0.1996÷0.1997 | 0.200÷0.2007 | 0.1996÷0.2001 | 0.1991÷0.2003 |

|  | Level 2/02 | Level 2/10 | Level 2/11 | Level 2/12 | Level 2/20 | Level 2/21 | Level 2/22 |
|---|---|---|---|---|---|---|---|
| $P_n(A_k)\in$ | 0.2974÷0.2982 | 0.2978÷0.2991 | 0.2979÷0.2979 | 0.2983÷0.2985 | 0.2975÷0.2976 | 0.2974÷0.2990 | 0.2975÷0.2979 |
| $P_n(T_k)\in$ | 0.3005÷0.3023 | 0.2997÷0.3006 | 0.3002÷0.3010 | 0.3006÷0.3008 | 0.3005÷0.3023 | 0.3005÷0.3010 | 0.3005÷0.3006 |
| $P_n(C_k)\in$ | 0.1994÷0.2005 | 0.1994÷0.1996 | 0.1995÷0.2004 | 0.1996÷0.2011 | 0.1993÷0.2003 | 0.1997÷0.2004 | 0.1993÷0.2002 |
| $P_n(G_k)\in$ | 0.1990÷0.1990 | 0.2000÷0.2008 | 0.1988÷0.2007 | 0.1991÷0.1996 | 0.1997÷0.1998 | 0.1996÷0.1996 | 0.1998÷0.2013 |

Fig. 29. Tables of fluctuation intervals of probabilities $P_n(A_k)$, $P_n(T_k)$, $P_n(C_k)$ and $P_n(G_k)$ for the set of all texts at each of levels of convolutions in the FGT-2 (upper table) and in the FGT-3 (bottom tables) in the case of the human sex chromosome Y (NCBI Reference Sequence: NC_000024.10). Here n = 1, 2, 3, 4, 5; $k \leq n$.

The same is true not only for each of human sex chromosomes X and Y (Fig. 28 and 29) but also for each of 22 human autosomes. Moreover, the same is true for long DNA-texts of all those model organisms and long DNA-texts, which are represented in the Appendixes 3-6 .

These our results allow putting forward the hypothesis about the validity of **the fifth rule of tetra-group symmetries, which concerns fractal genetic trees of long DNA-texts (including chromosomes) of different organisms**:

- For each of fractal genetic trees FGT-n (where n = 2, 3, 4, 5,… is not too large) of a long DNA-text, each of its daughter texts at different levels p of the tree (where p = 1, 2, 3, … is not too large) has aprroximately those values of probabilities $P_n(A_k)$, $P_n(T_k)$, $P_n(C_k)$ and $P_n(G_k)$ like the initial DNA-text.

Further researches are needed to check a degree of universality of this rule.

From Fig. 28 and 29 one can see also that fluctuation intervals of probabilities $P_n(A_k)$, $P_n(T_k)$, $P_n(C_k)$ and $P_n(G_k)$ are approximately equal each to other for all 20 (=7+13) texts at all considered levels of the fractal genetic net, which consists of the FGT-2 and FGT-3. It is true not only for each of human sex chromosomes X and Y but also for each of 22 human autosomes. Moreover, the same is true for long DNA-texts of all those model organisms and long DNA-texts, which are represented in the Appendixes 3-6.

Our results for long DNA-texts (about the implementation of the tetra-group symmetries in long convoluted texts at different levels of their FGT-n) testify in favor of existence of a fractal grammar of genetics in line with the Chargaff's words about a grammar of biology [Chargaff, 1971]. These results about a fractal grammar of long DNA-texts are additionally interesting by the following reasons:

- Many biological organisms have fractal-like inherited configurations in their bodies. This phenomenon can be considered as a consequence of the fractal-like organization of long DNA texts with the participation of tetra-group symmetries;
- As known, fractals allow a colossal compression of information (https://en.wikipedia.org/wiki/Fractal_compression). It is obvious that an opportunity of information compression is essential for genetic systems. Modern computer science knows a great number of methods of information compression including many methods of fractal compression. Our described results about fractal genetic nets can lead to a discovery of those «genetic» methods of information compression, which are used in genetic systems and in biological bodies in the whole.
- Many authors published their ideas and materials about relations of genomes with fractal structures in different aspects [Jeffrey, 1990; Lieberman-Aiden et al., 2009; Pellionisz, 2008; Pellionisz et al., 2012; Pellionisz, Ramanujam, Rajan, 2017; Peng et al., 1992; Perez, 2010]. For example, the work [Lieberman-Aiden et al., 2009] shown an existence of fractal globules in the three dimensional architecture of whole genomes, where spatial chromosome territories exist and where maximmally dense packing is provided on the basis of a special fractal packing, which provides the ability to easily fold and unfold any genomic locus. One should note that, by contrast to the work [Lieberman-Aiden et al., 2009], in our work we study not the spatial packing of whole genomes in a form of fractal globules but the quite another thing: we study the fractal organization of long DNA-texts, in particular, in the form of described fractal genetic trees and their nets of different kinds (FGT-n, where n = 2, 3, 4,…), which are connected with tetra-group symmetries of these texts (these symmetries and fractals were never studyed early);
- Fractals are actively used in study of cancer; some modern data testify that cancer processes are related with fractal patterns and their development [Baish, Jain, 2000; Bizzarri et al., 2011; Dokukin et al., 2015; Lennon et al., 2015; Pellionisz, Ramanujam, Rajan, 2017; Perez, 2017].
- Fractals are connected with theory of dynamic chaos, which has many applications in engineering technologies. We believe that the discovery of fractal-like properties of DNA-texts related with their tetra-group symmetries can lead to new ideas in theoretical and application areas, including problems of artificial intelligence and in-depth study of genetic phenomena for medical and biotechnological tasks.

Fig. 27 has shown above the example of a creation of daughter texts at different levels of FGN-n from a long DNA-text by the method of sequantel positional convolutions. But one can consider the inverse method, that is the method of the sequential positional assembly those texts, which belong to lower levels of FGN-n, into the extended texts, which belong to its higher levels up to the highest level $S_0$. If at a lower level (or at some lower levels) of FGN-n, its texts are permutated in their order, then the method of the sequental positional assembly from these texts leads to new extended texts at higher levels of this fractal tree. Such sequental positional assembly of permutated texts of lower levels of FGN-n can be used in biological evolution of DNA-texts of different organisms. As known, some biological organisms differ each from other by permutations of fragments of their DNA-texts. For example, mouse and human genomes can be viewed as a collection of about 200 fragments which are shuffled (rearranged) in mice as compared to humans; the chromosome 2 in humans is built from fragments that are similar to fragments from mouse DNA residing on chromosomes 1, 2, 6, 8, 11, 12, and 17 [Pevzner, 2000, Fig. 1.4 and 1.5, pp. 15, 16]. Genome rearrangements are a rather common chromosomal abnormality, which are associated with such genetic diseases as Down syndrome.

## 11. About letter-ordered representations of long DNA-texts conserving their collective probabilities in tetra-groups

The fifth and sixth rules of tetra-group symmetries, which are described in the previous Section about fractal genetic trees, are fulfilled at all not for all long texts of four letters C, G, A and T. To confirm this, lets us consider one of many examples of long texts, which don't satisfy the named rules. For example, take a long text, the first part of which contains only 100000 doublets CG and the second part contains only 100000 doublets AT: CG-CG-CG-…-CG-AT-AT-AT-...-AT. This text satisfies the second Chargaff's parity rule since %C=%G and %A=%T. But at the first level of the positional convolution of this text in its dichotomic fractal genetic tree (FGT-2), two texts are generated: the first of which contains only 50000 letters C and 50000 letters A, and the second of which contains only 50000 letters G and 50000 letters T. It is obvious that each of these two texts doesn't satisfy the second Chargaff's rule and other rules of tetra-group symmetries since %C≠%G and %A≠%T, etc.

Is there a variant of ordering rearrangement (shuffling) of all the letters of any long DNA-text, in which the characteristics of tetra-group symmetries of the text do not change? Let's demonstrate a possible variant of an ordered representation of any long DNA-text, which we call the "letter-ordered representation" (LO-representation). Such representation can be used as a convinient "canonical" form of representations of any long DNA-texts for some tasks of comparison analysis of such texts from the standpoint of their tetra-group symmetries.

Any long DNA-text, which contains $N_C$ letters C, $N_G$ letters G, $N_A$ letters A, $N_T$ letters T, can be formally represented by a text of the same length with an ordered sequence that contains four parts: its first part consists of $N_C$ letters C; its second part consists of $N_G$ letters G; its third part consists of $N_A$ letters A; its fourth part consists of $N_T$ letters T. Such letter-ordered sequence can be conditionally denoted $N_CC$-$N_GG$-$N_AA$-$N_TT$.

We have revealed that such letter-ordered representation of long DNA-texts possesses the same symmetric characteritics of tetra-group probabilities $P_n(C_k)$, $P_5(G_k)$, $P_n(A_k)$ and $P_n(T_k)$ as the original DNA-text (here n=1,2,3,4,... - is not too large, k≤n).

With a letter-ordered representation of a long DNA-text, a 4-dimensional metric space with a Cartesian coordinate system can be used, along the axes of which frequencies (numbers) of each of the four letters are plotted. In this case, different DNA-texts will be represented by different points of the given space, and distances between them can be calculated for a comparison analysis. (Another 4-dimensional metric space can be used for probabilities $P_n(C_k)$, $P_5(G_k)$, $P_n(A_k)$ and $P_n(T_k)$). Letter-ordered representations of long DNA-text are also convenient as beginning forms to study those permutations of fragments of long DNA-texts, which conserve characteristics of tetra-group symmetries of the texts.

## 12. On the biological sense of the symmetries of collective probabilities of tetra-groups in long DNA texts

The world of molecules, including DNA molecules, is subordinate to the principles of quantum mechanics. DNA molecules are carriers of quantum information that modern science analyzes and models on the basis of concepts and methods of quantum information and quantum computing: multi-qubit systems, pure and mixed quantum states, entangled states, edensity matrices, etc. The discovery of the symmetries of the collective probabilities of тетра-groups of oligomers in long DNA texts allowed us to propose a new model approach in which DNA alphabets and long DNA texts are represented by the academic formalisms of quantum informatics in the form of quantum-information multi-qubit systems. From the standpoint of this model approach, the existence of symmetries of these collective probabilities is a quantum-informational phenomenon and part of the general biological quantum-information system for ensuring the transfer of hereditary information in living organisms.

What is the biological sense of the quantum-information phenomenon of the symmetry of the collective probabilities of tetra groups in long DNA texts? A common property of all living organisms, consisting of quantum-mechanical molecular subsystems, is their ability to grow and develop on the basis of incorporation into their body of new and new molecules of nutrients becoming new parts of the whole organism. In the course of evolution, many species have acquired an even greater ability to unite parts to create single colonial and multicellular organisms. This means that biological evolution endowed organisms with a fundamental way of creating single quantum states of multicomponent quantum systems from the quantum states of their subsystems. But an analogous method in quantum mechanics and quantum informatics is built on the general principles of the mathematical operation of tensor multiplication of vectors and matrices. Figuratively speaking, a whole multicomponent organism is - to a certain extent - a product of tensor multiplication of the quantum states of its subsystems, including its molecular genetic systems (our work [Petoukhov, 2011] shows a connection of the tensor product of matrices with Punnett squares known in Mendelian genetics from 1905 year and included in text-books of genetics about Mendelian crosses of organisms). It is not for nothing that the symmetries of collective probabilities revealed in long DNA texts allow their modeling in the language of separable pure states of multi-qubit quantum systems that unite the quantum states of their subsystems. Our data testify that the biological meaning of symmetries of the collective probabilities of tetra groups in quantum-information long DNA-texts is related to the fundamental problem of the evolutionary quantum-information unification of the set of quantum subsystems into a single organism based on the principles of quantum mechanics. The genetic system, being an integral part of this quantum-information association, ensures the transfer of quantum information to descendants with the possibility of complicating the quantum-genetic messages in the process of biological evolution of organisms. Genetic symmetries of collective probabilities of tetra-groups in long quantum-information DNA texts designate the framework, inside which the evolutionary complication of quantum-genetic messages takes place. From the model standpoint, these symmetries of collective probabilities are a consequence of those quantum-information principles of creating the united states of quantum mechanical systems, which are modeled with the help of tensor products. To this we add that the reverse phenomenon - the appearance of independent "daughter" organisms or individual parts from integral biological objects - can be modeled as a result of the tensor factorization (decomposition) of the quantum states of an integral system upon its disintegration into two or more constituent parts.

The author hopes that the further usage in genetics the concepts and formalisms of quantum informatics, which was undertaken in this article in connection with symmetries of collective probabilities of tetra-groups, will lead to the development of substantial quantum-information genetics. This will promote the inclusion of genetics and all biology in the field of profound mathematical natural science. Consideration of biological phenomena, including the phenomena of inheritance of the intellectual abilities of biological bodies, from the standpoint of the theory of quantum computers, gives many valuable opportunities for their comprehension and also for development of artificial intelligence systems [Petoukhov et al., 2017]. For example, an adult human organism has around 10 trillion ($10^{14}$) human cells and each of cells containts an identical complect of DNA, genetic information from which is used for physiological functioning organism as the holistic system of cells. How such huge number of cells can reliably functioning as a cooperative whole? Quantum informatics and associations with quantum computers can help to model and understand such holistic biological systems with their ability to compute complex tasks and to transfer genetic information from one generation to another. The fundamental question about quantum computing was firstly touched upon in the book [Manin, 1980].

## 14. Hyperbolic rules of the oligomer collective organization of genomes.

The above described rules of collective probabilities of oligonucleotides (oligomers) from special tetra-groups can be reformulated into the hyperbolic rules of the oligomer collective organization of genomes. The hyperbolic rules say about total amounts of n-plets in the special tetra-groups instead of the collective probabilities of these n-plets. This new form seems to be more simple and comfortable for the perception of the phenomenological data about cooperative regularities in genomic sequences.

Let us explain this new form. The determination of the collective probability of one or another set of n-plets of a certain kind (for example, oligomers starting with the nucleotide A) is based on two follwing actions: 1) their total amount in the analyzed nucleotide sequence is calculated; 2) then this amount of the special kind of oligomers is divided by the sum of n-plets of all kinds, which is the whole DNA sequence length as a sequence of oligomers of the length n. Below, we restrict ourselves to only the first step and compare only the total amounts of all n-plets, which start with one of the nucleotides A, or T, or C, or G.

Let us turn to the human chromosome №1 and calculate - step by step - total amounts of its following n-plets at n=1, 2, 3, ..., 20: 1) all nucleotides A; 2) all 4 doublets starting with A (that is, AA, AT, AC, AG); 3) all 16 triplets starting with A (AAA, AAT, ..., GGG), ..., and so on (initial data on this chromosome are in the GenBank: https://www.ncbi.nlm.nih.gov/nuccore/NC_000001.11).

The result of calculations of such 20 total amounts is shown graphically in Fig. 30, at left (in blue). Here the abscissa axis represents the values of *n*, and the ordinate axis represents the values of the total amounts $\Sigma_{A,n,1}$ of *n*-plets, which start with the nucleotide A. The amazing result is that all 20 phenomenological points [$n$, $\Sigma_{A,n,1}$] lie - with a high level of accuracy - along the hyperbola $H_{A,1}$ = $S_A/n$ = $67070277/n$ shown in red in Fig. 30, middle. Deviations of phenomenological quantities $\Sigma_{A,n,1}$ from model values $S_A/n$ lie in the range -0.030%÷0.024%, that is, they comprise only hundredths of a percent (Fig. 30, at right).

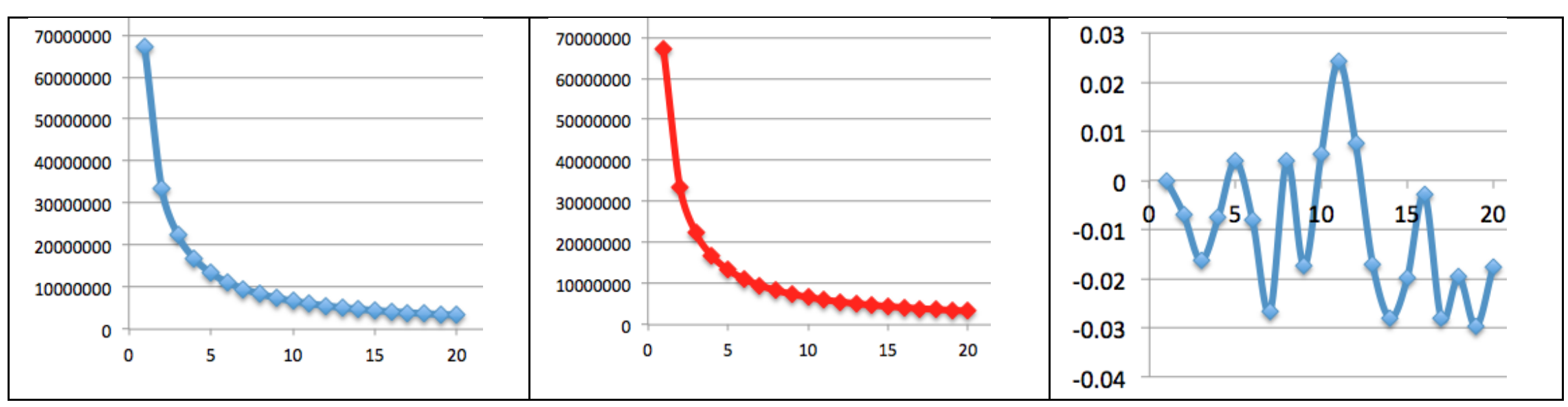

**Fig. 30**. The graphic data for the case of the OS-sequences of *n*-plets from the class $A_1$-oligomers of the human chromosome №1. In these graphs, the abscissa axis represents the values $n$ = 1, 2, 3, ..., 20. **Left**: the ordinate axis represents the set of phenomenological total amounts $\Sigma_{A,n,1}$ of *n*-plets beginning with the nucleotide A. **Middle**: the ordinate axis represents modeling values $S_A/n$ = $67070277/n$. The dots with coordinates [$n$, $S_A/n$] belong to the shown hyperbola $H_{A,1}$ = $S_A/n$ = $67070277/n$. **Right**: deviations of the real OS-sequence $\Sigma_{A,n,1}$ from the model hyperbolic progression $S_A/n$ in percentages.

This result is striking because it shows that knowing only the number of nucleotides A, that is, only one member of the number series shown in Fig. 30, at left, one can predict with the high accuracy all other 19 members, each of which is a sum of $4^{n-1}$ possible kinds of *n*-plets. The number of possible kinds of *n*-plets in these sums is growing rapidly, becoming astronomically huge: 4, 16, 64, 256, 1024, ..., $4^{10}$, ..., $4^{19}$. Of course, in the human chromosome №1, for example, not all $4^{19}$ kinds of the mentioned 20-plets exist but the total amount of all those kinds of 20-plets, which exist in this chromosome, is practically equal to $S_A/20$ with a high level of accuracy shown below.

One can remind here that genomic sequences in the GenBank sites usually contain some letters N, indicating that there can be any nucleotide in this place (https://www.ncbi.nlm.nih.gov/books/NBK21136/). By this reason, the total amount of all nucleotides A, T, C, G (that is the sum $S_A + S_T + S_C + S_G$, where $S_A$, $S_T$, $S_C$, and $S_G$ refer to the sums of appropriate nucleotides), calculated for the sequence from the GenBank, is slightly less than the complete length of the DNA sequence, which is indicated in the GenBank. But practically this is not essential for the results of the application of the OS-method to analyze genomic sequences.

Similar results were obtained when studying in this chromosome the total amounts of *n*-plets, which start with the nucleotide T (Fig. 31, at left), and the nucleotide C (Fig. 31, at middle), and the nucleotide G (Fig. 31, at right). The phenomenological values of the total amounts $\Sigma_{T,n,1}$, $\Sigma_{C,n,1}$, and $\Sigma_{G,n,1}$ of *n*-plets, which start with the corresponding nucleotide T, or C, or G, are also modeled effectively by appropriate hyperbolic progressions $H_{T,1}$, $H_{C,1}$, $H_{G,1}$, which differ from each other only by their numerators $S_T$, $S_C$, and $S_G$ (38):

$$H_{T,1}=S_T/n=67244164/n,\ \ H_{C,1}=S_C/n=48055043/n,\ \ H_{G,1}=S_G/n=48111528/n \quad (38)$$

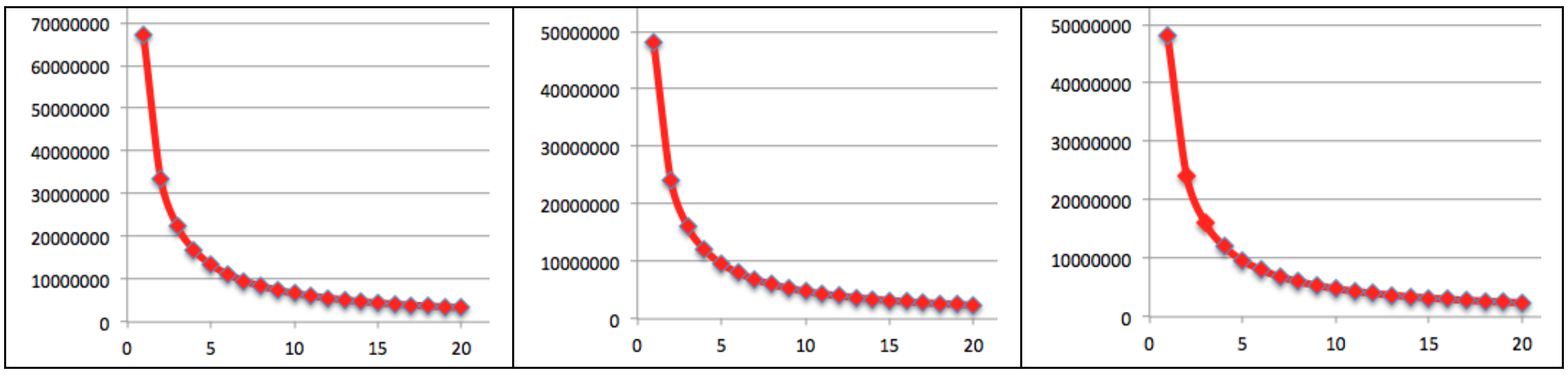

**Fig. 31**. Additional graph data on the human chromosome №1 regarding total amounts of n-plets, which start with the nucleotide T (at left), the nucleotide C (at middle), and the nucleotide G (at right). The abscissa axes represent the values $n$ = 1, 2, 3, ..., 20. The ordinate axes show model values $H_{T,1}(n)$, $H_{C,1}(n)$, and $H_{G,1}(n)$ (in red) from (38), which practically coincide phenomenological values $\Sigma_{T,n,1}$, $\Sigma_{C,n,1}$, and $\Sigma_{G,n,1}$ of the total amount of *n*-plets, which start with the nucleotide T (at the left graph), the nucleotide C (at the middle graph), and the nucleotide G (at the right graph). The numerical data on this coincidence is shown below.

Here one should note that this analysis of the oligomer cooperative organization of human chromosome №1 is made by the author's method of oligomer sums (abbreviation, the OS-method). The totality of data obtained by analyzing a nucleotide sequence by the OS-method is called its OS-representations. This method gives numeric sequences called oligomer sums sequences (or briefly, OS-sequences).

The application of the OS-method to the analysis of the human chromosome №1 and other chromosomes and genomes includes the following steps:

- Firstly, one should calculate phenomenological quantities $S_A$, $S_T$, $S_C$, and $S_G$ of monomers A, T, C, and G correspondingly in the considered nucleotide sequence. In the human chromosome № 1, the following quantities exist: $S_A = 67070277$, $S_T = 67244164$, $S_C = 48055043$, $S_G = 48111528$;
- Secondly, to construct the oligomer sums sequences, one should calculate the total amounts $\Sigma_{A,n,1}$, $\Sigma_{T,n,1}$, $\Sigma_{C,n,1}$, and $\Sigma_{G,n,1}$ of $n$-plets, which start with corresponding nucleotides A, or T, or C, or G at $n$ = 1, 2, 3, 4, … (here, for example, the symbol $\Sigma_{A,3,1}$ refers to the total amount of triplets, which start with the nucleotide A). These total amounts are members of the appropriate OS-sequences. For analysis of human chromosomes and various eukaryotic and prokaryotic genomes by this OS-method, the author usually takes $n$ = 1, 2, 3, … ,19, 20.

| *N* | 1 | 2 | 3 | 4 | 5 | 6 | 7 | 8 | 9 | 10 |
|---|---|---|---|---|---|---|---|---|---|---|
| **A** | | | | | | | | | | |
| Real | 67070277 | 33537501 | 22360413 | 16768845 | 13413532 | 11179286 | 9584038 | 8383461 | 7453552 | 6706672 |
| Model | 67070277 | 33535139 | 22356759 | 16767569 | 13414055 | 11178380 | 9581468 | 8383785 | 7452253 | 6707028 |
| Δ%A | 0.000 | -0.007 | -0.016 | -0.008 | 0.004 | -0.008 | -0.027 | 0.004 | -0.017 | 0.005 |
| **T** | | | | | | | | | | |
| Real | 67244164 | 33620498 | 22412993 | 16808862 | 13445360 | 11207274 | 9606748 | 8405040 | 7470145 | 6724359 |
| Model | 67244164 | 33622082 | 22414721 | 16811041 | 13448833 | 11207361 | 9606309 | 8405521 | 7471574 | 6724416 |
| Δ%T | 0.000 | 0.005 | 0.008 | 0.013 | 0.026 | 0.001 | -0.005 | 0.006 | 0.019 | 0.001 |
| **C** | | | | | | | | | | |
| Real | 48055043 | 24024903 | 16012711 | 12013624 | 9612227 | 8005708 | 6865944 | 6008215 | 5336968 | 4803919 |
| Model | 48055043 | 24027522 | 16018348 | 12013761 | 9611009 | 8009174 | 6865006 | 6006880 | 5339449 | 4805504 |
| Δ%C | 0.000 | 0.011 | 0.035 | 0.001 | -0.013 | 0.043 | -0.014 | -0.022 | 0.046 | 0.033 |
| **G** | | | | | | | | | | |
| Real | 48111528 | 24057606 | 16040889 | 12028924 | 9625086 | 8021235 | 6869132 | 6013412 | 5348337 | 4813156 |
| Model | 48111528 | 24055764 | 16037176 | 12027882 | 9622306 | 8018588 | 6873075 | 6013941 | 5345725 | 4811153 |
| Δ%G | 0.000 | -0.008 | -0.023 | -0.009 | -0.029 | -0.033 | 0.057 | 0.009 | -0.049 | -0.042 |

| *n* | 11 | 12 | 13 | 14 | 15 | 16 | 17 | 18 | 19 | 20 |
|---|---|---|---|---|---|---|---|---|---|---|
| **A** | | | | | | | | | | |
| Real | 6095821 | 5588773 | 5160139 | 4792078 | 4472245 | 4192017 | 3946422 | 3726860 | 3531067 | 3354107 |
| Model | 6097298 | 5589190 | 5159252 | 4790734 | 4471352 | 4191892 | 3945310 | 3726127 | 3530015 | 3353514 |
| Δ%A | 0.024 | 0.007 | -0.017 | -0.028 | -0.020 | -0.003 | -0.028 | -0.020 | -0.030 | -0.018 |
| **T** | | | | | | | | | | |
| Real | 6111970 | 5601854 | 5173904 | 4801395 | 4479492 | 4202773 | 3954021 | 3735327 | 3535288 | 3360459 |
| Model | 6113106 | 5603680 | 5172628 | 4803155 | 4482944 | 4202760 | 3955539 | 3735787 | 3539167 | 3362208 |
| Δ%T | 0.019 | 0.033 | -0.025 | 0.037 | 0.077 | 0.000 | 0.038 | 0.012 | 0.110 | 0.052 |
| **C** | | | | | | | | | | |
| Real | 4370502 | 4002753 | 3694018 | 3433636 | 3202830 | 3003511 | 2826568 | 2668499 | 2531448 | 2402186 |
| Model | 4368640 | 4004587 | 3696542 | 3432503 | 3203670 | 3003440 | 2826767 | 2669725 | 2529213 | 2402752 |
| Δ%C | -0.043 | 0.046 | 0.068 | -0.033 | 0.026 | -0.002 | 0.007 | 0.046 | -0.088 | 0.024 |
| **G** | | | | | | | | | | |
| Real | 4374518 | 4013372 | 3701250 | 3435824 | 3210839 | 3006763 | 2830698 | 2673815 | 2532772 | 2407301 |
| Model | 4373775 | 4009294 | 3700887 | 3436538 | 3207435 | 3006971 | 2830090 | 2672863 | 2532186 | 2405576 |
| Δ%G | -0.017 | -0.102 | -0.010 | 0.021 | -0.106 | 0.007 | -0.021 | -0.036 | -0.023 | -0.072 |

**Fig. 32.** Real and model values to the OS-representations of the classes of $A_1$-, $T_1$-, $C_1$-, and $G_1$-oligomers in human chromosome №1 are shown for $n$ = 1, 2, ..., 20. The real total amounts of $n$-plets, which start with a certain nucleotide (A, T, C, or G), are indicated (in blue) jointly with their model values $H_{A,1}(n)$, $H_{T,1}(n)$, $H_{C,1}(n)$, and $H_{G,1}(n)$ from (2.1) (in red). The symbol Δ% refers to deviations of real values from model values in percent (the model values are taken as 100%).

Fig. 32 shows real and model numeric values for the OS-representation of the human chromosome №1 at $n$ = 1, 2, 3,..., 20. The model values of the total amounts of $n$-plets, which start with a certain nucleotide (A, T, C, or G), are calculated correspondingly as values of the hyperbolic progressions $H_{A,1} = S_A/n$ = 67070277/$n$, $H_{T,1}=S_T/n$=67244164/$n$, $H_{C,1}=S_C/n$=48055043/$n$, and $H_{G,1}=S_G/n$=48111528/$n$. Deviations of real values from model values are also shown in percent in accordance with the expression: 100(1 – real value)/(model value). One can see that these deviations are much lesser than 0,2% in all cases.

The model hyperbolic progressions $H_{A,1} = S_A/n$, $H_{T,1} = S_T/n$, $H_{C,1} = S_C/n$, and $H_{G,1} = S_G/n$ serve as mathematical standards for the described phenomenological facts. These hyperbolic progressions differ from each other only in the magnitude of numerators in their expressions, and therefore they can be specified by the general expression (39):

$$H_{N,1}(n) = S_N/n, \qquad (39)$$

where N refers to any of nucleotides A, T, C, or G; $S_N$ refers to the number of corresponding nucleotides A, T, C, or G in the analyzed nucleotide sequence. If you know the total quantity $S_N$ of the monomer N, you can predict - with a high level of accuracy - the total amounts of $n$-plets belonging to the class $N_1$-oligomers by using the general expression (39). These phenomenological facts testify in favor of the cooperative entity of the nucleotide sequence in the human chromosome № 1.

By the corresponding compression of the ordinate axis in these cartesian coordinate systems (that is by appropriate scaling of numerators $S_A$, $S_T$, $S_C$, and $S_G$), each of these four hyperbolic sequences $H_{A,1}=S_A/n$, $H_{T,1}=S_T/n$, $H_{C,1}=S_C/n$, and $H_{G,1}=S_G/n$ reduces to the hyperbolic sequence (40):

$$y = 1/n, \qquad (40)$$

which we call the canonical (or reference) hyperbolic sequence of OS-representations (or the canonical OS-sequence) of nucleotide sequences. In mathematics, the sequence (41)

$$1/1,\ 1/2,\ 1/3,\ 1/4,\ldots,\ 1/n \qquad (41)$$

is known long ago as a harmonic progression (or a harmonic sequence) where each its term is the harmonic mean of the neighboring terms. For this reason, the revealed hyperbolic sequences in genomes can be also called genomic harmonic progressions and, in this mathematical sense, one can talk about the harmonic rules and the harmonious organization of genomes described below. The historically famous name "the harmonic progression" comes from the connection (41) with the series of harmonics in music. The sums of the first members of the harmonic progression (41) are called harmonic numbers. The rich centuries-old history of the study of harmonic progressions and harmonic series is associated with the names of Pythagoras, Orem (d'Oresme), Leibniz, Newton, Euler, Fourier, Dirichlet, Riemann and other researchers. The generalization of the harmonic series is known as the Riemann zeta function. Using musical terminology, where the term “timbre” refers to the totality of the set of sound frequencies in a prolonged sound, one can conditionally say that the oligomer sums method represents the analyzed nucleotide sequence as some "oligomer timbre". The series of harmonic numbers serves as the discrete analogue of the continuous function of natural logarithm ln $n$ [Graham, Knuth, Parashnik, 1994, p. 276]; this, in particular, connects the harmonic progression (41) with the Weber-Fechner logarithmic law, which is the main psychophysical law and dictates informatic peculiarities for all inherited sensory chanels - vision, hearing, smell, etc -, whose organs (eyes, ears, nose, etc.) very differ each other in appearance. It testifies that genetical and psychophysical levels of inherited biological informatics are structurally intercorrelated on the algebra-harmonical basis [Petoukhov, 2016, 2020].

Fig. 32 shows that the sequences of the total amounts of $n$-plets from the classes of $A_1$-oligomers and $T_1$-oligomers differ little from each other. The same is true for the sequences of the total amounts of $n$-plets from the classes of $C_1$- and $G_1$-oligomers. This fact is described by the expressions (42):

$$\Sigma_{A,n,1} \approx \Sigma_{T,n,1}, \qquad \Sigma_{C,n,1} \approx \Sigma_{G,n,1} \qquad (42)$$

In the particular case at $n$ = 1, expressions (41) demonstrate the second Chargaff's rule on the approximate equality between the amounts of nucleotides A and T, as well as C and G in long DNA sequences. Correspondingly the phenomenological fact, described by expressions (41), is a certain generalization of the second Chargaff's rule.

The results presented indicate, at least for the human chromosome №1, that there exists two general hyperbolic (or harmonic) rules regarding the total amounts of $n$-plets, which start with a certain nucleotide A, T, C, or G.

**The first hyperbolic rule** (about interlelations of oligomers in individual chromosomes):

- In individual chromosomes, for any of classes of oligomers, which start with the nucleotide A, or T, or C, or G, the total amounts $\Sigma_{N,n,1}(n)$ of their $n$-plets, corresponding different $n$, are interrelated each other through the general expression $\Sigma_{N,n,1} \approx S_N/n$ with a high level of accuracy (here N refers to any of nucleotides A, T, C, or G; $S_N$ refers to the quantity of monomers N; $n$ = 1, 2, 3, 4, ... is not too large compared to the full length of the nucleotide sequence). The phenomenological points with coordinates [$n$, $\Sigma_{N,n,1}$] practically lie on the hyperbola having points $H_{N,1} = S_N/n$.

**The second hyperbolic rule** (about the similarity in the pairs of OS-sequences):

- In individual chromosomes, two numeric OS-sequences expressing the total amounts of $n$-plets, which start with the nucleotide A and with the nucleotide T, are approximately identical. The same is true for two numeric OS-sequences expressing the total amounts of $n$-plets, which start with the nucleotide C and with the nucleotide G (in accordance with the expressions (42)). Here $n$ = 1, 2, 3, 4, ... is not too large compared to the full length of the nucleotide sequence.

Let us continue the description of obtained results of analysis of the human genome, which contains 22 autosomes and 2 sex chromosomes X and Y. These chromosomes are very different from each other in length, molecular weight, gene content, etc. What can be said about the other 23 human chromosomes?

| № | $S_A$ | Range % | $S_T$ | Range % | $S_C$ | Range % | $S_G$ | Range % |
|---|---|---|---|---|---|---|---|---|
| 1 | 67070277 | -0.030 ÷0.024 | 67244164 | -0.025 ÷0.110 | 48055043 | -0.088 ÷0.068 | 48111528 | -0.106 ÷0.057 |
| 2 | 71791213 | -0.079 ÷0.087 | 71987932 | -0.075 ÷0.095 | 48318180 | -0.097 ÷0.072 | 48450903 | -0.105 ÷0.141 |
| 3 | 59689091 | -0.021 ÷0.045 | 59833302 | -0.097 ÷0.098 | 39233483 | -0.130 ÷0.081 | 39344259 | -0.034 ÷0.088 |
| 4 | 58561236 | -0.065 ÷0.044 | 58623430 | -0.036 ÷0.128 | 36236976 | -0.039 ÷0.127 | 36331025 | -0.117 ÷0.075 |
| 5 | 54699094 | -0.052 ÷0.040 | 54955010 | -0.071 ÷0.078 | 35731600 | -0.012 ÷0.132 | 35879674 | -0.103 ÷0.085 |
| 6 | 51160489 | -0.039 ÷0.057 | 51151754 | -0.049 ÷0.022 | 33520786 | -0.092 ÷0.061 | 33516767 | -0.029 ÷0.069 |
| 7 | 47058248 | -0.104 ÷0.040 | 47215040 | -0.061 ÷0.030 | 32317984 | -0.086 ÷0.091 | 32378859 | -0.076 ÷0.069 |
| 8 | 42641072 | -0.061 ÷0.068 | 42581941 | -0.111 ÷0.071 | 28600559 | -0.110 ÷0.069 | 28600963 | -0.068 ÷0.050 |
| 9 | 31752642 | -0.134 ÷0.090 | 31733822 | -0.083 ÷0.065 | 22487631 | -0.099 ÷0.141 | 22470915 | -0.079 ÷0.143 |
| 10 | 38875926 | -0.081 ÷0.052 | 39027555 | -0.067 ÷0.099 | 27639505 | -0.058 ÷0.085 | 27719976 | -0.118 ÷0.085 |
| 11 | 39286730 | -0.032 ÷0.084 | 39361954 | -0.062 ÷0.042 | 27903257 | -0.139 ÷0.056 | 27981801 | -0.086 ÷0.112 |
| 12 | 39370109 | -0.096 ÷0.056 | 39492225 | -0.097 ÷0.094 | 27092804 | -0.076 ÷0.078 | 27182678 | -0.073 ÷0.105 |
| 13 | 29224840 | -0.067 ÷0.077 | 29320872 | -0.107 ÷0.069 | 18341128 | -0.107 ÷0.141 | 18346620 | -0.130 ÷0.065 |
| 14 | 25606393 | -0.109 ÷0.100 | 25819249 | -0.040 ÷0.086 | 17733667 | -0.137 ÷0.077 | 17782016 | -0.056 ÷0.142 |
| 15 | 24508669 | -0.085 ÷0.179 | 24553812 | -0.127 ÷0.088 | 17752941 | -0.090 ÷0.162 | 17825903 | -0.067 ÷0.113 |
| 16 | 22558319 | -0.122 ÷0.080 | 22774906 | -0.143 ÷0.104 | 18172742 | -0.146 ÷0.074 | 18299976 | -0.146 ÷0.173 |
| 17 | 22639499 | -0.141 ÷0.105 | 22705261 | -0.146 ÷0.070 | 18723944 | -0.134 ÷0.072 | 18851500 | -0.144 ÷0.105 |
| 18 | 22087028 | -0.160 ÷0.071 | 22109347 | -0.169 ÷0.121 | 14574701 | -0.090 ÷0.134 | 14594335 | -0.160 ÷0.210 |
| 19 | 15142293 | -0.160 ÷0.024 | 15282753 | -0.062 ÷0.062 | 13954580 | -0.103 ÷0.097 | 14061132 | -0.057 ÷0.226 |
| 20 | 16455618 | -0.106 ÷0.129 | 16643030 | -0.099 ÷0.089 | 13037092 | -0.062 ÷0.116 | 13098788 | -0.092 ÷0.155 |
| 21 | 9943435 | -0.161 ÷0.083 | 9882679 | -0.206 ÷0.173 | 6864570 | -0.134 ÷0.277 | 6852178 | -0.373 ÷0.219 |
| 22 | 10382214 | -0.175 ÷0.084 | 10370725 | -0.036 ÷0.209 | 9160652 | -0.258 ÷0.155 | 9246186 | -0.143 ÷0.235 |
| X | 46754807 | -0.078 ÷0.084 | 46916701 | -0.102 ÷0.055 | 30523780 | -0.116 ÷0.179 | 30697741 | -0.135 ÷0.067 |
| Y | 7886192 | -0.244 ÷0.097 | 7956168 | -0.063 ÷0.185 | 5285789 | -0.181 ÷0.407 | 5286894 | -0.247 ÷0.142 |

**Fig. 33.** Some results of analysis of all 24 human nuclear chromosomes by the oligomer sums method are represented. For each of the chromosomes, quantities $S_A$, $S_T$, $S_C$, and $S_G$ of monomers A, T, C, and G are shown to define the model hyperbolic progressions (39). The columns «Range %) show ranges of deviations of real OS-series of corresponding *n*-plets ($n$ = 1, 2, ..., 20) from their appropriate model values $S_A/n$, $S_T/n$, $S_C/n$, and $S_G/n$ in percentages (in each case, an appropriate model value is taken as 100%). The left column shows chromosome numbers.

Are there hyperbolic rules similar to formulated rules for the human chromosome №1 ? Yes, the author has got a positive answer to this question. For each of 24 human chromosomes, knowing its quantity $S_N$ of the nucleotide N (that is A, T, C, or G) allows you to calculate the total amounts of *n*-plets, which start with the oligomer N, with a high level of accuracy by using the general expression (39). Here $n$ = 1, 2, 3,... is not very large in comparison with the length of the DNA sequence. Fig. 33 shows general confirmational results of studying all 24 human chromosomes by the OS-method under $n$ =1, 2, 3, ..., 20.

These results show that both hyperbolic (or harmonic) rules № 1 and № 2 hold true for each of the human chromosomes with a high level of accuracy.

One can show that the obtained phenomenological data also lead to the third hyperbolic rule related with normalised versions of the OS-sequences $S_A/n$, $S_T/n$, $S_C/n$, and $S_G/n$. Scaling the numerators $S_A$, $S_T$, $S_C$, and $S_G$ by dividing by their total amount $S = S_A+S_T+S_C+S_G$, we obtain the corresponding scaling of all these OS-sequences, which are termed as "normalized OS-sequences " (43):

$$S_A/nS,\ S_T/nS,\ S_C/nS,\ S_G/nS \qquad (43)$$

It turns out that the normalized OS-sequences of all human chromosomes are similar to each other with a high degree of accuracy as Fig. 34 shows.

| **Chrom** | **$S_A/S$** | **$S_T/S$** | **$S_C/S$** | **$S_G/S$** |
|---|---|---|---|---|
| **1** | 0.2910 | 0.2918 | 0.2085 | 0.2087 |
| **2** | 0.2984 | 0.2993 | 0.2009 | 0.2014 |
| **3** | 0.3013 | 0.3020 | 0.1980 | 0.1986 |
| **4** | 0.3086 | 0.3089 | 0.1910 | 0.1915 |
| **5** | 0.3018 | 0.3032 | 0.1971 | 0.1979 |
| **6** | 0.3021 | 0.302 | 0.1979 | 0.197 |
| **7** | 0.2960 | 0.2970 | 0.2033 | 0.2037 |
| **8** | 0.2994 | 0.2990 | 0.2008 | 0.2008 |
| **9** | 0.2928 | 0.2926 | 0.2074 | 0.2072 |
| **10** | 0.2917 | 0.2929 | 0.2074 | 0.2080 |
| **11** | 0.2920 | 0.2926 | 0.2074 | 0.2080 |
| **12** | 0.2957 | 0.2966 | 0.2035 | 0.2042 |
| **13** | 0.3069 | 0.3079 | 0.1926 | 0.1926 |
| **14** | 0.2945 | 0.2970 | 0.2040 | 0.2045 |
| **15** | 0.2896 | 0.2901 | 0.2097 | 0.2106 |
| **16** | 0.2758 | 0.2784 | 0.2221 | 0.2237 |
| **17** | 0.2730 | 0.2738 | 0.2258 | 0.2273 |
| **18** | 0.3011 | 0.3014 | 0.1987 | 0.1989 |
| **19** | 0.2591 | 0.2615 | 0.2388 | 0.2406 |
| **20** | 0.2778 | 0.2810 | 0.2201 | 0.2211 |
| **21** | 0.2964 | 0.2946 | 0.2047 | 0.2043 |
| **22** | 0.2651 | 0.2648 | 0.2339 | 0.2361 |
| **X** | 0.3019 | 0.3029 | 0.1971 | 0.1982 |
| **Y** | 0.2985 | 0.3012 | 0.2001 | 0.2001 |

**Fig. 34.** Data of normalised OS-sequences $S_A/nS$, $S_T/nS$, $S_C/nS$, and $S_G/nS$ of all human chromosomes are shown for a comparison. Here $S = S_A+S_T+S_C+S_G$.

The same results on the similarity of normalized OS-sequences $S_A/nS$, $S_T/nS$, $S_C/nS$, and $S_G/nS$ in all chromosomes of a particular genome were obtained by the author when studying the genomes of a number of eukaryotes indicated above in Sections of this article These results allow proposing the third hyperbolic (or harmonic) rule on the total amounts of *n*-plets, which start with a certain nucleotide A, T, C, or G.

**The third hyperbolic rule** (about the similarity of chromosomes):

- All chromosomes of any individual eukaryotic genome have approximately the same normalized OS-sequences $S_A/nS$, $S_T/nS$, $S_C/nS$, and $S_G/nS$ representing classes of $A_1$-, $T_1$-, $C_1$-, and $G_1$-oligomers ($n$ = 1, 2, 3, 4, … is not too large compared to the full length of the nucleotide sequence).

The author has also received the confirmation that the hyperbolic rules №№ 1 and 2 are true for many prokaryotic genomes, which were analysed by the author on the basis of the OS-method (without no one exception from these rules till now). These results on prokaryotic genomes will be published sume later. The author suggests that the hyperbolic rules are universal genetic rules. But at this stage of the study, they are only candidates for the role of universal rules, since analysis of the widest variety of genomes is required to verify their universality.

As is known, mutations and the pressure of natural selection influence the genomic sequences of nucleotides. For these reasons, one can assume that as a result of many millions of years of biological evolution, genomic sequences, due to various influences, receive a completely random structure as a whole. This article provides evidence that, despite mutations, the pressure of natural selection, and other evolutionary factors, the nucleotide sequences of the eukaryotic and prokaryotic genomes have universal algebraic invariants. One can believe that the algebraic unity of living organisms is found (this should be tested further and further on more and more number of genomes). The discovery of these invariants gives new knowledge about the unity of the world of all living organisms and about the features of biological evolution. At the same time, new mathematical tools and approaches for in-depth study of this biological world and its evolution appear.

The genomic invariants, described in the article, are connected with hyperbolic sequences and transformations of hyperbolic rotations that shift the hyperbolic sequence along itself. Hyperbolic rotations, also called Lorentz transformations and known in the special theory of relativity, draw attention to the structural connection of genetic phenomena with the hyperbolic geometry of the Minkowski plane. One of the well-known models of two-dimensional hyperbolic geometry is the Poincaré disk model, also called the conformal disk model. The Poincaré disk model is connected with split-quaternions by J. Cockle and seems to be interesting for studying some genetic structures and inherited physiological phenomena as it was mentioned in previous author's publications on matrix genetics (see, for example, [Petoukhov, 2012]).

Living organisms are informational entities in which everything is subordinate to the task of reliably transmitting genetic information to descendants. All inherited physiological systems, as parts of a whole organism, must be structurally coupled with a genetic code for transmission to descendants in encoded form. Therefore, inherited physiological macrostructures can bear the imprint of structural features of the genetic code. For this reason, structural analogies exist between the genetic system and the properties of inherited physiological systems, for example, the unificated properties of different sensor systems, which are reflected in the main psychophysical Weber-Fechner law [Petoukhov, 2016].

It should be noted that the genomic hyperbolic rules are cardinally different from well-known hyperbolic Zipf's law. Zipf's law was originally formulated in terms of quantitative linguistics, stating that given some corpus of natural language utterances, the frequency of any word is inversely proportional to its rank in the frequency table (see, for example, [Fagan, Gençay, 2010]). In linguistics and in other fields, Zipf's law speaks on the frequency of encounter of separate words or other separate objects. In contrast, the hyperbolic rules of the genomes focuse on the OS-sequences of the total amounts of *n*-plets and the genomic entanglement, that is, on the relative number of not separate oligomers, but the whole sums of sets of different *n*-plets distributed inside the genomic sequence, where each separate nucleotide is a part of many oligomer sets existing simultaneously (each nucleotide is a distributed participant of many members of the appropriate genomic OS-sequence at once and makes a contribution to each of them). From the point of view of the quantum-information model, OS-sequences serve as quantum-information characteristics of genomic sequences.

The proposed oligomer sums method and the quantum-information model give new opportunities to study genetic systems and the inherited algebra-harmonic organization of living bodies.

## 15. Applications of the oligomer sums method to analysis of long genes

The application of the oligomer sums method to study long genes, whose sequences are much shorter that the complete DNA sequences in genomes, is also possible and gives interesting results. First of all, it unexpectedly reveals the phenomenon of regular rhythmic deviations of the sequences of real total sums of *n*-plets in the described genes from the corresponding model hyperbolic progressions.

Let us firstly consider human *TTN* gene encoding the largest known protein Titin. Titin, also known as connectin, is important in the contraction of striated muscle tissues. Figs. 33-37 show in graphical forms some results of analysis - by the oligomer sums method - of the nucleotide sequence of *TTN* gene (numeric results will be represented below). Initial data on its nucleotide sequence are taken in the GenBank https://www.ncbi.nlm.nih.gov/nuccore/X90568.1. This gene contains 26373 nucleotides A, 19569 nucleotides T, 17097 nucleotides C, and 18901 nucleotides G, that is $S_A$ = 26373, $S_T$ = 19569, $S_C$ = 17097, and $S_G$ = 18901 for the model hyperbolic progressions (39). It can be especially noted that, in this gene, the amounts of nucleotides A and T are significantly different (26373 and 19569), that is, the second Chargaff's rule on their approximate equality in long sequences is not satisfied here since this nucleotide sequence is not enough long for the Chargaff's rule.

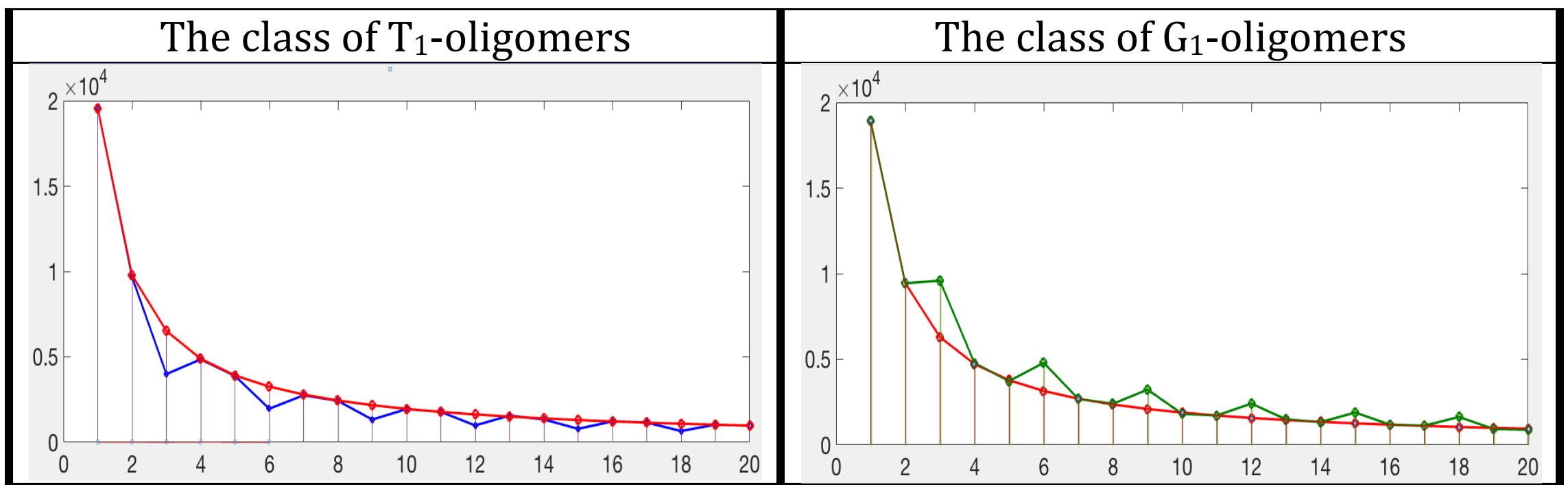

**Fig. 35**. Graphical representations of the results of analysis - by the oligomer sums method – of the human *TNT* gene. The OS-sequences of its total amounts of $n$-plets, which start with the nucleotide T (left) and the nucleotide G (right), are shown. The red lines refer to model hyperbolic progressions $S_T/n$ and $S_G/n$ correspondingly, where $S_T$ = 19569 and $S_G$ = 18901 are quantities of nucleotides T and G in the gene; $n$ = 1, 2, 3, ..., 20 as shown at the abscissa axes. The blue line (left) and the green line (right) with dots on them refer to the real OS-sequences of the total amounts of such $n$-plets. The ordinate axes indicate the total amounts of $n$-plets.

Fig. 35 shows the sequences of the highly regular significant deviations of the real total amounts of $n$-plets, which start with the nucleotide T and the nucleotide G, from model hyperbolic progressions $S_T/n$ = 19569/$n$ and $S_G/n$ = 18901/$n$. One should note that all these significant deviations happen only at $n$ = 3, 6, 9, ..., 3$m$, that is only for cases of 3$m$-plets (here $m$ = 1, 2, 3,...). Correspondingly these significant deviations can be called «triplet-deviations».

Fig. 36 shows the graph, which unites both graphs from Fig. 35 and demonstrates a few interesting features of the highly reqular series of these triplet-deviations.

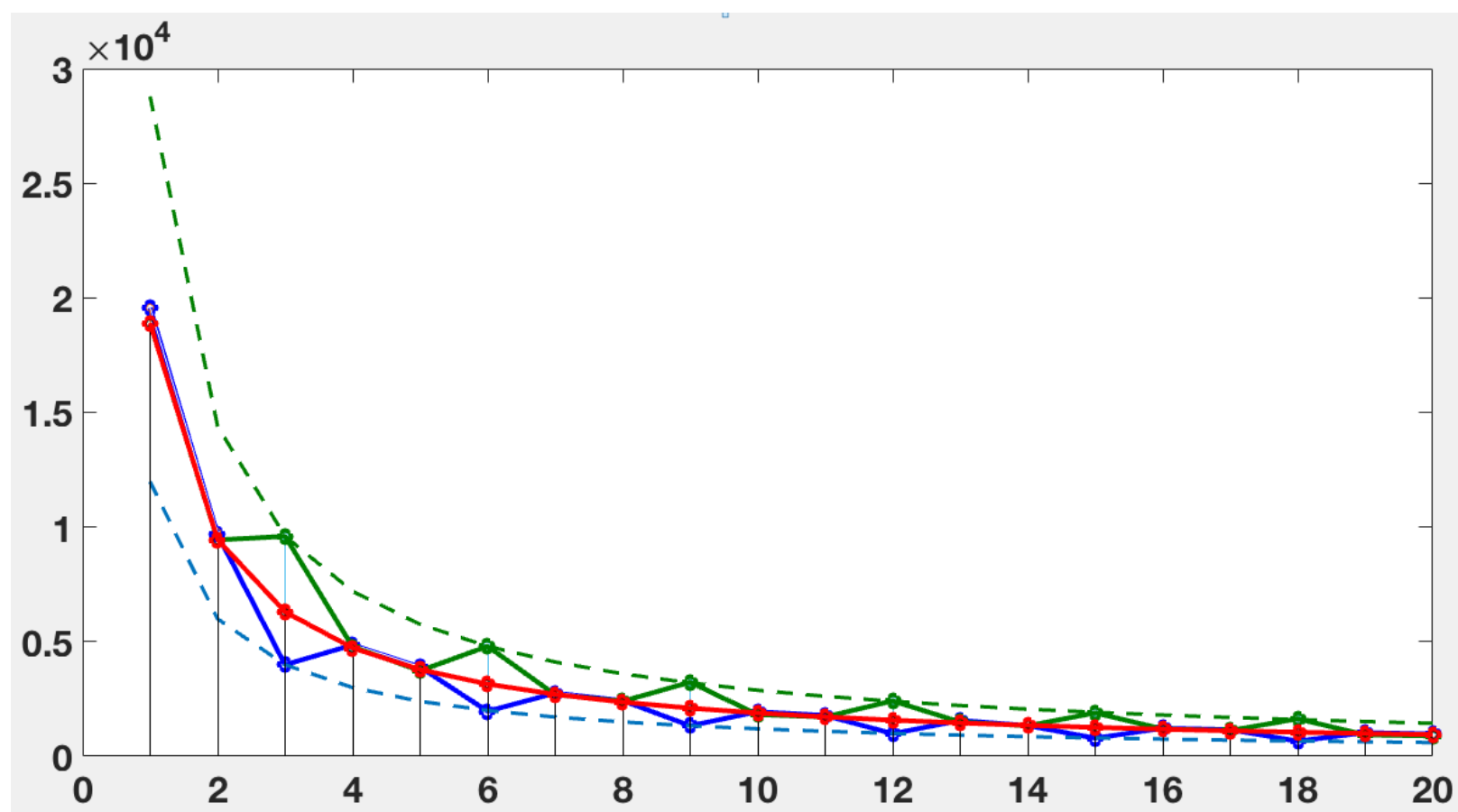

**Fig. 36.** The graph, uniting two graphs from Fig. 33 for the *TNT* gene, is shown. The blue dot line and the green dot lines correspond to those additional

hyperbolic progressions 11979/*n* and 28788/*n*, which model real total amounts of 3*m*-plets. Other parts of this united graph are the same as in Fig. 33.

Firstly, one can see in Fig. 36 that, in classes of $T_1$-oligomers and $G_1$-oligomers, the triplet-deviations happen in opposite directions (or, figuatively speaking, in antiphase):

- in the class of $T_1$-oligomers, they decrease real values compared with model values of the hyperbolic progression 19569/*n*;
- in the class of $G_1$-oligomers, they increase real values in comparison with model values of the hyperbolic progression 18901/*n.*

Secondly, under triplet-deviations, real total amounts of 3*m*-plets from the classes of $T_1$-oligomers and $G_1$-oligomers belong correspondingly to another hyperbolic progressions 11979/*n* and 28788/*n*. These hyperbolic progressions are indicated by the blue dot line and the green dot line in Fig. 36. Where did these numerators of model hyperbolas come from? Each of these numerators is associated with the total amount of triplets ($n$ = 3) in an appropriate class of oligomers in this gene: the total amount of triplets starting with nucleotide T is equal to 3993, and the total amount of triplets starting with nucleotide G is equal to 9596. To calculate the first values of the model hyperbolas, each of these amounts of triplets must be tripled, giving the shown numerators 11979 and 28788.

Similar triplet-deviations exist in the OS-representations not only of the *TNT* gene but also of other long genes, prokaryotic genomes, and viruses in different degrees as the author has discovered in analysis of a limited set of nucleotide series by the OS-method. In the genetic code system, triplets have an important meaning, which differ them from other *n*-plets: they encode amino acids and punctuations of protein synthesis. One can believe that the phenomenon of the triplet-deviations is related with this special meaning of triplets. For this reason, the deeper analysis of triplet-deviations in different species can be useful to study secrets of the genetic system and biological evolution.

Fig. 36 demonstrated the highly regular rhythmic triplet-deviations for $n$ = 1, 2, 3, …, 20, but similar rhythmic triplet-deviations exist in a much wider range of values *n*.

Fig. 37 shows in graphical forms percentage values of the highly regular rhythmic deviations of the real total amounts of *n*-plets, which start with the nucleotide T and with the nucleotide G in the *TNT* gene, from the appropriate model values 19569/*n* and 18901/*n*. Two cases of the range of values *n* are represented there: $n$ = 1, 2, 3,…, 20, and $n$ = 1, 2, 3, .., 100.

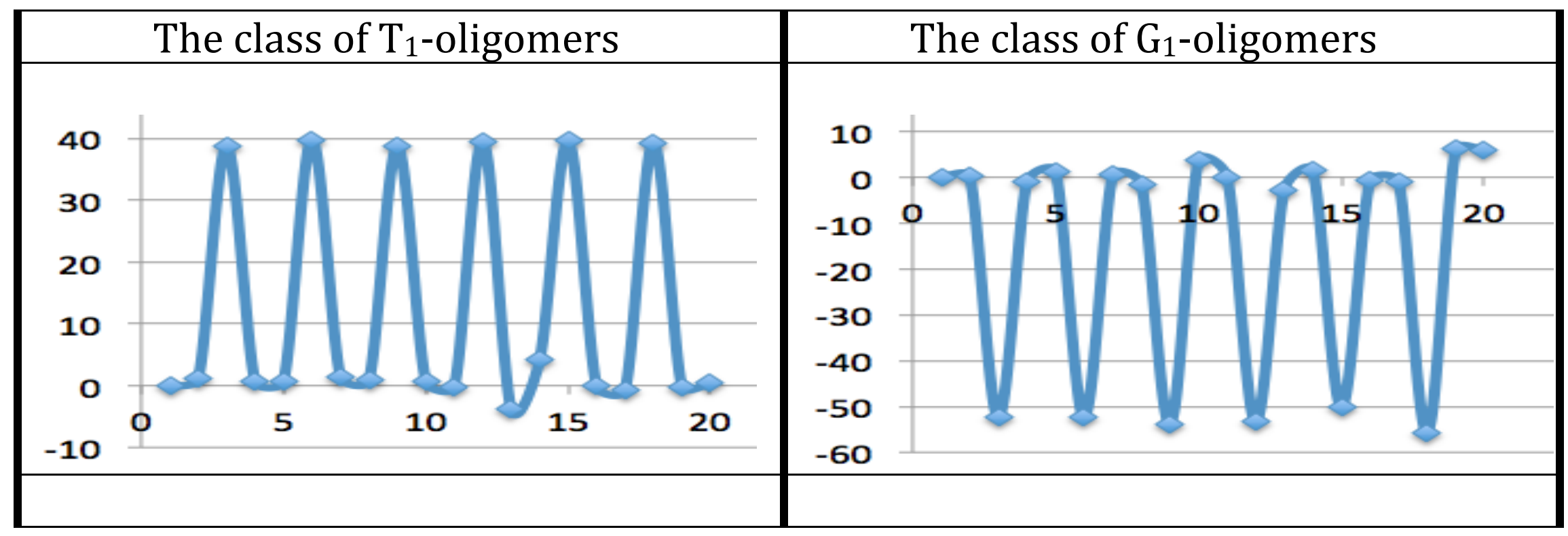

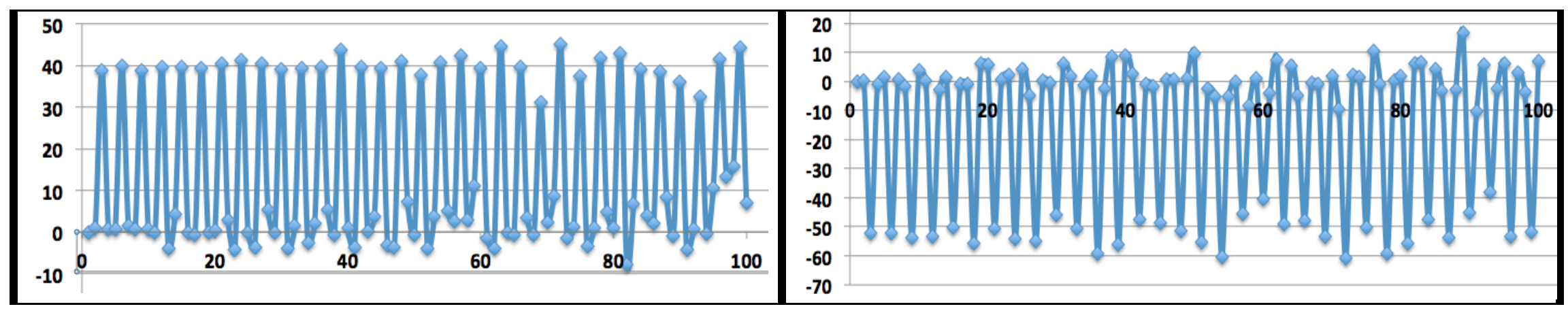

**Fig. 37.** Percentage representations of highly regular rhythmic sequences of the triplet-deviations of the real amounts of $n$-plets, which belong to classes of $T_1$-, and $G_1$-oligomers, from the appropriate model hyperbolic values $19569/n$ and $18901/n$ in the *TNT* gene. Here $n$ = 1, 2, 3, …, 20 (upper row) and $n$ = 1, 2, 3,…, 100 (bottom row) as shown at the abscissa axes. The ordinate axes show percentages of the deviations (the model values are taken as 100%).

The following Fig. 38 shows the OS-sequences of the total amounts of $n$-plets, which start with two other nucleotides A and C in the *TNT* gene. This gene contains 26373 nucleotides A and 17097 nucleotides C, that is $S_A$ = 26373 and $S_C$ = 17097 for the model hyperbolic progressions (39).

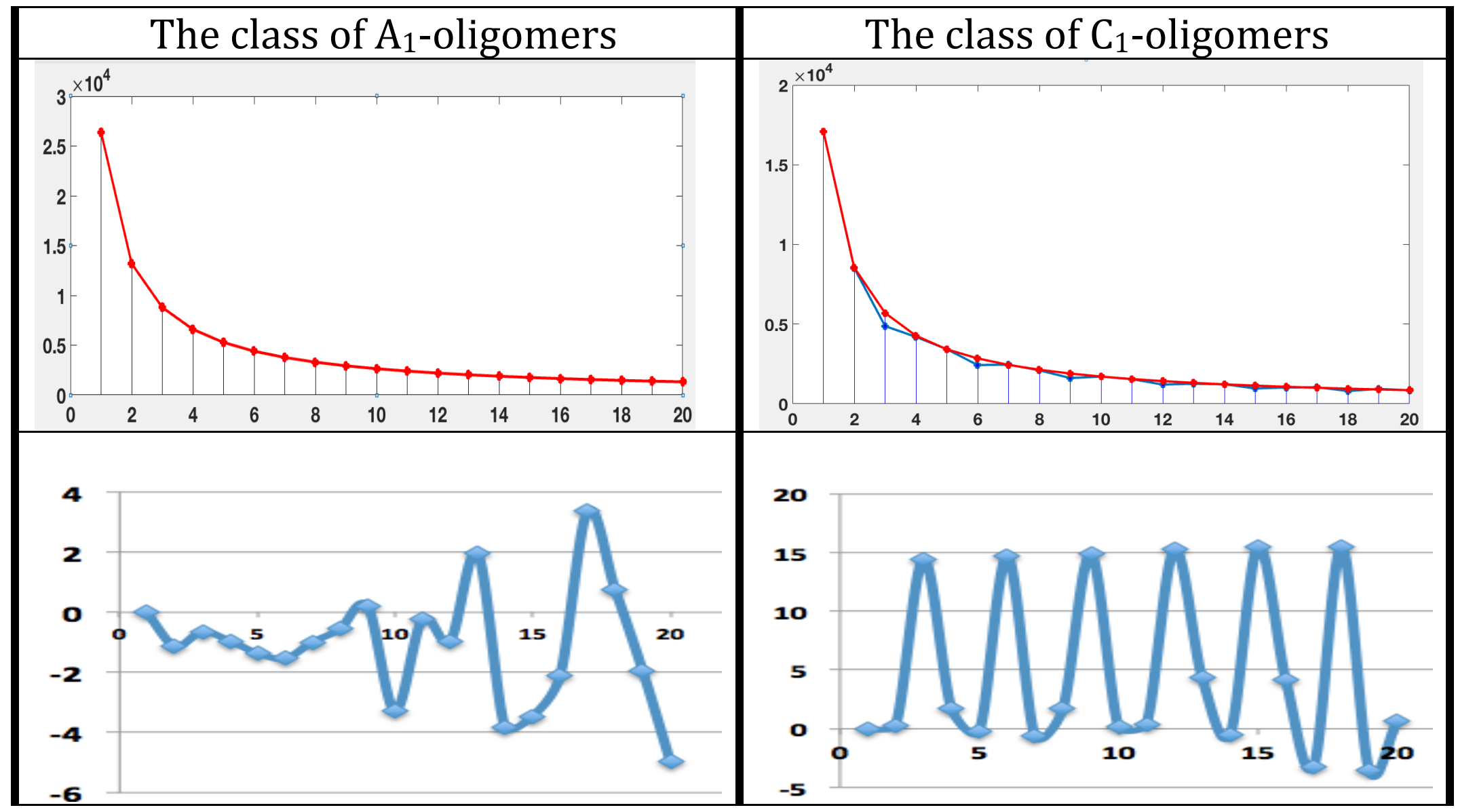

**Fig. 38.** Graphical representations of the results of analysis - by the oligomer sums method – of the human *TNT* gene regarding the sequences of the total amounts of $n$-plets, which start with the nucleotide A (left) and the nucleotide C (right). Here $n$ = 1, 2, 3,…, 20 (at the absciss axes). **Upper row**: the red lines refer to model hyperbolic progressions $S_A/n$ = $26373/n$ and $S_C/n$ =$17097/n$ correspondingly. The ordinate axes show the total amounts of appropriate $n$-plets. The class of oligomers starting with the nucleotide C has a regular sequences of the significant triplet-deviations at 3$m$-plets shown by the blue line. **Bottom row**: percentage representations of the sequences of deviations of the real total amounts of $n$-plets of these classes from the appropriate model hyperbolic values $26373/n$ and $17097/n$ (the ordinate axes show these percentages). The model values are taken as 100%.

One can see in Fig. 38 that the class of oligomers, which start with the nucleotide C, has a regular sequences of the significant triplet-deviations at 3*m*-plets shown by the blue line. The class of oligomers has not such regular sequences of significant deviations; in addition, its deviations are essentially less than deviations in the class of $C_1$-oligomers. In the class of $A_1$-oligomers, the real and model values differ little from each other, and therefore, in Fig. 38, the red line of model values covers the line of real values.

Fig. 39 shows numeric results of analysis of the *TNT* gene by the oligomer sums method.

| *n* | 1 | 2 | 3 | 4 | 5 | 6 | 7 | 8 | 9 | 10 |
|---|---|---|---|---|---|---|---|---|---|---|
| **A** | | | | | | | | | | |
| Real | 26373 | 13334 | 8848 | 6656 | 5346 | 4463 | 3805 | 3315 | 2924 | 2724 |
| Model | 26373 | 13187 | 8791 | 6593 | 5275 | 4396 | 3768 | 3297 | 2930 | 2637 |
| Δ% | 0 | -1.119 | -0.648 | -0.952 | -1.354 | -1.536 | -0.993 | -0.557 | 0.216 | -3.287 |
| **T** | | | | | | | | | | |
| Real | 19569 | 9677 | 3993 | 4857 | 3885 | 1964 | 2755 | 2426 | 1332 | 1943 |
| Model | 19569 | 9784.5 | 6523 | 4892 | 3914 | 3262 | 2796 | 2446 | 2174 | 1957 |
| Δ% | 0 | 1.099 | 38.786 | 0.721 | 0.736 | 39.782 | 1.451 | 0.823 | 38.740 | 0.710 |
| **C** | | | | | | | | | | |
| Real | 17097 | 8522 | 4876 | 4199 | 3426 | 2431 | 2458 | 2101 | 1617 | 1707 |
| Model | 17097 | 8549 | 5699 | 4274 | 3419 | 2850 | 2442 | 2137 | 1900 | 1710 |
| Δ% | 0 | 0.310 | 14.441 | 1.761 | -0.193 | 14.687 | -0.638 | 1.690 | 14.880 | 0.158 |
| **G** | | | | | | | | | | |
| Real | 18901 | 9437 | 9596 | 4773 | 3731 | 4798 | 2687 | 2400 | 3231 | 1820 |
| Model | 18901 | 9451 | 6300 | 4725 | 3780 | 3150 | 2700 | 2363 | 2100 | 1890 |
| Δ% | 0 | 0.143 | -52.309 | -1.011 | 1.302 | -52.309 | 0.487 | -1.582 | -53.849 | 3.709 |

| *n* | 11 | 12 | 13 | 14 | 15 | 16 | 17 | 18 | 19 | 20 |
|---|---|---|---|---|---|---|---|---|---|---|
| **A** | | | | | | | | | | |
| Real | 2403 | 2219 | 1989 | 1956 | 1819 | 1683 | 1499 | 1454 | 1415 | 1384 |
| Model | 2398 | 2198 | 2029 | 1884 | 1758 | 1648 | 1551 | 1465 | 1388 | 1319 |
| Δ% | -0.228 | -0.967 | 1.957 | -3.833 | -3.458 | -2.104 | 3.375 | 0.762 | -1.941 | -4.956 |
| **T** | | | | | | | | | | |
| Real | 1782 | 986 | 1563 | 1339 | 788 | 1224 | 1160 | 660 | 1032 | 974 |
| Model | 1779 | 1631 | 1505 | 1398 | 1305 | 1223 | 1151 | 1087 | 1030 | 978 |
| Δ% | -0.169 | 39.537 | -3.833 | 4.206 | 39.598 | -0.077 | -0.772 | 39.292 | -0.199 | 0.455 |
| **C** | | | | | | | | | | |
| Real | 1548 | 1207 | 1258 | 1227 | 963 | 1024 | 1038 | 803 | 932 | 849 |
| Model | 1554 | 1425 | 1315 | 1221 | 1140 | 1069 | 1006 | 950 | 900 | 855 |
| Δ% | 0.404 | 15.283 | 4.346 | -0.474 | 15.511 | 4.170 | -3.211 | 15.459 | -3.574 | 0.684 |
| **G** | | | | | | | | | | |
| Real | 1716 | 2416 | 1493 | 1330 | 1892 | 1190 | 1123 | 1635 | 933 | 890 |
| Model | 1718 | 1575 | 1454 | 1350 | 1260 | 1181 | 1112 | 1050 | 995 | 945 |
| Δ% | 0.132 | -53.389 | -2.688 | 1.487 | -50.151 | -0.735 | -1.005 | -55.706 | 6.211 | 5.825 |

**Fig. 39.** Real and model values to the OS-representations of the classes of $A_1$-, $T_1$-, $C_1$-, and $G_1$-oligomers in the human *TNT* gene are shown for $n$ = 1, 2, ..., 20. The real total amounts of *n*-plets, which start with a certain nucleotide (A, T, C, or G), are indicated jointly with their model values $H_{A,1}(n) = 26373/n$, $H_{T,1}(n) = 19569/n$, $H_{C,1}(n) = 17097/n$, and $H_{G,1}(n) = 18901/n$ (in red). The symbol Δ% refers to deviations of real values from model values in percent (the model values are taken as 100%).

Let us show briefly, for a comparison, also the OS-representations of another human *NEB* gene. Figs. 40 and 41 show graphs with the results of the *NEB* gene by the oligomer sums method. Initial data on this gene were taken from https://www.ncbi.nlm.nih.gov/nuccore/X83957. This gene contains 7071 nucleotides A, 4478 nucleotides T, 4578 nucleotides C, and 4754 nucleotides G, that is $S_A$ = 7071, $S_T$ = 4478, $S_C$ = 4578, and $S_G$ = 4754 for the model hyperbolic progressions (39). It can be especially noted that, in this gene, the amounts of nucleotides A and T are significantly different (7071 and 4478), that is, the second Chargaff's rule on their approximate equality in long sequences is not fulfilled here since this nucleotide sequence is not enough long for the Chargaff's rule.

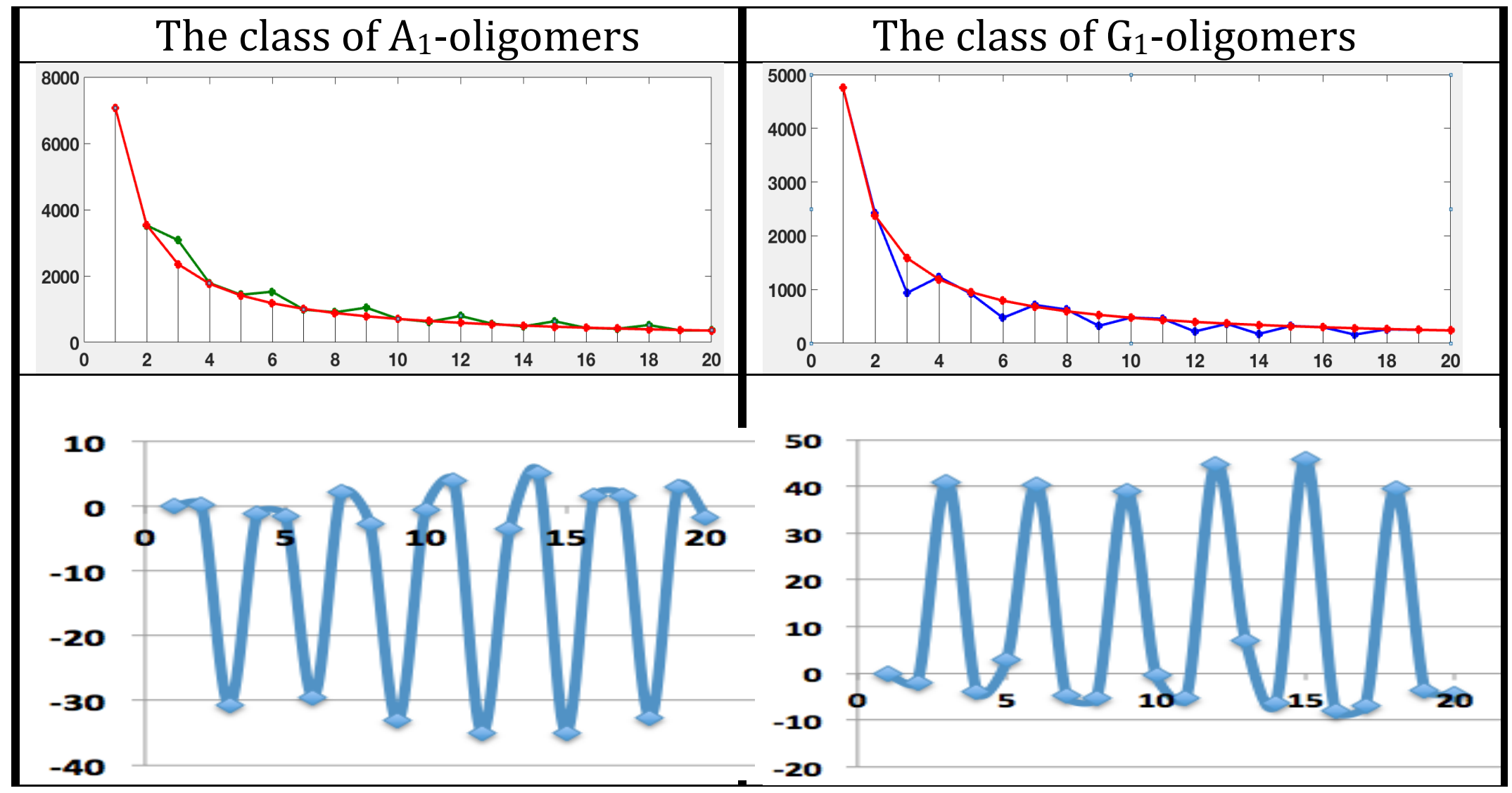

**Fig. 40.** Graphical representations of the results of analysis - by the oligomer sums method – of the human *NEB* gene: the sequences of the total amounts of *n*-plets, which start with the nucleotide A (left) and the nucleotide G (right) are shown. Here $n$ = 1, 2, 3,..., 20 (at the absciss axes). **Upper row**: the red lines refer to model hyperbolic progressions $S_A/n$ = 7071/*n* (left) and $S_G/n$ = 4754/*n* (at right). The ordinate axes show the total amounts of appropriate *n*-plets. The highly regular sequences of the significant triplet-deviations at 3*m*-plets shown by the green line (at left) and the blue line (right). **Bottom row**: percentage representations of the sequences of deviations of the real total amounts of *n*-plets of these classes from the appropriate model hyperbolic values 7071/*n* and 4754/*n* (the ordinate axes show these percentages). The model values are taken as 100%.

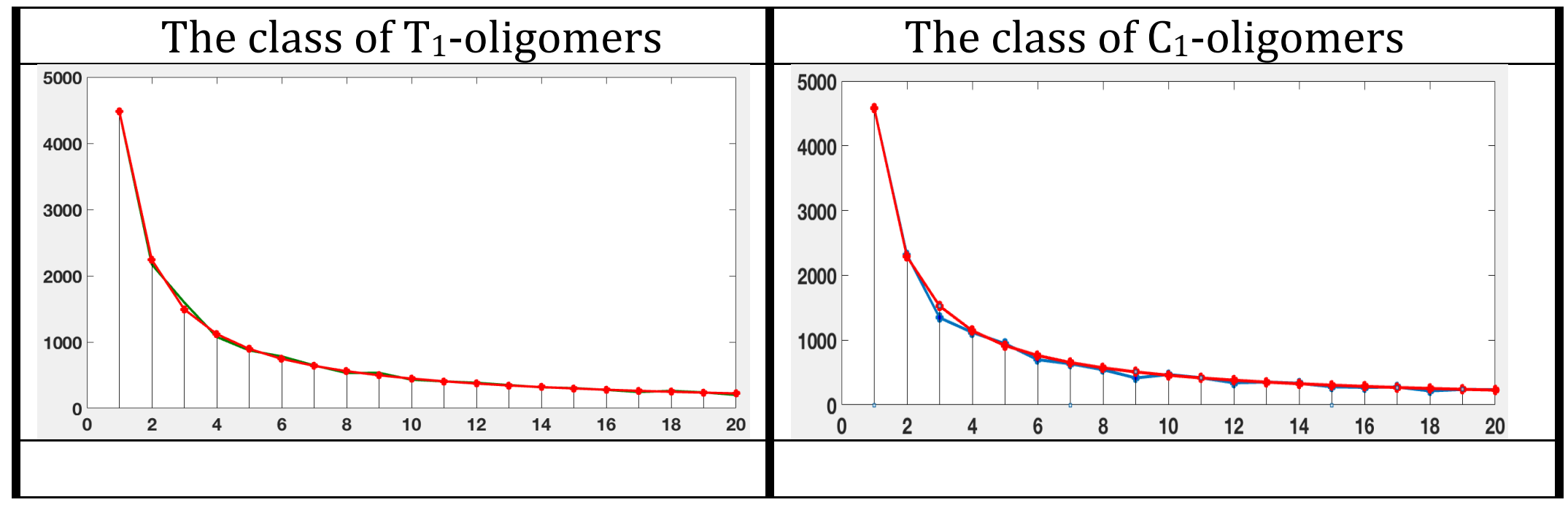

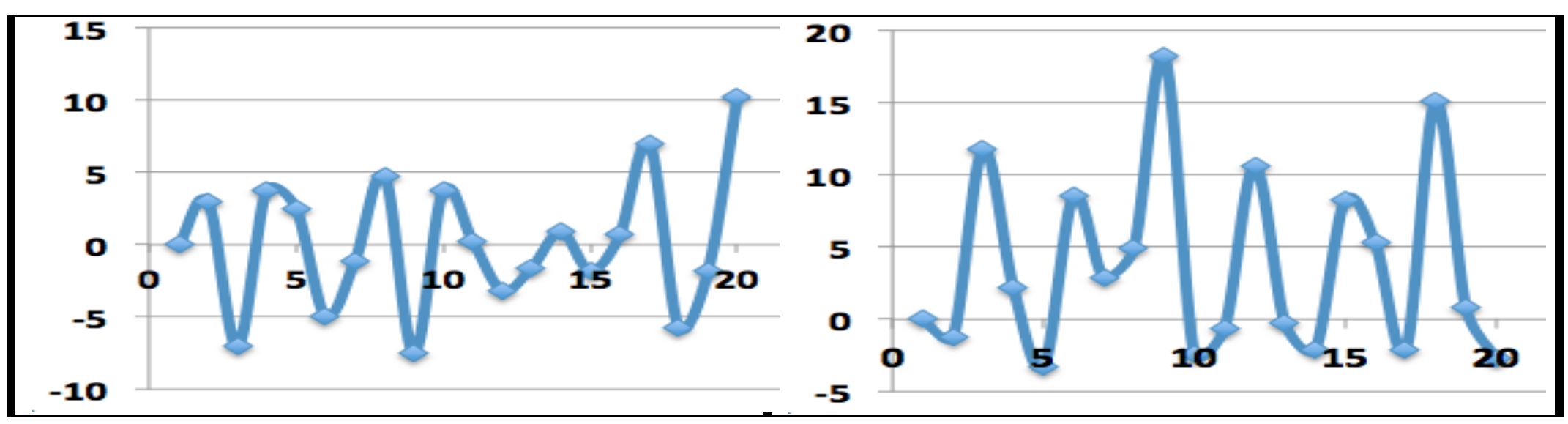

**Fig. 41.** Graphical representations of the results of analysis - by the oligomer sums method – of the human *NEB* gene: the sequences of the total amounts of *n*-plets, which start with the nucleotide T (left) and the nucleotide C (right) are shown. Here $n$ = 1, 2, 3,..., 20 (at the absciss axes). **Upper row**: the red lines refer to model hyperbolic progressions $S_T/n$ = 4478/$n$ (left) and $S_C/n$ = 4578/$n$ (right). The ordinate axes show the total amounts of appropriate *n*-plets. The weakly regular sequences of the significant triplet-deviations at 3*m*-plets shown by the green line (left) and the blue line (right). **Bottom row**: percentage representations of the sequences of deviations of the real total amounts of *n*-plets of these classes from the appropriate model hyperbolic values 4478/$n$ and 4578/$n$ (the ordinate axes show these percentages). The model values are taken as 100%.

In the author's laboratory, the oligomer sums method is now applied to systematic study of a wide list of eukaryotic and prokaryotic genomes, genes, viruses, bacteriophags and other DNA and RNA sequences. Results of this study will be published some later.

## 16. Some concluding remarks

The modern situation in the theoretic field of genetic informatics, where many millions of nucleotide sequences are described, can be characterized by the following citation: "*We are in the position of Johann Kepler when he first began looking for patterns in the volumes of data that Tycho Brahe had spent his life accumulating. We have the program that runs the cellular machinery, but we know very little about how to read it.*" [Fickett & Burks, 1989]. Kepler did not make his own astronomic observations, but he found – in the huge astronomic data of Tycho Brahe - his Kepler's laws of symmetric movements of planets relative to the Sun along ellipses. What are the hidden symmetries and rules in long DNA texts, in which the living nature for some reason records hereditary information about the most diverse organisms?

The Chargaff's parity rules played an important role in development of bioinformatics. We hope that the represented in this article rules of tetra-group symmetries in long texts of single stranded DNA will be also useful for further development of bioinformatics and theoretical biology.

The study of these tetra-group rules is now continued with including, in particular, relations of the tetra-groups of triplets with groups of amino acids and stop-codons, which are considered in many publications (see for example, [Dragovich, Dragovich, 2007; Dragovich, Khrennikov, Misic, 2017]).

In our opinion, characteristics of tetra-group symmetries of collective probabilities in long DNA-sequences and in sets of chromosomes of different organisms can be useful as new criteria of comparative analysis for taxonomic classification of biological organisms, for problems of evolutionary biology, etc.

The article [Trifonov, 1993] noted that the classical triplet code is not the only code carried by the sequences. DNA is a molecular language consisting of letters, words, and texts, carrying overlapping messages embodying multiple codes (as in our broadband communication channels of the Internet; on this, see additionally [Manin, Manin, 2017]). «*The nucleotide sequences are written in an unbroken manner. One way to detect "words" in such a continuous text is to evaluate the degree of internal correlation by calculating contrast values for the words. This technique allows one to derive vocabularies, which are species- and function-specific. The nucleotide sequences, thus, carry numerous superimposed messages. We do understand only a few of these messages while many more are waiting for their turn to be deciphered*» [Trifonov, 1993]. Our discover of rules in oligomer cooperative organization of long DNA sequences gives new materials to this topic about existence of many genetic codes of quantum-information types, about the geno-logic coding [Petoukhov, 2017a,b; Petoukhov, Petukhova, 2017a] and about the fractal grammar of DNA.

The presented study is a continuation of the author's researches on symmetries in biological objects described in his publications (see References below). This study further illustrates the effectiveness of symmetry analysis in natural systems. No wonder the theory of symmetries is one of the foundations of modern mathematical natural science. The presented results reveal the existence of a new broad class of symmetries in eukaryotic and prokaryotic genomes. The importance of these results were emphasized in the article "Petoukhov's rules on symmetries in long DNA-texts" [Darvas, 2018]. In this article, the head of the International Institute "Symmetrion" (Budapest, Hungary) proposed to launch a corresponding international project: "*Now, Petoukhov's above rules of symmetries are candidates for the role of universal rules of long DNA-texts in living bodies. Further researches are needed to determine the degree of universality of these rules. Taking into account the huge number of species and long DNA-texts to be tested in these relations, I propose to launch an international project to study these genetic symmetries. Symmetrion initiates and can take part as a centre of such an international project*" [Darvas, 2018].

## Appendix 1. Additional data about tetra-group rules and tetra-group symmetries in long DNA-texts

The Appendix 1 represents data about the fulfillment of the described tetra-group rules in a representative set of long sequences of oligonucleotides of single stranded DNA from GenBank. Kinds of studied sequences here are taken from articles of other authors to avoid a suspicion about a special choice of sequences. The results, presented below, confirm the fulfillment of the tetra-group rules and the existence of tetra-group symmetries in all the cases considered.

First of all, we represent results of analysis of tetra-group probabilities of 19 sequences from their list in the article about the second Chargaff's parity rule [Rapoport, Trifonov, 2012, p. 2]: "*Nucleotide disparities for prokaryotic coding sequences were taken from bacterial genomes of different groups both from Bacteria and Archea. All together 19 genomes were used: Aquifex aeolicus, Acidobacteria bacterium, Bradyrhizobium japonicum, Bacillus subtilis, Chlamydia trachomatis, Chromobacterium violaceum, Dehalococcoides ethenogenes, Escherichia coli, Flavobacterium psychrophilum, Gloeobacter violaceus, Helicobacter pilory, Methanosarcina acetivorans, Nanoarchaeum equitans, Syntrophus aciditrophicus, Streptomyces coelicolor, Sulfolobus solfataricus, Treponema denticola, Thermotoga maritima and Thermus thermophiles*". Fig. A1/1-A1/19 show results of analysis of tetra-group probabilities in these 19 DNA-sequences.

Then, Fig. A1/20-A1/35 show results of analysis of tetra-group symmetries in DNA-sequences, the length of which is which is approximately equal to 100000 nucleotides and more, from the article [Prahbu, 1993] devoted also to the second Chargaff's rule.

| NUCLEOTIDES | DOUBLETS | TRIPLETS | 4-PLETS | 5-PLETS |
|---|---|---|---|---|
| $P_1(A_1)$= 0,2841 | $P_2(A_1)$= 0,2834 | $P_3(A_1)$= 0,2820 | $P_4(A_1)$= 0,2834 | $P_5(A_1)$= 0,2843 |
| $P_1(T_1)$= 0,2811 | $P_2(T_1)$= 0,2818 | $P_3(T_1)$= 0,2815 | $P_4(T_1)$= 0,2820 | $P_5(T_1)$= 0,2821 |
| $P_1(C_1)$= 0,2168 | $P_2(C_1)$= 0,2173 | $P_3(C_1)$= 0,2167 | $P_4(C_1)$= 0,2170 | $P_5(C_1)$= 0,2160 |
| $P_1(G_1)$= 0,2179 | $P_2(G_1)$= 0,2176 | $P_3(G_1)$= 0,2198 | $P_4(G_1)$= 0,2174 | $P_5(G_1)$= 0,2178 |
| | $P_2(A_2)$= 0,2849 | $P_3(A_2)$= 0,2859 | $P_4(A_2)$= 0,2843 | $P_5(A_2)$= 0,2845 |
| | $P_2(T_2)$= 0,2805 | $P_3(T_2)$= 0,2827 | $P_4(T_2)$= 0,2807 | $P_5(T_2)$= 0,2814 |
| | $P_2(C_2)$= 0,2163 | $P_3(C_2)$= 0,2156 | $P_4(C_2)$= 0,2161 | $P_5(C_2)$= 0,2161 |
| | $P_2(G_2)$= 0,2183 | $P_3(G_2)$= 0,2157 | $P_4(G_2)$= 0,2188 | $P_5(G_2)$= 0,2180 |
| | | $P_3(A_3)$= 0,2845 | $P_4(A_3)$= 0,2831 | $P_5(A_3)$= 0,2844 |
| | | $P_3(T_3)$= 0,2791 | $P_4(T_3)$= 0,2815 | $P_5(T_3)$= 0,2806 |
| | | $P_3(C_3)$= 0,2181 | $P_4(C_3)$= 0,2176 | $P_5(C_3)$= 0,2171 |
| | | $P_3(G_3)$= 0,2183 | $P_4(G_3)$= 0,2177 | $P_5(G_3)$= 0,2178 |
| | | | $P_4(A_4)$= 0,2855 | $P_5(A_4)$= 0,2840 |
| | | | $P_4(T_4)$= 0,2802 | $P_5(T_4)$= 0,2804 |
| | | | $P_4(C_4)$= 0,2165 | $P_5(C_4)$= 0,2174 |
| | | | $P_4(G_4)$= 0,2178 | $P_5(G_4)$= 0,2182 |
| | | | | $P_5(A_5)$= 0,2835 |
| | | | | $P_5(T_5)$= 0,2810 |
| | | | | $P_5(C_5)$= 0,2177 |
| | | | | $P_5(G_5)$= 0,2180 |

Fig. A1/1. Probabilities of subgroups of tetra-groups in the sequence: Aquifex aeolicus VF5, complete genome
https://www.ncbi.nlm.nih.gov/nuccore/AE000657.1?report=fasta ;
https://www.ncbi.nlm.nih.gov/nuccore/AE000657.1?report=genbank :
LOCUS AE000657, 1551335 bp DNA circular BCT 30-JAN-2014 DEFINITION Aquifex aeolicus VF5, complete genome. ACCESSION AE000657 AE000669-AE000777 VERSION AE000657.1

| NUCLEOTIDES | DOUBLETS | TRIPLETS | 4-PLETS | 5-PLETS |
|---|---|---|---|---|
| $P_1(A_1)$= 0,2155 | $P_2(A_1)$= 0,2156 | $P_3(A_1)$= 0,2165 | $P_4(A_1)$= 0,2157 | $P_5(A_1)$= 0,2159 |
| $P_1(T_1)$= 0,2171 | $P_2(T_1)$= 0,2169 | $P_3(T_1)$= 0,2167 | $P_4(T_1)$= 0,2167 | $P_5(T_1)$= 0,2171 |

| $P_1(C_1)=$ 0,2855 | $P_2(C_1)=$ 0,2856 | $P_3(C_1)=$ 0,2836 | $P_4(C_1)=$ 0,2856 | $P_5(C_1)=$ 0,2850 |
|---|---|---|---|---|
| $P_1(G_1)=$ 0,2819 | $P_2(G_1)=$ 0,2819 | $P_3(G_1)=$ 0,2833 | $P_4(G_1)=$ 0,2820 | $P_5(G_1)=$ 0,2821 |
| | $P_2(A_2)=$ 0,2153 | $P_3(A_2)=$ 0,2154 | $P_4(A_2)=$ 0,2155 | $P_5(A_2)=$ 0,2153 |
| | $P_2(T_2)=$ 0,2173 | $P_3(T_2)=$ 0,2186 | $P_4(T_2)=$ 0,2171 | $P_5(T_2)=$ 0,2172 |
| | $P_2(C_2)=$ 0,2855 | $P_3(C_2)=$ 0,2863 | $P_4(C_2)=$ 0,2853 | $P_5(C_2)=$ 0,2855 |
| | $P_2(G_2)=$ 0,2818 | $P_3(G_2)=$ 0,2798 | $P_4(G_2)=$ 0,2821 | $P_5(G_2)=$ 0,2820 |
| | | $P_3(A_3)=$ 0,2146 | $P_4(A_3)=$ 0,2155 | $P_5(A_3)=$ 0,2159 |
| | | $P_3(T_3)=$ 0,2161 | $P_4(T_3)=$ 0,2171 | $P_5(T_3)=$ 0,2168 |
| | | $P_3(C_3)=$ 0,2868 | $P_4(C_3)=$ 0,2855 | $P_5(C_3)=$ 0,2855 |
| | | $P_3(G_3)=$ 0,2825 | $P_4(G_3)=$ 0,2819 | $P_5(G_3)=$ 0,2818 |
| | | | $P_4(A_4)=$ 0,2151 | $P_5(A_4)=$ 0,2154 |
| | | | $P_4(T_4)=$ 0,2175 | $P_5(T_4)=$ 0,2176 |
| | | | $P_4(C_4)=$ 0,2858 | $P_5(C_4)=$ 0,2858 |
| | | | $P_4(G_4)=$ 0,2815 | $P_5(G_4)=$ 0,2812 |
| | | | | $P_5(A_5)=$ 0,2149 |
| | | | | $P_5(T_5)=$ 0,21694 |
| | | | | $P_5(C_5)=$ 0,2859 |
| | | | | $P_5(G_5)=$ 0,2823 |

Fig. A1/2. Probabilities of subgroups of tetra-groups in the sequence: Acidobacteria bacterium KBS 146 M015DRAFT_scf7180000000004_quiver.1_C, whole genome shotgun sequence, GenBank: JHVA01000001.1, - https://www.ncbi.nlm.nih.gov/nuccore/JHVA01000001.1?report=fasta

https://www.ncbi.nlm.nih.gov/nuccore/JHVA01000001.1?report=genbank : LOCUS JHVA01000001, 4996384 bp , DNA linear BCT 08-APR-2014 DEFINITION Acidobacteria bacterium KBS 146 M015DRAFT_scf7180000000004_quiver.1_C, whole genome shotgun sequence. ACCESSION JHVA01000001 JHVA01000000 VERSION JHVA01000001.

| **NUCLEOTIDES** | **DOUBLETS** | **TRIPLETS** | **4-PLETS** | **5-PLETS** |
|---|---|---|---|---|
| $P_1(A_1)=$ 0,1819 | $P_2(A_1)=$ 0,1820 | $P_3(A_1)=$ 0,1806 | $P_4(A_1)=$ 0,1816 | $P_5(A_1)=$ 0,1814 |
| $P_1(T_1)=$ 0,1815 | $P_2(T_1)=$ 0,1814 | $P_3(T_1)=$ 0,1797 | $P_4(T_1)=$ 0,1817 | $P_5(T_1)=$ 0,1818 |
| $P_1(C_1)=$ 0,3184 | $P_2(C_1)=$ 0,3185 | $P_3(C_1)=$ 0,3197 | $P_4(C_1)=$ 0,3184 | $P_5(C_1)=$ 0,3182 |
| $P_1(G_1)=$ 0,3182 | $P_2(G_1)=$ 0,3182 | $P_3(G_1)=$ 0,3201 | $P_4(G_1)=$ 0,3183 | $P_5(G_1)=$ 0,3186 |
| | $P_2(A_2)=$ 0,1818 | $P_3(A_2)=$ 0,1836 | $P_4(A_2)=$ 0,1816 | $P_5(A_2)=$ 0,1819 |
| | $P_2(T_2)=$ 0,1816 | $P_3(T_2)=$ 0,1825 | $P_4(T_2)=$ 0,1815 | $P_5(T_2)=$ 0,1815 |
| | $P_2(C_2)=$ 0,3183 | $P_3(C_2)=$ 0,3166 | $P_4(C_2)=$ 0,3184 | $P_5(C_2)=$ 0,3186 |
| | $P_2(G_2)=$ 0,3182 | $P_3(G_2)=$ 0,3173 | $P_4(G_2)=$ 0,3185 | $P_5(G_2)=$ 0,3179 |
| | | $P_3(A_3)=$ 0,1815 | $P_4(A_3)=$ 0,1823 | $P_5(A_3)=$ 0,1820 |
| | | $P_3(T_3)=$ 0,1824 | $P_4(T_3)=$ 0,1811 | $P_5(T_3)=$ 0,1815 |
| | | $P_3(C_3)=$ 0,3189 | $P_4(C_3)=$ 0,3185 | $P_5(C_3)=$ 0,3186 |
| | | $P_3(G_3)=$ 0,3173 | $P_4(G_3)=$ 0,3181 | $P_5(G_3)=$ 0,3180 |
| | | | $P_4(A_4)=$ 0,1820 | $P_5(A_4)=$ 0,1817 |
| | | | $P_4(T_4)=$ 0,1818 | $P_5(T_4)=$ 0,1815 |
| | | | $P_4(C_4)=$ 0,3182 | $P_5(C_4)=$ 0,3183 |
| | | | $P_4(G_4)=$ 0,3180 | $P_5(G_4)=$ 0,3185 |
| | | | | $P_5(A_5)=$ 0,1824 |
| | | | | $P_5(T_5)=$ 0,1814 |
| | | | | $P_5(C_5)=$ 0,3183 |
| | | | | $P_5(G_5)=$ 0,3180 |

Fig. A1/3. Probabilities of subgroups of tetra-groups in the sequence: Bradyrhizobium japonicum strain E109, complete genome

GenBank: CP010313.1,

https://www.ncbi.nlm.nih.gov/nuccore/CP010313.1?report=fasta

https://www.ncbi.nlm.nih.gov/nuccore/CP010313.1?report=genbank :

LOCUS CP010313 , 9224208 bp, DNA circular BCT 24-FEB-2015

DEFINITION Bradyrhizobium japonicum strain E109, complete genome.

ACCESSION CP010313 VERSION CP010313.1

| NUCLEOTIDES | DOUBLETS | TRIPLETS | 4-PLETS | 5-PLETS |
|---|---|---|---|---|
| $P_1(A_1)$= 0,2805 | $P_2(A_1)$= 0,2802 | $P_3(A_1)$= 0,2813 | $P_4(A_1)$= 0,2801 | $P_5(A_1)$= 0,2800 |
| $P_1(T_1)$= 0,2806 | $P_2(T_1)$= 0,2806 | $P_3(T_1)$= 0,2809 | $P_4(T_1)$= 0,2807 | $P_5(T_1)$= 0,2805 |
| $P_1(C_1)$= 0,2192 | $P_2(C_1)$= 0,2196 | $P_3(C_1)$= 0,2181 | $P_4(C_1)$= 0,2195 | $P_5(C_1)$= 0,2197 |
| $P_1(G_1)$= 0,2197 | $P_2(G_1)$= 0,2197 | $P_3(G_1)$= 0,2196 | $P_4(G_1)$= 0,2198 | $P_5(G_1)$= 0,2198 |
| | $P_2(A_2)$= 0,2809 | $P_3(A_2)$= 0,2795 | $P_4(A_2)$= 0,2811 | $P_5(A_2)$= 0,2805 |
| | $P_2(T_2)$= 0,2805 | $P_3(T_2)$= 0,2803 | $P_4(T_2)$= 0,2801 | $P_5(T_2)$= 0,2807 |
| | $P_2(C_2)$= 0,2189 | $P_3(C_2)$= 0,2196 | $P_4(C_2)$= 0,2190 | $P_5(C_2)$= 0,2188 |
| | $P_2(G_2)$= 0,2197 | $P_3(G_2)$= 0,2206 | $P_4(G_2)$= 0,2198 | $P_5(G_2)$= 0,2200 |
| | | $P_3(A_3)$= 0,2807 | $P_4(A_3)$= 0,2802 | $P_5(A_3)$= 0,2811 |
| | | $P_3(T_3)$= 0,2805 | $P_4(T_3)$= 0,2806 | $P_5(T_3)$= 0,2805 |
| | | $P_3(C_3)$= 0,2200 | $P_4(C_3)$= 0,2197 | $P_5(C_3)$= 0,2195 |
| | | $P_3(G_3)$= 0,2189 | $P_4(G_3)$= 0,2196 | $P_5(G_3)$= 0,2190 |
| | | | $P_4(A_4)$= 0,2806 | $P_5(A_4)$= 0,2804 |
| | | | $P_4(T_4)$= 0,2810 | $P_5(T_4)$= 0,2812 |
| | | | $P_4(C_4)$= 0,2188 | $P_5(C_4)$= 0,2190 |
| | | | $P_4(G_4)$= 0,2195 | $P_5(G_4)$= 0,2194 |
| | | | | $P_5(A_5)$= 0,2805 |
| | | | | $P_5(T_5)$= 0,2800 |
| | | | | $P_5(C_5)$= 0,2193 |
| | | | | $P_5(G_5)$= 0,2201 |

Fig. A1/4. Probabilities of subgroups of tetra-groups in the sequence:

Bacillus subtilis strain UD1022, complete genome, GenBank: CP011534.1,

https://www.ncbi.nlm.nih.gov/nuccore/CP011534.1?report=fasta

https://www.ncbi.nlm.nih.gov/nuccore/CP011534.1?report=genbank : LOCUS CP011534 , 4025326 bp , DNA circular BCT 18-FEB-2016 DEFINITION Bacillus subtilis strain UD1022, complete genome. ACCESSION CP011534 VERSION CP011534.1

| NUCLEOTIDES | DOUBLETS | TRIPLETS | 4-PLETS | 5-PLETS |
|---|---|---|---|---|
| $P_1(A_1)$= 0,2942 | $P_2(A_1)$= 0,2939 | $P_3(A_1)$= 0,2927 | $P_4(A_1)$= 0,2939 | $P_5(A_1)$= 0,2933 |
| $P_1(T_1)$= 0,2930 | $P_2(T_1)$= 0,2931 | $P_3(T_1)$= 0,2921 | $P_4(T_1)$= 0,2933 | $P_5(T_1)$= 0,2932 |
| $P_1(C_1)$= 0,2067 | $P_2(C_1)$= 0,2066 | $P_3(C_1)$= 0,2074 | $P_4(C_1)$= 0,2059 | $P_5(C_1)$= 0,2066 |
| $P_1(G_1)$= 0,2061 | $P_2(G_1)$= 0,2063 | $P_3(G_1)$= 0,2078 | $P_4(G_1)$= 0,2069 | $P_5(G_1)$= 0,2068 |
| | $P_2(A_2)$= 0,2945 | $P_3(A_2)$= 0,2963 | $P_4(A_2)$= 0,2941 | $P_5(A_2)$= 0,2937 |
| | $P_2(T_2)$= 0,2930 | $P_3(T_2)$= 0,2935 | $P_4(T_2)$= 0,2922 | $P_5(T_2)$= 0,2937 |
| | $P_2(C_2)$= 0,2067 | $P_3(C_2)$= 0,2049 | $P_4(C_2)$= 0,2076 | $P_5(C_2)$= 0,2062 |
| | $P_2(G_2)$= 0,2058 | $P_3(G_2)$= 0,2053 | $P_4(G_2)$= 0,2061 | $P_5(G_2)$= 0,2064 |
| | | $P_3(A_3)$= 0,2936 | $P_4(A_3)$= 0,2939 | $P_5(A_3)$= 0,2943 |
| | | $P_3(T_3)$= 0,2935 | $P_4(T_3)$= 0,2930 | $P_5(T_3)$= 0,2932 |
| | | $P_3(C_3)$= 0,2077 | $P_4(C_3)$= 0,2073 | $P_5(C_3)$= 0,2070 |
| | | $P_3(G_3)$= 0,2051 | $P_4(G_3)$= 0,2058 | $P_5(G_3)$= 0,2055 |

| | | | |
|---|---|---|---|
| $P_4(A_4)$= | 0,2948 | $P_5(A_4)$= | 0,2943 |
| $P_4(T_4)$= | 0,2938 | $P_5(T_4)$= | 0,2925 |
| $P_4(C_4)$= | 0,2059 | $P_5(C_4)$= | 0,20743 |
| $P_4(G_4)$= | 0,2056 | $P_5(G_4)$= | 0,2058 |
| | | $P_5(A_5)$= | 0,2953 |
| | | $P_5(T_5)$= | 0,2926 |
| | | $P_5(C_5)$= | 0,2061 |
| | | $P_5(G_5)$= | 0,2059 |

Fig. A1/5. Probabilities of subgroups of tetra-groups in the sequence:

Chlamydia trachomatis strain QH111L, complete genome

GenBank: CP018052.1,

https://www.ncbi.nlm.nih.gov/nuccore/CP018052.1?report=fasta

https://www.ncbi.nlm.nih.gov/nuccore/CP018052.1?report=genbank :

LOCUS CP018052 , 1025839 bp , DNA circular BCT 22-NOV-2016

DEFINITION Chlamydia trachomatis strain QH111L, complete genome.

ACCESSION CP018052 VERSION CP018052.1

| **NUCLEOTIDES** | | **DOUBLETS** | | **TRIPLETS** | | **4-PLETS** | | **5-PLETS** | |
|---|---|---|---|---|---|---|---|---|---|
| $P_1(A_1)$= | 0,1670 | $P_2(A_1)$= | 0,1658 | $P_3(A_1)$= | 0,1758 | $P_4(A_1)$= | 0,1640 | $P_5(A_1)$= | 0,1678 |
| $P_1(T_1)$= | 0,1819 | $P_2(T_1)$= | 0,1812 | $P_3(T_1)$= | 0,1883 | $P_4(T_1)$= | 0,1809 | $P_5(T_1)$= | 0,1823 |
| $P_1(C_1)$= | 0,3049 | $P_2(C_1)$= | 0,3044 | $P_3(C_1)$= | 0,3009 | $P_4(C_1)$= | 0,3042 | $P_5(C_1)$= | 0,3055 |
| $P_1(G_1)$= | 0,3461 | $P_2(G_1)$= | 0,3486 | $P_3(G_1)$= | 0,3350 | $P_4(G_1)$= | 0,3509 | $P_5(G_1)$= | 0,3444 |
| | | $P_2(A_2)$= | 0,1682 | $P_3(A_2)$= | 0,1651 | $P_4(A_2)$= | 0,1683 | $P_5(A_2)$= | 0,1679 |
| | | $P_2(T_2)$= | 0,1827 | $P_3(T_2)$= | 0,1808 | $P_4(T_2)$= | 0,1805 | $P_5(T_2)$= | 0,1850 |
| | | $P_2(C_2)$= | 0,3055 | $P_3(C_2)$= | 0,3112 | $P_4(C_2)$= | 0,3062 | $P_5(C_2)$= | 0,2997 |
| | | $P_2(G_2)$= | 0,3437 | $P_3(G_2)$= | 0,3429 | $P_4(G_2)$= | 0,3450 | $P_5(G_2)$= | 0,3474 |
| | | | | $P_3(A_3)$= | 0,1602 | $P_4(A_3)$= | 0,1677 | $P_5(A_3)$= | 0,1697 |
| | | | | $P_3(T_3)$= | 0,1766 | $P_4(T_3)$= | 0,1814 | $P_5(T_3)$= | 0,1810 |
| | | | | $P_3(C_3)$= | 0,3027 | $P_4(C_3)$= | 0,3046 | $P_5(C_3)$= | 0,3044 |
| | | | | $P_3(G_3)$= | 0,3606 | $P_4(G_3)$= | 0,3463 | $P_5(G_3)$= | 0,3453 |
| | | | | | | $P_4(A_4)$= | 0,1681 | $P_5(A_4)$= | 0,1674 |
| | | | | | | $P_4(T_4)$= | 0,1849 | $P_5(T_4)$= | 0,1816 |
| | | | | | | $P_4(C_4)$= | 0,3047 | $P_5(C_4)$= | 0,3048 |
| | | | | | | $P_4(G_4)$= | 0,3424 | $P_5(G_4)$= | 0,3463 |
| | | | | | | | | $P_5(A_5)$= | 0,1627 |
| | | | | | | | | $P_5(T_5)$= | 0,1797 |
| | | | | | | | | $P_5(C_5)$= | 0,3103 |
| | | | | | | | | $P_5(G_5)$= | 0,3472 |

Fig. A1/6. Probabilities of subgroups of tetra-groups in the sequence:

Chromobacterium violaceum strain LK30 1, whole genome shotgun sequence

GenBank: LDUX01000001.1,

https://www.ncbi.nlm.nih.gov/nuccore/LDUX01000001.1?report=fasta

https://www.ncbi.nlm.nih.gov/nuccore/LDUX01000001.1?report=genbank :

LOCUS LDUX01000001, 127377 bp , DNA linear BCT 30-JUN-2015

DEFINITION Chromobacterium violaceum strain LK30 1, whole genome shotgun sequence. ACCESSION LDUX01000001 LDUX01000000 VERSION LDUX01000001.1

| **NUCLEOTIDES** | | **DOUBLETS** | | **TRIPLETS** | | **4-PLETS** | | **5-PLETS** | |
|---|---|---|---|---|---|---|---|---|---|
| $P_1(A_1)$= | 0,2664 | $P_2(A_1)$= | 0,2665 | $P_3(A_1)$= | 0,2706 | $P_4(A_1)$= | 0,2668 | $P_5(A_1)$= | 0,2673 |
| $P_1(T_1)$= | 0,2645 | $P_2(T_1)$= | 0,2647 | $P_3(T_1)$= | 0,2653 | $P_4(T_1)$= | 0,2649 | $P_5(T_1)$= | 0,2638 |
| $P_1(C_1)$= | 0,2338 | $P_2(C_1)$= | 0,2337 | $P_3(C_1)$= | 0,2295 | $P_4(C_1)$= | 0,2337 | $P_5(C_1)$= | 0,2340 |
| $P_1(G_1)$= | 0,2353 | $P_2(G_1)$= | 0,2351 | $P_3(G_1)$= | 0,2346 | $P_4(G_1)$= | 0,2346 | $P_5(G_1)$= | 0,2349 |
| | | $P_2(A_2)$= | 0,2663 | $P_3(A_2)$= | 0,2617 | $P_4(A_2)$= | 0,2659 | $P_5(A_2)$= | 0,2662 |
| | | $P_2(T_2)$= | 0,2643 | $P_3(T_2)$= | 0,2686 | $P_4(T_2)$= | 0,2642 | $P_5(T_2)$= | 0,2648 |
| | | $P_2(C_2)$= | 0,2339 | $P_3(C_2)$= | 0,2376 | $P_4(C_2)$= | 0,2335 | $P_5(C_2)$= | 0,2336 |
| | | $P_2(G_2)$= | 0,2355 | $P_3(G_2)$= | 0,2321 | $P_4(G_2)$= | 0,2364 | $P_5(G_2)$= | 0,2354 |
| | | | | $P_3(A_3)$= | 0,2668 | $P_4(A_3)$= | 0,2661 | $P_5(A_3)$= | 0,2659 |
| | | | | $P_3(T_3)$= | 0,2596 | $P_4(T_3)$= | 0,2644 | $P_5(T_3)$= | 0,2651 |
| | | | | $P_3(C_3)$= | 0,2343 | $P_4(C_3)$= | 0,2338 | $P_5(C_3)$= | 0,2339 |
| | | | | $P_3(G_3)$= | 0,2396 | $P_4(G_3)$= | 0,2356 | $P_5(G_3)$= | 0,2350 |
| | | | | | | $P_4(A_4)$= | 0,2666 | $P_5(A_4)$= | 0,2664 |
| | | | | | | $P_4(T_4)$= | 0,2644 | $P_5(T_4)$= | 0,2649 |
| | | | | | | $P_4(C_4)$= | 0,2342 | $P_5(C_4)$= | 0,2328 |
| | | | | | | $P_4(G_4)$= | 0,2347 | $P_5(G_4)$= | 0,2360 |
| | | | | | | | | $P_5(A_5)$= | 0,2660 |
| | | | | | | | | $P_5(T_5)$= | 0,2639 |
| | | | | | | | | $P_5(C_5)$= | 0,2346 |
| | | | | | | | | $P_5(G_5)$= | 0,2354 |

Fig. A1/7. Probabilities of subgroups of tetra-groups in the sequence:

Dehalococcoides mccartyi strain CG3, complete genome

NCBI Reference Sequence: NZ_CP013074.1,

https://www.ncbi.nlm.nih.gov/nuccore/NZ_CP013074.1?report=fasta

https://www.ncbi.nlm.nih.gov/nuccore/NZ_CP013074.1?report=genbank :

LOCUS NZ_CP013074 1521287 bp DNA circular CON 06-APR-2017

DEFINITION Dehalococcoides mccartyi strain CG3, complete genome.

ACCESSION NZ_CP013074 VERSION NZ_CP013074.1

| **NUCLEOTIDES** | | **DOUBLETS** | | **TRIPLETS** | | **4-PLETS** | | **5-PLETS** | |
|---|---|---|---|---|---|---|---|---|---|
| $P_1(A_1)$= | 0,2480 | $P_2(A_1)$= | 0,2481 | $P_3(A_1)$= | 0,2479 | $P_4(A_1)$= | 0,2479 | $P_5(A_1)$= | 0,2483 |
| $P_1(T_1)$= | 0,2472 | $P_2(T_1)$= | 0,2470 | $P_3(T_1)$= | 0,2489 | $P_4(T_1)$= | 0,2472 | $P_5(T_1)$= | 0,2472 |
| $P_1(C_1)$= | 0,2526 | $P_2(C_1)$= | 0,2523 | $P_3(C_1)$= | 0,2519 | $P_4(C_1)$= | 0,2522 | $P_5(C_1)$= | 0,2529 |
| $P_1(G_1)$= | 0,2522 | $P_2(G_1)$= | 0,2526 | $P_3(G_1)$= | 0,2514 | $P_4(G_1)$= | 0,2526 | $P_5(G_1)$= | 0,2516 |
| | | $P_2(A_2)$= | 0,2480 | $P_3(A_2)$= | 0,2468 | $P_4(A_2)$= | 0,2474 | $P_5(A_2)$= | 0,2479 |
| | | $P_2(T_2)$= | 0,2474 | $P_3(T_2)$= | 0,2461 | $P_4(T_2)$= | 0,2477 | $P_5(T_2)$= | 0,2475 |
| | | $P_2(C_2)$= | 0,2528 | $P_3(C_2)$= | 0,2544 | $P_4(C_2)$= | 0,2530 | $P_5(C_2)$= | 0,2522 |
| | | $P_2(G_2)$= | 0,2518 | $P_3(G_2)$= | 0,2527 | $P_4(G_2)$= | 0,2519 | $P_5(G_2)$= | 0,2523 |
| | | | | $P_3(A_3)$= | 0,2494 | $P_4(A_3)$= | 0,2483 | $P_5(A_3)$= | 0,2485 |
| | | | | $P_3(T_3)$= | 0,2466 | $P_4(T_3)$= | 0,2467 | $P_5(T_3)$= | 0,2470 |
| | | | | $P_3(C_3)$= | 0,2515 | $P_4(C_3)$= | 0,2525 | $P_5(C_3)$= | 0,2527 |
| | | | | $P_3(G_3)$= | 0,2525 | $P_4(G_3)$= | 0,2526 | $P_5(G_3)$= | 0,2518 |
| | | | | | | $P_4(A_4)$= | 0,2486 | $P_5(A_4)$= | 0,2479 |
| | | | | | | $P_4(T_4)$= | 0,2472 | $P_5(T_4)$= | 0,2472 |
| | | | | | | $P_4(C_4)$= | 0,2527 | $P_5(C_4)$= | 0,2523 |
| | | | | | | $P_4(G_4)$= | 0,2516 | $P_5(G_4)$= | 0,2526 |
| | | | | | | | | $P_5(A_5)$= | 0,2475 |
| | | | | | | | | $P_5(T_5)$= | 0,2471 |
| | | | | | | | | $P_5(C_5)$= | 0,2529 |
| | | | | | | | | $P_5(G_5)$= | 0,2525 |

Fig. A1/8. Probabilities of subgroups of tetra-groups in the sequence:

Escherichia coli CFT073, complete genome, GenBank: AE014075.1,

https://www.ncbi.nlm.nih.gov/nuccore/AE014075.1?report=fasta

https://www.ncbi.nlm.nih.gov/nuccore/AE014075.1?report=genbank:

LOCUS AE014075 5231428 bp DNA circular BCT 31-JAN-2014
DEFINITION Escherichia coli CFT073, complete genome. ACCESSION
AE014075 AE016755 AE016756 AE016757 AE016758 AE016759 AE016760
AE016761 AE016762 AE016763 AE016764 AE016765 AE016766 AE016767
AE016768 AE016769 AE016770 AE016771 AE016772 VERSION AE014075.1

| **NUCLEOTIDES** | | **DOUBLETS** | | **TRIPLETS** | | **4-PLETS** | | **5-PLETS** | |
|---|---|---|---|---|---|---|---|---|---|
| $P_1(A_1)=$ | 0,3326 | $P_2(A_1)=$ | 0,3326 | $P_3(A_1)=$ | 0,3335 | $P_4(A_1)=$ | 0,3323 | $P_5(A_1)=$ | 0,3329 |
| $P_1(T_1)=$ | 0,3420 | $P_2(T_1)=$ | 0,3416 | $P_3(T_1)=$ | 0,3411 | $P_4(T_1)=$ | 0,3419 | $P_5(T_1)=$ | 0,3421 |
| $P_1(C_1)=$ | 0,1640 | $P_2(C_1)=$ | 0,1643 | $P_3(C_1)=$ | 0,1616 | $P_4(C_1)=$ | 0,1645 | $P_5(C_1)=$ | 0,1637 |
| $P_1(G_1)=$ | 0,1614 | $P_2(G_1)=$ | 0,1615 | $P_3(G_1)=$ | 0,1638 | $P_4(G_1)=$ | 0,1613 | $P_5(G_1)=$ | 0,1612 |
| | | $P_2(A_2)=$ | 0,3326 | $P_3(A_2)=$ | 0,3306 | $P_4(A_2)=$ | 0,3334 | $P_5(A_2)=$ | 0,3324 |
| | | $P_2(T_2)=$ | 0,3425 | $P_3(T_2)=$ | 0,3425 | $P_4(T_2)=$ | 0,3413 | $P_5(T_2)=$ | 0,3425 |
| | | $P_2(C_2)=$ | 0,1637 | $P_3(C_2)=$ | 0,1665 | $P_4(C_2)=$ | 0,1638 | $P_5(C_2)=$ | 0,1640 |
| | | $P_2(G_2)=$ | 0,1612 | $P_3(G_2)=$ | 0,1606 | $P_4(G_2)=$ | 0,1615 | $P_5(G_2)=$ | 0,1611 |
| | | | | $P_3(A_3)=$ | 0,3338 | $P_4(A_3)=$ | 0,3330 | $P_5(A_3)=$ | 0,3328 |
| | | | | $P_3(T_3)=$ | 0,3427 | $P_4(T_3)=$ | 0,3413 | $P_5(T_3)=$ | 0,3422 |
| | | | | $P_3(C_3)=$ | 0,1638 | $P_4(C_3)=$ | 0,1640 | $P_5(C_3)=$ | 0,1635 |
| | | | | $P_3(G_3)=$ | 0,1597 | $P_4(G_3)=$ | 0,1617 | $P_5(G_3)=$ | 0,1614 |
| | | | | | | $P_4(A_4)=$ | 0,3319 | $P_5(A_4)=$ | 0,3326 |
| | | | | | | $P_4(T_4)=$ | 0,3436 | $P_5(T_4)=$ | 0,3414 |
| | | | | | | $P_4(C_4)=$ | 0,1636 | $P_5(C_4)=$ | 0,1639 |
| | | | | | | $P_4(G_4)=$ | 0,1610 | $P_5(G_4)=$ | 0,1620 |
| | | | | | | | | $P_5(A_5)=$ | 0,3324 |
| | | | | | | | | $P_5(T_5)=$ | 0,3420 |
| | | | | | | | | $P_5(C_5)=$ | 0,1647 |
| | | | | | | | | $P_5(G_5)=$ | 0,16095 |

Fig. A1/9. Probabilities of subgroups of tetra-groups in the sequence: Flavobacterium psychrophilum JIP02/86, https://www.ncbi.nlm.nih.gov/nuccore/NC_009613.3 :
LOCUS NC_009613, 2860382 bp, DNA circular CON 03-AUG-2016
DEFINITION Flavobacterium psychrophilum JIP02/86 complete genome.
ACCESSION NC_009613 VERSION NC_009613.3

| **NUCLEOTIDES** | | **DOUBLETS** | | **TRIPLETS** | | **4-PLETS** | | **5-PLETS** | |
|---|---|---|---|---|---|---|---|---|---|
| $P_1(A_1)=$ | 0,1906 | $P_2(A_1)=$ | 0,1905 | $P_3(A_1)=$ | 0,1892 | $P_4(A_1)=$ | 0,1910 | $P_5(A_1)=$ | 0,1906 |
| $P_1(T_1)=$ | 0,1894 | $P_2(T_1)=$ | 0,1899 | $P_3(T_1)=$ | 0,1881 | $P_4(T_1)=$ | 0,1898 | $P_5(T_1)=$ | 0,1895 |
| $P_1(C_1)=$ | 0,3101 | $P_2(C_1)=$ | 0,3098 | $P_3(C_1)=$ | 0,3120 | $P_4(C_1)=$ | 0,3096 | $P_5(C_1)=$ | 0,3104 |
| $P_1(G_1)=$ | 0,3099 | $P_2(G_1)=$ | 0,3099 | $P_3(G_1)=$ | 0,3107 | $P_4(G_1)=$ | 0,3095 | $P_5(G_1)=$ | 0,3095 |
| | | $P_2(A_2)=$ | 0,1907 | $P_3(A_2)=$ | 0,1925 | $P_4(A_2)=$ | 0,1906 | $P_5(A_2)=$ | 0,1907 |
| | | $P_2(T_2)=$ | 0,1890 | $P_3(T_2)=$ | 0,1899 | $P_4(T_2)=$ | 0,1892 | $P_5(T_2)=$ | 0,1894 |
| | | $P_2(C_2)=$ | 0,3104 | $P_3(C_2)=$ | 0,3074 | $P_4(C_2)=$ | 0,3105 | $P_5(C_2)=$ | 0,3100 |
| | | $P_2(G_2)=$ | 0,3099 | $P_3(G_2)=$ | 0,3103 | $P_4(G_2)=$ | 0,3097 | $P_5(G_2)=$ | 0,3100 |
| | | | | $P_3(A_3)=$ | 0,1901 | $P_4(A_3)=$ | 0,1900 | $P_5(A_3)=$ | 0,1908 |
| | | | | $P_3(T_3)=$ | 0,1903 | $P_4(T_3)=$ | 0,1899 | $P_5(T_3)=$ | 0,1892 |
| | | | | $P_3(C_3)=$ | 0,3108 | $P_4(C_3)=$ | 0,3099 | $P_5(C_3)=$ | 0,3097 |
| | | | | $P_3(G_3)=$ | 0,3088 | $P_4(G_3)=$ | 0,3103 | $P_5(G_3)=$ | 0,3104 |
| | | | | | | $P_4(A_4)=$ | 0,1908 | $P_5(A_4)=$ | 0,1905 |
| | | | | | | $P_4(T_4)=$ | 0,1889 | $P_5(T_4)=$ | 0,1896 |
| | | | | | | $P_4(C_4)=$ | 0,3102 | $P_5(C_4)=$ | 0,3095 |
| | | | | | | $P_4(G_4)=$ | 0,3102 | $P_5(G_4)=$ | 0,3105 |
| | | | | | | | | $P_5(A_5)=$ | 0,1904 |
| | | | | | | | | $P_5(T_5)=$ | 0,1896 |
| | | | | | | | | $P_5(C_5)=$ | 0,3107 |
| | | | | | | | | $P_5(G_5)=$ | 0,3093 |

Fig. A1/10. Probabilities of subgroups of tetra-groups in the sequence:

Gloeobacter violaceus PCC 7421 DNA, complete genome, GenBank: BA000045.2,
https://www.ncbi.nlm.nih.gov/nuccore/BA000045.2?report=fasta

https://www.ncbi.nlm.nih.gov/nuccore/BA000045.2?report=genbank :

LOCUS BA000045 4659019 bp DNA circular BCT 07-OCT-2016
DEFINITION Gloeobacter violaceus PCC 7421 DNA, complete genome.
ACCESSION BA000045 AP006568-AP006583 VERSION BA000045.2

| **NUCLEOTIDES** | | **DOUBLETS** | | **TRIPLETS** | | **4-PLETS** | | **5-PLETS** | |
|---|---|---|---|---|---|---|---|---|---|
| $P_1(A_1)=$ | 0,3033 | $P_2(A_1)=$ | 0,3034 | $P_3(A_1)=$ | 0,3033 | $P_4(A_1)=$ | 0,3031 | $P_5(A_1)=$ | 0,3040 |
| $P_1(T_1)=$ | 0,3048 | $P_2(T_1)=$ | 0,3049 | $P_3(T_1)=$ | 0,3037 | $P_4(T_1)=$ | 0,3047 | $P_5(T_1)=$ | 0,3051 |
| $P_1(C_1)=$ | 0,1970 | $P_2(C_1)=$ | 0,1969 | $P_3(C_1)=$ | 0,1989 | $P_4(C_1)=$ | 0,1971 | $P_5(C_1)=$ | 0,1963 |
| $P_1(G_1)=$ | 0,1949 | $P_2(G_1)=$ | 0,1949 | $P_3(G_1)=$ | 0,1941 | $P_4(G_1)=$ | 0,1951 | $P_5(G_1)=$ | 0,1946 |
| | | $P_2(A_2)=$ | 0,3032 | $P_3(A_2)=$ | 0,3042 | $P_4(A_2)=$ | 0,3036 | $P_5(A_2)=$ | 0,3029 |
| | | $P_2(T_2)=$ | 0,3048 | $P_3(T_2)=$ | 0,3028 | $P_4(T_2)=$ | 0,3051 | $P_5(T_2)=$ | 0,3062 |
| | | $P_2(C_2)=$ | 0,1970 | $P_3(C_2)=$ | 0,1958 | $P_4(C_2)=$ | 0,1968 | $P_5(C_2)=$ | 0,1963 |
| | | $P_2(G_2)=$ | 0,1950 | $P_3(G_2)=$ | 0,1972 | $P_4(G_2)=$ | 0,1945 | $P_5(G_2)=$ | 0,1946 |
| | | | | $P_3(A_3)=$ | 0,3022 | $P_4(A_3)=$ | 0,3037 | $P_5(A_3)=$ | 0,3045 |
| | | | | $P_3(T_3)=$ | 0,3080 | $P_4(T_3)=$ | 0,3050 | $P_5(T_3)=$ | 0,3028 |
| | | | | $P_3(C_3)=$ | 0,1962 | $P_4(C_3)=$ | 0,1967 | $P_5(C_3)=$ | 0,1974 |
| | | | | $P_3(G_3)=$ | 0,1936 | $P_4(G_3)=$ | 0,1946 | $P_5(G_3)=$ | 0,1954 |
| | | | | | | $P_4(A_4)=$ | 0,3028 | $P_5(A_4)=$ | 0,3027 |
| | | | | | | $P_4(T_4)=$ | 0,3045 | $P_5(T_4)=$ | 0,3052 |
| | | | | | | $P_4(C_4)=$ | 0,1973 | $P_5(C_4)=$ | 0,1974 |
| | | | | | | $P_4(G_4)=$ | 0,1954 | $P_5(G_4)=$ | 0,1946 |
| | | | | | | | | $P_5(A_5)=$ | 0,3022 |
| | | | | | | | | $P_5(T_5)=$ | 0,3049 |
| | | | | | | | | $P_5(C_5)=$ | 0,1974 |
| | | | | | | | | $P_5(G_5)=$ | 0,1955 |

Fig. A1/11. Probabilities of subgroups of tetra-groups in the sequence:

Helicobacter pilory, NCBI Reference Sequence: NC_000921.1

https://www.ncbi.nlm.nih.gov/nuccore/NC_000921.1

LOCUS NC_000921 1643831 bp DNA circular CON 22-MAR-2017
DEFINITION Helicobacter pylori J99, complete genome. ACCESSION
NC_000921 NZ_AE001440-NZ_AE001571 VERSION NC_000921.1

| **NUCLEOTIDES** | | **DOUBLETS** | | **TRIPLETS** | | **4-PLETS** | | **5-PLETS** | |
|---|---|---|---|---|---|---|---|---|---|
| $P_1(A_1)=$ | 0,2848 | $P_2(A_1)=$ | 0,2848 | $P_3(A_1)=$ | 0,2830 | $P_4(A_1)=$ | 0,2850 | $P_5(A_1)=$ | 0,2843 |
| $P_1(T_1)=$ | 0,2884 | $P_2(T_1)=$ | 0,2881 | $P_3(T_1)=$ | 0,2890 | $P_4(T_1)=$ | 0,2881 | $P_5(T_1)=$ | 0,2882 |
| $P_1(C_1)=$ | 0,2136 | $P_2(C_1)=$ | 0,2139 | $P_3(C_1)=$ | 0,2148 | $P_4(C_1)=$ | 0,2134 | $P_5(C_1)=$ | 0,2137 |
| $P_1(G_1)=$ | 0,2132 | $P_2(G_1)=$ | 0,2132 | $P_3(G_1)=$ | 0,2132 | $P_4(G_1)=$ | 0,2134 | $P_5(G_1)=$ | 0,2138 |
| | | $P_2(A_2)=$ | 0,2848 | $P_3(A_2)=$ | 0,2851 | $P_4(A_2)=$ | 0,2846 | $P_5(A_2)=$ | 0,2850 |
| | | $P_2(T_2)=$ | 0,2887 | $P_3(T_2)=$ | 0,2875 | $P_4(T_2)=$ | 0,2888 | $P_5(T_2)=$ | 0,2879 |
| | | $P_2(C_2)=$ | 0,2132 | $P_3(C_2)=$ | 0,2133 | $P_4(C_2)=$ | 0,2133 | $P_5(C_2)=$ | 0,2133 |
| | | $P_2(G_2)=$ | 0,2133 | $P_3(G_2)=$ | 0,2141 | $P_4(G_2)=$ | 0,2134 | $P_5(G_2)=$ | 0,2137 |
| | | | | $P_3(A_3)=$ | 0,2863 | $P_4(A_3)=$ | 0,2847 | $P_5(A_3)=$ | 0,2852 |
| | | | | $P_3(T_3)=$ | 0,2887 | $P_4(T_3)=$ | 0,2880 | $P_5(T_3)=$ | 0,2886 |
| | | | | $P_3(C_3)=$ | 0,2126 | $P_4(C_3)=$ | 0,2144 | $P_5(C_3)=$ | 0,2137 |
| | | | | $P_3(G_3)=$ | 0,2124 | $P_4(G_3)=$ | 0,2129 | $P_5(G_3)=$ | 0,2126 |
| | | | | | | $P_4(A_4)=$ | 0,2849 | $P_5(A_4)=$ | 0,2846 |
| | | | | | | $P_4(T_4)=$ | 0,2887 | $P_5(T_4)=$ | 0,2889 |
| | | | | | | $P_4(C_4)=$ | 0,2132 | $P_5(C_4)=$ | 0,2133 |
| | | | | | | $P_4(G_4)=$ | 0,2131 | $P_5(G_4)=$ | 0,2132 |
| | | | | | | | | $P_5(A_5)=$ | 0,2848 |
| | | | | | | | | $P_5(T_5)=$ | 0,2883 |
| | | | | | | | | $P_5(C_5)=$ | 0,2140 |
| | | | | | | | | $P_5(G_5)=$ | 0,2129 |

Fig. A1/12. Probabilities of subgroups of tetra-groups in the sequence:

Methanosarcina acetivorans str. C2A, complete genome,
GenBank: AE010299.1,
https://www.ncbi.nlm.nih.gov/nuccore/AE010299.1?report=fasta
https://www.ncbi.nlm.nih.gov/nuccore/AE01029 -
LOCUS AE010299 5751492 bp, DNA circular BCT 01-OCT-2014
DEFINITION Methanosarcina acetivorans str. C2A, complete genome.
ACCESSION AE010299 AE010656-AE011189 VERSION AE010299.1

| **NUCLEOTIDES** | | **DOUBLETS** | | **TRIPLETS** | | **4-PLETS** | | **5-PLETS** | |
|---|---|---|---|---|---|---|---|---|---|
| $P_1(A_1)=$ | 0,3422 | $P_2(A_1)=$ | 0,3424 | $P_3(A_1)=$ | 0,3413 | $P_4(A_1)=$ | 0,3422 | $P_5(A_1)=$ | 0,3413 |
| $P_1(T_1)=$ | 0,3422 | $P_2(T_1)=$ | 0,3419 | $P_3(T_1)=$ | 0,3420 | $P_4(T_1)=$ | 0,3422 | $P_5(T_1)=$ | 0,3426 |
| $P_1(C_1)=$ | 0,1576 | $P_2(C_1)=$ | 0,1581 | $P_3(C_1)=$ | 0,1593 | $P_4(C_1)=$ | 0,1572 | $P_5(C_1)=$ | 0,1585 |
| $P_1(G_1)=$ | 0,1580 | $P_2(G_1)=$ | 0,1577 | $P_3(G_1)=$ | 0,1573 | $P_4(G_1)=$ | 0,1584 | $P_5(G_1)=$ | 0,1576 |
| | | $P_2(A_2)=$ | 0,3420 | $P_3(A_2)=$ | 0,3423 | $P_4(A_2)=$ | 0,3408 | $P_5(A_2)=$ | 0,3433 |
| | | $P_2(T_2)=$ | 0,3425 | $P_3(T_2)=$ | 0,3413 | $P_4(T_2)=$ | 0,3425 | $P_5(T_2)=$ | 0,3411 |
| | | $P_2(C_2)=$ | 0,1571 | $P_3(C_2)=$ | 0,1571 | $P_4(C_2)=$ | 0,1572 | $P_5(C_2)=$ | 0,1575 |
| | | $P_2(G_2)=$ | 0,1583 | $P_3(G_2)=$ | 0,1594 | $P_4(G_2)=$ | 0,1596 | $P_5(G_2)=$ | 0,1582 |
| | | | | $P_3(A_3)=$ | 0,3431 | $P_4(A_3)=$ | 0,3425 | $P_5(A_3)=$ | 0,3430 |
| | | | | $P_3(T_3)=$ | 0,3433 | $P_4(T_3)=$ | 0,3416 | $P_5(T_3)=$ | 0,3439 |
| | | | | $P_3(C_3)=$ | 0,1563 | $P_4(C_3)=$ | 0,1590 | $P_5(C_3)=$ | 0,1586 |
| | | | | $P_3(G_3)=$ | 0,1573 | $P_4(G_3)=$ | 0,1570 | $P_5(G_3)=$ | 0,1545 |
| | | | | | | $P_4(A_4)=$ | 0,3433 | $P_5(A_4)=$ | 0,3417 |
| | | | | | | $P_4(T_4)=$ | 0,3426 | $P_5(T_4)=$ | 0,3432 |
| | | | | | | $P_4(C_4)=$ | 0,1571 | $P_5(C_4)=$ | 0,1574 |
| | | | | | | $P_4(G_4)=$ | 0,1570 | $P_5(G_4)=$ | 0,1577 |
| | | | | | | | | $P_5(A_5)=$ | 0,3418 |
| | | | | | | | | $P_5(T_5)=$ | 0,3403 |
| | | | | | | | | $P_5(C_5)=$ | 0,1560 |
| | | | | | | | | $P_5(G_5)=$ | 0,1619 |

Fig. A1/13. Probabilities of subgroups of tetra-groups in the sequence:

Nanoarchaeum equitans Kin4-M, complete genome, GenBank: AE017199.1,
https://www.ncbi.nlm.nih.gov/nuccore/AE017199.1?report=fasta
https://www.ncbi.nlm.nih.gov/nuccore/AE017199.1?report=genbank -
LOCUS AE017199 490885 bp DNA circular BCT 30-JAN-2014
DEFINITION Nanoarchaeum equitans Kin4-M, complete genome. ACCESSION AE017199 AACL01000000 AACL01000001 VERSION AE017199.1

| **NUCLEOTIDES** | | **DOUBLETS** | | **TRIPLETS** | | **4-PLETS** | | **5-PLETS** | |
|---|---|---|---|---|---|---|---|---|---|
| $P_1(A_1)=$ | 0,2431 | $P_2(A_1)=$ | 0,2431 | $P_3(A_1)=$ | 0,2441 | $P_4(A_1)=$ | 0,2435 | $P_5(A_1)=$ | 0,2425 |

| $P_1(T_1)=$ | 0,2423 | $P_2(T_1)=$ | 0,2423 | $P_3(T_1)=$ | 0,2446 | $P_4(T_1)=$ | 0,2423 | $P_5(T_1)=$ | 0,2425 |
|---|---|---|---|---|---|---|---|---|---|
| $P_1(C_1)=$ | 0,2556 | $P_2(C_1)=$ | 0,2556 | $P_3(C_1)=$ | 0,2546 | $P_4(C_1)=$ | 0,2557 | $P_5(C_1)=$ | 0,2560 |
| $P_1(G_1)=$ | 0,2590 | $P_2(G_1)=$ | 0,2590 | $P_3(G_1)=$ | 0,2567 | $P_4(G_1)=$ | 0,2585 | $P_5(G_1)=$ | 0,2590 |
| | | $P_2(A_2)=$ | 0,2431 | $P_3(A_2)=$ | 0,2418 | $P_4(A_2)=$ | 0,2432 | $P_5(A_2)=$ | 0,2435 |
| | | $P_2(T_2)=$ | 0,2424 | $P_3(T_2)=$ | 0,2413 | $P_4(T_2)=$ | 0,2423 | $P_5(T_2)=$ | 0,2418 |
| | | $P_2(C_2)=$ | 0,2557 | $P_3(C_2)=$ | 0,2573 | $P_4(C_2)=$ | 0,2557 | $P_5(C_2)=$ | 0,2554 |
| | | $P_2(G_2)=$ | 0,2589 | $P_3(G_2)=$ | 0,2597 | $P_4(G_2)=$ | 0,2588 | $P_5(G_2)=$ | 0,2592 |
| | | | | $P_3(A_3)=$ | 0,2433 | $P_4(A_3)=$ | 0,2426 | $P_5(A_3)=$ | 0,2434 |
| | | | | $P_3(T_3)=$ | 0,2412 | $P_4(T_3)=$ | 0,2422 | $P_5(T_3)=$ | 0,2422 |
| | | | | $P_3(C_3)=$ | 0,2551 | $P_4(C_3)=$ | 0,2556 | $P_5(C_3)=$ | 0,2557 |
| | | | | $P_3(G_3)=$ | 0,2604 | $P_4(G_3)=$ | 0,2596 | $P_5(G_3)=$ | 0,2587 |
| | | | | | | $P_4(A_4)=$ | 0,2429 | $P_5(A_4)=$ | 0,2427 |
| | | | | | | $P_4(T_4)=$ | 0,2425 | $P_5(T_4)=$ | 0,2427 |
| | | | | | | $P_4(C_4)=$ | 0,2556 | $P_5(C_4)=$ | 0,2561 |
| | | | | | | $P_4(G_4)=$ | 0,2590 | $P_5(G_4)=$ | 0,2584 |
| | | | | | | | | $P_5(A_5)=$ | 0,2431 |
| | | | | | | | | $P_5(T_5)=$ | 0,2425 |
| | | | | | | | | $P_5(C_5)=$ | 0,2550 |
| | | | | | | | | $P_5(G_5)=$ | 0,2595 |

Fig. A1/14. Probabilities of subgroups of tetra-groups in the sequence:

Syntrophus aciditrophicus SB, complete genome, GenBank: CP000252.1, https://www.ncbi.nlm.nih.gov/nuccore/CP000252.1?report=fasta

https://www.ncbi.nlm.nih.gov/nuccore/CP000252.1?report=genbank - LOCUS CP000252, 3179300 bp, DNA circular BCT 31-JAN-2014 DEFINITION Syntrophus aciditrophicus SB, complete genome. ACCESSION CP000252 VERSION CP000252.1

| **NUCLEOTIDES** | | **DOUBLETS** | | **TRIPLETS** | | **4-PLETS** | | **5-PLETS** | |
|---|---|---|---|---|---|---|---|---|---|
| $P_1(A_1)=$ | 0,1389 | $P_2(A_1)=$ | 0,1386 | $P_3(A_1)=$ | 0,1366 | $P_4(A_1)=$ | 0,1385 | $P_5(A_1)=$ | 0,1392 |
| $P_1(T_1)=$ | 0,1400 | $P_2(T_1)=$ | 0,1400 | $P_3(T_1)=$ | 0,1381 | $P_4(T_1)=$ | 0,1399 | $P_5(T_1)=$ | 0,1401 |
| $P_1(C_1)=$ | 0,3601 | $P_2(C_1)=$ | 0,3600 | $P_3(C_1)=$ | 0,3614 | $P_4(C_1)=$ | 0,3606 | $P_5(C_1)=$ | 0,3598 |
| $P_1(G_1)=$ | 0,3611 | $P_2(G_1)=$ | 0,3613 | $P_3(G_1)=$ | 0,3639 | $P_4(G_1)=$ | 0,3611 | $P_5(G_1)=$ | 0,3609 |
| | | $P_2(A_2)=$ | 0,1391 | $P_3(A_2)=$ | 0,1408 | $P_4(A_2)=$ | 0,1388 | $P_5(A_2)=$ | 0,1388 |
| | | $P_2(T_2)=$ | 0,1399 | $P_3(T_2)=$ | 0,1414 | $P_4(T_2)=$ | 0,1402 | $P_5(T_2)=$ | 0,1400 |
| | | $P_2(C_2)=$ | 0,3602 | $P_3(C_2)=$ | 0,3580 | $P_4(C_2)=$ | 0,3601 | $P_5(C_2)=$ | 0,3597 |
| | | $P_2(G_2)=$ | 0,3608 | $P_3(G_2)=$ | 0,3598 | $P_4(G_2)=$ | 0,3610 | $P_5(G_2)=$ | 0,3615 |
| | | | | $P_3(A_3)=$ | 0,1392 | $P_4(A_3)=$ | 0,1388 | $P_5(A_3)=$ | 0,1389 |
| | | | | $P_3(T_3)=$ | 0,1403 | $P_4(T_3)=$ | 0,1401 | $P_5(T_3)=$ | 0,1397 |
| | | | | $P_3(C_3)=$ | 0,3609 | $P_4(C_3)=$ | 0,3595 | $P_5(C_3)=$ | 0,3610 |
| | | | | $P_3(G_3)=$ | 0,3596 | $P_4(G_3)=$ | 0,3616 | $P_5(G_3)=$ | 0,3605 |
| | | | | | | $P_4(A_4)=$ | 0,1394 | $P_5(A_4)=$ | 0,1389 |
| | | | | | | $P_4(T_4)=$ | 0,1397 | $P_5(T_4)=$ | 0,1398 |
| | | | | | | $P_4(C_4)=$ | 0,3603 | $P_5(C_4)=$ | 0,3600 |
| | | | | | | $P_4(G_4)=$ | 0,3606 | $P_5(G_4)=$ | 0,3614 |
| | | | | | | | | $P_5(A_5)=$ | 0,1386 |
| | | | | | | | | $P_5(T_5)=$ | 0,1403 |
| | | | | | | | | $P_5(C_5)=$ | 0,3601 |
| | | | | | | | | $P_5(G_5)=$ | 0,3611 |

Fig. A1/15. Probabilities of subgroups of tetra-groups in the sequence:

Streptomyces coelicolor A3(2) complete genome, GenBank: AL645882.2, https://www.ncbi.nlm.nih.gov/nuccore/AL645882.2?report=fasta

https://www.ncbi.nlm.nih.gov/nuccore/AL645882.2?report=genbank :

LOCUS AL645882, 8667507 bp, DNA linear CON 06-FEB-2015
DEFINITION Streptomyces coelicolor A3(2) complete genome. ACCESSION
AL645882 VERSION AL645882.2

| NUCLEOTIDES | | DOUBLETS | | TRIPLETS | | 4-PLETS | | 5-PLETS | |
|---|---|---|---|---|---|---|---|---|---|
| $P_1(A_1)$= | 0,3181 | $P_2(A_1)$= | 0,3179 | $P_3(A_1)$= | 0,3182 | $P_4(A_1)$= | 0,3177 | $P_5(A_1)$= | 0,3185 |
| $P_1(T_1)$= | 0,3233 | $P_2(T_1)$= | 0,3231 | $P_3(T_1)$= | 0,3225 | $P_4(T_1)$= | 0,3234 | $P_5(T_1)$= | 0,3236 |
| $P_1(C_1)$= | 0,1798 | $P_2(C_1)$= | 0,1797 | $P_3(C_1)$= | 0,1798 | $P_4(C_1)$= | 0,1792 | $P_5(C_1)$= | 0,1793 |
| $P_1(G_1)$= | 0,1788 | $P_2(G_1)$= | 0,1793 | $P_3(G_1)$= | 0,1795 | $P_4(G_1)$= | 0,1798 | $P_5(G_1)$= | 0,1787 |
| | | $P_2(A_2)$= | 0,3184 | $P_3(A_2)$= | 0,3185 | $P_4(A_2)$= | 0,3183 | $P_5(A_2)$= | 0,3183 |
| | | $P_2(T_2)$= | 0,3234 | $P_3(T_2)$= | 0,3230 | $P_4(T_2)$= | 0,3237 | $P_5(T_2)$= | 0,3228 |
| | | $P_2(C_2)$= | 0,1800 | $P_3(C_2)$= | 0,1782 | $P_4(C_2)$= | 0,1801 | $P_5(C_2)$= | 0,1801 |
| | | $P_2(G_2)$= | 0,1783 | $P_3(G_2)$= | 0,1803 | $P_4(G_2)$= | 0,1779 | $P_5(G_2)$= | 0,1787 |
| | | | | $P_3(A_3)$= | 0,3177 | $P_4(A_3)$= | 0,3181 | $P_5(A_3)$= | 0,3165 |
| | | | | $P_3(T_3)$= | 0,3243 | $P_4(T_3)$= | 0,3229 | $P_5(T_3)$= | 0,3236 |
| | | | | $P_3(C_3)$= | 0,1814 | $P_4(C_3)$= | 0,1803 | $P_5(C_3)$= | 0,1809 |
| | | | | $P_3(G_3)$= | 0,1765 | $P_4(G_3)$= | 0,1788 | $P_5(G_3)$= | 0,1790 |
| | | | | | | $P_4(A_4)$= | 0,3185 | $P_5(A_4)$= | 0,3187 |
| | | | | | | $P_4(T_4)$= | 0,3232 | $P_5(T_4)$= | 0,3232 |
| | | | | | | $P_4(C_4)$= | 0,1797 | $P_5(C_4)$= | 0,1794 |
| | | | | | | $P_4(G_4)$= | 0,1786 | $P_5(G_4)$= | 0,1788 |
| | | | | | | | | $P_5(A_5)$= | 0,3187 |
| | | | | | | | | $P_5(T_5)$= | 0,3232 |
| | | | | | | | | $P_5(C_5)$= | 0,1795 |
| | | | | | | | | $P_5(G_5)$= | 0,1786 |

Fig. A1/16. Probabilities of subgroups of tetra-groups in the sequence:

Sulfolobus solfataricus strain SULA, complete genome, GenBank: CP011057.1,
https://www.ncbi.nlm.nih.gov/nuccore/CP011057.1?report=fasta
https://www.ncbi.nlm.nih.gov/nuccore/CP011057.1?report=genbank -
LOCUS CP011057, 2727337 bp, DNA circular BCT 10-JUN-2015
DEFINITION Sulfolobus solfataricus strain SULA, complete genome. ACCESSION
CP011057 VERSION CP011057.1

| **NUCLEOTIDES** | | **DOUBLETS** | | **TRIPLETS** | | **4-PLETS** | | **5-PLETS** | |
|---|---|---|---|---|---|---|---|---|---|
| $P_1(A_1)$= | 0,3095 | $P_2(A_1)$= | 0,3095 | $P_3(A_1)$= | 0,3079 | $P_4(A_1)$= | 0,3095 | $P_5(A_1)$= | 0,3089 |
| $P_1(T_1)$= | 0,3106 | $P_2(T_1)$= | 0,3104 | $P_3(T_1)$= | 0,3120 | $P_4(T_1)$= | 0,3106 | $P_5(T_1)$= | 0,3106 |
| $P_1(C_1)$= | 0,1880 | $P_2(C_1)$= | 0,1877 | $P_3(C_1)$= | 0,1882 | $P_4(C_1)$= | 0,1871 | $P_5(C_1)$= | 0,1885 |
| $P_1(G_1)$= | 0,1919 | $P_2(G_1)$= | 0,1923 | $P_3(G_1)$= | 0,1918 | $P_4(G_1)$= | 0,1928 | $P_5(G_1)$= | 0,1920 |
| | | $P_2(A_2)$= | 0,3096 | $P_3(A_2)$= | 0,3097 | $P_4(A_2)$= | 0,3102 | $P_5(A_2)$= | 0,3095 |
| | | $P_2(T_2)$= | 0,3107 | $P_3(T_2)$= | 0,3089 | $P_4(T_2)$= | 0,3108 | $P_5(T_2)$= | 0,3110 |
| | | $P_2(C_2)$= | 0,1882 | $P_3(C_2)$= | 0,1879 | $P_4(C_2)$= | 0,1879 | $P_5(C_2)$= | 0,1871 |
| | | $P_2(G_2)$= | 0,1915 | $P_3(G_2)$= | 0,1934 | $P_4(G_2)$= | 0,1911 | $P_5(G_2)$= | 0,1924 |
| | | | | $P_3(A_3)$= | 0,3109 | $P_4(A_3)$= | 0,3095 | $P_5(A_3)$= | 0,3102 |
| | | | | $P_3(T_3)$= | 0,3108 | $P_4(T_3)$= | 0,3102 | $P_5(T_3)$= | 0,3103 |
| | | | | $P_3(C_3)$= | 0,1877 | $P_4(C_3)$= | 0,1883 | $P_5(C_3)$= | 0,1876 |
| | | | | $P_3(G_3)$= | 0,1906 | $P_4(G_3)$= | 0,1919 | $P_5(G_3)$= | 0,1918 |
| | | | | | | $P_4(A_4)$= | 0,3089 | $P_5(A_4)$= | 0,3092 |
| | | | | | | $P_4(T_4)$= | 0,3107 | $P_5(T_4)$= | 0,3103 |
| | | | | | | $P_4(C_4)$= | 0,1884 | $P_5(C_4)$= | 0,1882 |

| | | | |
|---|---|---|---|
| $P_4(G_4)=$ | 0,1920 | $P_5(G_4)=$ | 0,1923 |
| | | $P_5(A_5)=$ | 0,3099 |
| | | $P_5(T_5)=$ | 0,3106 |
| | | $P_5(C_5)=$ | 0,1883 |
| | | $P_5(G_5)=$ | 0,1911 |

Fig. A1/17. Probabilities of subgroups of tetra-groups in the sequence:

Treponema denticola SP33 supercont1.1, whole genome shotgun sequence
NCBI Reference Sequence: NZ_KB442453.1,
https://www.ncbi.nlm.nih.gov/nuccore/NZ_KB442453.1?report=fasta
https://www.ncbi.nlm.nih.gov/nuccore/NZ_KB442453.1?report=genbank -
LOCUS NZ_KB442453, 1850823 bp, DNA linear CON 02-APR-2017
DEFINITION Treponema denticola SP33 supercont1.1, whole genome shotgun sequence. ACCESSION NZ_KB442453 NZ_AGDZ01000000 VERSION NZ_KB442453.1

| **NUCLEOTIDES** | | **DOUBLETS** | | **TRIPLETS** | | **4-PLETS** | | **5-PLETS** | |
|---|---|---|---|---|---|---|---|---|---|
| $P_1(A_1)=$ | 0,2695 | $P_2(A_1)=$ | 0,2700 | $P_3(A_1)=$ | 0,2703 | $P_4(A_1)=$ | 0,2700 | $P_5(A_1)=$ | 0,2692 |
| $P_1(T_1)=$ | 0,2678 | $P_2(T_1)=$ | 0,2681 | $P_3(T_1)=$ | 0,2689 | $P_4(T_1)=$ | 0,2680 | $P_5(T_1)=$ | 0,2672 |
| $P_1(C_1)=$ | 0,2281 | $P_2(C_1)=$ | 0,2277 | $P_3(C_1)=$ | 0,2271 | $P_4(C_1)=$ | 0,2276 | $P_5(C_1)=$ | 0,2291 |
| $P_1(G_1)=$ | 0,2347 | $P_2(G_1)=$ | 0,2341 | $P_3(G_1)=$ | 0,2337 | $P_4(G_1)=$ | 0,2345 | $P_5(G_1)=$ | 0,2346 |
| | | $P_2(A_2)=$ | 0,2689 | $P_3(A_2)=$ | 0,2697 | $P_4(A_2)=$ | 0,2688 | $P_5(A_2)=$ | 0,2691 |
| | | $P_2(T_2)=$ | 0,2675 | $P_3(T_2)=$ | 0,2694 | $P_4(T_2)=$ | 0,2671 | $P_5(T_2)=$ | 0,2678 |
| | | $P_2(C_2)=$ | 0,2284 | $P_3(C_2)=$ | 0,2289 | $P_4(C_2)=$ | 0,2295 | $P_5(C_2)=$ | 0,2275 |
| | | $P_2(G_2)=$ | 0,2352 | $P_3(G_2)=$ | 0,2320 | $P_4(G_2)=$ | 0,2345 | $P_5(G_2)=$ | 0,2356 |
| | | | | $P_3(A_3)=$ | 0,2684 | $P_4(A_3)=$ | 0,2701 | $P_5(A_3)=$ | 0,2703 |
| | | | | $P_3(T_3)=$ | 0,2651 | $P_4(T_3)=$ | 0,2683 | $P_5(T_3)=$ | 0,2674 |
| | | | | $P_3(C_3)=$ | 0,2282 | $P_4(C_3)=$ | 0,2279 | $P_5(C_3)=$ | 0,2285 |
| | | | | $P_3(G_3)=$ | 0,2383 | $P_4(G_3)=$ | 0,23371 | $P_5(G_3)=$ | 0,2339 |
| | | | | | | $P_4(A_4)=$ | 0,2690 | $P_5(A_4)=$ | 0,2700 |
| | | | | | | $P_4(T_4)=$ | 0,2679 | $P_5(T_4)=$ | 0,2683 |
| | | | | | | $P_4(C_4)=$ | 0,2272 | $P_5(C_4)=$ | 0,2272 |
| | | | | | | $P_4(G_4)=$ | 0,2359 | $P_5(G_4)=$ | 0,2345 |
| | | | | | | | | $P_5(A_5)=$ | 0,2689 |
| | | | | | | | | $P_5(T_5)=$ | 0,2684 |
| | | | | | | | | $P_5(C_5)=$ | 0,2280 |
| | | | | | | | | $P_5(G_5)=$ | 0,2347 |

Fig. A1/18. Probabilities of subgroups of tetra-groups in the sequence:

Thermotoga maritima strain Tma200, complete genome, GenBank: CP010967.1,
https://www.ncbi.nlm.nih.gov/nuccore/CP010967.1?report=fasta
https://www.ncbi.nlm.nih.gov/nuccore/CP010967.1?report=genbank -
LOCUS CP010967, 1859582 bp , DNA circular BCT 10-JUN-2015
DEFINITION Thermotoga maritima strain Tma200, complete genome.
ACCESSION CP010967 VERSION CP010967.1

| **NUCLEOTIDES** | | **DOUBLETS** | | **TRIPLETS** | | **4-PLETS** | | **5-PLETS** | |
|---|---|---|---|---|---|---|---|---|---|
| $P_1(A_1)=$ | 0,1543 | $P_2(A_1)=$ | 0,1543 | $P_3(A_1)=$ | 0,1515 | $P_4(A_1)=$ | 0,1541 | $P_5(A_1)=$ | 0,1539 |
| $P_1(T_1)=$ | 0,1557 | $P_2(T_1)=$ | 0,1556 | $P_3(T_1)=$ | 0,1556 | $P_4(T_1)=$ | 0,1549 | $P_5(T_1)=$ | 0,1559 |
| $P_1(C_1)=$ | 0,3461 | $P_2(C_1)=$ | 0,3458 | $P_3(C_1)=$ | 0,3490 | $P_4(C_1)=$ | 0,3460 | $P_5(C_1)=$ | 0,3461 |
| $P_1(G_1)$= | 0,3439 | $P_2(G_1)$= | 0,3443 | $P_3(G_1)$= | 0,3439 | $P_4(G_1)$= | 0,3450 | $P_5(G_1)$= | 0,3441 |
| | | $P_2(A_2)$= | 0,1542 | $P_3(A_2)$= | 0,1542 | $P_4(A_2)$= | 0,1541 | $P_5(A_2)$= | 0,1547 |
| | | $P_2(T_2)$= | 0,1558 | $P_3(T_2)$= | 0,1529 | $P_4(T_2)$= | 0,1555 | $P_5(T_2)$= | 0,1561 |
| | | $P_2(C_2)$= | 0,3464 | $P_3(C_2)$= | 0,3465 | $P_4(C_2)$= | 0,3467 | $P_5(C_2)$= | 0,3447 |
| | | $P_2(G_2)$= | 0,3435 | $P_3(G_2)$= | 0,3464 | $P_4(G_2)$= | 0,3436 | $P_5(G_2)$= | 0,3445 |
| | | | | $P_3(A_3)$= | 0,1571 | $P_4(A_3)$= | 0,1545 | $P_5(A_3)$= | 0,1544 |
| | | | | $P_3(T_3)$= | 0,1586 | $P_4(T_3)$= | 0,1563 | $P_5(T_3)$= | 0,1555 |
| | | | | $P_3(C_3)$= | 0,3428 | $P_4(C_3)$= | 0,3456 | $P_5(C_3)$= | 0,3465 |
| | | | | $P_3(G_3)$= | 0,3414 | $P_4(G_3)$= | 0,3436 | $P_5(G_3)$= | 0,3436 |
| | | | | | | $P_4(A_4)$= | 0,1543 | $P_5(A_4)$= | 0,1546 |
| | | | | | | $P_4(T_4)$= | 0,1562 | $P_5(T_4)$= | 0,1551 |
| | | | | | | $P_4(C_4)$= | 0,3461 | $P_5(C_4)$= | 0,3465 |
| | | | | | | $P_4(G_4)$= | 0,3434 | $P_5(G_4)$= | 0,3438 |
| | | | | | | | | $P_5(A_5)$= | 0,1536 |
| | | | | | | | | $P_5(T_5)$= | 0,1560 |
| | | | | | | | | $P_5(C_5)$= | 0,3468 |
| | | | | | | | | $P_5(G_5)$= | 0,3436 |

Fig. A1/19. Probabilities of subgroups of tetra-groups in the sequence:

Thermus thermophilus DNA, complete genome, strain: TMY

GenBank: AP017920.1, -

https://www.ncbi.nlm.nih.gov/nuccore/AP017920.1?report=fasta

https://www.ncbi.nlm.nih.gov/nuccore/AP017920.1?report=genbank

LOCUS AP017920, 2121526 bp, DNA circular BCT 26-NOV-2016
DEFINITION Thermus thermophilus DNA, complete genome, strain: TMY.
ACCESSION AP017920 VERSION AP017920.1

| **NUCLEOTIDES** | **DOUBLETS** | **TRIPLETS** | **4-PLETS** | **5-PLETS** |
|---|---|---|---|---|
| $\Sigma_1$ = 229354 | $\Sigma_2$ =114677 | $\Sigma_3$ =76451 | $\Sigma_4$ =57338 | $\Sigma_5$ =45870 |
| $F_1(A_1)$=49475 | $F_2(A_1)$=24800 | $F_3(A_1)$=16825 | $F_4(A_1)$=12423 | $F_5(A_1)$=9981 |
| $P_1(A_1)$=0,2157 | $P_2(A_1)$=0,2163 | $P_3(A_1)$=0,2201 | $P_4(A_1)$=0,2167 | $P_5(A_1)$=0,2176 |
| $F_1(T_1)$=48776 | $F_2(T_1)$=24340 | $F_3(T_1)$=16601 | $F_4(T_1)$=12266 | $F_5(T_1)$=9660 |
| $P_1(T_1)$=0,2127 | $P_2(T_1)$=0,2122 | $P_3(T_1)$=0,2171 | $P_4(T_1)$=0,2139 | $P_5(T_1)$=0,2106 |
| $F_1(C_1)$=64911 | $F_2(C_1)$=32407 | $F_3(C_1)$=21303 | $F_4(C_1)$=16132 | $F_5(C_1)$=12992 |
| $P_1(C_1)$=0,2830 | $P_2(C_1)$=0,2826 | $P_3(C_1)$=0,2786 | $P_4(C_1)$=0,2813 | $P_5(C_1)$=0,2832 |
| $F_1(G_1)$=66192 | $F_2(G_1)$=33130 | $F_3(G_1)$=21722 | $F_4(G_1)$=16517 | $F_5(G_1)$=13237 |
| $P_1(G_1)$=0,2886 | $P_2(G_1)$=0,2889 | $P_3(G_1)$=0,2841 | $P_4(G_1)$=0,2881 | $P_5(G_1)$=0,2886 |
| | $F_2(A_2)$=24675 | $F_3(A_2)$=16002 | $F_4(A_2)$= 12396 | $F_5(A_2)$= 9846 |
| | $P_2(A_2)$=0,2152 | $P_3(A_2)$=0,2093 | $P_4(A_2)$= 0,2161 | $P_5(A_2)$= 0,2146 |
| | $F_2(T_2)$=24436 | $F_3(T_2)$=15964 | $F_4(T_2)$= 12116 | $F_5(T_2)$= 9707 |
| | $P_2(T_2)$=0,2131 | $P_3(T_2)$=0,2088 | $P_4(T_2)$= 0,2113 | $P_5(T_2)$= 0,2116 |
| | $F_2(C_2)$=32504 | $F_3(C_2)$=22117 | $F_4(C_2)$= 16329 | $F_5(C_2)$= 13119 |
| | $P_2(C_2)$=0,2834 | $P_3(C_2)$=0,2893 | $P_4(C_2)$= 0,2848 | $P_5(C_2)$= 0,2860 |
| | $F_2(G_2)$=33062 | $F_3(G_2)$=22368 | $F_4(G_2)$= 16497 | $F_5(G_2)$= 13198 |
| | $P_2(G_2)$=0,2883 | $P_3(G_2)$=0,2926 | $P_4(G_2)$= 0,2877 | $P_5(G_2)$= 0,2877 |
| | | $F_3(A_3)$=16648 | $F_4(A_3)$= 12377 | $F_5(A_3)$= 9747 |
| | | $P_3(A_3)$=0,2178 | $P_4(A_3)$= 0,2159 | $P_5(A_3)$= 0,2125 |
| | | $F_3(T_3)$=16211 | $F_4(T_3)$= 12074 | $F_5(T_3)$= 9842 |
| | | $P_3(T_3)$=0,2120 | $P_4(T_3)$= 0,2106 | $P_5(T_3)$= 0,2146 |
| | | $F_3(C_3)$=21491 | $F_4(C_3)$= 16275 | $F_5(C_3)$= 12970 |
| | | $P_3(C_3)$=0,2811 | $P_4(C_3)$= 0,2838 | $P_5(C_3)$= 0,2828 |

| | | |
|---|---|---|
| $F_3(G_3)=22101$ | $F_4(G_3)= 16612$ | $F_5(G_3)= 13311$ |
| $P_3(G_3)=0,2891$ | $P_4(G_3)= 0,2897$ | $P_5(G_3)= 0,2902$ |
| | $F_4(A_4)=12279$ | $F_5(A_4)= 9843$ |
| | $P_4(A_4)= 0,2142$ | $P_5(A_4)= 0,2146$ |
| | $F_4(T_4)=12320$ | $F_5(T_4)= 9778$ |
| | $P_4(T_4)= 0,2149$ | $P_5(T_4)= 0,2132$ |
| | $F_4(C_4)=16175$ | $F_5(C_4)= 12916$ |
| | $P_4(C_4)= 0,2821$ | $P_5(C_4)= 0,2816$ |
| | $F_4(G_4)=16564$ | $F_5(G_4)= 13333$ |
| | $P_4(G_4)= 0,2889$ | $P_5(G_4)= 0,2907$ |
| | | $F_5(A_5)=10057$ |
| | | $P_5(A_5)=0,2193$ |
| | | $F_5(T_5)=9789$ |
| | | $P_5(T_5)=0,2134$ |
| | | $F_5(C_5)=12914$ |
| | | $P_5(C_5)=0,2815$ |
| | | $F_5(G_5)=13110$ |
| | | $P_5(G_5)=0,2858$ |

Fig. A1/20. Collective frequencies $F_n(A_k)$, $F_n(T_k)$, $F_n(C_k)$, $F_n(G_k)$ and collective probabilties $P_n(A_k)$, $P_n(T_k)$, $P_n(C_k)$ and $P_n(G_k)$ ($n$ = 1, 2, 3, 4, 5 and $k \le n$) of tetra-group subgroups in sequences of n-plets in the case of the Human cytomegalovirus strain AD169 complete genome, 229354 bp, GenBank, accession X17403.1. The index k denotes a position of the letter inside n-plets.

| **NUCLEOTIDES** $\Sigma_1$ = 191737 | **DOUBLETS** $\Sigma_2$ = 95868 | **TRIPLETS** $\Sigma_3$ = 63912 | **4-PLETS** $\Sigma_4$ = 47934 | **5-PLETS** $\Sigma_5$ = 38347 |
|---|---|---|---|---|
| $F_1(A_1)=63921$ | $F_2(A_1)=31832$ | $F_3(A_1)=21310$ | $F_4(A_1)=15932$ | $F_5(A_1)=12872$ |
| $P_1(A_1)=0,3334$ | $P_2(A_1)=0,3320$ | $P_3(A_1)=0,3334$ | $P_4(A_1)=0,3324$ | $P_5(A_1)=0,3357$ |
| $F_1(T_1)=63776$ | $F_2(T_1)=31540$ | $F_3(T_1)=21513$ | $F_4(T_1)=15720$ | $F_5(T_1)=12796$ |
| $P_1(T_1)=0,3326$ | $P_2(T_1)=0,3290$ | $P_3(T_1)=0,3366$ | $P_4(T_1)=0,3280$ | $P_5(T_1)=0,3337$ |
| $F_1(C_1)= 32010$ | $F_2(C_1)=16257$ | $F_3(C_1)=10452$ | $F_4(C_1)=8137$ | $F_5(C_1)=6355$ |
| $P_1(C_1)=0,1669$ | $P_2(C_1)=0,1696$ | $P_3(C_1)=0,1635$ | $P_4(C_1)=0,1698$ | $P_5(C_1)=0,1657$ |
| $F_1(G_1)= 32030$ | $F_2(G_1)=16239$ | $F_3(G_1)=10637$ | $F_4(G_1)=8145$ | $F_5(G_1)=6324$ |
| $P_1(G_1)=0,1671$ | $P_2(G_1)=0,1694$ | $P_3(G_1)=0,1664$ | $P_4(G_1)=0,1699$ | $P_5(G_1)=0,1649$ |
| | $F_2(A_2)=32088$ | $F_3(A_2)= 21273$ | $F_4(A_2)=16064$ | $F_5(A_2)=12661$ |
| | $P_2(A_2)=0,3347$ | $P_3(A_2)= 0,3328$ | $P_4(A_2)=0,3351$ | $P_5(A_2)=0,3302$ |
| | $F_2(T_2)=32236$ | $F_3(T_2)=21024$ | $F_4(T_2)=16132$ | $F_5(T_2)=12832$ |
| | $P_2(T_2)=0,3363$ | $P_3(T_2)=0,3290$ | $P_4(T_2)=0,3365$ | $P_5(T_2)=0,3346$ |
| | $F_2(C_2)=15753$ | $F_3(C_2)=10551$ | $F_4(C_2)=7882$ | $F_5(C_2)=6439$ |
| | $P_2(C_2)=0,1643$ | $P_3(C_2)=0,1651$ | $P_4(C_2)=0,1644$ | $P_5(C_2)=0,1679$ |
| | $F_2(G_2)=15791$ | $F_3(G_2)=11064$ | $F_4(G_2)=7856$ | $F_5(G_2)=6415$ |
| | $P_2(G_2)=0,1647$ | $P_3(G_2)=0,1731$ | $P_4(G_2)=0,1639$ | $P_5(G_2)=0,1673$ |
| | | $F_3(A_3)=21337$ | $F_4(A_3)=15900$ | $F_5(A_3)=12785$ |
| | | $P_3(A_3)=0,3338$ | $P_4(A_3)=0,3317$ | $P_5(A_3)=0,3334$ |
| | | $F_3(T_3)=21239$ | $F_4(T_3)=15820$ | $F_5(T_3)=12666$ |
| | | $P_3(T_3)=0,3323$ | $P_4(T_3)=0,3300$ | $P_5(T_3)=0,3303$ |
| | | $F_3(C_3)=11007$ | $F_4(C_3)=8120$ | $F_5(C_3)=6480$ |
| | | $P_3(C_3)=0,1722$ | $P_4(C_3)=0,1694$ | $P_5(C_3)=0,1690$ |
| | | $F_3(G_3)=10329$ | $F_4(G_3)=8094$ | $F_5(G_3)=6416$ |
| | | $P_3(G_3)=0,1616$ | $P_4(G_3)=0,1689$ | $P_5(G_3)=0,1673$ |

| | |
|---|---|
| $F_4(A_4)$=16024 | $F_5(A_4)$=12890 |
| $P_4(A_4)$=0,3343 | $P_5(A_4)$=0,3361 |
| $F_4(T_4)$=16104 | $F_5(T_4)$=12635 |
| $P_4(T_4)$=0,3360 | $P_5(T_4)$=0,3295 |
| $F_4(C_4)$=7871 | $F_5(C_4)$=6429 |
| $P_4(C_4)$=0,1642 | $P_5(C_4)$=0,1677 |
| $F_4(G_4)$=7935 | $F_5(G_4)$=6393 |
| $P_4(G_4)$=0,1655 | $P_5(G_4)$=0,1667 |
| | $F_5(A_5)$=12711 |
| | $P_5(A_5)$=0,3315 |
| | $F_5(T_5)$=12847 |
| | $P_5(T_5)$=0,3350 |
| | $F_5(C_5)$=6307 |
| | $P_5(C_5)$=0,1645 |
| | $F_5(G_5)$=6482 |
| | $P_5(G_5)$=0,1690 |

Fig. A1/21. Collective frequencies $F_n(A_k)$, $F_n(T_k)$, $F_n(C_k)$, $F_n(G_k)$ and collective probabilties $P_n(A_k)$, $P_n(T_k)$, $P_n(C_k)$ and $P_n(G_k)$ (n = 1, 2, 3, 4, 5 and $k \leq n$) of tetra-group subgroups in sequences of n-plets in the case of the VACCG, Vaccinia virus Copenhagen, complete genome, 191737 bp., accession M35027.1, https://www.ncbi.nlm.nih.gov/nuccore/335317

| **NUCLEOTIDES** $\Sigma_1$ = 186609 | **DOUBLETS** $\Sigma_2$ = 93304 | **TRIPLETS** $\Sigma_3$ =62203 | **4-PLETS** $\Sigma_4$ =46652 | **5-PLETS** $\Sigma_5$ =37321 |
|---|---|---|---|---|
| $F_1(A_1)$= 53206 | $F_2(A_1)$= 26575 | $F_3(A_1)$= 17874 | $F_4(A_1)$= 13285 | $F_5(A_1)$= 10665 |
| $P_1(A_1)$= 0,2851 | $P_2(A_1)$=0,2848 | $P_3(A_1)$=0,2873 | $P_4(A_1)$=0,2848 | $P_5(A_1)$=0,2858 |
| $F_1(T_1)$= 54264 | $F_2(T_1)$= 27190 | $F_3(T_1)$= 17831 | $F_4(T_1)$= 13477 | $F_5(T_1)$= 10782 |
| $P_1(T_1)$=0,2908 | $P_2(T_1)$=0,2914 | $P_3(T_1)$=0,2867 | $P_4(T_1)$=0,2889 | $P_5(T_1)$=0,2889 |
| $F_1(C_1)$= 39215 | $F_2(C_1)$= 19603 | $F_3(C_1)$= 12979 | $F_4(C_1)$= 9882 | $F_5(C_1)$= 7742 |
| $P_1(C_1)$=0,2101 | $P_2(C_1)$=0,2101 | $P_3(C_1)$=0,2087 | $P_4(C_1)$=0,2118 | $P_5(C_1)$=0,2074 |
| $F_1(G_1)$= 39924 | $F_2(G_1)$= 19936 | $F_3(G_1)$= 12979 | $F_4(G_1)$= 9882 | $F_5(G_1)$= 7742 |
| $P_1(G_1)$=0,2139 | $P_2(G_1)$=0,2137 | $P_3(G_1)$=0,2087 | $P_4(G_1)$=0,2118 | $P_5(G_1)$=0,2074 |
| | $F_2(A_2)$= 26631 | $F_3(A_2)$=17725 | $F_4(A_2)$=13424 | $F_5(A_2)$=10815 |
| | $P_2(A_2)$=0,2854 | $P_3(A_2)$=0,2850 | $P_4(A_2)$=0,2877 | $P_5(A_2)$=0,2898 |
| | $F_2(T_2)$= 27073 | $F_3(T_2)$=18094 | $F_4(T_2)$=13473 | $F_5(T_2)$=10781 |
| | $P_2(T_2)$=0,2902 | $P_3(T_2)$=0,2909 | $P_4(T_2)$=0,2888 | $P_5(T_2)$=0,2889 |
| | $F_2(C_2)$= 19612 | $F_3(C_2)$=13077 | $F_4(C_2)$= 9746 | $F_5(C_2)$=7721 |
| | $P_2(C_2)$=0,2102 | $P_3(C_2)$=0,2102 | $P_4(C_2)$=0,2089 | $P_5(C_2)$=0,2069 |
| | $F_2(G_2)$= 19988 | $F_3(G_2)$=13307 | $F_4(G_2)$=10009 | $F_5(G_2)$=8004 |
| | $P_2(G_2)$=0,2142 | $P_3(G_2)$=0,2139 | $P_4(G_2)$=0,2145 | $P_5(G_2)$=0,2144 |
| | | $F_3(A_3)$= 17607 | $F_4(A_3)$= 13290 | $F_5(A_3)$=10608 |
| | | $P_3(A_3)$=0,2831 | $P_4(A_3)$=0,2849 | $P_5(A_3)$=0,2842 |
| | | $F_3(T_3)$= 18339 | $F_4(T_3)$=13713 | $F_5(T_3)$=10949 |
| | | $P_3(T_3)$=0,2948 | $P_4(T_3)$=0,2939 | $P_5(T_3)$=0,2934 |
| | | $F_3(C_3)$= 13159 | $F_4(C_3)$=9721 | $F_5(C_3)$=7829 |
| | | $P_3(C_3)$=0,2115 | $P_4(C_3)$=0,2084 | $P_5(C_3)$=0,2098 |
| | | $F_3(G_3)$= 13098 | $F_4(G_3)$=9928 | $F_5(G_3)$=7935 |
| | | $P_3(G_3)$=0,2106 | $P_4(G_3)$=0,2128 | $P_5(G_3)$=0,2126 |
| | | | $F_4(A_4)$= 13207 | $F_5(A_4)$=10611 |
| | | | $P_4(A_4)$=0,2831 | $P_5(A_4)$=0,2843 |
| | | | $F_4(T_4)$= 13600 | $F_5(T_4)$=10891 |

| | |
|---|---|
| $P_4(T_4)$=0,2915 | $P_5(T_4)$=0,2918 |
| $F_4(C_4)$= 9866 | $F_5(C_4)$= 7902 |
| $P_4(C_4)$=0,2115 | $P_5(C_4)$=0,2117 |
| $F_4(G_4)$= 9979 | $F_5(G_4)$= 7917 |
| $P_4(G_4)$=0,2139 | $P_5(G_4)$= 0,2121 |
| | $F_5(A_5)$= 10504 |
| | $P_5(A_5)$=0,2815 |
| | $F_5(T_5)$= 10860 |
| | $P_5(T_5)$=0,2910 |
| | $F_5(C_5)$= 8021 |
| | $P_5(C_5)$=0,2149 |
| | $F_5(G_5)$= 7936 |
| | $P_5(G_5)$=0,2126 |

Fig. A1/22. Collective frequencies $F_n(A_k)$, $F_n(T_k)$, $F_n(C_k)$, $F_n(G_k)$ and collective probabilties $P_n(A_k)$, $P_n(T_k)$, $P_n(C_k)$ and $P_n(G_k)$ (n = 1, 2, 3, 4, 5 and k ≤ n) of tetra-group subgroups in sequences of n-plets in the case of the MPOMTCG, Marchantia paleacea isolate A 18 mitochondrion, complete genome, 186609 bp, accession M68929.1, https://www.ncbi.nlm.nih.gov/nuccore/786182

| **NUCLEOTIDES** | **DOUBLETS** | **TRIPLETS** | **4-PLETS** | **5-PLETS** |
|---|---|---|---|---|
| $\Sigma_1$ = 184113 | $\Sigma_2$ = 92056 | $\Sigma_3$ =61371 | $\Sigma_4$ =46028 | $\Sigma_5$ =36822 |
| $F_1(A_1)$= 36002 | $F_2(A_1)$= 18098 | $F_3(A_1)$= 11910 | $F_4(A_1)$= 9044 | $F_5(A_1)$= 7210 |
| $P_1(A_1)$=0,1955 | $P_2(A_1)$=0,1966 | $P_3(A_1)$=0,1941 | $P_4(A_1)$=0,1965 | $P_5(A_1)$=0,1958 |
| $F_1(T_1)$= 37665 | $F_2(T_1)$= 18887 | $F_3(T_1)$= 12160 | $F_4(T_1)$= 9394 | $F_5(T_1)$= 7571 |
| $P_1(T_1)$=0,2046 | $P_2(T_1)$=0,2052 | $P_3(T_1)$=0,1981 | $P_4(T_1)$=0,2041 | $P_5(T_1)$=0,2056 |
| $F_1(C_1)$= 55824 | $F_2(C_1)$= 27823 | $F_3(C_1)$= 19144 | $F_4(C_1)$= 13990 | $F_5(C_1)$= 11162 |
| $P_1(C_1)$=0,3032 | $P_2(C_1)$=0,3022 | $P_3(C_1)$=0,3119 | $P_4(C_1)$=0,3039 | $P_5(C_1)$=0,3031 |
| $F_1(G_1)$= 54622 | $F_2(G_1)$= 27248 | $F_3(G_1)$= 18157 | $F_4(G_1)$= 13600 | $F_5(G_1)$= 10879 |
| $P_1(G_1)$=0,2967 | $P_2(G_1)$=0,2960 | $P_3(G_1)$=0,2959 | $P_4(G_1)$=0,2955 | $P_5(G_1)$=0,2954 |
| | $F_2(A_2)$= 17903 | $F_3(A_2)$= 11597 | $F_4(A_2)$=8758 | $F_5(A_2)$=7145 |
| | $P_2(A_2)$=0,1945 | $P_3(A_2)$=0,1890 | $P_4(A_2)$=0,1903 | $P_5(A_2)$=0,1940 |
| | $F_2(T_2)$= 18778 | $F_3(T_2)$= 12565 | $F_4(T_2)$=9279 | $F_5(T_2)$=7464 |
| | $P_2(T_2)$=0,2040 | $P_3(T_2)$=0,2047 | $P_4(T_2)$=0,2016 | $P_5(T_2)$=0,2027 |
| | $F_2(C_2)$= 28001 | $F_3(C_2)$= 18517 | $F_4(C_2)$=14361 | $F_5(C_2)$= 11163 |
| | $P_2(C_2)$=0,3042 | $P_3(C_2)$=0,3017 | $P_4(C_2)$=0,3120 | $P_5(C_2)$=0,3032 |
| | $F_2(G_2)$= 27374 | $F_3(G_2)$=18692 | $F_4(G_2)$=13630 | $F_5(G_2)$= 11050 |
| | $P_2(G_2)$=0,2974 | $P_3(G_2)$=0,3046 | $P_4(G_2)$=0,2961 | $P_5(G_2)$=0,3001 |
| | | $F_3(A_3)$= 12495 | $F_4(A_3)$= 9054 | $F_5(A_3)$= 7232 |
| | | $P_3(A_3)$= 0,2036 | $P_4(A_3)$=0,1967 | $P_5(A_3)$=0,1964 |
| | | $F_3(T_3)$= 12940 | $F_4(T_3)$= 9493 | $F_5(T_3)$= 7570 |
| | | $P_3(T_3)$=0,2108 | $P_4(T_3)$=0,2062 | $P_5(T_3)$= 0,2056 |
| | | $F_3(C_3)$= 18163 | $F_4(C_3)$= 13833 | $F_5(C_3)$=11189 |
| | | $P_3(C_3)$=0,2960 | $P_4(C_3)$= 0,3005 | $P_5(C_3)$=0,3039 |
| | | $F_3(G_3)$= 17773 | $F_4(G_3)$=13648 | $F_5(G_3)$=10831 |
| | | $P_3(G_3)$=0,2896 | $P_4(G_3)$= 0,2965 | $P_5(G_3)$=0,2941 |
| | | | $F_4(A_4)$= 9145 | $F_5(A_4)$= 7236 |
| | | | $P_4(A_4)$=0,1987 | $P_5(A_4)$=0,1965 |
| | | | $F_4(T_4)$= 9499 | $F_5(T_4)$=7475 |
| | | | $P_4(T_4)$=0,2064 | $P_5(T_4)$=0,2030 |
| | | | $F_4(C_4)$= 13640 | $F_5(C_4)$=11165 |

| | |
|---|---|
| $P_4(C_4)$=0,2963 | $P_5(C_4)$=0,3032 |
| $F_4(G_4)$= 13744 | $F_5(G_4)$=10946 |
| $P_4(G_4)$=0,2986 | $P_5(G_4)$=0,2973 |
| | $F_5(A_5)$= 7178 |
| | $P_5(A_5)$=0,1949 |
| | $F_5(T_5)$= 7583 |
| | $P_5(T_5)$=0,2059 |
| | $F_5(C_5)$= 11145 |
| | $P_5(C_5)$=0,3027 |
| | $F_5(G_5)$= 10916 |
| | $P_5(G_5)$=0,2965 |

Fig. A1/23. Collective frequencies $F_n(A_k)$, $F_n(T_k)$, $F_n(C_k)$, $F_n(G_k)$ and collective probabilties $P_n(A_k)$, $P_n(T_k)$, $P_n(C_k)$ and $P_n(G_k)$ ($n$ = 1, 2, 3, 4, 5 and $k \le n$) of tetra-group subgroups in sequences of n-plets in the case of the HS4B958RAJ, Epstein-Barr virus, 184113 bp, accession M80517.1, https://www.ncbi.nlm.nih.gov/nuccore/330330

| **NUCLEOTIDES** $\Sigma_1$ = 155943 | **DOUBLETS** $\Sigma_2$ = 77971 | **TRIPLETS** $\Sigma_3$ =51981 | **4-PLETS** $\Sigma_4$ =38985 | **5-PLETS** $\Sigma_5$ =31188 |
|---|---|---|---|---|
| $F_1(A_1)$= 47860 | $F_2(A_1)$= 23952 | $F_3(A_1)$= 16025 | $F_4(A_1)$= 12043 | $F_5(A_1)$= 9661 |
| $P_1(A_1)$=0,3069 | $P_2(A_1)$=0,3072 | $P_3(A_1)$=0,3083 | $P_4(A_1)$=0,3089 | $P_5(A_1)$=0,3098 |
| $F_1(T_1)$= 49064 | $F_2(T_1)$= 24497 | $F_3(T_1)$= 16460 | $F_4(T_1)$= 12313 | $F_5(T_1)$= 9743 |
| $P_1(T_1)$=0,3146 | $P_2(T_1)$=0,3142 | $P_3(T_1)$=0,3167 | $P_4(T_1)$=0,3158 | $P_5(T_1)$=0,3124 |
| $F_1(C_1)$= 30014 | $F_2(C_1)$= 15095 | $F_3(C_1)$= 9974 | $F_4(C_1)$= 7452 | $F_5(C_1)$= 5997 |
| $P_1(C_1)$=0,1925 | $P_2(C_1)$=0,1936 | $P_3(C_1)$=0,1919 | $P_4(C_1)$=0,1912 | $P_5(C_1)$=0,1923 |
| $F_1(G_1)$= 29005 | $F_2(G_1)$= 14427 | $F_3(G_1)$= 9522 | $F_4(G_1)$= 7177 | $F_5(G_1)$= 5787 |
| $P_1(G_1)$=0,1860 | $P_2(G_1)$=0,1850 | $P_3(G_1)$=0,1832 | $P_4(G_1)$=0,1841 | $P_5(G_1)$=0,1856 |
| | $F_2(A_2)$= 23907 | $F_3(A_2)$=16233 | $F_4(A_2)$= 12039 | $F_5(A_2)$=9605 |
| | $P_2(A_2)$=0,3066 | $P_3(A_2)$=0,3123 | $P_4(A_2)$=0,3088 | $P_5(A_2)$=0,3080 |
| | $F_2(T_2)$= 24567 | $F_3(T_2)$=16124 | $F_4(T_2)$=12192 | $F_5(T_2)$=9680 |
| | $P_2(T_2)$=0,3151 | $P_3(T_2)$=0,3102 | $P_4(T_2)$=0,3127 | $P_5(T_2)$=0,3104 |
| | $F_2(C_2)$= 14919 | $F_3(C_2)$=9963 | $F_4(C_2)$=7486 | $F_5(C_2)$=6021 |
| | $P_2(C_2)$=0,1913 | $P_3(C_2)$=0,1917 | $P_4(C_2)$=0,1920 | $P_5(C_2)$=0,1930 |
| | $F_2(G_2)$= 14578 | $F_3(G_2)$=9661 | $F_4(G_2)$=7268 | $F_5(G_2)$=5882 |
| | $P_2(G_2)$=0,1870 | $P_3(G_2)$=0,1859 | $P_4(G_2)$=0,1864 | $P_5(G_2)$=0,1886 |
| | | $F_3(A_3)$= 15602 | $F_4(A_3)$=11909 | $F_5(A_3)$=9461 |
| | | $P_3(A_3)$=0,3001 | $P_4(A_3)$=0,3055 | $P_5(A_3)$=0,3034 |
| | | $F_3(T_3)$= 16480 | $F_4(T_3)$=12183 | $F_5(T_3)$=9958 |
| | | $P_3(T_3)$=0,3170 | $P_4(T_3)$=0,3125 | $P_5(T_3)$=0,3193 |
| | | $F_3(C_3)$= 10077 | $F_4(C_3)$=7643 | $F_5(C_3)$=6022 |
| | | $P_3(C_3)$=0,1939 | $P_4(C_3)$=0,1960 | $P_5(C_3)$=0,1931 |
| | | $F_3(G_3)$= 9822 | $F_4(G_3)$= 7250 | $F_5(G_3)$= 5747 |
| | | $P_3(G_3)$=0,1890 | $P_4(G_3)$=0,1860 | $P_5(G_3)$=0,1843 |
| | | | $F_4(A_4)$= 11867 | $F_5(A_4)$= 9641 |
| | | | $P_4(A_4)$=0,3044 | $P_5(A_4)$=0,3091 |
| | | | $F_4(T_4)$= 12375 | $F_5(T_4)$= 9791 |
| | | | $P_4(T_4)$=0,3174 | $P_5(T_4)$=0,3139 |
| | | | $F_4(C_4)$= 7433 | $F_5(C_4)$=6004 |

| | |
|---|---|
| $P_4(C_4)$=0,1907 | $P_5(C_4)$=0,1925 |
| $F_4(G_4)$= 7310 | $F_5(G_4)$=5752 |
| $P_4(G_4)$=0,1875 | $P_5(G_4)$=0,1844 |
| | $F_5(A_5)$= 9490 |
| | $P_5(A_5)$=0,3043 |
| | $F_5(T_5)$= 9891 |
| | $P_5(T_5)$=0,3171 |
| | $F_5(C_5)$= 5970 |
| | $P_5(C_5)$=0,1914 |
| | $F_5(G_5)$= 5837 |
| | $P_5(G_5)$=0,1872 |

Fig. A1/24. Collective frequencies $F_n(A_k)$, $F_n(T_k)$, $F_n(C_k)$, $F_n(G_k)$ and collective probabilties $P_n(A_k)$, $P_n(T_k)$, $P_n(C_k)$ and $P_n(G_k)$ ($n$ = 1, 2, 3, 4, 5 and $k \leq n$) of tetra-group subgroups in sequences of n-plets in the case of the Nicotiana tabacum chloroplast genome DNA,155943 bp, accession Z00044.2, https://www.ncbi.nlm.nih.gov/nuccore/Z00044

| **NUCLEOTIDES** | **DOUBLETS** | **TRIPLETS** | **4-PLETS** | **5-PLETS** |
|---|---|---|---|---|
| $\Sigma_1$ = 150224 | $\Sigma_2$ = 75112 | $\Sigma_3$ =50074 | $\Sigma_4$ =37556 | $\Sigma_5$ =30044 |
| $F_1(A_1)$= 32616 | $F_2(A_1)$= 16273 | $F_3(A_1)$= 11331 | $F_4(A_1)$= 8082 | $F_5(A_1)$= 6480 |
| $P_1(A_1)$=0,2171 | $P_2(A_1)$=0,2166 | $P_3(A_1)$=0,2263 | $P_4(A_1)$= 0,2152 | $P_5(A_1)$=0,2157 |
| $F_1(T_1)$= 32482 | $F_2(T_1)$= 16191 | $F_3(T_1)$= 10871 | $F_4(T_1)$= 8106 | $F_5(T_1)$= 6514 |
| $P_1(T_1)$=0,2162 | $P_2(T_1)$=0,2156 | $P_3(T_1)$=0,2171 | $P_4(T_1)$=0,2158 | $P_5(T_1)$=0,2168 |
| $F_1(C_1)$= 43173 | $F_2(C_1)$= 21746 | $F_3(C_1)$= 14287 | $F_4(C_1)$= 10735 | $F_5(C_1)$= 8646 |
| $P_1(C_1)$=0,2874 | $P_2(C_1)$=0,2895 | $P_3(C_1)$=0,2853 | $P_4(C_1)$=0,2858 | $P_5(C_1)$=0,2878 |
| $F_1(G_1)$= 41953 | $F_2(G_1)$= 20902 | $F_3(G_1)$= 13585 | $F_4(G_1)$= 10633 | $F_5(G_1)$= 8404 |
| $P_1(G_1)$=0,2793 | $P_2(G_1)$=0,2783 | $P_3(G_1)$=0,2713 | $P_4(G_1)$=0,2831 | $P_5(G_1)$=0,2797 |
| | $F_2(A_2)$= 16343 | $F_3(A_2)$=10347 | $F_4(A_2)$=8121 | $F_5(A_2)$=6551 |
| | $P_2(A_2)$=0,2176 | $P_3(A_2)$=0,2066 | $P_4(A_2)$=0,2162 | $P_5(A_2)$=0,2180 |
| | $F_2(T_2)$= 16291 | $F_3(T_2)$=10789 | $F_4(T_2)$=8144 | $F_5(T_2)$=6599 |
| | $P_2(T_2)$=0,2169 | $P_3(T_2)$=0,2155 | $P_4(T_2)$=0,2168 | $P_5(T_2)$=0,2196 |
| | $F_2(C_2)$= 21427 | $F_3(C_2)$=14889 | $F_4(C_2)$=10703 | $F_5(C_2)$=8569 |
| | $P_2(C_2)$=0,2853 | $P_3(C_2)$=0,2973 | $P_4(C_2)$=0,2850 | $P_5(C_2)$=0,2852 |
| | $F_2(G_2)$= 21051 | $F_3(G_2)$=14049 | $F_4(G_2)$=10588 | $F_5(G_2)$=8325 |
| | $P_2(G_2)$=0,2803 | $P_3(G_2)$=0,2806 | $P_4(G_2)$=0,2819 | $P_5(G_2)$=0,2771 |
| | | $F_3(A_3)$= 10938 | $F_4(A_3)$=8191 | $F_5(A_3)$=6492 |
| | | $P_3(A_3)$=0,2184 | $P_4(A_3)$=0,2181 | $P_5(A_3)$=0,2161 |
| | | $F_3(T_3)$= 10822 | $F_4(T_3)$=8085 | $F_5(T_3)$=6475 |
| | | $P_3(T_3)$=0,2161 | $P_4(T_3)$=0,2153 | $P_5(T_3)$=0,2155 |
| | | $F_3(C_3)$= 13997 | $F_4(C_3)$=11011 | $F_5(C_3)$=8519 |
| | | $P_3(C_3)$=0,2795 | $P_4(C_3)$=0,2932 | $P_5(C_3)$=0,2835 |
| | | $F_3(G_3)$= 14317 | $F_4(G_3)$=10269 | $F_5(G_3)$=8558 |
| | | $P_3(G_3)$=0,2859 | $P_4(G_3)$=0,2734 | $P_5(G_3)$=0,2848 |
| | | | $F_4(A_4)$= 8222 | $F_5(A_4)$=6551 |
| | | | $P_4(A_4)$=0,2189 | $P_5(A_4)$=0,2180 |
| | | | $F_4(T_4)$= 8147 | $F_5(T_4)$=6423 |
| | | | $P_4(T_4)$=0,2169 | $P_5(T_4)$=0,2138 |
| | | | $F_4(C_4)$= 10724 | $F_5(C_4)$=8716 |
| | | | $P_4(C_4)$=0,2855 | $P_5(C_4)$=0,2901 |

| $F_4(G_4)$= 10463 | $F_5(G_4)$=8354 |
| --- | --- |
| $P_4(G_4)$=0,2786 | $P_5(G_4)$=0,2781 |
| | $F_5(A_5)$= 6542 |
| | $P_5(A_5)$=0,2177 |
| | $F_5(T_5)$= 6471 |
| | $P_5(T_5)$=0,2154 |
| | $F_5(C_5)$= 8723 |
| | $P_5(C_5)$=0,2903 |
| | $F_5(G_5)$= 8308 |
| | $P_5(G_5)$=0,2765 |

Fig. A1/25. Collective frequencies $F_n(A_k)$, $F_n(T_k)$, $F_n(C_k)$, $F_n(G_k)$ and collective probabilties $P_n(A_k)$, $P_n(T_k)$, $P_n(C_k)$ and $P_n(G_k)$ ($n$ = 1, 2, 3, 4, 5 and $k \leq n$) of tetra-group subgroups in sequences of n-plets in the case of the Equine herpesvirus 1 strain Ab4, complete genome, 150224 bp, accession AY665713.1, https://www.ncbi.nlm.nih.gov/nuccore/AY665713.1

| **NUCLEOTIDES** $\Sigma_1$ = 134502 | **DOUBLETS** $\Sigma_2$ = 67251 | **TRIPLETS** $\Sigma_3$ =44834 | **4-PLETS** $\Sigma_4$ =33625 | **5-PLETS** $\Sigma_5$ =26900 |
| --- | --- | --- | --- | --- |
| $F_1(A_1)$= 41231 | $F_2(A_1)$= 20572 | $F_3(A_1)$= 13637 | $F_4(A_1)$= 10209 | $F_5(A_1)$= 8302 |
| $P_1(A_1)$=0,3065 | $P_2(A_1)$=0,3059 | $P_3(A_1)$=0,3042 | $P_4(A_1)$=0,3036 | $P_5(A_1)$=0,3086 |
| $F_1(T_1)$= 40818 | $F_2(T_1)$= 20391 | $F_3(T_1)$= 13675 | $F_4(T_1)$= 10159 | $F_5(T_1)$= 8094 |
| $P_1(T_1)$=0,3035 | $P_2(T_1)$=0,3032 | $P_3(T_1)$=0,3050 | $P_4(T_1)$=0,3021 | $P_5(T_1)$=0,3009 |
| $F_1(C_1)$= 26129 | $F_2(C_1)$= 13035 | $F_3(C_1)$= 8815 | $F_4(C_1)$= 6576 | $F_5(C_1)$= 5341 |
| $P_1(C_1)$=0,1943 | $P_2(C_1)$=0,1938 | $P_3(C_1)$=0,1966 | $P_4(C_1)$=0,1956 | $P_5(C_1)$=0,1986 |
| $F_1(G_1)$= 26324 | $F_2(G_1)$= 13253 | $F_3(G_1)$= 8707 | $F_4(G_1)$= 6681 | $F_5(G_1)$= 5163 |
| $P_1(G_1)$=0,1957 | $P_2(G_1)$=0,1971 | $P_3(G_1)$=0,1942 | $P_4(G_1)$=0,1987 | $P_5(G_1)$=0,1919 |
| | $F_2(A_2)$= 20659 | $F_3(A_2)$=13872 | $F_4(A_2)$=10372 | $F_5(A_2)$=8240 |
| | $P_2(A_2)$=0,3072 | $P_3(A_2)$=0,3094 | $P_4(A_2)$=0,3085 | $P_5(A_2)$=0,3063 |
| | $F_2(T_2)$= 20427 | $F_3(T_2)$=13761 | $F_4(T_2)$=10178 | $F_5(T_2)$=8209 |
| | $P_2(T_2)$=0,3037 | $P_3(T_2)$=0,3069 | $P_4(T_2)$=0,3027 | $P_5(T_2)$=0,3052 |
| | $F_2(C_2)$= 13094 | $F_3(C_2)$=8610 | $F_4(C_2)$=6498 | $F_5(C_2)$=5195 |
| | $P_2(C_2)$=0,1947 | $P_3(C_2)$=0,1920 | $P_4(C_2)$=0,1932 | $P_5(C_2)$=0,1931 |
| | $F_2(G_2)$= 13071 | $F_3(G_2)$=8591 | $F_4(G_2)$=6577 | $F_5(G_2)$=5256 |
| | $P_2(G_2)$=0,1944 | $P_3(G_2)$=0,1916 | $P_4(G_2)$=0,1956 | $P_5(G_2)$=0,1954 |
| | | $F_3(A_3)$= 13722 | $F_4(A_3)$=10363 | $F_5(A_3)$=8236 |
| | | $P_3(A_3)$= 0,3061 | $P_4(A_3)$=0,3082 | $P_5(A_3)$=0,3062 |
| | | $F_3(T_3)$=13382 | $F_4(T_3)$=10231 | $F_5(T_3)$=8206 |
| | | $P_3(T_3)$=0,2985 | $P_4(T_3)$=0,3043 | $P_5(T_3)$=0,3051 |
| | | $F_3(C_3)$= 8704 | $F_4(C_3)$=6459 | $F_5(C_3)$=5129 |
| | | $P_3(C_3)$=0,1941 | $P_4(C_3)$=0,1921 | $P_5(C_3)$=0,1907 |
| | | $F_3(G_3)$= 9026 | $F_4(G_3)$=6572 | $F_5(G_3)$=5329 |
| | | $P_3(G_3)$=0,2013 | $P_4(G_3)$=0,1954 | $P_5(G_3)$=0,1981 |
| | | | $F_4(A_4)$= 10286 | $F_5(A_4)$=8255 |
| | | | $P_4(A_4)$=0,3059 | $P_5(A_4)$=0,3069 |
| | | | $F_4(T_4)$= 10249 | $F_5(T_4)$=8179 |
| | | | $P_4(T_4)$=0,3048 | $P_5(T_4)$=0,3041 |
| | | | $F_4(C_4)$= 6596 | $F_5(C_4)$= 5157 |
| | | | $P_4(C_4)$=0,1962 | $P_5(C_4)$=0,1917 |

| | |
|---|---|
| $F_4(G_4)$= 6494<br>$P_4(G_4)$=0,1931 | $F_5(G_4)$= 5309<br>$P_5(G_4)$=0,1974 |
| | $F_5(A_5)$= 8197<br>$P_5(A_5)$=0,3047 |
| | $F_5(T_5)$= 8129<br>$P_5(T_5)$=0,3022 |
| | $F_5(C_5)$= 5307<br>$P_5(C_5)$=0,1973 |
| | $F_5(G_5)$= 5267<br>$P_5(G_5)$=0,1958 |

Fig. A1/26. Collective frequencies $F_n(A_k)$, $F_n(T_k)$, $F_n(C_k)$, $F_n(G_k)$ and collective probabilties $P_n(A_k)$, $P_n(T_k)$, $P_n(C_k)$ and $P_n(G_k)$ ($n$ = 1, 2, 3, 4, 5 and $k \le n$) of tetra-group subgroups in sequences of n-plets in the case of the Oryza sativa cultivar TN1 chloroplast, complete genome, 134502 bp, accession NC_031333.1, https://www.ncbi.nlm.nih.gov/nuccore/NC_031333.1

| **NUCLEOTIDES**<br>**$\Sigma_1$ =** 134226 | **DOUBLETS**<br>**$\Sigma_2$ =** 67113 | **TRIPLETS**<br>**$\Sigma_3$ =**44742 | **4-PLETS**<br>**$\Sigma_4$ =**33556 | **5-PLETS**<br>**$\Sigma_5$ =**26845 |
|---|---|---|---|---|
| $F_1(A_1)$= 28727<br>$P_1(A_1)$=0,2140 | $F_2(A_1)$= 14299<br>$P_2(A_1)$=0,2131 | $F_3(A_1)$= 9855<br>$P_3(A_1)$=0,2203 | $F_4(A_1)$= 7148<br>$P_4(A_1)$=0,2130 | $F_5(A_1)$= 5795<br>$P_5(A_1)$= 0,2159 |
| $F_1(T_1)$= 30025<br>$P_1(T_1)$=0,2237 | $F_2(T_1)$= 15047<br>$P_2(T_1)$=0,2242 | $F_3(T_1)$= 10633<br>$P_3(T_1)$=0,2377 | $F_4(T_1)$= 7535<br>$P_4(T_1)$=0,2246 | $F_5(T_1)$= 5969<br>$P_5(T_1)$=0,2224 |
| $F_1(C_1)$= 37767<br>$P_1(C_1)$=0,2814 | $F_2(C_1)$= 18943<br>$P_2(C_1)$=0,2823 | $F_3(C_1)$= 12066<br>$P_3(C_1)$=0,2697 | $F_4(C_1)$= 9436<br>$P_4(C_1)$=0,2812 | $F_5(C_1)$= 7480<br>$P_5(C_1)$=0,2786 |
| $F_1(G_1)$= 37707<br>$P_1(G_1)$=0,2809 | $F_2(G_1)$= 18824<br>$P_2(G_1)$=0,2805 | $F_3(G_1)$= 12188<br>$P_3(G_1)$=0,2724 | $F_4(G_1)$= 9437<br>$P_4(G_1)$=0,2812 | $F_5(G_1)$= 7601<br>$P_5(G_1)$=0,2831 |
| | $F_2(A_2)$= 14428<br>$P_2(A_2)$=0,2150 | $F_3(A_2)$=8918<br>$P_3(A_2)$=0,1993 | $F_4(A_2)$=7195<br>$P_4(A_2)$=0,2144 | $F_5(A_2)$=5812<br>$P_5(A_2)$=0,2165 |
| | $F_2(T_2)$= 14978<br>$P_2(T_2)$=0,2232 | $F_3(T_2)$=9597<br>$P_3(T_2)$=0,2145 | $F_4(T_2)$=7472<br>$P_4(T_2)$=0,2227 | $F_5(T_2)$=6090<br>$P_5(T_2)$=0,2266 |
| | $F_2(C_2)$= 18824<br>$P_2(C_2)$=0,2805 | $F_3(C_2)$=13595<br>$P_3(C_2)$=0,3038 | $F_4(C_2)$=9411<br>$P_4(C_2)$=0,2805 | $F_5(C_2)$=7444<br>$P_5(C_2)$=0,2793 |
| | $F_2(G_2)$= 18883<br>$P_2(G_2)$=0,2814 | $F_3(G_2)$=12632<br>$P_3(G_2)$=0,2824 | $F_4(G_2)$=9478<br>$P_4(G_2)$=0,2825 | $F_5(G_2)$=7499<br>$P_5(G_2)$=0,2793 |
| | | $F_3(A_3)$= 9954<br>$P_3(A_3)$=0,2225 | $F_4(A_3)$=7151<br>$P_4(A_3)$=0,2131 | $F_5(A_3)$=5737<br>$P_5(A_3)$=0,2137 |
| | | $F_3(T_3)$= 9795<br>$P_3(T_3)$=0,2189 | $F_4(T_3)$=7512<br>$P_4(T_3)$=0,2239 | $F_5(T_3)$=6059<br>$P_5(T_3)$=0,2257 |
| | | $F_3(C_3)$= 12106<br>$P_3(C_3)$=0,2706 | $F_4(C_3)$=9506<br>$P_4(C_3)$=0,2833 | $F_5(C_3)$=7527<br>$P_5(C_3)$=0,2804 |
| | | $F_3(G_3)$= 12887<br>$P_3(G_3)$=0,2880 | $F_4(G_3)$=9387<br>$P_4(G_3)$=0,2797 | $F_5(G_3)$=7522<br>$P_5(G_3)$=0,2802 |
| | | | $F_4(A_4)$= 7233<br>$P_4(A_4)$=0,2156 | $F_5(A_4)$=5627<br>$P_5(A_4)$=0,2096 |
| | | | $F_4(T_4)$= 7506<br>$P_4(T_4)$=0,2237 | $F_5(T_4)$=6042<br>$P_5(T_4)$=0,2251 |
| | | | $F_4(C_4)$= 9413<br>$P_4(C_4)$=0,2805 | $F_5(C_4)$=7634<br>$P_5(C_4)$=0,2844 |

| | |
|---|---|
| $F_4(G_4)$= 9404 | $F_5(G_4)$=7542 |
| $P_4(G_4)$=0,2802 | $P_5(G_4)$=0,2809 |
| | $F_5(A_5)$= 5756 |
| | $P_5(A_5)$=0,2144 |
| | $F_5(T_5)$= 5865 |
| | $P_5(T_5)$=0,2185 |
| | $F_5(C_5)$= 7682 |
| | $P_5(C_5)$=0,2862 |
| | $F_5(G_5)$= 7542 |
| | $P_5(G_5)$=0,2809 |

Fig. A1/27. Collective frequencies $F_n(A_k)$, $F_n(T_k)$, $F_n(C_k)$, $F_n(G_k)$ and collective probabilties $P_n(A_k)$, $P_n(T_k)$, $P_n(C_k)$ and $P_n(G_k)$ (n = 1, 2, 3, 4, 5 and k ≤ n) of tetra-group subgroups in sequences of n-plets in the case of the IH1CG, Ictalurid herpesvirus 1 strain Auburn 1, complete genome, 134226 bp, accession M75136.2, https://www.ncbi.nlm.nih.gov/nuccore/519868059

| **NUCLEOTIDES** $\Sigma_1$ = 124884 | **DOUBLETS** $\Sigma_2$ = 62442 | **TRIPLETS** $\Sigma_3$ =41628 | **4-PLETS** $\Sigma_4$ =31221 | **5-PLETS** $\Sigma_5$ =24976 |
|---|---|---|---|---|
| $F_1(A_1)$= 33782 | $F_2(A_1)$= 16995 | $F_3(A_1)$= 11327 | $F_4(A_1)$= 8467 | $F_5(A_1)$= 6724 |
| $P_1(A_1)$=0,2705 | $P_2(A_1)$=0,2722 | $P_3(A_1)$=0,2721 | $P_4(A_1)$=0,2712 | $P_5(A_1)$=0,2692 |
| $F_1(T_1)$= 33623 | $F_2(T_1)$= 16785 | $F_3(T_1)$= 11586 | $F_4(T_1)$= 8445 | $F_5(T_1)$= 6767 |
| $P_1(T_1)$=0,2692 | $P_2(T_1)$=0,2688 | $P_3(T_1)$=0,2783 | $P_4(T_1)$=0,2705 | $P_5(T_1)$=0,2709 |
| $F_1(C_1)$= 29295 | $F_2(C_1)$= 14590 | $F_3(C_1)$= 9512 | $F_4(C_1)$= 7336 | $F_5(C_1)$= 5826 |
| $P_1(C_1)$=0,2346 | $P_2(C_1)$=0,2337 | $P_3(C_1)$=0,2285 | $P_4(C_1)$=0,2350 | $P_5(C_1)$=0,2333 |
| $F_1(G_1)$= 28184 | $F_2(G_1)$= 14072 | $F_3(G_1)$= 9203 | $F_4(G_1)$= 6973 | $F_5(G_1)$= 5659 |
| $P_1(G_1)$=0,2257 | $P_2(G_1)$=0,2254 | $P_3(G_1)$=0,2211 | $P_4(G_1)$=0,2233 | $P_5(G_1)$=0,2266 |
| | $F_2(A_2)$= 16787 | $F_3(A_2)$=11097 | $F_4(A_2)$=8363 | $F_5(A_2)$= 6808 |
| | $P_2(A_2)$=0,2688 | $P_3(A_2)$=0,2666 | $P_4(A_2)$=0,2679 | $P_5(A_2)$=0,2726 |
| | $F_2(T_2)$= 16838 | $F_3(T_2)$=11042 | $F_4(T_2)$= 8448 | $F_5(T_2)$=6689 |
| | $P_2(T_2)$=0,2697 | $P_3(T_2)$=0,2653 | $P_4(T_2)$=0,2706 | $P_5(T_2)$=0,2678 |
| | $F_2(C_2)$= 14705 | $F_3(C_2)$=9814 | $F_4(C_2)$=7387 | $F_5(C_2)$=5797 |
| | $P_2(C_2)$=0,2355 | $P_3(C_2)$=0,2358 | $P_4(C_2)$=0,2366 | $P_5(C_2)$=0,2321 |
| | $F_2(G_2)$= 14112 | $F_3(G_2)$=9675 | $F_4(G_2)$=7023 | $F_5(G_2)$=5682 |
| | $P_2(G_2)$=0,2260 | $P_3(G_2)$=0,2324 | $P_4(G_2)$=0,2249 | $P_5(G_2)$=0,2275 |
| | | $F_3(A_3)$= 11358 | $F_4(A_3)$=8528 | $F_5(A_3)$=6678 |
| | | $P_3(A_3)$=0,2728 | $P_4(A_3)$=0,2731 | $P_5(A_3)$=0,2674 |
| | | $F_3(T_3)$= 10995 | $F_4(T_3)$=8340 | $F_5(T_3)$=6737 |
| | | $P_3(T_3)$=0,2641 | $P_4(T_3)$=0,2671 | $P_5(T_3)$=0,2697 |
| | | $F_3(C_3)$= 9969 | $F_4(C_3)$=7254 | $F_5(C_3)$=5980 |
| | | $P_3(C_3)$=0,2395 | $P_4(C_3)$=0,2323 | $P_5(C_3)$=0,2394 |
| | | $F_3(G_3)$= 9306 | $F_4(G_3)$=7099 | $F_5(G_3)$= 5581 |
| | | $P_3(G_3)$=0,2236 | $P_4(G_3)$=0,2274 | $P_5(G_3)$=0,2235 |
| | | | $F_4(A_4)$= 8424 | $F_5(A_4)$=6818 |
| | | | $P_4(A_4)$=0,2698 | $P_5(A_4)$=0,2730 |
| | | | $F_4(T_4)$= 8390 | $F_5(T_4)$= 6623 |
| | | | $P_4(T_4)$=0,2687 | $P_5(T_4)$=0,2652 |
| | | | $F_4(C_4)$= 7318 | $F_5(C_4)$=5846 |
| | | | $P_4(C_4)$=0,2344 | $P_5(C_4)$=0,2341 |
| | | | $F_4(G_4)$= 7089 | $F_5(G_4)$=5689 |
| | | | $P_4(G_4)$=0,2271 | $P_5(G_4)$=0,2278 |
| | | | | $F_5(A_5)$= 6753 |
| | | | | $P_5(A_5)$=0,2704 |
| | | | | $F_5(T_5)$= 6807 |
| | | | | $P_5(T_5)$=0,2725 |
| | | | | $F_5(C_5)$= 5846 |
| | | | | $P_5(C_5)$=0,2341 |
| | | | | $F_5(G_5)$= 5570 |
| | | | | $P_5(G_5)$=0,2230 |

Fig. A1/28. Collective frequencies $F_n(A_k)$, $F_n(T_k)$, $F_n(C_k)$, $F_n(G_k)$ and collective probabilties $P_n(A_k)$, $P_n(T_k)$, $P_n(C_k)$ and $P_n(G_k)$ (n = 1, 2, 3, 4, 5 and k ≤ n) of tetra-group subgroups in sequences of n-plets in the case of the Human herpesvirus 3 isolate 667/2005, complete genome, 124884 bp, accession JN704693.1, https://www.ncbi.nlm.nih.gov/nuccore/JN704693.1

| **NUCLEOTIDES** | **DOUBLETS** | **TRIPLETS** | **4-PLETS** | **5-PLETS** |
|---|---|---|---|---|
| $\Sigma_1$ = 121024 | $\Sigma_2$ = 60512 | $\Sigma_3$ =40341 | $\Sigma_4$ =30256 | $\Sigma_5$ =24204 |
| $F_1(A_1)$= 42896 | $F_2(A_1)$= 21385 | $F_3(A_1)$= 14349 | $F_4(A_1)$= 10739 | $F_5(A_1)$= 8517 |
| $P_1(A_1)$=0,3544 | $P_2(A_1)$=0,3534 | $P_3(A_1)$=0,3557 | $P_4(A_1)$=0,3549 | $P_5(A_1)$=0,3519 |
| $F_1(T_1)$= 43263 | $F_2(T_1)$= 21703 | $F_3(T_1)$= 14637 | $F_4(T_1)$= 10870 | $F_5(T_1)$= 8677 |
| $P_1(T_1)$=0,3575 | $P_2(T_1)$=0,3587 | $P_3(T_1)$=0,3628 | $P_4(T_1)$=0,3593 | $P_5(T_1)$=0,3585 |
| $F_1(C_1)$= 17309 | $F_2(C_1)$= 8677 | $F_3(C_1)$= 5687 | $F_4(C_1)$= 4340 | $F_5(C_1)$= 3448 |
| $P_1(C_1)$=0,1430 | $P_2(C_1)$=0,1434 | $P_3(C_1)$=0,1410 | $P_4(C_1)$=0,1434 | $P_5(C_1)$=0,1425 |
| $F_1(G_1)$= 17556 | $F_2(G_1)$= 8747 | $F_3(G_1)$= 5668 | $F_4(G_1)$= 4307 | $F_5(G_1)$= 3562 |
| $P_1(G_1)$=0,1451 | $P_2(G_1)$=0,1445 | $P_3(G_1)$=0,1405 | $P_4(G_1)$=0,1424 | $P_5(G_1)$=0,1472 |
| | $F_2(A_2)$= 21511 | $F_3(A_2)$=14274 | $F_4(A_2)$=10782 | $F_5(A_2)$=8550 |
| | $P_2(A_2)$=0,3555 | $P_3(A_2)$=0,3538 | $P_4(A_2)$=0,3564 | $P_5(A_2)$=0,3532 |
| | $F_2(T_2)$= 21560 | $F_3(T_2)$=14746 | $F_4(T_2)$=10815 | $F_5(T_2)$=8658 |
| | $P_2(T_2)$=0,3563 | $P_3(T_2)$=0,3655 | $P_4(T_2)$=0,3574 | $P_5(T_2)$=0,3577 |
| | $F_2(C_2)$= 8632 | $F_3(C_2)$=5657 | $F_4(C_2)$=4288 | $F_5(C_2)$=3533 |
| | $P_2(C_2)$=0,1426 | $P_3(C_2)$=0,1402 | $P_4(C_2)$=0,1417 | $P_5(C_2)$=0,1460 |
| | $F_2(G_2)$= 8809 | $F_3(G_2)$=5664 | $F_4(G_2)$=4371 | $F_5(G_2)$=3463 |
| | $P_2(G_2)$=0,1456 | $P_3(G_2)$=0,1404 | $P_4(G_2)$=0,1445 | $P_5(G_2)$=0,1431 |
| | | $F_3(A_3)$= 14273 | $F_4(A_3)$=10646 | $F_5(A_3)$=8596 |
| | | $P_3(A_3)$=0,3538 | $P_4(A_3)$=0,3519 | $P_5(A_3)$=0,3551 |
| | | $F_3(T_3)$= 13879 | $F_4(T_3)$=10833 | $F_5(T_3)$=8694 |
| | | $P_3(T_3)$=0,3440 | $P_4(T_3)$=0,3580 | $P_5(T_3)$=0,3592 |
| | | $F_3(C_3)$= 5965 | $F_4(C_3)$=4337 | $F_5(C_3)$=3416 |
| | | $P_3(C_3)$=0,1479 | $P_4(C_3)$=0,1433 | $P_5(C_3)$=0,1411 |
| | | $F_3(G_3)$= 6224 | $F_4(G_3)$=4440 | $F_5(G_3)$=3498 |
| | | $P_3(G_3)$=0,1543 | $P_4(G_3)$=0,1467 | $P_5(G_3)$=0,1445 |
| | | | $F_4(A_4)$= 10729 | $F_5(A_4)$=8680 |
| | | | $P_4(A_4)$=0,3546 | $P_5(A_4)$=0,3586 |
| | | | $F_4(T_4)$= 10745 | $F_5(T_4)$=8552 |
| | | | $P_4(T_4)$=0,3551 | $P_5(T_4)$=0,3533 |
| | | | $F_4(C_4)$= 4344 | $F_5(C_4)$=3475 |
| | | | $P_4(C_4)$=0,1436 | $P_5(C_4)$=0,1436 |
| | | | $F_4(G_4)$= 4438 | $F_5(G_4)$=3497 |
| | | | $P_4(G_4)$=0,1467 | $P_5(G_4)$=0,1445 |
| | | | | $F_5(A_5)$= 8552 |
| | | | | $P_5(A_5)$=0,3533 |

| |
|---|
| $F_5(T_5)$= 8680 |
| $P_5(T_5)$=0,3586 |
| $F_5(C_5)$= 3436 |
| $P_5(C_5)$=0,1420 |
| $F_5(G_5)$= 3536 |
| $P_5(G_5)$=0,1461 |

Fig. A1/29. Collective frequencies $F_n(A_k)$, $F_n(T_k)$, $F_n(C_k)$, $F_n(G_k)$ and collective probabilties $P_n(A_k)$, $P_n(T_k)$, $P_n(C_k)$ and $P_n(G_k)$ (n = 1, 2, 3, 4, 5 and k ≤ n) of tetra-group subgroups in sequences of n-plets in the case of the Marchantia paleacea chloroplast genome DNA, 121024 bp, accession X04465.1, https://www.ncbi.nlm.nih.gov/nuccore/X04465

| **NUCLEOTIDES** $\Sigma_1$ = 132464 | **DOUBLETS** $\Sigma_2$ = 66232 | **TRIPLETS** $\Sigma_3$ =44154 | **4-PLETS** $\Sigma_4$ =33116 | **5-PLETS** $\Sigma_5$ =26492 |
|---|---|---|---|---|
| $F_1(A_1)$= 33030 | $F_2(A_1)$= 16477 | $F_3(A_1)$= 11106 | $F_4(A_1)$= 8304 | $F_5(A_1)$= 6592 |
| $P_1(A_1)$=0,2494 | $P_2(A_1)$=0,2488 | $P_3(A_1)$=0,2515 | $P_4(A_1)$=0,2508 | $P_5(A_1)$=0,2488 |
| $F_1(T_1)$= 33824 | $F_2(T_1)$= 16910 | $F_3(T_1)$= 11340 | $F_4(T_1)$= 8526 | $F_5(T_1)$= 6907 |
| $P_1(T_1)$=0,2553 | $P_2(T_1)$=0,2553 | $P_3(T_1)$=0,2568 | $P_4(T_1)$=0,2575 | $P_5(T_1)$=0,2607 |
| $F_1(C_1)$= 34062 | $F_2(C_1)$= 17075 | $F_3(C_1)$= 11321 | $F_4(C_1)$= 8428 | $F_5(C_1)$= 6813 |
| $P_1(C_1)$=0,2571 | $P_2(C_1)$=0,2578 | $P_3(C_1)$=0,2564 | $P_4(C_1)$=0,2545 | $P_5(C_1)$=0,2572 |
| $F_1(G_1)$= 31548 | $F_2(G_1)$= 15770 | $F_3(G_1)$= 10387 | $F_4(G_1)$= 7858 | $F_5(G_1)$= 6180 |
| $P_1(G_1)$=0,2382 | $P_2(G_1)$=0,2381 | $P_3(G_1)$=0,2352 | $P_4(G_1)$=0,2373 | $P_5(G_1)$=0,2333 |
| | $F_2(A_2)$= 16553 | $F_3(A_2)$=11103 | $F_4(A_2)$=8196 | $F_5(A_2)$=6647 |
| | $P_2(A_2)$=0,2499 | $P_3(A_2)$=0,2515 | $P_4(A_2)$=0,2475 | $P_5(A_2)$=0,2509 |
| | $F_2(T_2)$= 16914 | $F_3(T_2)$=11245 | $F_4(T_2)$=8491 | $F_5(T_2)$=26492 |
| | $P_2(T_2)$=0,2554 | $P_3(T_2)$=0,2547 | $P_4(T_2)$=0,2564 | $P_5(T_2)$=0,2530 |
| | $F_2(C_2)$= 16987 | $F_3(C_2)$=11149 | $F_4(C_2)$=8478 | $F_5(C_2)$=6817 |
| | $P_2(C_2)$=0,2565 | $P_3(C_2)$=0,2525 | $P_4(C_2)$=0,2560 | $P_5(C_2)$=0,2573 |
| | $F_2(G_2)$= 15778 | $F_3(G_2)$=10657 | $F_4(G_2)$=7951 | $F_5(G_2)$=6326 |
| | $P_2(G_2)$=0,2382 | $P_3(G_2)$=0,2414 | $P_4(G_2)$=0,2401 | $P_5(G_2)$=0,2388 |
| | | $F_3(A_3)$= 10821 | $F_4(A_3)$=8173 | $F_5(A_3)$=6649 |
| | | $P_3(A_3)$=0,2451 | $P_4(A_3)$=0,2468 | $P_5(A_3)$=0,2510 |
| | | $F_3(T_3)$= 11239 | $F_4(T_3)$=8384 | $F_5(T_3)$=6801 |
| | | $P_3(T_3)$=0,2545 | $P_4(T_3)$=0,2532 | $P_5(T_3)$=0,2567 |
| | | $F_3(C_3)$= 11592 | $F_4(C_3)$=8647 | $F_5(C_3)$=6815 |
| | | $P_3(C_3)$=0,2625 | $P_4(C_3)$=0,2611 | $P_5(C_3)$=0,2572 |
| | | $F_3(G_3)$= 10502 | $F_4(G_3)$=7912 | $F_5(G_3)$=6227 |
| | | $P_3(G_3)$=0,2378 | $P_4(G_3)$=0,2389 | $P_5(G_3)$=0,2351 |
| | | | $F_4(A_4)$= 8357 | $F_5(A_4)$=6547 |
| | | | $P_4(A_4)$=0,2524 | $P_5(A_4)$=0,2471 |
| | | | $F_4(T_4)$= 8423 | $F_5(T_4)$=6723 |
| | | | $P_4(T_4)$=0,2543 | $P_5(T_4)$=0,2538 |
| | | | $F_4(C_4)$= 8509 | $F_5(C_4)$=6812 |
| | | | $P_4(C_4)$=0,2569 | $P_5(C_4)$=0,2571 |
| | | | $F_4(G_4)$= 7827 | $F_5(G_4)$=6410 |
| | | | $P_4(G_4)$=0,2364 | $P_5(G_4)$=0,2420 |
| | | | | $F_5(A_5)$= 6595 |
| | | | | $P_5(A_5)$=0,2489 |

| 5-PLETS |
| --- |
| $F_5(T_5)$= 6691 |
| $P_5(T_5)$=0,2526 |
| $F_5(C_5)$= 6805 |
| $P_5(C_5)$=0,2569 |
| $F_5(G_5)$= 6401 |
| $P_5(G_5)$=0,2416 |

Fig. A1/30. Collective frequencies $F_n(A_k)$, $F_n(T_k)$, $F_n(C_k)$, $F_n(G_k)$ and collective probabilties $P_n(A_k)$, $P_n(T_k)$, $P_n(C_k)$ and $P_n(G_k)$ (n = 1, 2, 3, 4, 5 and $k \leq n$) of tetra-group subgroups in sequences of n-plets in the case of the Escherichia coli strain PSUO78 plasmid pPSUO78_1, complete sequence, 132464 bp, accession CP012113.1, https://www.ncbi.nlm.nih.gov/nuccore/CP012113.1

| **NUCLEOTIDES**<br>$\Sigma_1$ = 152261 | **DOUBLETS**<br>$\Sigma_2$ = 76130 | **TRIPLETS**<br>$\Sigma_3$ =50753 | **4-PLETS**<br>$\Sigma_4$ =38065 | **5-PLETS**<br>$\Sigma_5$ =30452 |
| --- | --- | --- | --- | --- |
| $F_1(A_1)$= 24240 | $F_2(A_1)$= 12192 | $F_3(A_1)$= 7949 | $F_4(A_1)$= 6121 | $F_5(A_1)$= 4926 |
| $P_1(A_1)$=0,1592 | $P_2(A_1)$=0,1601 | $P_3(A_1)$=0,1566 | $P_4(A_1)$=0,1608 | $P_5(A_1)$=0,1618 |
| $F_1(T_1)$= 24050 | $F_2(T_1)$= 11941 | $F_3(T_1)$= 7848 | $F_4(T_1)$= 5945 | $F_5(T_1)$= 4862 |
| $P_1(T_1)$=0,1580 | $P_2(T_1)$=0,1569 | $P_3(T_1)$=0,1546 | $P_4(T_1)$=0,1562 | $P_5(T_1)$=0,1597 |
| $F_1(C_1)$= 51458 | $F_2(C_1)$= 25758 | $F_3(C_1)$= 17447 | $F_4(C_1)$= 12906 | $F_5(C_1)$= 10344 |
| $P_1(C_1)$=0,3380 | $P_2(C_1)$=0,3383 | $P_3(C_1)$=0,3438 | $P_4(C_1)$=0,3391 | $P_5(C_1)$=0,3397 |
| $F_1(G_1)$= 52513 | $F_2(G_1)$= 26239 | $F_3(G_1)$= 17509 | $F_4(G_1)$= 13093 | $F_5(G_1)$= 10320 |
| $P_1(G_1)$=0,3449 | $P_2(G_1)$=0,3447 | $P_3(G_1)$=0,3450 | $P_4(G_1)$=0,3440 | $P_5(G_1)$=0,3389 |
| | $F_2(A_2)$= 12048 | $F_3(A_2)$=8057 | $F_4(A_2)$=6079 | $F_5(A_2)$=4882 |
| | $P_2(A_2)$=0,1583 | $P_3(A_2)$=0,1587 | $P_4(A_2)$=0,1597 | $P_5(A_2)$=0,1603 |
| | $F_2(T_2)$= 12109 | $F_3(T_2)$=7837 | $F_4(T_2)$=6061 | $F_5(T_2)$=4809 |
| | $P_2(T_2)$=0,1591 | $P_3(T_2)$=0,1544 | $P_4(T_2)$=0,1592 | $P_5(T_2)$=0,1579 |
| | $F_2(C_2)$= 25699 | $F_3(C_2)$=17182 | $F_4(C_2)$=12881 | $F_5(C_2)$=10219 |
| | $P_2(C_2)$=0,3376 | $P_3(C_2)$=0,3385 | $P_4(C_2)$=0,3384 | $P_5(C_2)$=0,3356 |
| | $F_2(G_2)$= 26274 | $F_3(G_2)$=17677 | $F_4(G_2)$=13044 | $F_5(G_2)$=10542 |
| | $P_2(G_2)$=0,3451 | $P_3(G_2)$=0,3483 | $P_4(G_2)$=0,3427 | $P_5(G_2)$=0,3462 |
| | | $F_3(A_3)$= 8234 | $F_4(A_3)$=6071 | $F_5(A_3)$=4785 |
| | | $P_3(A_3)$=0,1622 | $P_4(A_3)$=0,1595 | $P_5(A_3)$=0,1571 |
| | | $F_3(T_3)$= 8365 | $F_4(T_3)$=5996 | $F_5(T_3)$=4806 |
| | | $P_3(T_3)$=0,1648 | $P_4(T_3)$=0,1575 | $P_5(T_3)$=0,1578 |
| | | $F_3(C_3)$= 16828 | $F_4(C_3)$=12852 | $F_5(C_3)$=10270 |
| | | $P_3(C_3)$=0,3316 | $P_4(C_3)$=0,3376 | $P_5(C_3)$=0,3373 |
| | | $F_3(G_3)$= 17326 | $F_4(G_3)$=13146 | $F_5(G_3)$=10591 |
| | | $P_3(G_3)$=0,3414 | $P_4(G_3)$=0,3455 | $P_5(G_3)$=0,3478 |
| | | | $F_4(A_4)$= 5969 | $F_5(A_4)$=4772 |
| | | | $P_4(A_4)$=0,1568 | $P_5(A_4)$=0,1567 |
| | | | $F_4(T_4)$= 6048 | $F_5(T_4)$=4771 |
| | | | $P_4(T_4)$=0,1589 | $P_5(T_4)$=0,1567 |
| | | | $F_4(C_4)$= 12818 | $F_5(C_4)$=10352 |
| | | | $P_4(C_4)$=0,3367 | $P_5(C_4)$=0,3399 |
| | | | $F_4(G_4)$= 13230 | $F_5(G_4)$=10557 |
| | | | $P_4(G_4)$=0,3476 | $P_5(G_4)$=0,34677 |
| | | | | $F_5(A_5)$= 4875 |
| | | | | $P_5(A_5)$=0,1601 |
| | | | | $F_5(T_5)$= 4802 |
| | | | | $P_5(T_5)$=0,1577 |

|  |
|---|
| $F_5(C_5)$= 10272 |
| $P_5(C_5)$=0,3373 |
| $F_5(G_5)$= 10503 |
| $P_5(G_5)$=0,3449 |

Fig. A1/31. Collective frequencies $F_n(A_k)$, $F_n(T_k)$, $F_n(C_k)$, $F_n(G_k)$ and collective probabilties $P_n(A_k)$, $P_n(T_k)$, $P_n(C_k)$ and $P_n(G_k)$ (n = 1, 2, 3, 4, 5 and $k \le n$) of tetra-group subgroups in sequences of n-plets in the case of the HSIULR, Human herpesvirus 1 complete genome, 152261 bp, accession X14112.1, [https://www.ncbi.nlm.nih.gov/nuccore/X14112.1](https://www.ncbi.nlm.nih.gov/nuccore/X14112.1)

| **NUCLEOTIDES** | **DOUBLETS** | **TRIPLETS** | **4-PLETS** | **5-PLETS** |
|---|---|---|---|---|
| $\Sigma_1$ = 100849 | $\Sigma_2$ = 50424 | $\Sigma_3$ =33616 | $\Sigma_4$ =25212 | $\Sigma_5$ =20169 |
| $F_1(A_1)$= 30346 | $F_2(A_1)$= 15271 | $F_3(A_1)$= 10157 | $F_4(A_1)$= 7636 | $F_5(A_1)$= 5978 |
| $P_1(A_1)$=0,3009 | $P_2(A_1)$=0,3029 | $P_3(A_1)$=0,3021 | $P_4(A_1)$=0,3029 | $P_5(A_1)$=0,2964 |
| $F_1(T_1)$= 32481 | $F_2(T_1)$= 16175 | $F_3(T_1)$= 10746 | $F_4(T_1)$= 8030 | $F_5(T_1)$= 6570 |
| $P_1(T_1)$=0,3221 | $P_2(T_1)$=0,3208 | $P_3(T_1)$=0,3197 | $P_4(T_1)$=0,3185 | $P_5(T_1)$=0,3257 |
| $F_1(C_1)$= 18635 | $F_2(C_1)$= 9390 | $F_3(C_1)$= 6302 | $F_4(C_1)$= 4727 | $F_5(C_1)$= 3709 |
| $P_1(C_1)$=0,1848 | $P_2(C_1)$=0,1862 | $P_3(C_1)$=0,1875 | $P_4(C_1)$=0,1875 | $P_5(C_1)$=0,1839 |
| $F_1(G_1)$= 19387 | $F_2(G_1)$= 9588 | $F_3(G_1)$= 6411 | $F_4(G_1)$= 4819 | $F_5(G_1)$= 3912 |
| $P_1(G_1)$=0,1922 | $P_2(G_1)$=0,1901 | $P_3(G_1)$=0,1907 | $P_4(G_1)$=0,1911 | $P_5(G_1)$=0,1940 |
| | $F_2(A_2)$= 15075 | $F_3(A_2)$=10055 | $F_4(A_2)$=7552 | $F_5(A_2)$= 6095 |
| | $P_2(A_2)$=0,2990 | $P_3(A_2)$=0,2991 | $P_4(A_2)$=0,2995 | $P_5(A_2)$=0,3022 |
| | $F_2(T_2)$= 16306 | $F_3(T_2)$=10870 | $F_4(T_2)$=8147 | $F_5(T_2)$=6461 |
| | $P_2(T_2)$=0,3234 | $P_3(T_2)$=0,3234 | $P_4(T_2)$=0,3231 | $P_5(T_2)$=0,3203 |
| | $F_2(C_2)$= 9245 | $F_3(C_2)$=6166 | $F_4(C_2)$=4624 | $F_5(C_2)$=3795 |
| | $P_2(C_2)$=0,1833 | $P_3(C_2)$=0,1834 | $P_4(C_2)$=0,1834 | $P_5(C_2)$=0,1882 |
| | $F_2(G_2)$= 9798 | $F_3(G_2)$=6525 | $F_4(G_2)$=4889 | $F_5(G_2)$=3818 |
| | $P_2(G_2)$=0,1943 | $P_3(G_2)$=0,1941 | $P_4(G_2)$=0,1939 | $P_5(G_2)$=0,1893 |
| | | $F_3(A_3)$= 10134 | $F_4(A_3)$=25212 | $F_5(A_3)$=6045 |
| | | $P_3(A_3)$=0,3015 | $P_4(A_3)$=0,3028 | $P_5(A_3)$=0,2997 |
| | | $F_3(T_3)$= 10865 | $F_4(T_3)$=8145 | $F_5(T_3)$=6413 |
| | | $P_3(T_3)$=0,3232 | $P_4(T_3)$=0,3231 | $P_5(T_3)$=0,3180 |
| | | $F_3(C_3)$= 6167 | $F_4(C_3)$=4663 | $F_5(C_3)$=3798 |
| | | $P_3(C_3)$=0,1835 | $P_4(C_3)$=0,1850 | $P_5(C_3)$=0,1883 |
| | | $F_3(G_3)$= 6450 | $F_4(G_3)$=4769 | $F_5(G_3)$=3913 |
| | | $P_3(G_3)$=0,1919 | $P_4(G_3)$=0,1892 | $P_5(G_3)$=0,1940 |
| | | | $F_4(A_4)$= 7523 | $F_5(A_4)$= 6102 |
| | | | $P_4(A_4)$=0,2984 | $P_5(A_4)$=0,3025 |
| | | | $F_4(T_4)$= 8159 | $F_5(T_4)$=6534 |
| | | | $P_4(T_4)$=0,3236 | $P_5(T_4)$=0,3240 |
| | | | $F_4(C_4)$= 4621 | $F_5(C_4)$=3702 |
| | | | $P_4(C_4)$=0,1833 | $P_5(C_4)$=0,1835 |
| | | | $F_4(G_4)$= 4909 | $F_5(G_4)$=3831 |
| | | | $P_4(G_4)$=0,1947 | $P_5(G_4)$=0,1899 |
| | | | | $F_5(A_5)$= 6125 |
| | | | | $P_5(A_5)$=0,3037 |
| | | | | $F_5(T_5)$= 6502 |
| | | | | $P_5(T_5)$=0,3224 |
| | | | | $F_5(C_5)$= 3630 |
| | | | | $P_5(C_5)$=0,1800 |

$F_5(G_5)$= 3912
$P_5(G_5)$=0,1940

Fig. A1/32. Collective frequencies $F_n(A_k)$, $F_n(T_k)$, $F_n(C_k)$, $F_n(G_k)$ and collective probabilties $P_n(A_k)$, $P_n(T_k)$, $P_n(C_k)$ and $P_n(G_k)$ ($n$ = 1, 2, 3, 4, 5 and $k \le n$) of tetra-group subgroups in sequences of n-plets in the following case: HUMNEUROF, Human oligodendrocyte myelin glycoprotein (OMG) exons 1-2; neurofibromatosis 1 (NF1) exons 28-49; ecotropic viral integration site 2B (EVI2B) exons 1-2; ecotropic viral integration site 2A (EVI2A) exons 1-2; adenylate kinase (AK3) exons 1-2, 100849 bp, accession L05367.1, https://www.ncbi.nlm.nih.gov/nuccore/189152

| **NUCLEOTIDES** | **DOUBLETS** | **TRIPLETS** | **4-PLETS** | **5-PLETS** |
|---|---|---|---|---|
| $\Sigma_1$ = 100314 | $\Sigma_2$ = 50157 | $\Sigma_3$ =33438 | $\Sigma_4$ =25078 | $\Sigma_5$ =20062 |
| $F_1(A_1)$= 35804 | $F_2(A_1)$= 17906 | $F_3(A_1)$= 11917 | $F_4(A_1)$= 8814 | $F_5(A_1)$= 7101 |
| $P_1(A_1)$=0,3569 | $P_2(A_1)$=0,3570 | $P_3(A_1)$=0,3564 | $P_4(A_1)$=0,3515 | $P_5(A_1)$=0,3540 |
| $F_1(T_1)$= 34358 | $F_2(T_1)$= 17236 | $F_3(T_1)$= 11802 | $F_4(T_1)$= 8693 | $F_5(T_1)$= 6967 |
| $P_1(T_1)$=0,3425 | $P_2(T_1)$=0,3436 | $P_3(T_1)$=0,3530 | $P_4(T_1)$=0,3466 | $P_5(T_1)$=0,3473 |
| $F_1(C_1)$= 13428 | $F_2(C_1)$= 6669 | $F_3(C_1)$= 4415 | $F_4(C_1)$= 3413 | $F_5(C_1)$= 2614 |
| $P_1(C_1)$=0,1339 | $P_2(C_1)$=0,1330 | $P_3(C_1)$=0,1320 | $P_4(C_1)$=0,1361 | $P_5(C_1)$=0,1303 |
| $F_1(G_1)$= 16724 | $F_2(G_1)$= 8346 | $F_3(G_1)$= 5304 | $F_4(G_1)$= 4158 | $F_5(G_1)$= 3380 |
| $P_1(G_1)$=0,1667 | $P_2(G_1)$=0,1664 | $P_3(G_1)$=0,1586 | $P_4(G_1)$=0,1658 | $P_5(G_1)$=0,1685 |
| | $F_2(A_2)$= 17898 | $F_3(A_2)$=12161 | $F_4(A_2)$=8975 | $F_5(A_2)$=7134 |
| | $P_2(A_2)$=0,3568 | $P_3(A_2)$=0,3637 | $P_4(A_2)$=0,3579 | $P_5(A_2)$=0,3556 |
| | $F_2(T_2)$= 17122 | $F_3(T_2)$=11010 | $F_4(T_2)$=8550 | $F_5(T_2)$=6849 |
| | $P_2(T_2)$=0,3414 | $P_3(T_2)$=0,3293 | $P_4(T_2)$=0,3409 | $P_5(T_2)$=0,3414 |
| | $F_2(C_2)$= 6759 | $F_3(C_2)$=4448 | $F_4(C_2)$=3344 | $F_5(C_2)$=2707 |
| | $P_2(C_2)$=0,1348 | $P_3(C_2)$=0,1330 | $P_4(C_2)$=0,1333 | $P_5(C_2)$=0,1349 |
| | $F_2(G_2)$= 8378 | $F_3(G_2)$=5819 | $F_4(G_2)$=4209 | $F_5(G_2)$=3372 |
| | $P_2(G_2)$=0,1670 | $P_3(G_2)$=0,1740 | $P_4(G_2)$=0,1678 | $P_5(G_2)$=0,1681 |
| | | $F_3(A_3)$= 11726 | $F_4(A_3)$=9092 | $F_5(A_3)$=7130 |
| | | $P_3(A_3)$=0,3507 | $P_4(A_3)$=0,3625 | $P_5(A_3)$=0,3554 |
| | | $F_3(T_3)$= 11546 | $F_4(T_3)$=8542 | $F_5(T_3)$=6833 |
| | | $P_3(T_3)$=0,3453 | $P_4(T_3)$=0,3406 | $P_5(T_3)$=0,3406 |
| | | $F_3(C_3)$= 4565 | $F_4(C_3)$=3256 | $F_5(C_3)$=2677 |
| | | $P_3(C_3)$=0,1365 | $P_4(C_3)$=0,1298 | $P_5(C_3)$=0,1334 |
| | | $F_3(G_3)$= 5601 | $F_4(G_3)$=4188 | $F_5(G_3)$=3422 |
| | | $P_3(G_3)$=0,1675 | $P_4(G_3)$=0,1670 | $P_5(G_3)$=0,1706 |
| | | | $F_4(A_4)$= 8922 | $F_5(A_4)$=7177 |
| | | | $P_4(A_4)$=0,3558 | $P_5(A_4)$=0,3577 |
| | | | $F_4(T_4)$= 8572 | $F_5(T_4)$=6914 |
| | | | $P_4(T_4)$=0,3418 | $P_5(T_4)$=0,3446 |
| | | | $F_4(C_4)$= 3415 | $F_5(C_4)$=2730 |
| | | | $P_4(C_4)$=0,1362 | $P_5(C_4)$=0,1361 |
| | | | $F_4(G_4)$= 4169 | $F_5(G_4)$=3241 |
| | | | $P_4(G_4)$=0,1662 | $P_5(G_4)$=0,1615 |
| | | | | $F_5(A_5)$= 7261 |
| | | | | $P_5(A_5)$=0,3619 |
| | | | | $F_5(T_5)$= 6792 |
| | | | | $P_5(T_5)$=0,3386 |
| | | | | $F_5(C_5)$= 2700 |
| | | | | $P_5(C_5)$=0,1346 |

| $F_5(G_5)$= 3309<br>$P_5(G_5)$=0,1649 |
|---|

Fig. A1/33. Collective frequencies $F_n(A_k)$, $F_n(T_k)$, $F_n(C_k)$, $F_n(G_k)$ and collective probabilties $P_n(A_k)$, $P_n(T_k)$, $P_n(C_k)$ and $P_n(G_k)$ ($n$ = 1, 2, 3, 4, 5 and $k \leq n$) of tetra-group subgroups in sequences of n-plets in the case of Podospora anserina mitochondrion, complete genome, 100314 bp, accession NC_001329.3, https://www.ncbi.nlm.nih.gov/nuccore/NC_001329.3

| **NUCLEOTIDES**<br>$\mathbf{\Sigma_1}$ **=** 97630 | **DOUBLETS**<br>$\mathbf{\Sigma_2}$ **=** 48815 | **TRIPLETS**<br>$\mathbf{\Sigma_3}$ **=**32543 | **4-PLETS**<br>$\mathbf{\Sigma_4}$ **=**24407 | **5-PLETS**<br>$\mathbf{\Sigma_5}$ **=**19526 |
|---|---|---|---|---|
| $F_1(A_1)$= 28063<br>$P_1(A_1)$=0,2874 | $F_2(A_1)$= 13947<br>$P_2(A_1)$=0,2857 | $F_3(A_1)$= 9319<br>$P_3(A_1)$=0,2864 | $F_4(A_1)$= 6869<br>$P_4(A_1)$=0,2814 | $F_5(A_1)$= 5662<br>$P_5(A_1)$=0,2900 |
| $F_1(T_1)$= 26383<br>$P_1(T_1)$=0,2702 | $F_2(T_1)$= 13185<br>$P_2(T_1)$=0,2701 | $F_3(T_1)$= 8723<br>$P_3(T_1)$=0,2680 | $F_4(T_1)$= 6622<br>$P_4(T_1)$=0,2713 | $F_5(T_1)$= 5189<br>$P_5(T_1)$=0,2657 |
| $F_1(C_1)$= 20949<br>$P_1(C_1)$=0,2146 | $F_2(C_1)$= 10479<br>$P_2(C_1)$=0,2147 | $F_3(C_1)$= 6987<br>$P_3(C_1)$=0,2147 | $F_4(C_1)$= 5254<br>$P_4(C_1)$=0,2153 | $F_5(C_1)$= 4242<br>$P_5(C_1)$=0,2172 |
| $F_1(G_1)$= 22235<br>$P_1(G_1)$=0,2277 | $F_2(G_1)$= 11204<br>$P_2(G_1)$=0,2295 | $F_3(G_1)$= 7514<br>$P_3(G_1)$=0,2309 | $F_4(G_1)$= 5662<br>$P_4(G_1)$=0,2320 | $F_5(G_1)$= 4433<br>$P_5(G_1)$=0,2270 |
| | $F_2(A_2)$= 14116<br>$P_2(A_2)$=0,2892 | $F_3(A_2)$=9440<br>$P_3(A_2)$=0,2901 | $F_4(A_2)$=7077<br>$P_4(A_2)$=0,2900 | $F_5(A_2)$=5677<br>$P_5(A_2)$=0,2907 |
| | $F_2(T_2)$= 13198<br>$P_2(T_2)$=0,2704 | $F_3(T_2)$=8787<br>$P_3(T_2)$=0,2700 | $F_4(T_2)$=6630<br>$P_4(T_2)$=0,2716 | $F_5(T_2)$=5202<br>$P_5(T_2)$=0,2664 |
| | $F_2(C_2)$= 10470<br>$P_2(C_2)$=0,2145 | $F_3(C_2)$=6991<br>$P_3(C_2)$=0,2148 | $F_4(C_2)$=5212<br>$P_4(C_2)$=0,2135 | $F_5(C_2)$=4153<br>$P_5(C_2)$=0,2127 |
| | $F_2(G_2)$= 11031<br>$P_2(G_2)$=0,2260 | $F_3(G_2)$=6991<br>$P_3(G_2)$=0,2251 | $F_4(G_2)$=5488<br>$P_4(G_2)$=0,2249 | $F_5(G_2)$=4494<br>$P_5(G_2)$=0,2302 |
| | | $F_3(A_3)$= 9304<br>$P_3(A_3)$=0,2859 | $F_4(A_3)$=7078<br>$P_4(A_3)$=0,2900 | $F_5(A_3)$=5570<br>$P_5(A_3)$=0,2853 |
| | | $F_3(T_3)$= 8873<br>$P_3(T_3)$=0,2727 | $F_4(T_3)$=6563<br>$P_4(T_3)$=0,2689 | $F_5(T_3)$=5250<br>$P_5(T_3)$=0,2689 |
| | | $F_3(C_3)$= 6970<br>$P_3(C_3)$=0,2142 | $F_4(C_3)$=5224<br>$P_4(C_3)$=0,2140 | $F_5(C_3)$=4184<br>$P_5(C_3)$=0,2143 |
| | | $F_3(G_3)$= 7396<br>$P_3(G_3)$=0,2273 | $F_4(G_3)$=5542<br>$P_4(G_3)$=0,2271 | $F_5(G_3)$=4522<br>$P_5(G_3)$=0,2316 |
| | | | $F_4(A_4)$= 7039<br>$P_4(A_4)$=0,2884 | $F_5(A_4)$=5561<br>$P_5(A_4)$=0,2848 |
| | | | $F_4(T_4)$= 6568<br>$P_4(T_4)$=0,2691 | $F_5(T_4)$=5356<br>$P_5(T_4)$=0,2743 |
| | | | $F_4(C_4)$= 5257<br>$P_4(C_4)$=0,2154 | $F_5(C_4)$=4207<br>$P_5(C_4)$=0,2155 |
| | | | $F_4(G_4)$= 5543<br>$P_4(G_4)$=0,2271 | $F_5(G_4)$=4402<br>$P_5(G_4)$=0,2254 |
| | | | | $F_5(A_5)$= 5593<br>$P_5(A_5)$=0,2864 |
| | | | | $F_5(T_5)$= 5386<br>$P_5(T_5)$=0,2758 |
| | | | | $F_5(C_5)$= 4163<br>$P_5(C_5)$=0,2132 |

$F_5(G_5)$= 4384<br>$P_5(G_5)$=0,2245

Fig. A1/34. Collective frequencies $F_n(A_k)$, $F_n(T_k)$, $F_n(C_k)$, $F_n(G_k)$ and collective probabilties $P_n(A_k)$, $P_n(T_k)$, $P_n(C_k)$ and $P_n(G_k)$ ($n$ = 1, 2, 3, 4, 5 and $k \leq n$) of tetra-group subgroups in sequences of n-plets in the case of the HUMTCRADCV, Human T-cell receptor genes (Human Tcr-C-delta gene, exons 1-4; Tcr-V-delta gene, exons 1-2; T-cell receptor alpha (Tcr-alpha) gene, J1-J61 segments; and Tcr-C-alpha gene, exons 1-4), 97630 bp, accession M94081.1, https://www.ncbi.nlm.nih.gov/nuccore/2627263

| **NUCLEOTIDES**<br>**$\Sigma_1$ =** 94647 | **DOUBLETS**<br>**$\Sigma_2$ =** 47323 | **TRIPLETS**<br>**$\Sigma_3$ =**31549 | **4-PLETS**<br>**$\Sigma_4$ =**23661 | **5-PLETS**<br>**$\Sigma_5$ =**18929 |
|---|---|---|---|---|
| $F_1(A_1)$= 26359<br>$P_1(A_1)$=0,2785 | $F_2(A_1)$= 13093<br>$P_2(A_1)$=0,2767 | $F_3(A_1)$= 8879<br>$P_3(A_1)$= 0,2814 | $F_4(A_1)$= 6527<br>$P_4(A_1)$=0,2759 | $F_5(A_1)$= 5303<br>$P_5(A_1)$=0,2802 |
| $F_1(T_1)$= 25769<br>$P_1(T_1)$=0,2723 | $F_2(T_1)$= 13008<br>$P_2(T_1)$=0,2749 | $F_3(T_1)$= 8621<br>$P_3(T_1)$=0,2733 | $F_4(T_1)$= 6476<br>$P_4(T_1)$=0,2737 | $F_5(T_1)$= 5124<br>$P_5(T_1)$=0,2707 |
| $F_1(C_1)$= 20790<br>$P_1(C_1)$=0,2197 | $F_2(C_1)$= 10284<br>$P_2(C_1)$=0,2173 | $F_3(C_1)$= 6911<br>$P_3(C_1)$=0,2191 | $F_4(C_1)$= 5166<br>$P_4(C_1)$=0,2183 | $F_5(C_1)$= 4108<br>$P_5(C_1)$=0,2170 |
| $F_1(G_1)$= 21729<br>$P_1(G_1)$=0,2296 | $F_2(G_1)$= 10938<br>$P_2(G_1)$=0,2311 | $F_3(G_1)$= 7138<br>$P_3(G_1)$=0,2263 | $F_4(G_1)$= 5492<br>$P_4(G_1)$=0,2321 | $F_5(G_1)$= 4394<br>$P_5(G_1)$=0,2321 |
| | $F_2(A_2)$= 13265<br>$P_2(A_2)$=0,2803 | $F_3(A_2)$=8763<br>$P_3(A_2)$=0,2778 | $F_4(A_2)$=6632<br>$P_4(A_2)$=0,2803 | $F_5(A_2)$=5317<br>$P_5(A_2)$=0,2809 |
| | $F_2(T_2)$= 12761<br>$P_2(T_2)$= 0,2697 | $F_3(T_2)$=8621<br>$P_3(T_2)$=0,2733 | $F_4(T_2)$=6381<br>$P_4(T_2)$=0,2697 | $F_5(T_2)$=5184<br>$P_5(T_2)$=0,2739 |
| | $F_2(C_2)$= 10506<br>$P_2(C_2)$=0,2220 | $F_3(C_2)$=6875<br>$P_3(C_2)$=0,2179 | $F_4(C_2)$=5247<br>$P_4(C_2)$=0,2218 | $F_5(C_2)$=4152<br>$P_5(C_2)$=0,2193 |
| | $F_2(G_2)$= 10791<br>$P_2(G_2)$=0,2280 | $F_3(G_2)$=7290<br>$P_3(G_2)$=0,2311 | $F_4(G_2)$=5401<br>$P_4(G_2)$=0,2283 | $F_5(G_2)$=4276<br>$P_5(G_2)$=0,2259 |
| | | $F_3(A_3)$= 8717<br>$P_3(A_3)$=0,2763 | $F_4(A_3)$=6566<br>$P_4(A_3)$=0,2775 | $F_5(A_3)$=5242<br>$P_5(A_3)$=0,2769 |
| | | $F_3(T_3)$= 8527<br>$P_3(T_3)$=0,2703 | $F_4(T_3)$=6532<br>$P_4(T_3)$=0,2761 | $F_5(T_3)$=5160<br>$P_5(T_3)$=0,2726 |
| | | $F_3(C_3)$= 7004<br>$P_3(C_3)$=0,2220 | $F_4(C_3)$=5117<br>$P_4(C_3)$=0,2163 | $F_5(C_3)$=4163<br>$P_5(C_3)$=0,2199 |
| | | $F_3(G_3)$= 7301<br>$P_3(G_3)$=0,2314 | $F_4(G_3)$=5446<br>$P_4(G_3)$=0,2302 | $F_5(G_3)$=4364<br>$P_5(G_3)$=0,2305 |
| | | | $F_4(A_4)$= 6633<br>$P_4(A_4)$=0,2803 | $F_5(A_4)$=5261<br>$P_5(A_4)$=0,2779 |
| | | | $F_4(T_4)$= 6380<br>$P_4(T_4)$=0,2696 | $F_5(T_4)$=5137<br>$P_5(T_4)$=0,2714 |
| | | | $F_4(C_4)$= 5259<br>$P_4(C_4)$=0,2223 | $F_5(C_4)$=4217<br>$P_5(C_4)$=0,2228 |
| | | | $F_4(G_4)$= 5389<br>$P_4(G_4)$=0,2278 | $F_5(G_4)$=4314<br>$P_5(G_4)$=0,2279 |
| | | | | $F_5(A_5)$= 5235<br>$P_5(A_5)$=0,2766 |
| | | | | $F_5(T_5)$= 5164<br>$P_5(T_5)$=0,2728 |
| | | | | $F_5(C_5)$= 4150<br>$P_5(C_5)$=0,2192 |

| $F_5(G_5)$= 4380 |
|---|
| $P_5(G_5)$=0,2314 |

Fig. A1/35. Collective frequencies $F_n(A_k)$, $F_n(T_k)$, $F_n(C_k)$, $F_n(G_k)$ and collective probabilties $P_n(A_k)$, $P_n(T_k)$, $P_n(C_k)$ and $P_n(G_k)$ ($n$ = 1, 2, 3, 4, 5 and $k \leq n$) of tetra-group subgroups in sequences of n-plets in the case of MUSTCRA, Mouse T-cell receptor alpha/delta chain locus, 94647 bp, accession M64239.1, https://www.ncbi.nlm.nih.gov/nuccore/201744

One can see from the data of the long DNA-sequences in Fig. A1/1-A1/35 that all these sequences satisfy the three tetra-group rules and demonstrate the existence of tetra-group symmetries in them.

## Appendix 2. Symmetries of tetra-group probabilities in the compete set of human chromosomes

The Appendix 2 represents data about the fulfillment of the described tetra-group rules in the complete set of 24 chromosomes of human genome. All initial data about chromosomes are taken from the CenBank.

| **NUCLEOTIDES** | | **DOUBLETS** | | **TRIPLETS** | | **4-PLETS** | | **5-PLETS** | |
|---|---|---|---|---|---|---|---|---|---|
| $P_1(A_1)$= | 0,2910 | $P_2(A_1)$= | 0,2910 | $P_3(A_1)$= | 0,2910 | $P_4(A_1)$= | 0,2910 | $P_5(A_1)$= | 0,2910 |
| $P_1(T_1)$= | 0,2918 | $P_2(T_1)$= | 0,2917 | $P_3(T_1)$= | 0,2917 | $P_4(T_1)$= | 0,2917 | $P_5(T_1)$= | 0,2917 |
| $P_1(C_1)$= | 0,2085 | $P_2(C_1)$= | 0,2085 | $P_3(C_1)$= | 0,2084 | $P_4(C_1)$= | 0,2085 | $P_5(C_1)$= | 0,2085 |
| $P_1(G_1)$= | 0,2087 | $P_2(G_1)$= | 0,2088 | $P_3(G_1)$= | 0,2088 | $P_4(G_1)$= | 0,2088 | $P_5(G_1)$= | 0,2088 |
| | | $P_2(A_2)$= | 0,2910 | $P_3(A_2)$= | 0,2910 | $P_4(A_2)$= | 0,2910 | $P_5(A_2)$= | 0,2910 |
| | | $P_2(T_2)$= | 0,2918 | $P_3(T_2)$= | 0,2917 | $P_4(T_2)$= | 0,2918 | $P_5(T_2)$= | 0,2917 |
| | | $P_2(C_2)$= | 0,2085 | $P_3(C_2)$= | 0,2085 | $P_4(C_2)$= | 0,2085 | $P_5(C_2)$= | 0,2086 |
| | | $P_2(G_2)$= | 0,2087 | $P_3(G_2)$= | 0,2088 | $P_4(G_2)$= | 0,2087 | $P_5(G_2)$= | 0,2087 |
| | | | | $P_3(A_3)$= | 0,2910 | $P_4(A_3)$= | 0,2910 | $P_5(A_3)$= | 0,2910 |
| | | | | $P_3(T_3)$= | 0,2918 | $P_4(T_3)$= | 0,2918 | $P_5(T_3)$= | 0,2918 |
| | | | | $P_3(C_3)$= | 0,2085 | $P_4(C_3)$= | 0,2085 | $P_5(C_3)$= | 0,2085 |
| | | | | $P_3(G_3)$= | 0,2087 | $P_4(G_3)$= | 0,2088 | $P_5(G_3)$= | 0,2087 |
| | | | | | | $P_4(A_4)$= | 0,2910 | $P_5(A_4)$= | 0,2910 |
| | | | | | | $P_4(T_4)$= | 0,2918 | $P_5(T_4)$= | 0,2919 |
| | | | | | | $P_4(C_4)$= | 0,2085 | $P_5(C_4)$= | 0,2084 |
| | | | | | | $P_4(G_4)$= | 0,2087 | $P_5(G_4)$= | 0,2087 |
| | | | | | | | | $P_5(A_5)$= | 0,2910 |
| | | | | | | | | $P_5(T_5)$= | 0,2917 |
| | | | | | | | | $P_5(C_5)$= | 0,2085 |
| | | | | | | | | $P_5(G_5)$= | 0,2088 |

Fig. A2/1. The table of probabilities of subgroups of tetra-groups in the sequence: Homo sapiens chromosome 1, GRCh38.p7 Primary Assembly
NCBI Reference Sequence: NC_000001.11
LOCUS NC_000001 248956422 bp DNA linear CON 06-JUN-2016
DEFINITION Homo sapiens chromosome 1, GRCh38.p7 Primary Assembly.
ACCESSION NC_000001 GPC_000001293 VERSION NC_000001.11
https://www.ncbi.nlm.nih.gov/nuccore/NC_000001.11
https://www.ncbi.nlm.nih.gov/nuccore/NC_000001.11?report=fasta

| NUCLEOTIDES | | DOUBLETS | | TRIPLETS | | 4-PLETS | | 5-PLETS | |
|---|---|---|---|---|---|---|---|---|---|
| $P_1(A_1)$= | 0,2984 | $P_2(A_1)$= | 0,2985 | $P_3(A_1)$= | 0,2983 | $P_4(A_1)$= | 0,2985 | $P_5(A_1)$= | 0,2986 |
| $P_1(T_1)$= | 0,2993 | $P_2(T_1)$= | 0,2992 | $P_3(T_1)$= | 0,2993 | $P_4(T_1)$= | 0,2991 | $P_5(T_1)$= | 0,2992 |
| $P_1(C_1)$= | 0,2009 | $P_2(C_1)$= | 0,2009 | $P_3(C_1)$= | 0,2009 | $P_4(C_1)$= | 0,2008 | $P_5(C_1)$= | 0,2008 |
| $P_1(G_1)$= | 0,2014 | $P_2(G_1)$= | 0,2014 | $P_3(G_1)$= | 0,2014 | $P_4(G_1)$= | 0,2015 | $P_5(G_1)$= | 0,2014 |
| | | $P_2(A_2)$= | 0,2988 | $P_3(A_2)$= | 0,2985 | $P_4(A_2)$= | 0,2984 | $P_5(A_2)$= | 0,2985 |
| | | $P_2(T_2)$= | 0,2993 | $P_3(T_2)$= | 0,2992 | $P_4(T_2)$= | 0,2993 | $P_5(T_2)$= | 0,2993 |
| | | $P_2(C_2)$= | 0,2009 | $P_3(C_2)$= | 0,2009 | $P_4(C_2)$= | 0,2009 | $P_5(C_2)$= | 0,2008 |
| | | $P_2(G_2)$= | 0,2014 | $P_3(G_2)$= | 0,2014 | $P_4(G_2)$= | 0,2014 | $P_5(G_2)$= | 0,2014 |
| | | | | $P_3(A_3)$= | 0,2985 | $P_4(A_3)$= | 0,2985 | $P_5(A_3)$= | 0,2984 |
| | | | | $P_3(T_3)$= | 0,2993 | $P_4(T_3)$= | 0,2993 | $P_5(T_3)$= | 0,2992 |
| | | | | $P_3(C_3)$= | 0,2008 | $P_4(C_3)$= | 0,2009 | $P_5(C_3)$= | 0,2010 |
| | | | | $P_3(G_3)$= | 0,2014 | $P_4(G_3)$= | 0,2014 | $P_5(G_3)$= | 0,2014 |
| | | | | | | $P_4(A_4)$= | 0,2984 | $P_5(A_4)$= | 0,2984 |
| | | | | | | $P_4(T_4)$= | 0,2994 | $P_5(T_4)$= | 0,2993 |
| | | | | | | $P_4(C_4)$= | 0,2008 | $P_5(C_4)$= | 0,2009 |
| | | | | | | $P_4(G_4)$= | 0,2014 | $P_5(G_4)$= | 0,2014 |
| | | | | | | | | $P_5(A_5)$= | 0,2984 |
| | | | | | | | | $P_5(T_5)$= | 0,2992 |
| | | | | | | | | $P_5(C_5)$= | 0,2009 |
| | | | | | | | | $P_5(G_5)$= | 0,2015 |

Fig. A2/2. The table of probabilities of subgroups of tetra-groups in the sequence: Homo sapiens chromosome 2, GRCh38.p7 Primary Assembly
NCBI Reference Sequence: NC_000002.12
LOCUS NC_000002 242193529 bp DNA linear CON 06-JUN-2016
DEFINITION Homo sapiens chromosome 2, GRCh38.p7 Primary Assembly.
ACCESSION NC_000002 GPC_000001294 VERSION NC_000002.12
https://www.ncbi.nlm.nih.gov/nuccore/NC_000002.12
https://www.ncbi.nlm.nih.gov/nuccore/NC_000002.12?report=fasta

| NUCLEOTIDES | | DOUBLETS | | TRIPLETS | | 4-PLETS | | 5-PLETS | |
|---|---|---|---|---|---|---|---|---|---|
| $P_1(A_1)$= | 0,3013 | $P_2(A_1)$= | 0,3012 | $P_3(A_1)$= | 0,3013 | $P_4(A_1)$= | 0,3013 | $P_5(A_1)$= | 0,3013 |
| $P_1(T_1)$= | 0,3020 | $P_2(T_1)$= | 0,3020 | $P_3(T_1)$= | 0,3022 | $P_4(T_1)$= | 0,3020 | $P_5(T_1)$= | 0,3021 |
| $P_1(C_1)$= | 0,1980 | $P_2(C_1)$= | 0,1981 | $P_3(C_1)$= | 0,1979 | $P_4(C_1)$= | 0,1981 | $P_5(C_1)$= | 0,1980 |
| $P_1(G_1)$= | 0,1986 | $P_2(G_1)$= | 0,1986 | $P_3(G_1)$= | 0,1986 | $P_4(G_1)$= | 0,1985 | $P_5(G_1)$= | 0,1986 |
| | | $P_2(A_2)$= | 0,3013 | $P_3(A_2)$= | 0,3013 | $P_4(A_2)$= | 0,3013 | $P_5(A_2)$= | 0,3013 |
| | | $P_2(T_2)$= | 0,3020 | $P_3(T_2)$= | 0,3020 | $P_4(T_2)$= | 0,3021 | $P_5(T_2)$= | 0,3019 |
| | | $P_2(C_2)$= | 0,1980 | $P_3(C_2)$= | 0,1981 | $P_4(C_2)$= | 0,1980 | $P_5(C_2)$= | 0,1981 |
| | | $P_2(G_2)$= | 0,1986 | $P_3(G_2)$= | 0,1986 | $P_4(G_2)$= | 0,1986 | $P_5(G_2)$= | 0,1987 |
| | | | | $P_3(A_3)$= | 0,3014 | $P_4(A_3)$= | 0,3012 | $P_5(A_3)$= | 0,3013 |
| | | | | $P_3(T_3)$= | 0,3019 | $P_4(T_3)$= | 0,3021 | $P_5(T_3)$= | 0,3021 |
| | | | | $P_3(C_3)$= | 0,1981 | $P_4(C_3)$= | 0,1980 | $P_5(C_3)$= | 0,1980 |
| | | | | $P_3(G_3)$= | 0,1986 | $P_4(G_3)$= | 0,1987 | $P_5(G_3)$= | 0,1986 |
| | | | | | | $P_4(A_4)$= | 0,3014 | $P_5(A_4)$= | 0,3012 |
| | | | | | | $P_4(T_4)$= | 0,3020 | $P_5(T_4)$= | 0,3021 |
| | | | | | | $P_4(C_4)$= | 0,1980 | $P_5(C_4)$= | 0,1981 |
| | | | | | | $P_4(G_4)$= | 0,1986 | $P_5(G_4)$= | 0,1986 |
| | | | | | | | | $P_5(A_5)$= | 0,3014 |
| | | | | | | | | $P_5(T_5)$= | 0,3020 |
| | | | | | | | | $P_5(C_5)$= | 0,1981 |
| | | | | | | | | $P_5(G_5)$= | 0,1985 |

Fig. A2/3. The table of probabilities of subgroups of tetra-groups in the sequence: Homo sapiens chromosome 3, GRCh38.p7 Primary Assembly

NCBI Reference Sequence: NC_000003.12
LOCUS NC_000003 198295559 bp DNA linear CON 06-JUN-2016
DEFINITION Homo sapiens chromosome 3, GRCh38.p7 Primary Assembly.
ACCESSION NC_000003 GPC_000001295 VERSION NC_000003.12
https://www.ncbi.nlm.nih.gov/nuccore/NC_000003.12

| NUCLEOTIDES | | DOUBLETS | | TRIPLETS | | 4-PLETS | | 5-PLETS | |
|---|---|---|---|---|---|---|---|---|---|
| $P_1(A_1)=$ | 0,3086 | $P_2(A_1)=$ | 0,3086 | $P_3(A_1)=$ | 0,3087 | $P_4(A_1)=$ | 0,3086 | $P_5(A_1)=$ | 0,3088 |
| $P_1(T_1)=$ | 0,3089 | $P_2(T_1)=$ | 0,3090 | $P_3(T_1)=$ | 0,3089 | $P_4(T_1)=$ | 0,3089 | $P_5(T_1)=$ | 0,3089 |
| $P_1(C_1)=$ | 0,1910 | $P_2(C_1)=$ | 0,1909 | $P_3(C_1)=$ | 0,1909 | $P_4(C_1)=$ | 0,1909 | $P_5(C_1)=$ | 0,1910 |
| $P_1(G_1)=$ | 0,1915 | $P_2(G_1)=$ | 0,1915 | $P_3(G_1)=$ | 0,1915 | $P_4(G_1)=$ | 0,1916 | $P_5(G_1)=$ | 0,1913 |
| | | $P_2(A_2)=$ | 0,3086 | $P_3(A_2)=$ | 0,3086 | $P_4(A_2)=$ | 0,3087 | $P_5(A_2)=$ | 0,3086 |
| | | $P_2(T_2)=$ | 0,3089 | $P_3(T_2)=$ | 0,3090 | $P_4(T_2)=$ | 0,3089 | $P_5(T_2)=$ | 0,3091 |
| | | $P_2(C_2)=$ | 0,1910 | $P_3(C_2)=$ | 0,1910 | $P_4(C_2)=$ | 0,1910 | $P_5(C_2)=$ | 0,1908 |
| | | $P_2(G_2)=$ | 0,1914 | $P_3(G_2)=$ | 0,1914 | $P_4(G_2)=$ | 0,1914 | $P_5(G_2)=$ | 0,1915 |
| | | | | $P_3(A_3)=$ | 0,3086 | $P_4(A_3)=$ | 0,3086 | $P_5(A_3)=$ | 0,3087 |
| | | | | $P_3(T_3)=$ | 0,3090 | $P_4(T_3)=$ | 0,3090 | $P_5(T_3)=$ | 0,3088 |
| | | | | $P_3(C_3)=$ | 0,1910 | $P_4(C_3)=$ | 0,1909 | $P_5(C_3)=$ | 0,1910 |
| | | | | $P_3(G_3)=$ | 0,1914 | $P_4(G_3)=$ | 0,1914 | $P_5(G_3)=$ | 0,1915 |
| | | | | | | $P_4(A_4)=$ | 0,3085 | $P_5(A_4)=$ | 0,3085 |
| | | | | | | $P_4(T_4)=$ | 0,3090 | $P_5(T_4)=$ | 0,3090 |
| | | | | | | $P_4(C_4)=$ | 0,1910 | $P_5(C_4)=$ | 0,1910 |
| | | | | | | $P_4(G_4)=$ | 0,1915 | $P_5(G_4)=$ | 0,1915 |
| | | | | | | | | $P_5(A_5)=$ | 0,3085 |
| | | | | | | | | $P_5(T_5)=$ | 0,3090 |
| | | | | | | | | $P_5(C_5)=$ | 0,1910 |
| | | | | | | | | $P_5(G_5)=$ | 0,1915 |

Fig. A2/4. The table of probabilities of subgroups of tetra-groups in the sequence: Homo sapiens chromosome 4, GRCh38.p7 Primary Assembly
NCBI Reference Sequence: NC_000004.12
LOCUS NC_000004 190214555 bp DNA linear CON 06-JUN-2016
DEFINITION Homo sapiens chromosome 4, GRCh38.p7 Primary Assembly.
ACCESSION NC_000004 GPC_000001296 VERSION NC_000004.12
https://www.ncbi.nlm.nih.gov/nuccore/NC_000004.12

| NUCLEOTIDES | | DOUBLETS | | TRIPLETS | | 4-PLETS | | 5-PLETS | |
|---|---|---|---|---|---|---|---|---|---|
| $P_1(A_1)=$ | 0,3018 | $P_2(A_1)=$ | 0,3018 | $P_3(A_1)=$ | 0,3018 | $P_4(A_1)=$ | 0,3017 | $P_5(A_1)=$ | 0,3018 |
| $P_1(T_1)=$ | 0,3032 | $P_2(T_1)=$ | 0,3032 | $P_3(T_1)=$ | 0,3031 | $P_4(T_1)=$ | 0,3032 | $P_5(T_1)=$ | 0,3031 |
| $P_1(C_1)=$ | 0,1971 | $P_2(C_1)=$ | 0,1971 | $P_3(C_1)=$ | 0,1971 | $P_4(C_1)=$ | 0,1972 | $P_5(C_1)=$ | 0,1971 |
| $P_1(G_1)=$ | 0,1979 | $P_2(G_1)=$ | 0,1979 | $P_3(G_1)=$ | 0,1989 | $P_4(G_1)=$ | 0,1979 | $P_5(G_1)=$ | 0,1980 |
| | | $P_2(A_2)=$ | 0,3018 | $P_3(A_2)=$ | 0,3017 | $P_4(A_2)=$ | 0,3018 | $P_5(A_2)=$ | 0,3019 |
| | | $P_2(T_2)=$ | 0,3032 | $P_3(T_2)=$ | 0,3032 | $P_4(T_2)=$ | 0,3032 | $P_5(T_2)=$ | 0,3031 |
| | | $P_2(C_2)=$ | 0,1971 | $P_3(C_2)=$ | 0,1972 | $P_4(C_2)=$ | 0,1971 | $P_5(C_2)=$ | 0,1972 |
| | | $P_2(G_2)=$ | 0,1980 | $P_3(G_2)=$ | 0,1979 | $P_4(G_2)=$ | 0,1979 | $P_5(G_2)=$ | 0,1978 |
| | | | | $P_3(A_3)=$ | 0,3018 | $P_4(A_3)=$ | 0,3018 | $P_5(A_3)=$ | 0,3018 |
| | | | | $P_3(T_3)=$ | 0,3032 | $P_4(T_3)=$ | 0,3032 | $P_5(T_3)=$ | 0,3032 |
| | | | | $P_3(C_3)=$ | 0,1971 | $P_4(C_3)=$ | 0,1971 | $P_5(C_3)=$ | 0,1972 |
| | | | | $P_3(G_3)=$ | 0,1979 | $P_4(G_3)=$ | 0,1979 | $P_5(G_3)=$ | 0,1978 |
| | | | | | | $P_4(A_4)=$ | 0,3017 | $P_5(A_4)=$ | 0,3016 |
| | | | | | | $P_4(T_4)=$ | 0,3031 | $P_5(T_4)=$ | 0,3032 |
| | | | | | | $P_4(C_4)=$ | 0,1971 | $P_5(C_4)=$ | 0,1971 |
| | | | | | | $P_4(G_4)=$ | 0,1980 | $P_5(G_4)=$ | 0,1981 |
| | | | | | | | | $P_5(A_5)=$ | 0,3017 |
| | | | | | | | | $P_5(T_5)=$ | 0,3032 |

| | | | | |
|---|---|---|---|---|
| | | | | $P_5(C_5)=$ 0,1972 |
| | | | | $P_5(G_5)=$ 0,1979 |

Fig. A2/5. The table of probabilities of subgroups of tetra-groups in the sequence: Homo sapiens chromosome 5, GRCh38.p7 Primary Assembly
NCBI Reference Sequence: NC_000005.10
LOCUS NC_000005 181538259 bp DNA linear CON 06-JUN-2016
DEFINITION Homo sapiens chromosome 5, GRCh38.p7 Primary Assembly.
ACCESSION NC_000005 GPC_000001297 VERSION NC_000005.10
https://www.ncbi.nlm.nih.gov/nuccore/NC_000005.10

| **NUCLEOTIDES** | **DOUBLETS** | **TRIPLETS** | **4-PLETS** | **5-PLETS** |
|---|---|---|---|---|
| $P_1(A_1)=$ 0,3021 | $P_2(A_1)=$ 0,3021 | $P_3(A_1)=$ 0,3021 | $P_4(A_1)=$ 0,3021 | $P_5(A_1)=$ 0,3021 |
| $P_1(T_1)=$ 0,3020 | $P_2(T_1)=$ 0,3020 | $P_3(T_1)=$ 0,3021 | $P_4(T_1)=$ 0,3020 | $P_5(T_1)=$ 0,3020 |
| $P_1(C_1)=$ 0,1979 | $P_2(C_1)=$ 0,1979 | $P_3(C_1)=$ 0,1979 | $P_4(C_1)=$ 0,1980 | $P_5(C_1)=$ 0,1980 |
| $P_1(G_1)=$ 0,1979 | $P_2(G_1)=$ 0,1979 | $P_3(G_1)=$ 0,1980 | $P_4(G_1)=$ 0,1979 | $P_5(G_1)=$ 0,1979 |
| | $P_2(A_2)=$ 0,3021 | $P_3(A_2)=$ 0,3021 | $P_4(A_2)=$ 0,3023 | $P_5(A_2)=$ 0,3021 |
| | $P_2(T_2)=$ 0,3021 | $P_3(T_2)=$ 0,3020 | $P_4(T_2)=$ 0,3020 | $P_5(T_2)=$ 0,3021 |
| | $P_2(C_2)=$ 0,1979 | $P_3(C_2)=$ 0,1980 | $P_4(C_2)=$ 0,1979 | $P_5(C_2)=$ 0,1979 |
| | $P_2(G_2)=$ 0,1979 | $P_3(G_2)=$ 0,1979 | $P_4(G_2)=$ 0,1979 | $P_5(G_2)=$ 0,1979 |
| | | $P_3(A_3)=$ 0,3022 | $P_4(A_3)=$ 0,3021 | $P_5(A_3)=$ 0,3020 |
| | | $P_3(T_3)=$ 0,3020 | $P_4(T_3)=$ 0,3021 | $P_5(T_3)=$ 0,3021 |
| | | $P_3(C_3)=$ 0,1979 | $P_4(C_3)=$ 0,1978 | $P_5(C_3)=$ 0,1980 |
| | | $P_3(G_3)=$ 0,1979 | $P_4(G_3)=$ 0,1979 | $P_5(G_3)=$ 0,1979 |
| | | | $P_4(A_4)=$ 0,3019 | $P_5(A_4)=$ 0,3022 |
| | | | $P_4(T_4)=$ 0,3021 | $P_5(T_4)=$ 0,3020 |
| | | | $P_4(C_4)=$ 0,1980 | $P_5(C_4)=$ 0,1978 |
| | | | $P_4(G_4)=$ 0,1979 | $P_5(G_4)=$ 0,1980 |
| | | | | $P_5(A_5)=$ 0,3020 |
| | | | | $P_5(T_5)=$ 0,3020 |
| | | | | $P_5(C_5)=$ 0,1980 |
| | | | | $P_5(G_5)=$ 0,1980 |

Fig. A2/6. The table of probabilities of subgroups of tetra-groups in the sequence: Homo sapiens chromosome 6, whole genome shotgun sequence
GenBank: CM000257.1
LOCUS CM000257 171718000 bp DNA linear CON 23-MAR-2015
DEFINITION Homo sapiens chromosome 6, whole genome shotgun sequence.
ACCESSION CM000257 AADB02000000 CH003477 VERSION CM000257.1
https://www.ncbi.nlm.nih.gov/nuccore/CM000257.1
https://www.ncbi.nlm.nih.gov/nuccore/CM000257.1?report=fasta

| **NUCLEOTIDES** | **DOUBLETS** | **TRIPLETS** | **4-PLETS** | **5-PLETS** |
|---|---|---|---|---|
| $P_1(A_1)=$ 0,2960 | $P_2(A_1)=$ 0,2960 | $P_3(A_1)=$ 0,2960 | $P_4(A_1)=$ 0,2960 | $P_5(A_1)=$ 0,2960 |
| $P_1(T_1)=$ 0,2970 | $P_2(T_1)=$ 0,2970 | $P_3(T_1)=$ 0,2971 | $P_4(T_1)=$ 0,2970 | $P_5(T_1)=$ 0,2970 |
| $P_1(C_1)=$ 0,2033 | $P_2(C_1)=$ 0,2033 | $P_3(C_1)=$ 0,2032 | $P_4(C_1)=$ 0,2032 | $P_5(C_1)=$ 0,2032 |
| $P_1(G_1)=$ 0,2037 | $P_2(G_1)=$ 0,2037 | $P_3(G_1)=$ 0,2037 | $P_4(G_1)=$ 0,2037 | $P_5(G_1)=$ 0,2037 |
| | $P_2(A_2)=$ 0,2960 | $P_3(A_2)=$ 0,2960 | $P_4(A_2)=$ 0,2959 | $P_5(A_2)=$ 0,2961 |
| | $P_2(T_2)=$ 0,2970 | $P_3(T_2)=$ 0,2970 | $P_4(T_2)=$ 0,2970 | $P_5(T_2)=$ 0,2969 |
| | $P_2(C_2)=$ 0,2033 | $P_3(C_2)=$ 0,2034 | $P_4(C_2)=$ 0,2033 | $P_5(C_2)=$ 0,2032 |

| | | | | | | | | | |
|---|---|---|---|---|---|---|---|---|---|
| | | $P_2(G_2)=$ | 0,2037 | $P_3(G_2)=$ | 0,2037 | $P_4(G_2)=$ | 0,2037 | $P_5(G_2)=$ | 0,2038 |
| | | | | $P_3(A_3)=$ | 0,2960 | $P_4(A_3)=$ | 0,2960 | $P_5(A_3)=$ | 0,2961 |
| | | | | $P_3(T_3)=$ | 0,2970 | $P_4(T_3)=$ | 0,2970 | $P_5(T_3)=$ | 0,2969 |
| | | | | $P_3(C_3)=$ | 0,2033 | $P_4(C_3)=$ | 0,2033 | $P_5(C_3)=$ | 0,2033 |
| | | | | $P_3(G_3)=$ | 0,2037 | $P_4(G_3)=$ | 0,2036 | $P_5(G_3)=$ | 0,2036 |
| | | | | | | $P_4(A_4)=$ | 0,2961 | $P_5(A_4)=$ | 0,2959 |
| | | | | | | $P_4(T_4)=$ | 0,2970 | $P_5(T_4)=$ | 0,2971 |
| | | | | | | $P_4(C_4)=$ | 0,2033 | $P_5(C_4)=$ | 0,2034 |
| | | | | | | $P_4(G_4)=$ | 0,2037 | $P_5(G_4)=$ | 0,2036 |
| | | | | | | | | $P_5(A_5)=$ | 0,2960 |
| | | | | | | | | $P_5(T_5)=$ | 0,2971 |
| | | | | | | | | $P_5(C_5)=$ | 0,2033 |
| | | | | | | | | $P_5(G_5)=$ | 0,2036 |

Fig. A2/7. The table of probabilities of subgroups of tetra-groups in the sequence: Homo sapiens chromosome 7, GRCh38.p7 Primary Assembly

NCBI Reference Sequence: NC_000007.14

LOCUS NC_000007 159345973 bp DNA linear CON 06-JUN-2016

DEFINITION Homo sapiens chromosome 7, GRCh38.p7 Primary Assembly.

ACCESSION NC_000007 GPC_000001299 VERSION NC_000007.14

https://www.ncbi.nlm.nih.gov/nuccore/NC_000007.14

https://www.ncbi.nlm.nih.gov/nuccore/NC_000007.14?report=fasta

| **NUCLEOTIDES** | | **DOUBLETS** | | **TRIPLETS** | | **4-PLETS** | | **5-PLETS** | |
|---|---|---|---|---|---|---|---|---|---|
| $P_1(A_1)=$ | 0,2994 | $P_2(A_1)=$ | 0,2994 | $P_3(A_1)=$ | 0,2994 | $P_4(A_1)=$ | 0,2993 | $P_5(A_1)=$ | 0,2993 |
| $P_1(T_1)=$ | 0,2990 | $P_2(T_1)=$ | 0,2990 | $P_3(T_1)=$ | 0,2990 | $P_4(T_1)=$ | 0,2991 | $P_5(T_1)=$ | 0,2991 |
| $P_1(C_1)=$ | 0,2008 | $P_2(C_1)=$ | 0,2008 | $P_3(C_1)=$ | 0,2008 | $P_4(C_1)=$ | 0,2008 | $P_5(C_1)=$ | 0,2008 |
| $P_1(G_1)=$ | 0,2008 | $P_2(G_1)=$ | 0,2008 | $P_3(G_1)=$ | 0,2008 | $P_4(G_1)=$ | 0,2008 | $P_5(G_1)=$ | 0,2008 |
| | | $P_2(A_2)=$ | 0,2994 | $P_3(A_2)=$ | 0,2994 | $P_4(A_2)=$ | 0,2994 | $P_5(A_2)=$ | 0,2994 |
| | | $P_2(T_2)=$ | 0,2989 | $P_3(T_2)=$ | 0,2990 | $P_4(T_2)=$ | 0,2990 | $P_5(T_2)=$ | 0,2989 |
| | | $P_2(C_2)=$ | 0,2008 | $P_3(C_2)=$ | 0,2008 | $P_4(C_2)=$ | 0,2008 | $P_5(C_2)=$ | 0,2008 |
| | | $P_2(G_2)=$ | 0,2009 | $P_3(G_2)=$ | 0,2008 | $P_4(G_2)=$ | 0,2008 | $P_5(G_2)=$ | 0,2009 |
| | | | | $P_3(A_3)=$ | 0,2994 | $P_4(A_3)=$ | 0,2994 | $P_5(A_3)=$ | 0,2994 |
| | | | | $P_3(T_3)=$ | 0,2990 | $P_4(T_3)=$ | 0,2990 | $P_5(T_3)=$ | 0,2990 |
| | | | | $P_3(C_3)=$ | 0,2008 | $P_4(C_3)=$ | 0,2008 | $P_5(C_3)=$ | 0,2008 |
| | | | | $P_3(G_3)=$ | 0,2008 | $P_4(G_3)=$ | 0,2007 | $P_5(G_3)=$ | 0,2008 |
| | | | | | | $P_4(A_4)=$ | 0,2994 | $P_5(A_4)=$ | 0,2994 |
| | | | | | | $P_4(T_4)=$ | 0,2989 | $P_5(T_4)=$ | 0,2990 |
| | | | | | | $P_4(C_4)=$ | 0,2008 | $P_5(C_4)=$ | 0,2008 |
| | | | | | | $P_4(G_4)=$ | 0,2009 | $P_5(G_4)=$ | 0,2008 |
| | | | | | | | | $P_5(A_5)=$ | 0,2994 |
| | | | | | | | | $P_5(T_5)=$ | 0,2990 |
| | | | | | | | | $P_5(C_5)=$ | 0,2008 |
| | | | | | | | | $P_5(G_5)=$ | 0,2007 |

Fig. A2/8. The table of probabilities of subgroups of tetra-groups in the sequence: Homo sapiens chromosome 8, alternate assembly CHM1_1.1, whole genome shotgun sequence

NCBI Reference Sequence: NC_018919.2

LOCUS NC_018919 146399655 bp DNA linear CON 06-JUN-2016

DEFINITION Homo sapiens chromosome 8, alternate assembly CHM1_1.1, whole genome shotgun sequence. ACCESSION NC_018919 GPC_000001167 VERSION NC_018919.2

https://www.ncbi.nlm.nih.gov/nuccore/NC_018919.2

https://www.ncbi.nlm.nih.gov/nuccore/NC_018919.2?report=fasta

| NUCLEOTIDES | | DOUBLETS | | TRIPLETS | | 4-PLETS | | 5-PLETS | |
|---|---|---|---|---|---|---|---|---|---|
| $P_1(A_1)=$ | 0,2928 | $P_2(A_1)=$ | 0,2928 | $P_3(A_1)=$ | 0,2927 | $P_4(A_1)=$ | 0,2929 | $P_5(A_1)=$ | 0,2927 |
| $P_1(T_1)=$ | 0,2926 | $P_2(T_1)=$ | 0,2927 | $P_3(T_1)=$ | 0,2926 | $P_4(T_1)=$ | 0,2928 | $P_5(T_1)=$ | 0,2926 |
| $P_1(C_1)=$ | 0,2074 | $P_2(C_1)=$ | 0,2073 | $P_3(C_1)=$ | 0,2074 | $P_4(C_1)=$ | 0,2072 | $P_5(C_1)=$ | 0,2074 |
| $P_1(G_1)=$ | 0,2072 | $P_2(G_1)=$ | 0,2072 | $P_3(G_1)=$ | 0,2073 | $P_4(G_1)=$ | 0,2071 | $P_5(G_1)=$ | 0,2073 |
| | | $P_2(A_2)=$ | 0,2928 | $P_3(A_2)=$ | 0,2928 | $P_4(A_2)=$ | 0,2927 | $P_5(A_2)=$ | 0,2927 |
| | | $P_2(T_2)=$ | 0,2925 | $P_3(T_2)=$ | 0,2926 | $P_4(T_2)=$ | 0,2925 | $P_5(T_2)=$ | 0,2927 |
| | | $P_2(C_2)=$ | 0,2074 | $P_3(C_2)=$ | 0,2074 | $P_4(C_2)=$ | 0,2074 | $P_5(C_2)=$ | 0,2075 |
| | | $P_2(G_2)=$ | 0,2073 | $P_3(G_2)=$ | 0,2072 | $P_4(G_2)=$ | 0,2073 | $P_5(G_2)=$ | 0,2072 |
| | | | | $P_3(A_3)=$ | 0,2929 | $P_4(A_3)=$ | 0,2928 | $P_5(A_3)=$ | 0,2930 |
| | | | | $P_3(T_3)=$ | 0,2927 | $P_4(T_3)=$ | 0,2926 | $P_5(T_3)=$ | 0,2926 |
| | | | | $P_3(C_3)=$ | 0,2073 | $P_4(C_3)=$ | 0,2074 | $P_5(C_3)=$ | 0,2074 |
| | | | | $P_3(G_3)=$ | 0,2072 | $P_4(G_3)=$ | 0,2072 | $P_5(G_3)=$ | 0,2071 |
| | | | | | | $P_4(A_4)=$ | 0,2928 | $P_5(A_4)=$ | 0,2927 |
| | | | | | | $P_4(T_4)=$ | 0,2926 | $P_5(T_4)=$ | 0,2927 |
| | | | | | | $P_4(C_4)=$ | 0,2075 | $P_5(C_4)=$ | 0,2073 |
| | | | | | | $P_4(G_4)=$ | 0,2072 | $P_5(G_4)=$ | 0,2072 |
| | | | | | | | | $P_5(A_5)=$ | 0,2929 |
| | | | | | | | | $P_5(T_5)=$ | 0,2925 |
| | | | | | | | | $P_5(C_5)=$ | 0,2073 |
| | | | | | | | | $P_5(G_5)=$ | 0,2072 |

Fig. A2/9. The table of probabilities of subgroups of tetra-groups in the sequence: Homo sapiens chromosome 9, whole genome shotgun sequence
GenBank: CM000260.1
LOCUS CM000260 111583154 bp DNA linear CON 23-MAR-2015
DEFINITION Homo sapiens chromosome 9, whole genome shotgun sequence.
ACCESSION CM000260 AADB02000000 CH003480 VERSION CM000260.1
https://www.ncbi.nlm.nih.gov/nuccore/CM000260.1
https://www.ncbi.nlm.nih.gov/nuccore/CM000260.1?report=fasta

| NUCLEOTIDES | | DOUBLETS | | TRIPLETS | | 4-PLETS | | 5-PLETS | |
|---|---|---|---|---|---|---|---|---|---|
| $P_1(A_1)=$ | 0,2917 | $P_2(A_1)=$ | 0,2918 | $P_3(A_1)=$ | 0,2917 | $P_4(A_1)=$ | 0,2918 | $P_5(A_1)=$ | 0,2917 |
| $P_1(T_1)=$ | 0,2928 | $P_2(T_1)=$ | 0,2928 | $P_3(T_1)=$ | 0,2929 | $P_4(T_1)=$ | 0,2928 | $P_5(T_1)=$ | 0,2928 |
| $P_1(C_1)=$ | 0,2074 | $P_2(C_1)=$ | 0,2073 | $P_3(C_1)=$ | 0,2074 | $P_4(C_1)=$ | 0,2074 | $P_5(C_1)=$ | 0,2074 |
| $P_1(G_1)=$ | 0,2080 | $P_2(G_1)=$ | 0,2080 | $P_3(G_1)=$ | 0,2080 | $P_4(G_1)=$ | 0,2080 | $P_5(G_1)=$ | 0,2081 |
| | | $P_2(A_2)=$ | 0,2916 | $P_3(A_2)=$ | 0,2917 | $P_4(A_2)=$ | 0,2916 | $P_5(A_2)=$ | 0,2918 |
| | | $P_2(T_2)=$ | 0,2929 | $P_3(T_2)=$ | 0,2928 | $P_4(T_2)=$ | 0,2930 | $P_5(T_2)=$ | 0,2928 |
| | | $P_2(C_2)=$ | 0,2075 | $P_3(C_2)=$ | 0,2075 | $P_4(C_2)=$ | 0,2075 | $P_5(C_2)=$ | 0,2075 |
| | | $P_2(G_2)=$ | 0,2080 | $P_3(G_2)=$ | 0,2080 | $P_4(G_2)=$ | 0,2079 | $P_5(G_2)=$ | 0,2079 |
| | | | | $P_3(A_3)=$ | 0,2918 | $P_4(A_3)=$ | 0,2918 | $P_5(A_3)=$ | 0,2918 |
| | | | | $P_3(T_3)=$ | 0,2929 | $P_4(T_3)=$ | 0,2929 | $P_5(T_3)=$ | 0,2929 |
| | | | | $P_3(C_3)=$ | 0,2073 | $P_4(C_3)=$ | 0,2073 | $P_5(C_3)=$ | 0,2074 |
| | | | | $P_3(G_3)=$ | 0,2080 | $P_4(G_3)=$ | 0,2080 | $P_5(G_3)=$ | 0,2079 |
| | | | | | | $P_4(A_4)=$ | 0,2917 | $P_5(A_4)=$ | 0,2916 |
| | | | | | | $P_4(T_4)=$ | 0,2928 | $P_5(T_4)=$ | 0,2930 |
| | | | | | | $P_4(C_4)=$ | 0,2074 | $P_5(C_4)=$ | 0,2074 |
| | | | | | | $P_4(G_4)=$ | 0,2081 | $P_5(G_4)=$ | 0,2080 |
| | | | | | | | | $P_5(A_5)=$ | 0,2917 |
| | | | | | | | | $P_5(T_5)=$ | 0,2928 |
| | | | | | | | | $P_5(C_5)=$ | 0,2073 |
| | | | | | | | | $P_5(G_5)=$ | 0,2082 |

Fig. A2/10. The table of probabilities of subgroups of tetra-groups in the sequence: Homo sapiens chromosome 10, GRCh38.p7 Primary Assembly
NCBI Reference Sequence: NC_000010.11
LOCUS NC_000010 133797422 bp DNA linear CON 06-JUN-2016
DEFINITION Homo sapiens chromosome 10, GRCh38.p7 Primary Assembly.
ACCESSION NC_000010 GPC_000001302 VERSION NC_000010.11
https://www.ncbi.nlm.nih.gov/nuccore/NC_000010.11
https://www.ncbi.nlm.nih.gov/nuccore/NC_000010.11?report=fasta

| **NUCLEOTIDES** | | **DOUBLETS** | | **TRIPLETS** | | **4-PLETS** | | **5-PLETS** | |
|---|---|---|---|---|---|---|---|---|---|
| $P_1(A_1)$= | 0,2920 | $P_2(A_1)$= | 0,2920 | $P_3(A_1)$= | 0,2920 | $P_4(A_1)$= | 0,2919 | $P_5(A_1)$= | 0,2920 |
| $P_1(T_1)$= | 0,2926 | $P_2(T_1)$= | 0,2926 | $P_3(T_1)$= | 0,2925 | $P_4(T_1)$= | 0,2926 | $P_5(T_1)$= | 0,2926 |
| $P_1(C_1)$= | 0,2074 | $P_2(C_1)$= | 0,2074 | $P_3(C_1)$= | 0,2074 | $P_4(C_1)$= | 0,2075 | $P_5(C_1)$= | 0,2073 |
| $P_1(G_1)$= | 0,2080 | $P_2(G_1)$= | 0,2080 | $P_3(G_1)$= | 0,2081 | $P_4(G_1)$= | 0,2080 | $P_5(G_1)$= | 0,2081 |
| | | $P_2(A_2)$= | 0,2921 | $P_3(A_2)$= | 0,2921 | $P_4(A_2)$= | 0,2920 | $P_5(A_2)$= | 0,2921 |
| | | $P_2(T_2)$= | 0,2926 | $P_3(T_2)$= | 0,2926 | $P_4(T_2)$= | 0,2926 | $P_5(T_2)$= | 0,2924 |
| | | $P_2(C_2)$= | 0,2074 | $P_3(C_2)$= | 0,2074 | $P_4(C_2)$= | 0,2074 | $P_5(C_2)$= | 0,2075 |
| | | $P_2(G_2)$= | 0,2080 | $P_3(G_2)$= | 0,2079 | $P_4(G_2)$= | 0,2080 | $P_5(G_2)$= | 0,2080 |
| | | | | $P_3(A_3)$= | 0,2920 | $P_4(A_3)$= | 0,2920 | $P_5(A_3)$= | 0,2920 |
| | | | | $P_3(T_3)$= | 0,2927 | $P_4(T_3)$= | 0,2926 | $P_5(T_3)$= | 0,2927 |
| | | | | $P_3(C_3)$= | 0,2073 | $P_4(C_3)$= | 0,2074 | $P_5(C_3)$= | 0,2074 |
| | | | | $P_3(G_3)$= | 0,2080 | $P_4(G_3)$= | 0,2080 | $P_5(G_3)$= | 0,2079 |
| | | | | | | $P_4(A_4)$= | 0,2922 | $P_5(A_4)$= | 0,2920 |
| | | | | | | $P_4(T_4)$= | 0,2926 | $P_5(T_4)$= | 0,2926 |
| | | | | | | $P_4(C_4)$= | 0,2073 | $P_5(C_4)$= | 0,2075 |
| | | | | | | $P_4(G_4)$= | 0,2080 | $P_5(G_4)$= | 0,2079 |
| | | | | | | | | $P_5(A_5)$= | 0,2919 |
| | | | | | | | | $P_5(T_5)$= | 0,2927 |
| | | | | | | | | $P_5(C_5)$= | 0,2073 |
| | | | | | | | | $P_5(G_5)$= | 0,2081 |

Fig. A2/11. The table of probabilities of subgroups of tetra-groups in the sequence Homo sapiens chromosome 11, GRCh38.p7 Primary Assembly
NCBI Reference Sequence: NC_000011.10
LOCUS NC_000011 135086622 bp DNA linear CON 06-JUN-2016
DEFINITION Homo sapiens chromosome 11, GRCh38.p7 Primary Assembly.
ACCESSION NC_000011 GPC_000001303 VERSION NC_000011.10
https://www.ncbi.nlm.nih.gov/nuccore/NC_000011.10
https://www.ncbi.nlm.nih.gov/nuccore/NC_000011.10?report=fasta

| **NUCLEOTIDES** | | **DOUBLETS** | | **TRIPLETS** | | **4-PLETS** | | **5-PLETS** | |
|---|---|---|---|---|---|---|---|---|---|
| $P_1(A_1)$= | 0,2957 | $P_2(A_1)$= | 0,2957 | $P_3(A_1)$= | 0,2958 | $P_4(A_1)$= | 0,2958 | $P_5(A_1)$= | 0,2957 |
| $P_1(T_1)$= | 0,2966 | $P_2(T_1)$= | 0,2966 | $P_3(T_1)$= | 0,2966 | $P_4(T_1)$= | 0,2966 | $P_5(T_1)$= | 0,2967 |
| $P_1(C_1)$= | 0,2035 | $P_2(C_1)$= | 0,2035 | $P_3(C_1)$= | 0,2035 | $P_4(C_1)$= | 0,2035 | $P_5(C_1)$= | 0,2035 |
| $P_1(G_1)$= | 0,2042 | $P_2(G_1)$= | 0,2042 | $P_3(G_1)$= | 0,2041 | $P_4(G_1)$= | 0,2042 | $P_5(G_1)$= | 0,2041 |
| | | $P_2(A_2)$= | 0,2958 | $P_3(A_2)$= | 0,2956 | $P_4(A_2)$= | 0,2958 | $P_5(A_2)$= | 0,2956 |
| | | $P_2(T_2)$= | 0,2966 | $P_3(T_2)$= | 0,2968 | $P_4(T_2)$= | 0,2966 | $P_5(T_2)$= | 0,2966 |
| | | $P_2(C_2)$= | 0,2035 | $P_3(C_2)$= | 0,2035 | $P_4(C_2)$= | 0,2034 | $P_5(C_2)$= | 0,2035 |
| | | $P_2(G_2)$= | 0,2041 | $P_3(G_2)$= | 0,2042 | $P_4(G_2)$= | 0,2042 | $P_5(G_2)$= | 0,2043 |
| | | | | $P_3(A_3)$= | 0,2958 | $P_4(A_3)$= | 0,2955 | $P_5(A_3)$= | 0,2956 |
| | | | | $P_3(T_3)$= | 0,2966 | $P_4(T_3)$= | 0,2967 | $P_5(T_3)$= | 0,2966 |
| | | | | $P_3(C_3)$= | 0,2035 | $P_4(C_3)$= | 0,2035 | $P_5(C_3)$= | 0,2036 |
| | | | | $P_3(G_3)$= | 0,2042 | $P_4(G_3)$= | 0,2042 | $P_5(G_3)$= | 0,2042 |
| | | | | | | $P_4(A_4)$= | 0,2957 | $P_5(A_4)$= | 0,2958 |

| | | | |
|---|---|---|---|
| $P_4(T_4)=$ | 0,2966 | $P_5(T_4)=$ | 0,2966 |
| $P_4(C_4)=$ | 0,2035 | $P_5(C_4)=$ | 0,2034 |
| $P_4(G_4)=$ | 0,2041 | $P_5(G_4)=$ | 0,2041 |
| | | $P_5(A_5)=$ | 0,2957 |
| | | $P_5(T_5)=$ | 0,2966 |
| | | $P_5(C_5)=$ | 0,2035 |
| | | $P_5(G_5)=$ | 0,2042 |

Fig. A2/12. Probabilities of subgroups of tetra-groups in the sequence: Homo sapiens chromosome 12, GRCh38.p7 Primary Assembly

NCBI Reference Sequence: NC_000012.12

LOCUS NC_000012 133275309 bp DNA linear CON 06-JUN-2016

DEFINITION Homo sapiens chromosome 12, GRCh38.p7 Primary Assembly.

ACCESSION NC_000012 GPC_000001304 VERSION NC_000012.12

https://www.ncbi.nlm.nih.gov/nuccore/NC_000012.12

https://www.ncbi.nlm.nih.gov/nuccore/NC_000012.12?report=fasta

| **NUCLEOTIDES** | | **DOUBLETS** | | **TRIPLETS** | | **4-PLETS** | | **5-PLETS** | |
|---|---|---|---|---|---|---|---|---|---|
| $P_1(A_1)=$ | 0,3069 | $P_2(A_1)=$ | 0,3068 | $P_3(A_1)=$ | 0,3069 | $P_4(A_1)=$ | 0,3068 | $P_5(A_1)=$ | 0,3071 |
| $P_1(T_1)=$ | 0,3079 | $P_2(T_1)=$ | 0,3078 | $P_3(T_1)=$ | 0,3079 | $P_4(T_1)=$ | 0,3078 | $P_5(T_1)=$ | 0,3077 |
| $P_1(C_1)=$ | 0,1926 | $P_2(C_1)=$ | 0,1927 | $P_3(C_1)=$ | 0,1926 | $P_4(C_1)=$ | 0,1926 | $P_5(C_1)=$ | 0,1925 |
| $P_1(G_1)=$ | 0,1926 | $P_2(G_1)=$ | 0,1927 | $P_3(G_1)=$ | 0,1926 | $P_4(G_1)=$ | 0,1927 | $P_5(G_1)=$ | 0,1927 |
| | | $P_2(A_2)=$ | 0,3069 | $P_3(A_2)=$ | 0,3069 | $P_4(A_2)=$ | 0,3068 | $P_5(A_2)=$ | 0,3068 |
| | | $P_2(T_2)=$ | 0,3079 | $P_3(T_2)=$ | 0,3079 | $P_4(T_2)=$ | 0,3079 | $P_5(T_2)=$ | 0,3081 |
| | | $P_2(C_2)=$ | 0,1925 | $P_3(C_2)=$ | 0,1925 | $P_4(C_2)=$ | 0,1926 | $P_5(C_2)=$ | 0,1926 |
| | | $P_2(G_2)=$ | 0,1926 | $P_3(G_2)=$ | 0,1926 | $P_4(G_2)=$ | 0,1927 | $P_5(G_2)=$ | 0,1925 |
| | | | | $P_3(A_3)=$ | 0,3068 | $P_4(A_3)=$ | 0,3068 | $P_5(A_3)=$ | 0,3067 |
| | | | | $P_3(T_3)=$ | 0,3078 | $P_4(T_3)=$ | 0,3078 | $P_5(T_3)=$ | 0,3079 |
| | | | | $P_3(C_3)=$ | 0,1927 | $P_4(C_3)=$ | 0,1927 | $P_5(C_3)=$ | 0,1927 |
| | | | | $P_3(G_3)=$ | 0,19274 | $P_4(G_3)=$ | 0,1926 | $P_5(G_3)=$ | 0,1927 |
| | | | | | | $P_4(A_4)=$ | 0,3071 | $P_5(A_4)=$ | 0,3069 |
| | | | | | | $P_4(T_4)=$ | 0,3080 | $P_5(T_4)=$ | 0,3079 |
| | | | | | | $P_4(C_4)=$ | 0,1924 | $P_5(C_4)=$ | 0,1927 |
| | | | | | | $P_4(G_4)=$ | 0,1925 | $P_5(G_4)=$ | 0,1926 |
| | | | | | | | | $P_5(A_5)=$ | 0,3070 |
| | | | | | | | | $P_5(T_5)=$ | 0,3079 |
| | | | | | | | | $P_5(C_5)=$ | 0,1925 |
| | | | | | | | | $P_5(G_5)=$ | 0,1927 |

Fig. A2/13. The table of probabilities of subgroups of tetra-groups in the sequence: Homo sapiens chromosome 13, whole genome shotgun sequence

GenBank: CM000264.1

LOCUS CM000264 95789532 bp DNA linear CON 23-MAR-2015

DEFINITION Homo sapiens chromosome 13, whole genome shotgun sequence.

ACCESSION CM000264 AADB02000000 CH003484 VERSION CM000264.1

https://www.ncbi.nlm.nih.gov/nuccore/CM000264.1

https://www.ncbi.nlm.nih.gov/nuccore/CM000264.1?report=fasta

| **NUCLEOTIDES** | | **DOUBLETS** | | **TRIPLETS** | | **4-PLETS** | | **5-PLETS** | |
|---|---|---|---|---|---|---|---|---|---|
| $P_1(A_1)=$ | 0,2945 | $P_2(A_1)=$ | 0,2944 | $P_3(A_1)=$ | 0,2946 | $P_4(A_1)=$ | 0,2944 | $P_5(A_1)=$ | 0,2945 |
| $P_1(T_1)=$ | 0,2970 | $P_2(T_1)=$ | 0,2970 | $P_3(T_1)=$ | 0,2970 | $P_4(T_1)=$ | 0,2970 | $P_5(T_1)=$ | 0,2970 |
| $P_1(C_1)=$ | 0,2040 | $P_2(C_1)=$ | 0,2041 | $P_3(C_1)=$ | 0,2040 | $P_4(C_1)=$ | 0,2041 | $P_5(C_1)=$ | 0,2039 |
| $P_1(G_1)=$ | 0,2045 | $P_2(G_1)=$ | 0,2045 | $P_3(G_1)=$ | 0,2045 | $P_4(G_1)=$ | 0,2045 | $P_5(G_1)=$ | 0,2046 |
| | | $P_2(A_2)=$ | 0,2946 | $P_3(A_2)=$ | 0,2946 | $P_4(A_2)=$ | 0,2946 | $P_5(A_2)=$ | 0,2945 |

| | | | | | | | |
|---|---|---|---|---|---|---|---|
| $P_2(T_2)=$ | 0,2970 | $P_3(T_2)=$ | 0,2970 | $P_4(T_2)=$ | 0,2970 | $P_5(T_2)=$ | 0,2971 |
| $P_2(C_2)=$ | 0,2039 | $P_3(C_2)=$ | 0,2039 | $P_4(C_2)=$ | 0,2038 | $P_5(C_2)=$ | 0,2039 |
| $P_2(G_2)=$ | 0,2045 | $P_3(G_2)=$ | 0,2045 | $P_4(G_2)=$ | 0,2045 | $P_5(G_2)=$ | 0,2045 |
| | | $P_3(A_3)=$ | 0,2944 | $P_4(A_3)=$ | 0,2945 | $P_5(A_3)=$ | 0,2945 |
| | | $P_3(T_3)=$ | 0,2970 | $P_4(T_3)=$ | 0,2969 | $P_5(T_3)=$ | 0,2968 |
| | | $P_3(C_3)=$ | 0,2040 | $P_4(C_3)=$ | 0,2041 | $P_5(C_3)=$ | 0,2040 |
| | | $P_3(G_3)=$ | 0,2046 | $P_4(G_3)=$ | 0,2045 | $P_5(G_3)=$ | 0,2047 |
| | | | | $P_4(A_4)=$ | 0,2946 | $P_5(A_4)=$ | 0,2946 |
| | | | | $P_4(T_4)=$ | 0,2970 | $P_5(T_4)=$ | 0,2968 |
| | | | | $P_4(C_4)=$ | 0,2039 | $P_5(C_4)=$ | 0,2041 |
| | | | | $P_4(G_4)=$ | 0,2046 | $P_5(G_4)=$ | 0,2045 |
| | | | | | | $P_5(A_5)=$ | 0,2945 |
| | | | | | | $P_5(T_5)=$ | 0,2971 |
| | | | | | | $P_5(C_5)=$ | 0,2039 |
| | | | | | | $P_5(G_5)=$ | 0,2044 |

Fig. A2/14. The table of probabilities of subgroups of tetra-groups in the sequence: Homo sapiens chromosome 14, whole genome shotgun sequence
GenBank: CM000265.1
LOCUS CM000265 87316725 bp DNA linear CON 23-MAR-2015
DEFINITION Homo sapiens chromosome 14, whole genome shotgun sequence.
ACCESSION CM000265 AADB02000000 CH003485 VERSION CM000265.1
https://www.ncbi.nlm.nih.gov/nuccore/CM000265.1
https://www.ncbi.nlm.nih.gov/nuccore/CM000265.1?report=fasta

| **NUCLEOTIDES** | | **DOUBLETS** | | **TRIPLETS** | | **4-PLETS** | | **5-PLETS** | |
|---|---|---|---|---|---|---|---|---|---|
| $P_1(A_1)=$ | 0,2896 | $P_2(A_1)=$ | 0,2896 | $P_3(A_1)=$ | 0,2897 | $P_4(A_1)=$ | 0,2896 | $P_5(A_1)=$ | 0,2895 |
| $P_1(T_1)=$ | 0,2901 | $P_2(T_1)=$ | 0,2901 | $P_3(T_1)=$ | 0,2900 | $P_4(T_1)=$ | 0,2901 | $P_5(T_1)=$ | 0,2902 |
| $P_1(C_1)=$ | 0,2097 | $P_2(C_1)=$ | 0,2097 | $P_3(C_1)=$ | 0,2098 | $P_4(C_1)=$ | 0,2098 | $P_5(C_1)=$ | 0,2098 |
| $P_1(G_1)=$ | 0,2106 | $P_2(G_1)=$ | 0,2106 | $P_3(G_1)=$ | 0,2105 | $P_4(G_1)=$ | 0,2105 | $P_5(G_1)=$ | 0,2105 |
| | | $P_2(A_2)=$ | 0,2895 | $P_3(A_2)=$ | 0,2894 | $P_4(A_2)=$ | 0,2895 | $P_5(A_2)=$ | 0,2894 |
| | | $P_2(T_2)=$ | 0,2901 | $P_3(T_2)=$ | 0,2902 | $P_4(T_2)=$ | 0,2901 | $P_5(T_2)=$ | 0,2901 |
| | | $P_2(C_2)=$ | 0,2098 | $P_3(C_2)=$ | 0,2097 | $P_4(C_2)=$ | 0,2098 | $P_5(C_2)=$ | 0,2098 |
| | | $P_2(G_2)=$ | 0,2106 | $P_3(G_2)=$ | 0,2106 | $P_4(G_2)=$ | 0,2106 | $P_5(G_2)=$ | 0,2107 |
| | | | | $P_3(A_3)=$ | 0,2896 | $P_4(A_3)=$ | 0,2897 | $P_5(A_3)=$ | 0,2895 |
| | | | | $P_3(T_3)=$ | 0,2900 | $P_4(T_3)=$ | 0,2901 | $P_5(T_3)=$ | 0,2902 |
| | | | | $P_3(C_3)=$ | 0,2097 | $P_4(C_3)=$ | 0,2096 | $P_5(C_3)=$ | 0,2097 |
| | | | | $P_3(G_3)=$ | 0,2107 | $P_4(G_3)=$ | 0,2106 | $P_5(G_3)=$ | 0,2106 |
| | | | | | | $P_4(A_4)=$ | 0,2895 | $P_5(A_4)=$ | 0,2898 |
| | | | | | | $P_4(T_4)=$ | 0,2901 | $P_5(T_4)=$ | 0,2901 |
| | | | | | | $P_4(C_4)=$ | 0,2098 | $P_5(C_4)=$ | 0,2097 |
| | | | | | | $P_4(G_4)=$ | 0,2107 | $P_5(G_4)=$ | 0,2104 |
| | | | | | | | | $P_5(A_5)=$ | 0,2896 |
| | | | | | | | | $P_5(T_5)=$ | 0,2899 |
| | | | | | | | | $P_5(C_5)=$ | 0,2097 |
| | | | | | | | | $P_5(G_5)=$ | 0,2108 |

Fig. A2/15. The table of probabilities of subgroups of tetra-groups in the sequence: Homo sapiens chromosome 15, GRCh38.p7 Primary Assembly
NCBI Reference Sequence: NC_000015.10
LOCUS NC_000015 101991189 bp DNA linear CON 06-JUN-2016
DEFINITION Homo sapiens chromosome 15, GRCh38.p7 Primary Assembly.
ACCESSION NC_000015 GPC_000001307 VERSION NC_000015.10
https://www.ncbi.nlm.nih.gov/nuccore/NC_000015.10
https://www.ncbi.nlm.nih.gov/nuccore/NC_000015.10?report=fasta

| NUCLEOTIDES | | DOUBLETS | | TRIPLETS | | 4-PLETS | | 5-PLETS | |
|---|---|---|---|---|---|---|---|---|---|
| $P_1(A_1)=$ | 0,2758 | $P_2(A_1)=$ | 0,2757 | $P_3(A_1)=$ | 0,2758 | $P_4(A_1)=$ | 0,2757 | $P_5(A_1)=$ | 0,2757 |
| $P_1(T_1)=$ | 0,2784 | $P_2(T_1)=$ | 0,2785 | $P_3(T_1)=$ | 0,2784 | $P_4(T_1)=$ | 0,2785 | $P_5(T_1)=$ | 0,2784 |
| $P_1(C_1)=$ | 0,2221 | $P_2(C_1)=$ | 0,2222 | $P_3(C_1)=$ | 0,2221 | $P_4(C_1)=$ | 0,2222 | $P_5(C_1)=$ | 0,2221 |
| $P_1(G_1)=$ | 0,2237 | $P_2(G_1)=$ | 0,2236 | $P_3(G_1)=$ | 0,2237 | $P_4(G_1)=$ | 0,2236 | $P_5(G_1)=$ | 0,2237 |
| | | $P_2(A_2)=$ | 0,2758 | $P_3(A_2)=$ | 0,2757 | $P_4(A_2)=$ | 0,2758 | $P_5(A_2)=$ | 0,2758 |
| | | $P_2(T_2)=$ | 0,2783 | $P_3(T_2)=$ | 0,2784 | $P_4(T_2)=$ | 0,2781 | $P_5(T_2)=$ | 0,2783 |
| | | $P_2(C_2)=$ | 0,2221 | $P_3(C_2)=$ | 0,2222 | $P_4(C_2)=$ | 0,2222 | $P_5(C_2)=$ | 0,2222 |
| | | $P_2(G_2)=$ | 0,2238 | $P_3(G_2)=$ | 0,2237 | $P_4(G_2)=$ | 0,2238 | $P_5(G_2)=$ | 0,2237 |
| | | | | $P_3(A_3)=$ | 0,2758 | $P_4(A_3)=$ | 0,2756 | $P_5(A_3)=$ | 0,2760 |
| | | | | $P_3(T_3)=$ | 0,2784 | $P_4(T_3)=$ | 0,2785 | $P_5(T_3)=$ | 0,2784 |
| | | | | $P_3(C_3)=$ | 0,2222 | $P_4(C_3)=$ | 0,2222 | $P_5(C_3)=$ | 0,2221 |
| | | | | $P_3(G_3)=$ | 0,2237 | $P_4(G_3)=$ | 0,2237 | $P_5(G_3)=$ | 0,2236 |
| | | | | | | $P_4(A_4)=$ | 0,2759 | $P_5(A_4)=$ | 0,2756 |
| | | | | | | $P_4(T_4)=$ | 0,2784 | $P_5(T_4)=$ | 0,2783 |
| | | | | | | $P_4(C_4)=$ | 0,2220 | $P_5(C_4)=$ | 0,2222 |
| | | | | | | $P_4(G_4)=$ | 0,2237 | $P_5(G_4)=$ | 0,2239 |
| | | | | | | | | $P_5(A_5)=$ | 0,2757 |
| | | | | | | | | $P_5(T_5)=$ | 0,2785 |
| | | | | | | | | $P_5(C_5)=$ | 0,2222 |
| | | | | | | | | $P_5(G_5)=$ | 0,2236 |

Fig. A2/16. The table of probabilities of subgroups of tetra-groups in the sequence: Homo sapiens chromosome 16, GRCh38.p7 Primary Assembly
NCBI Reference Sequence: NC_000016.10
LOCUS NC_000016 90338345 bp DNA linear CON 06-JUN-2016
DEFINITION Homo sapiens chromosome 16, GRCh38.p7 Primary Assembly.
ACCESSION NC_000016 GPC_000001308 VERSION NC_000016.10
https://www.ncbi.nlm.nih.gov/nuccore/NC_000016.10
https://www.ncbi.nlm.nih.gov/nuccore/NC_000016.10?report=fasta

| NUCLEOTIDES | | DOUBLETS | | TRIPLETS | | 4-PLETS | | 5-PLETS | |
|---|---|---|---|---|---|---|---|---|---|
| $P_1(A_1)=$ | 0,2730 | $P_2(A_1)=$ | 0,2730 | $P_3(A_1)=$ | 0,2730 | $P_4(A_1)=$ | 0,2729 | $P_5(A_1)=$ | 0,2732 |
| $P_1(T_1)=$ | 0,2738 | $P_2(T_1)=$ | 0,2739 | $P_3(T_1)=$ | 0,2739 | $P_4(T_1)=$ | 0,2738 | $P_5(T_1)=$ | 0,2738 |
| $P_1(C_1)=$ | 0,2258 | $P_2(C_1)=$ | 0,2257 | $P_3(C_1)=$ | 0,2258 | $P_4(C_1)=$ | 0,2258 | $P_5(C_1)=$ | 0,2257 |
| $P_1(G_1)=$ | 0,2273 | $P_2(G_1)=$ | 0,2274 | $P_3(G_1)=$ | 0,2272 | $P_4(G_1)=$ | 0,2275 | $P_5(G_1)=$ | 0,2273 |
| | | $P_2(A_2)=$ | 0,2730 | $P_3(A_2)=$ | 0,2730 | $P_4(A_2)=$ | 0,2731 | $P_5(A_2)=$ | 0,2729 |
| | | $P_2(T_2)=$ | 0,2737 | $P_3(T_2)=$ | 0,2738 | $P_4(T_2)=$ | 0,2737 | $P_5(T_2)=$ | 0,2739 |
| | | $P_2(C_2)=$ | 0,2259 | $P_3(C_2)=$ | 0,2257 | $P_4(C_2)=$ | 0,2259 | $P_5(C_2)=$ | 0,2257 |
| | | $P_2(G_2)=$ | 0,2273 | $P_3(G_2)=$ | 0,2274 | $P_4(G_2)=$ | 0,2273 | $P_5(G_2)=$ | 0,2275 |
| | | | | $P_3(A_3)=$ | 0,2730 | $P_4(A_3)=$ | 0,2731 | $P_5(A_3)=$ | 0,2731 |
| | | | | $P_3(T_3)=$ | 0,2737 | $P_4(T_3)=$ | 0,2740 | $P_5(T_3)=$ | 0,2738 |
| | | | | $P_3(C_3)=$ | 0,2259 | $P_4(C_3)=$ | 0,2256 | $P_5(C_3)=$ | 0,2258 |
| | | | | $P_3(G_3)=$ | 0,2274 | $P_4(G_3)=$ | 0,2273 | $P_5(G_3)=$ | 0,2273 |
| | | | | | | $P_4(A_4)=$ | 0,2730 | $P_5(A_4)=$ | 0,2731 |
| | | | | | | $P_4(T_4)=$ | 0,2738 | $P_5(T_4)=$ | 0,2738 |
| | | | | | | $P_4(C_4)=$ | 0,2259 | $P_5(C_4)=$ | 0,2258 |
| | | | | | | $P_4(G_4)=$ | 0,2273 | $P_5(G_4)=$ | 0,2273 |
| | | | | | | | | $P_5(A_5)=$ | 0,2729 |
| | | | | | | | | $P_5(T_5)=$ | 0,2739 |
| | | | | | | | | $P_5(C_5)=$ | 0,2260 |
| | | | | | | | | $P_5(G_5)=$ | 0,2272 |

Fig. A2/17. The table of probabilities of subgroups of tetra-groups in the sequence: Homo sapiens chromosome 17, GRCh38.p7 Primary Assembly
NCBI Reference Sequence: NC_000017.11
LOCUS NC_000017 83257441 bp DNA linear CON 06-JUN-2016
DEFINITION Homo sapiens chromosome 17, GRCh38.p7 Primary Assembly.
ACCESSION NC_000017 GPC_000001309 VERSION NC_000017.11
https://www.ncbi.nlm.nih.gov/nuccore/NC_000017.11
https://www.ncbi.nlm.nih.gov/nuccore/NC_000017.11?report=fasta

| NUCLEOTIDES | | DOUBLETS | | TRIPLETS | | 4-PLETS | | 5-PLETS | |
|---|---|---|---|---|---|---|---|---|---|
| $P_1(A_1)=$ | 0,3011 | $P_2(A_1)=$ | 0,3012 | $P_3(A_1)=$ | 0,3011 | $P_4(A_1)=$ | 0,3012 | $P_5(A_1)=$ | 0,3009 |
| $P_1(T_1)=$ | 0,3014 | $P_2(T_1)=$ | 0,3013 | $P_3(T_1)=$ | 0,3014 | $P_4(T_1)=$ | 0,3014 | $P_5(T_1)=$ | 0,3017 |
| $P_1(C_1)=$ | 0,1987 | $P_2(C_1)=$ | 0,1985 | $P_3(C_1)=$ | 0,1987 | $P_4(C_1)=$ | 0,1986 | $P_5(C_1)=$ | 0,1986 |
| $P_1(G_1)=$ | 0,1989 | $P_2(G_1)=$ | 0,1989 | $P_3(G_1)=$ | 0,1988 | $P_4(G_1)=$ | 0,1988 | $P_5(G_1)=$ | 0,1988 |
| | | $P_2(A_2)=$ | 0,3009 | $P_3(A_2)=$ | 0,3011 | $P_4(A_2)=$ | 0,3009 | $P_5(A_2)=$ | 0,3012 |
| | | $P_2(T_2)=$ | 0,3014 | $P_3(T_2)=$ | 0,3013 | $P_4(T_2)=$ | 0,3014 | $P_5(T_2)=$ | 0,3013 |
| | | $P_2(C_2)=$ | 0,1988 | $P_3(C_2)=$ | 0,1987 | $P_4(C_2)=$ | 0,1987 | $P_5(C_2)=$ | 0,1986 |
| | | $P_2(G_2)=$ | 0,1989 | $P_3(G_2)=$ | 0,1989 | $P_4(G_2)=$ | 0,1990 | $P_5(G_2)=$ | 0,1989 |
| | | | | $P_3(A_3)=$ | 0,3010 | $P_4(A_3)=$ | 0,3012 | $P_5(A_3)=$ | 0,3014 |
| | | | | $P_3(T_3)=$ | 0,3014 | $P_4(T_3)=$ | 0,3012 | $P_5(T_3)=$ | 0,3011 |
| | | | | $P_3(C_3)=$ | 0,1986 | $P_4(C_3)=$ | 0,1985 | $P_5(C_3)=$ | 0,1988 |
| | | | | $P_3(G_3)=$ | 0,1990 | $P_4(G_3)=$ | 0,1990 | $P_5(G_3)=$ | 0,1988 |
| | | | | | | $P_4(A_4)=$ | 0,3009 | $P_5(A_4)=$ | 0,3010 |
| | | | | | | $P_4(T_4)=$ | 0,3013 | $P_5(T_4)=$ | 0,3014 |
| | | | | | | $P_4(C_4)=$ | 0,1988 | $P_5(C_4)=$ | 0,1986 |
| | | | | | | $P_4(G_4)=$ | 0,1989 | $P_5(G_4)=$ | 0,1990 |
| | | | | | | | | $P_5(A_5)=$ | 0,3008 |
| | | | | | | | | $P_5(T_5)=$ | 0,3014 |
| | | | | | | | | $P_5(C_5)=$ | 0,1988 |
| | | | | | | | | $P_5(G_5)=$ | 0,1991 |

Fig. A2/18. The table of probabilities of subgroups of tetra-groups in the sequence: Homo sapiens chromosome 18, whole genome shotgun sequence
GenBank: CM000269.1
LOCUS CM000269 74792881 bp DNA linear CON 23-MAR-2015
DEFINITION Homo sapiens chromosome 18, whole genome shotgun sequence.
ACCESSION CM000269 AADB02000000 CH003489 VERSION CM000269.1
https://www.ncbi.nlm.nih.gov/nuccore/CM000269.1
https://www.ncbi.nlm.nih.gov/nuccore/CM000269.1?report=fasta

| NUCLEOTIDES | | DOUBLETS | | TRIPLETS | | 4-PLETS | | 5-PLETS | |
|---|---|---|---|---|---|---|---|---|---|
| $P_1(A_1)=$ | 0,2591 | $P_2(A_1)=$ | 0,2592 | $P_3(A_1)=$ | 0,2592 | $P_4(A_1)=$ | 0,2591 | $P_5(A_1)=$ | 0,2592 |
| $P_1(T_1)=$ | 0,2615 | $P_2(T_1)=$ | 0,2615 | $P_3(T_1)=$ | 0,2616 | $P_4(T_1)=$ | 0,2615 | $P_5(T_1)=$ | 0,2615 |
| $P_1(C_1)=$ | 0,2388 | $P_2(C_1)=$ | 0,2388 | $P_3(C_1)=$ | 0,2387 | $P_4(C_1)=$ | 0,2388 | $P_5(C_1)=$ | 0,2386 |
| $P_1(G_1)=$ | 0,2406 | $P_2(G_1)=$ | 0,2405 | $P_3(G_1)=$ | 0,2405 | $P_4(G_1)=$ | 0,2405 | $P_5(G_1)=$ | 0,2407 |
| | | $P_2(A_2)=$ | 0,2590 | $P_3(A_2)=$ | 0,2591 | $P_4(A_2)=$ | 0,2589 | $P_5(A_2)=$ | 0,2591 |
| | | $P_2(T_2)=$ | 0,2615 | $P_3(T_2)=$ | 0,2616 | $P_4(T_2)=$ | 0,2616 | $P_5(T_2)=$ | 0,2615 |
| | | $P_2(C_2)=$ | 0,2388 | $P_3(C_2)=$ | 0,2388 | $P_4(C_2)=$ | 0,2389 | $P_5(C_2)=$ | 0,2387 |
| | | $P_2(G_2)=$ | 0,2407 | $P_3(G_2)=$ | 0,2405 | $P_4(G_2)=$ | 0,2407 | $P_5(G_2)=$ | 0,2408 |
| | | | | $P_3(A_3)=$ | 0,2591 | $P_4(A_3)=$ | 0,2593 | $P_5(A_3)=$ | 0,2590 |
| | | | | $P_3(T_3)=$ | 0,2613 | $P_4(T_3)=$ | 0,2615 | $P_5(T_3)=$ | 0,2615 |
| | | | | $P_3(C_3)$= | 0,2388 | $P_4(C_3)$= | 0,2387 | $P_5(C_3)$= | 0,2390 |
| | | | | $P_3(G_3)$= | 0,2408 | $P_4(G_3)$= | 0,2405 | $P_5(G_3)$= | 0,2405 |
| | | | | | | $P_4(A_4)$= | 0,2591 | $P_5(A_4)$= | 0,2592 |
| | | | | | | $P_4(T_4)$= | 0,2614 | $P_5(T_4)$= | 0,2615 |
| | | | | | | $P_4(C_4)$= | 0,2387 | $P_5(C_4)$= | 0,2388 |
| | | | | | | $P_4(G_4)$= | 0,2408 | $P_5(G_4)$= | 0,2404 |
| | | | | | | | | $P_5(A_5)$= | 0,2591 |
| | | | | | | | | $P_5(T_5)$= | 0,2615 |
| | | | | | | | | $P_5(C_5)$= | 0,2388 |
| | | | | | | | | $P_5(G_5)$= | 0,2406 |

Fig. A2/19. The table of probabilities of subgroups of tetra-groups in the sequence: Homo sapiens chromosome 19, GRCh38.p7 Primary Assembly
NCBI Reference Sequence: NC_000019.10
LOCUS NC_000019 58617616 bp DNA linear CON 06-JUN-2016
DEFINITION Homo sapiens chromosome 19, GRCh38.p7 Primary Assembly.
ACCESSION NC_000019 GPC_000001311 VERSION NC_000019.10
https://www.ncbi.nlm.nih.gov/nuccore/NC_000019.10
https://www.ncbi.nlm.nih.gov/nuccore/NC_000019.10?report=fasta

| NUCLEOTIDES | | DOUBLETS | | TRIPLETS | | 4-PLETS | | 5-PLETS | |
|---|---|---|---|---|---|---|---|---|---|
| $P_1(A_1)$= | 0,2778 | $P_2(A_1)$= | 0,2779 | $P_3(A_1)$= | 0,2778 | $P_4(A_1)$= | 0,2779 | $P_5(A_1)$= | 0,2777 |
| $P_1(T_1)$= | 0,2810 | $P_2(T_1)$= | 0,2808 | $P_3(T_1)$= | 0,2810 | $P_4(T_1)$= | 0,2809 | $P_5(T_1)$= | 0,2811 |
| $P_1(C_1)$= | 0,2201 | $P_2(C_1)$= | 0,2202 | $P_3(C_1)$= | 0,2201 | $P_4(C_1)$= | 0,2202 | $P_5(C_1)$= | 0,2200 |
| $P_1(G_1)$= | 0,2211 | $P_2(G_1)$= | 0,2212 | $P_3(G_1)$= | 0,2211 | $P_4(G_1)$= | 0,2211 | $P_5(G_1)$= | 0,2212 |
| | | $P_2(A_2)$= | 0,2777 | $P_3(A_2)$= | 0,2778 | $P_4(A_2)$= | 0,2776 | $P_5(A_2)$= | 0,2780 |
| | | $P_2(T_2)$= | 0,2811 | $P_3(T_2)$= | 0,2810 | $P_4(T_2)$= | 0,2813 | $P_5(T_2)$= | 0,2806 |
| | | $P_2(C_2)$= | 0,2200 | $P_3(C_2)$= | 0,2201 | $P_4(C_2)$= | 0,2200 | $P_5(C_2)$= | 0,2201 |
| | | $P_2(G_2)$= | 0,2211 | $P_3(G_2)$= | 0,2212 | $P_4(G_2)$= | 0,2211 | $P_5(G_2)$= | 0,2212 |
| | | | | $P_3(A_3)$= | 0,2779 | $P_4(A_3)$= | 0,2779 | $P_5(A_3)$= | 0,2777 |
| | | | | $P_3(T_3)$= | 0,2809 | $P_4(T_3)$= | 0,2807 | $P_5(T_3)$= | 0,2811 |
| | | | | $P_3(C_3)$= | 0,2201 | $P_4(C_3)$= | 0,2202 | $P_5(C_3)$= | 0,2203 |
| | | | | $P_3(G_3)$= | 0,2211 | $P_4(G_3)$= | 0,2212 | $P_5(G_3)$= | 0,2209 |
| | | | | | | $P_4(A_4)$= | 0,2778 | $P_5(A_4)$= | 0,2778 |
| | | | | | | $P_4(T_4)$= | 0,2810 | $P_5(T_4)$= | 0,2810 |
| | | | | | | $P_4(C_4)$= | 0,2201 | $P_5(C_4)$= | 0,2199 |
| | | | | | | $P_4(G_4)$= | 0,2211 | $P_5(G_4)$= | 0,2213 |
| | | | | | | | | $P_5(A_5)$= | 0,2779 |
| | | | | | | | | $P_5(T_5)$= | 0,2810 |
| | | | | | | | | $P_5(C_5)$= | 0,2201 |
| | | | | | | | | $P_5(G_5)$= | 0,2211 |

Fig. A2/20. The table of probabilities of subgroups of tetra-groups in the sequence: Homo sapiens chromosome 20, whole genome shotgun sequence
GenBank: CM000271.1
LOCUS CM000271 59605541 bp DNA linear CON 23-MAR-2015
DEFINITION Homo sapiens chromosome 20, whole genome shotgun sequence.
ACCESSION CM000271 AADB02000000 CH003491 VERSION CM000271.1
https://www.ncbi.nlm.nih.gov/nuccore/CM000271.1
https://www.ncbi.nlm.nih.gov/nuccore/CM000271.1?report=fasta

| NUCLEOTIDES | | DOUBLETS | | TRIPLETS | | 4-PLETS | | 5-PLETS | |
|---|---|---|---|---|---|---|---|---|---|
| $P_1(A_1)$= | 0,2964 | $P_2(A_1)$= | 0,2964 | $P_3(A_1)$= | 0,2963 | $P_4(A_1)$= | 0,2966 | $P_5(A_1)$= | 0,2965 |
| $P_1(T_1)$= | 0,2946 | $P_2(T_1)$= | 0,2947 | $P_3(T_1)$= | 0,2946 | $P_4(T_1)$= | 0,2946 | $P_5(T_1)$= | 0,2943 |
| $P_1(C_1)=$ | 0,2047 | $P_2(C_1)=$ | 0,2046 | $P_3(C_1)=$ | 0,2046 | $P_4(C_1)=$ | 0,2046 | $P_5(C_1)=$ | 0,2045 |
| $P_1(G_1)=$ | 0,2043 | $P_2(G_1)=$ | 0,2043 | $P_3(G_1)=$ | 0,2045 | $P_4(G_1)=$ | 0,2042 | $P_5(G_1)=$ | 0,2046 |
| | | $P_2(A_2)=$ | 0,2965 | $P_3(A_2)=$ | 0,2964 | $P_4(A_2)=$ | 0,2967 | $P_5(A_2)=$ | 0,2966 |
| | | $P_2(T_2)=$ | 0,2945 | $P_3(T_2)=$ | 0,2946 | $P_4(T_2)=$ | 0,2945 | $P_5(T_2)=$ | 0,2948 |
| | | $P_2(C_2)=$ | 0,2047 | $P_3(C_2)=$ | 0,2049 | $P_4(C_2)=$ | 0,2046 | $P_5(C_2)=$ | 0,2047 |
| | | $P_2(G_2)=$ | 0,2043 | $P_3(G_2)=$ | 0,2041 | $P_4(G_2)=$ | 0,2042 | $P_5(G_2)=$ | 0,2040 |
| | | | | $P_3(A_3)=$ | 0,2966 | $P_4(A_3)=$ | 0,2962 | $P_5(A_3)=$ | 0,2964 |
| | | | | $P_3(T_3)=$ | 0,2947 | $P_4(T_3)=$ | 0,2949 | $P_5(T_3)=$ | 0,2947 |
| | | | | $P_3(C_3)=$ | 0,2044 | $P_4(C_3)=$ | 0,2045 | $P_5(C_3)=$ | 0,2047 |
| | | | | $P_3(G_3)=$ | 0,2043 | $P_4(G_3)=$ | 0,2044 | $P_5(G_3)=$ | 0,2042 |
| | | | | | | $P_4(A_4)=$ | 0,2963 | $P_5(A_4)=$ | 0,2964 |
| | | | | | | $P_4(T_4)=$ | 0,2945 | $P_5(T_4)=$ | 0,2946 |
| | | | | | | $P_4(C_4)=$ | 0,2049 | $P_5(C_4)=$ | 0,2048 |
| | | | | | | $P_4(G_4)=$ | 0,2043 | $P_5(G_4)=$ | 0,2042 |
| | | | | | | | | $P_5(A_5)=$ | 0,2964 |
| | | | | | | | | $P_5(T_5)=$ | 0,2947 |
| | | | | | | | | $P_5(C_5)=$ | 0,2045 |
| | | | | | | | | $P_5(G_5)=$ | 0,2044 |

Fig. A2/21. The table of probabilities of subgroups of tetra-groups in the sequence: Homo sapiens genomic DNA, chromosome 21q
LOCUS BA000005 33543332 bp DNA linear CON 12-JUL-2008
DEFINITION Homo sapiens genomic DNA,
chromosome 21q. ACCESSION BA000005 VERSION BA000005.3
https://www.ncbi.nlm.nih.gov/nuccore/BA000005
https://www.ncbi.nlm.nih.gov/nuccore/BA000005.3?report=fasta

| **NUCLEOTIDES** | | **DOUBLETS** | | **TRIPLETS** | | **4-PLETS** | | **5-PLETS** | |
|---|---|---|---|---|---|---|---|---|---|
| $P_1(A_1)=$ | 0,2651 | $P_2(A_1)=$ | 0,2651 | $P_3(A_1)=$ | 0,2651 | $P_4(A_1)=$ | 0,2651 | $P_5(A_1)=$ | 0,2652 |
| $P_1(T_1)=$ | 0,2648 | $P_2(T_1)=$ | 0,2648 | $P_3(T_1)=$ | 0,2648 | $P_4(T_1)=$ | 0,2649 | $P_5(T_1)=$ | 0,2647 |
| $P_1(C_1)=$ | 0,2339 | $P_2(C_1)=$ | 0,2339 | $P_3(C_1)=$ | 0,2340 | $P_4(C_1)=$ | 0,2339 | $P_5(C_1)=$ | 0,2340 |
| $P_1(G_1)=$ | 0,2361 | $P_2(G_1)=$ | 0,2362 | $P_3(G_1)=$ | 0,2362 | $P_4(G_1)=$ | 0,2361 | $P_5(G_1)=$ | 0,2361 |
| | | $P_2(A_2)=$ | 0,2651 | $P_3(A_2)=$ | 0,2652 | $P_4(A_2)=$ | 0,2651 | $P_5(A_2)=$ | 0,2651 |
| | | $P_2(T_2)=$ | 0,2648 | $P_3(T_2)=$ | 0,2647 | $P_4(T_2)=$ | 0,2647 | $P_5(T_2)=$ | 0,2645 |
| | | $P_2(C_2)=$ | 0,2340 | $P_3(C_2)=$ | 0,2341 | $P_4(C_2)=$ | 0,2339 | $P_5(C_2)=$ | 0,2340 |
| | | $P_2(G_2)=$ | 0,2360 | $P_3(G_2)=$ | 0,2360 | $P_4(G_2)=$ | 0,2362 | $P_5(G_2)=$ | 0,2364 |
| | | | | $P_3(A_3)=$ | 0,2651 | $P_4(A_3)=$ | 0,2651 | $P_5(A_3)=$ | 0,2652 |
| | | | | $P_3(T_3)=$ | 0,2650 | $P_4(T_3)=$ | 0,2648 | $P_5(T_3)=$ | 0,2650 |
| | | | | $P_3(C_3)=$ | 0,2337 | $P_4(C_3)=$ | 0,2338 | $P_5(C_3)=$ | 0,2339 |
| | | | | $P_3(G_3)=$ | 0,2362 | $P_4(G_3)=$ | 0,2363 | $P_5(G_3)=$ | 0,2359 |
| | | | | | | $P_4(A_4)=$ | 0,2652 | $P_5(A_4)=$ | 0,2650 |
| | | | | | | $P_4(T_4)=$ | 0,2649 | $P_5(T_4)=$ | 0,2649 |
| | | | | | | $P_4(C_4)=$ | 0,2340 | $P_5(C_4)=$ | 0,2340 |
| | | | | | | $P_4(G_4)=$ | 0,2358 | $P_5(G_4)=$ | 0,2362 |
| | | | | | | | | $P_5(A_5)=$ | 0,2652 |
| | | | | | | | | $P_5(T_5)=$ | 0,2651 |
| | | | | | | | | $P_5(C_5)=$ | 0,2337 |
| | | | | | | | | $P_5(G_5)=$ | 0,2360 |

Fig. A2/22. The table of probabilities of subgroups of tetra-groups in the sequence: Homo sapiens chromosome 22, GRCh38.p7 Primary Assembly
NCBI Reference Sequence: NC_000022.11
LOCUS NC_000022 50818468 bp DNA linear CON 06-JUN-2016
DEFINITION Homo sapiens chromosome 22, GRCh38.p7 Primary Assembly.
ACCESSION NC_000022 GPC_000001314 VERSION NC_000022.11
https://www.ncbi.nlm.nih.gov/nuccore/NC_000022.11

https://www.ncbi.nlm.nih.gov/nuccore/NC_000022.11?report=fasta

| NUCLEOTIDES | | DOUBLETS | | TRIPLETS | | 4-PLETS | | 5-PLETS | |
|---|---|---|---|---|---|---|---|---|---|
| $P_1(A_1)$= | 0,3019 | $P_2(A_1)$= | 0,3019 | $P_3(A_1)$= | 0,3019 | $P_4(A_1)$= | 0,3019 | $P_5(A_1)$= | 0,3017 |
| $P_1(T_1)$= | 0,3029 | $P_2(T_1)$= | 0,3029 | $P_3(T_1)$= | 0,3029 | $P_4(T_1)$= | 0,3029 | $P_5(T_1)$= | 0,3030 |
| $P_1(C_1)$= | 0,1971 | $P_2(C_1)$= | 0,1970 | $P_3(C_1)$= | 0,1971 | $P_4(C_1)$= | 0,1970 | $P_5(C_1)$= | 0,1971 |
| $P_1(G_1)$= | 0,1982 | $P_2(G_1)$= | 0,1982 | $P_3(G_1)$= | 0,1982 | $P_4(G_1)$= | 0,1982 | $P_5(G_1)$= | 0,1982 |
| | | $P_2(A_2)$= | 0,3018 | $P_3(A_2)$= | 0,3018 | $P_4(A_2)$= | 0,3019 | $P_5(A_2)$= | 0,3019 |
| | | $P_2(T_2)$= | 0,3029 | $P_3(T_2)$= | 0,3029 | $P_4(T_2)$= | 0,3029 | $P_5(T_2)$= | 0,3029 |
| | | $P_2(C_2)$= | 0,1971 | $P_3(C_2)$= | 0,19711 | $P_4(C_2)$= | 0,1970 | $P_5(C_2)$= | 0,1970 |
| | | $P_2(G_2)$= | 0,1982 | $P_3(G_2)$= | 0,1982 | $P_4(G_2)$= | 0,1982 | $P_5(G_2)$= | 0,1982 |
| | | | | $P_3(A_3)$= | 0,3019 | $P_4(A_3)$= | 0,3018 | $P_5(A_3)$= | 0,3019 |
| | | | | $P_3(T_3)$= | 0,3029 | $P_4(T_3)$= | 0,3029 | $P_5(T_3)$= | 0,3029 |
| | | | | $P_3(C_3)$= | 0,1970 | $P_4(C_3)$= | 0,1971 | $P_5(C_3)$= | 0,1971 |
| | | | | $P_3(G_3)$= | 0,1982 | $P_4(G_3)$= | 0,1981 | $P_5(G_3)$= | 0,1982 |
| | | | | | | $P_4(A_4)$= | 0,3017 | $P_5(A_4)$= | 0,3019 |
| | | | | | | $P_4(T_4)$= | 0,3029 | $P_5(T_4)$= | 0,3028 |
| | | | | | | $P_4(C_4)$= | 0,1972 | $P_5(C_4)$= | 0,1971 |
| | | | | | | $P_4(G_4)$= | 0,1982 | $P_5(G_4)$= | 0,1982 |
| | | | | | | | | $P_5(A_5)$= | 0,3019 |
| | | | | | | | | $P_5(T_5)$= | 0,3029 |
| | | | | | | | | $P_5(C_5)$= | 0,1971 |
| | | | | | | | | $P_5(G_5)$= | 0,1982 |

Fig. A2/23. The table of probabilities of subgroups of tetra-groups in the sequence: Homo sapiens chromosome X, GRCh38.p7 Primary Assembly
NCBI Reference Sequence: NC_000023.11
LOCUS NC_000023 156040895 bp DNA linear CON 06-JUN-2016
DEFINITION Homo sapiens chromosome X, GRCh38.p7 Primary Assembly.
ACCESSION NC_000023 GPC_000001315
VERSION NC_000023.11
https://www.ncbi.nlm.nih.gov/nuccore/NC_000023.11
https://www.ncbi.nlm.nih.gov/nuccore/NC_000023.11?report=fasta

| NUCLEOTIDES | | DOUBLETS | | TRIPLETS | | 4-PLETS | | 5-PLETS | |
|---|---|---|---|---|---|---|---|---|---|
| $P_1(A_1)$= | 0,2985 | $P_2(A_1)$= | 0,2986 | $P_3(A_1)$= | 0,2985 | $P_4(A_1)$= | 0,2987 | $P_5(A_1)$= | 0,2983 |
| $P_1(T_1)$= | 0,3012 | $P_2(T_1)$= | 0,3013 | $P_3(T_1)$= | 0,3013 | $P_4(T_1)$= | 0,3014 | $P_5(T_1)$= | 0,3009 |
| $P_1(C_1)$= | 0,2001 | $P_2(C_1)$= | 0,2001 | $P_3(C_1)$= | 0,2001 | $P_4(C_1)$= | 0,2000 | $P_5(C_1)$= | 0,2004 |
| $P_1(G_1)$= | 0,2001 | $P_2(G_1)$= | 0,2000 | $P_3(G_1)$= | 0,2001 | $P_4(G_1)$= | 0,2000 | $P_5(G_1)$= | 0,2004 |
| | | $P_2(A_2)$= | 0,2985 | $P_3(A_2)$= | 0,2987 | $P_4(A_2)$= | 0,2984 | $P_5(A_2)$= | 0,2986 |
| | | $P_2(T_2)$= | 0,3011 | $P_3(T_2)$= | 0,3011 | $P_4(T_2)$= | 0,3011 | $P_5(T_2)$= | 0,3009 |
| | | $P_2(C_2)$= | 0,2001 | $P_3(C_2)$= | 0,2001 | $P_4(C_2)$= | 0,2002 | $P_5(C_2)$= | 0,2002 |
| | | $P_2(G_2)$= | 0,2003 | $P_3(G_2)$= | 0,2001 | $P_4(G_2)$= | 0,2003 | $P_5(G_2)$= | 0,2002 |
| | | | | $P_3(A_3)$= | 0,2984 | $P_4(A_3)$= | 0,2985 | $P_5(A_3)$= | 0,2988 |
| | | | | $P_3(T_3)$= | 0,3012 | $P_4(T_3)$= | 0,3012 | $P_5(T_3)$= | 0,3013 |
| | | | | $P_3(C_3)$= | 0,2001 | $P_4(C_3)$= | 0,2002 | $P_5(C_3)$= | 0,2002 |
| | | | | $P_3(G_3)$= | 0,2003 | $P_4(G_3)$= | 0,2000 | $P_5(G_3)$= | 0,1998 |
| | | | | | | $P_4(A_4)$= | 0,2986 | $P_5(A_4)$= | 0,2984 |
| | | | | | | $P_4(T_4)$= | 0,3012 | $P_5(T_4)$= | 0,3016 |
| | | | | | | $P_4(C_4)$= | 0,2000 | $P_5(C_4)$= | 0,2000 |
| | | | | | | $P_4(G_4)$= | 0,2003 | $P_5(G_4)$= | 0,2000 |
| | | | | | | | | $P_5(A_5)$= | 0,2987 |
| | | | | | | | | $P_5(T_5)$= | 0,3012 |
| | | | | | | | | $P_5(C_5)$= | 0,1998 |

| $P_5(G_5)=$ 0,2003 |
|---|

Fig. A2/24. The table of probabilities of subgroups of tetra-groups in the sequence: Homo sapiens chromosome Y, GRCh38.p7 Primary Assembly
NCBI Reference Sequence: NC_000024.10
LOCUS NC_000024 57227415 bp DNA linear CON 06-JUN-2016
DEFINITION Homo sapiens chromosome Y, GRCh38.p7 Primary Assembly.
ACCESSION NC_000024 GPC_000001316 VERSION NC_000024.10
https://www.ncbi.nlm.nih.gov/nuccore/NC_000024.10
https://www.ncbi.nlm.nih.gov/nuccore/NC_000024.10?report=fasta

## Appendix 3. Symmetries of tetra-group probabilities in the complete set of chromosomes of a nematode *Caenorhabditis elegans*

The appendix 3 represents data about the fulfillment of the described tetra-group rules in the complete set of chromosomes of *Caenorhabditis elegans*. This free-living soil nematode is widely used as a model organism in genetics for a long time. The *Caenorhabditis elegans* nuclear genome is approximately 100 Mb, distributed among six chromosomes. All initial data are taken from the CenBank (https://www.ncbi.nlm.nih.gov/genome?term=caenorhabditis%20elegans).

| NUCLEOTIDES | DOUBLETS | TRIPLETS | 4-PLETS | 5-PLETS |
|---|---|---|---|---|
| $P_1(A_1)=$ 0.3208 | $P_2(A_1)=$ 0.3207 | $P_3(A_1)=$ 0.3205 | $P_4(A_1)=$ 0.3207 | $P_5(A_1)=$ 0.3205 |
| $P_1(T_1)=$ 0.3217 | $P_2(T_1)=$ 0.3218 | $P_3(T_1)=$ 0.3216 | $P_4(T_1)=$ 0.3219 | $P_5(T_1)=$ 0.3219 |
| $P_1(C_1)=$ 0.1789 | $P_2(C_1)=$ 0.1789 | $P_3(C_1)=$ 0.1788 | $P_4(C_1)=$ 0.1789 | $P_5(C_1)=$ 0.1789 |
| $P_1(G_1)=$ 0.1786 | $P_2(G_1)=$ 0.1787 | $P_3(G_1)=$ 0.1791 | $P_4(G_1)=$ 0.1784 | $P_5(G_1)=$ 0.1787 |
| | $P_2(A_2)=$ 0.321 | $P_3(A_2)=$ 0.3212 | $P_4(A_2)=$ 0.321 | $P_5(A_2)=$ 0.321 |
| | $P_2(T_2)=$ 0.3216 | $P_3(T_2)=$ 0.3216 | $P_4(T_2)=$ 0.3218 | $P_5(T_2)=$ 0.3215 |
| | $P_2(C_2)=$ 0.1788 | $P_3(C_2)=$ 0.1793 | $P_4(C_2)=$ 0.1785 | $P_5(C_2)=$ 0.179 |
| | $P_2(G_2)=$ 0.1786 | $P_3(G_2)=$ 0.1779 | $P_4(G_2)=$ 0.1788 | $P_5(G_2)=$ 0.1785 |
| | | $P_3(A_3)=$ 0.3209 | $P_4(A_3)=$ 0.3206 | $P_5(A_3)=$ 0.3208 |
| | | $P_3(T_3)=$ 0.3218 | $P_4(T_3)=$ 0.3217 | $P_5(T_3)=$ 0.3218 |
| | | $P_3(C_3)=$ 0.1784 | $P_4(C_3)=$ 0.1788 | $P_5(C_3)=$ 0.1789 |
| | | $P_3(G_3)=$ 0.1789 | $P_4(G_3)=$ 0.1789 | $P_5(G_3)=$ 0.1785 |
| | | | $P_4(A_4)=$ 0.3211 | $P_5(A_4)=$ 0.321 |
| | | | $P_4(T_4)=$ 0.3213 | $P_5(T_4)=$ 0.3218 |
| | | | $P_4(C_4)=$ 0.1792 | $P_5(C_4)=$ 0.1786 |
| | | | $P_4(G_4)=$ 0.1784 | $P_5(G_4)=$ 0.1787 |
| | | | | $P_5(A_5)=$ 0.3211 |
| | | | | $P_5(T_5)=$ 0.3213 |
| | | | | $P_5(C_5)=$ 0.179 |
| | | | | $P_5(G_5)=$ 0.1787 |

Fig. A3/1. The table of probabilities of subgroups of tetra-groups in the sequence: Caenorhabditis elegans chromosome I, NCBI Reference Sequence: NC_003279.8

LOCUS NC_003279 15072434 bp DNA linear CON 11-OCT-2017

DEFINITION Caenorhabditis elegans chromosome I. ACCESSION NC_003279

VERSION NC_003279.8

https://www.ncbi.nlm.nih.gov/nuccore/NC_003279.8

| NUCLEOTIDES | DOUBLETS | TRIPLETS | 4-PLETS | 5-PLETS |
|---|---|---|---|---|
| $P_1(A_1)=$ 0.3193 | $P_2(A_1)=$ 0.3193 | $P_3(A_1)=$ 0.319 | $P_4(A_1)=$ 0.3193 | $P_5(A_1)=$ 0.3191 |
| $P_1(T_1)=$ 0.3187 | $P_2(T_1)=$ 0.3187 | $P_3(T_1)=$ 0.3189 | $P_4(T_1)=$ 0.3187 | $P_5(T_1)=$ 0.3189 |
| $P_1(C_1)=$ 0.1812 | $P_2(C_1)=$ 0.1813 | $P_3(C_1)=$ 0.1814 | $P_4(C_1)=$ 0.1813 | $P_5(C_1)=$ 0.1814 |
| $P_1(G_1)=$ 0.1808 | $P_2(G_1)=$ 0.1807 | $P_3(G_1)=$ 0.1807 | $P_4(G_1)=$ 0.1806 | $P_5(G_1)=$ 0.1806 |
| | $P_2(A_2)=$ 0.3193 | $P_3(A_2)=$ 0.3192 | $P_4(A_2)=$ 0.3193 | $P_5(A_2)=$ 0.3193 |
| | $P_2(T_2)=$ 0.3187 | $P_3(T_2)=$ 0.318 | $P_4(T_2)=$ 0.3187 | $P_5(T_2)=$ 0.319 |
| | $P_2(C_2)=$ 0.1812 | $P_3(C_2)=$ 0.1814 | $P_4(C_2)=$ 0.181 | $P_5(C_2)=$ 0.1809 |
| | $P_2(G_2)=$ 0.1809 | $P_3(G_2)=$ 0.1814 | $P_4(G_2)=$ 0.181 | $P_5(G_2)=$ 0.1809 |
| | | $P_3(A_3)=$ 0.3196 | $P_4(A_3)=$ 0.3192 | $P_5(A_3)=$ 0.3193 |
| | | $P_3(T_3)=$ 0.3192 | $P_4(T_3)=$ 0.3187 | $P_5(T_3)=$ 0.3183 |
| | | $P_3(C_3)=$ 0.1809 | $P_4(C_3)=$ 0.1813 | $P_5(C_3)=$ 0.1813 |
| | | $P_3(G_3)=$ 0.1803 | $P_4(G_3)=$ 0.1808 | $P_5(G_3)=$ 0.181 |
| | | | $P_4(A_4)=$ 0.3192 | $P_5(A_4)=$ 0.3194 |
| | | | $P_4(T_4)=$ 0.3187 | $P_5(T_4)=$ 0.3185 |
| | | | $P_4(C_4)=$ 0.1813 | $P_5(C_4)=$ 0.1813 |
| | | | $P_4(G_4)=$ 0.1807 | $P_5(G_4)=$ 0.1808 |
| | | | | $P_5(A_5)=$ 0.3193 |
| | | | | $P_5(T_5)=$ 0.3189 |
| | | | | $P_5(C_5)=$ 0.1813 |
| | | | | $P_5(G_5)=$ 0.1805 |

Fig. A3/2. The table of probabilities of subgroups of tetra-groups in the sequence: Caenorhabditis elegans chromosome II,

NCBI Reference Sequence: NC_003280.10

LOCUS NC_003280 15279421 bp DNA linear CON 11-OCT-2017
DEFINITION Caenorhabditis elegans chromosome II. ACCESSION NC_003280
VERSION NC_003280.10
https://www.ncbi.nlm.nih.gov/nuccore/NC_003280.10

| NUCLEOTIDES | DOUBLETS | TRIPLETS | 4-PLETS | 5-PLETS |
|---|---|---|---|---|
| $P_1(A_1)$= 0.3225 | $P_2(A_1)$= 0.3226 | $P_3(A_1)$= 0.3223 | $P_4(A_1)$= 0.3225 | $P_5(A_1)$= 0.3226 |
| $P_1(T_1)$= 0.3209 | $P_2(T_1)$= 0.3209 | $P_3(T_1)$= 0.3206 | $P_4(T_1)$= 0.3209 | $P_5(T_1)$= 0.3211 |
| $P_1(C_1)$= 0.1777 | $P_2(C_1)$= 0.1778 | $P_3(C_1)$= 0.178 | $P_4(C_1)$= 0.1778 | $P_5(C_1)$= 0.1777 |
| $P_1(G_1)$= 0.1789 | $P_2(G_1)$= 0.1787 | $P_3(G_1)$= 0.1791 | $P_4(G_1)$= 0.1788 | $P_5(G_1)$= 0.1787 |
| | $P_2(A_2)$= 0.3223 | $P_3(A_2)$= 0.3227 | $P_4(A_2)$= 0.3223 | $P_5(A_2)$= 0.3223 |
| | $P_2(T_2)$= 0.3209 | $P_3(T_2)$= 0.3213 | $P_4(T_2)$= 0.3209 | $P_5(T_2)$= 0.321 |
| | $P_2(C_2)$= 0.1776 | $P_3(C_2)$= 0.1776 | $P_4(C_2)$= 0.1776 | $P_5(C_2)$= 0.178 |
| | $P_2(G_2)$= 0.1792 | $P_3(G_2)$= 0.1784 | $P_4(G_2)$= 0.1792 | $P_5(G_2)$= 0.1788 |
| | | $P_3(A_3)$= 0.3223 | $P_4(A_3)$= 0.3227 | $P_5(A_3)$= 0.3222 |
| | | $P_3(T_3)$= 0.3209 | $P_4(T_3)$= 0.321 | $P_5(T_3)$= 0.3208 |
| | | $P_3(C_3)$= 0.1775 | $P_4(C_3)$= 0.1778 | $P_5(C_3)$= 0.1779 |
| | | $P_3(G_3)$= 0.1792 | $P_4(G_3)$= 0.1785 | $P_5(G_3)$= 0.179 |
| | | | $P_4(A_4)$= 0.3224 | $P_5(A_4)$= 0.3225 |
| | | | $P_4(T_4)$= 0.321 | $P_5(T_4)$= 0.3209 |
| | | | $P_4(C_4)$= 0.1775 | $P_5(C_4)$= 0.1775 |
| | | | $P_4(G_4)$= 0.1791 | $P_5(G_4)$= 0.179 |
| | | | | $P_5(A_5)$= 0.3227 |
| | | | | $P_5(T_5)$= 0.3208 |
| | | | | $P_5(C_5)$= 0.1773 |
| | | | | $P_5(G_5)$= 0.1792 |

Fig. A3/3. The table of probabilities of subgroups of tetra-groups in the sequence: Caenorhabditis elegans chromosome III,

NCBI Reference Sequence: NC_003281.10

LOCUS NC_003281 13783801 bp DNA linear CON 11-OCT-2017
DEFINITION Caenorhabditis elegans chromosome III. ACCESSION NC_003281
VERSION NC_003281.10
https://www.ncbi.nlm.nih.gov/nuccore/NC_003281.10

| NUCLEOTIDES | DOUBLETS | TRIPLETS | 4-PLETS | 5-PLETS |
|---|---|---|---|---|
| $P_1(A_1)$= 0.3265 | $P_2(A_1)$= 0.3263 | $P_3(A_1)$= 0.3264 | $P_4(A_1)$= 0.3265 | $P_5(A_1)$= 0.3263 |
| $P_1(T_1)$= 0.3276 | $P_2(T_1)$= 0.3277 | $P_3(T_1)$= 0.3273 | $P_4(T_1)$= 0.3276 | $P_5(T_1)$= 0.328 |
| $P_1(C_1)$= 0.1735 | $P_2(C_1)$= 0.1735 | $P_3(C_1)$= 0.1735 | $P_4(C_1)$= 0.1736 | $P_5(C_1)$= 0.1733 |
| $P_1(G_1)$= 0.1725 | $P_2(G_1)$= 0.1724 | $P_3(G_1)$= 0.1728 | $P_4(G_1)$= 0.1723 | $P_5(G_1)$= 0.1723 |
| | $P_2(A_2)$= 0.3266 | $P_3(A_2)$= 0.3265 | $P_4(A_2)$= 0.3266 | $P_5(A_2)$= 0.3265 |
| | $P_2(T_2)$= 0.3275 | $P_3(T_2)$= 0.3275 | $P_4(T_2)$= 0.3275 | $P_5(T_2)$= 0.3277 |
| | $P_2(C_2)$= 0.1734 | $P_3(C_2)$= 0.1737 | $P_4(C_2)$= 0.1732 | $P_5(C_2)$= 0.1736 |
| | $P_2(G_2)$= 0.1725 | $P_3(G_2)$= 0.1724 | $P_4(G_2)$= 0.1727 | $P_5(G_2)$= 0.1722 |
| | | $P_3(A_3)$= 0.3265 | $P_4(A_3)$= 0.3262 | $P_5(A_3)$= 0.3268 |
| | | $P_3(T_3)$= 0.328 | $P_4(T_3)$= 0.3278 | $P_5(T_3)$= 0.3272 |
| | | $P_3(C_3)$= 0.1732 | $P_4(C_3)$= 0.1735 | $P_5(C_3)$= 0.1735 |
| | | $P_3(G_3)$= 0.1723 | $P_4(G_3)$= 0.1725 | $P_5(G_3)$= 0.1725 |
| | | | $P_4(A_4)$= 0.3266 | $P_5(A_4)$= 0.3265 |
| | | | $P_4(T_4)$= 0.3275 | $P_5(T_4)$= 0.3276 |
| | | | $P_4(C_4)$= 0.1736 | $P_5(C_4)$= 0.1733 |
| | | | $P_4(G_4)$= 0.1724 | $P_5(G_4)$= 0.1726 |
| | | | | $P_5(A_5)$= 0.3262 |
| | | | | $P_5(T_5)$= 0.3275 |
| | | | | $P_5(C_5)$= 0.1736 |
| | | | | $P_5(G_5)$= 0.1726 |

Fig. A3/4. The table of probabilities of subgroups of tetra-groups in the sequence: Caenorhabditis elegans chromosome IV,
NCBI Reference Sequence: NC_003282.8

LOCUS NC_003282 17493829 bp DNA linear CON 11-OCT-2017
DEFINITION Caenorhabditis elegans chromosome IV. ACCESSION NC_003282
VERSION NC_003282.8
https://www.ncbi.nlm.nih.gov/nuccore/NC_003282.8

| NUCLEOTIDES | DOUBLETS | TRIPLETS | 4-PLETS | 5-PLETS |
|---|---|---|---|---|
| $P_1(A_1)=$ 0.3226 | $P_2(A_1)=$ 0.3225 | $P_3(A_1)=$ 0.3227 | $P_4(A_1)=$ 0.3224 | $P_5(A_1)=$ 0.3227 |
| $P_1(T_1)=$ 0.3231 | $P_2(T_1)=$ 0.3232 | $P_3(T_1)=$ 0.3229 | $P_4(T_1)=$ 0.3233 | $P_5(T_1)=$ 0.3232 |
| $P_1(C_1)=$ 0.1774 | $P_2(C_1)=$ 0.1775 | $P_3(C_1)=$ 0.1772 | $P_4(C_1)=$ 0.1776 | $P_5(C_1)=$ 0.177 |
| $P_1(G_1)=$ 0.1769 | $P_2(G_1)=$ 0.1768 | $P_3(G_1)=$ 0.1772 | $P_4(G_1)=$ 0.1768 | $P_5(G_1)=$ 0.1771 |
| | $P_2(A_2)=$ 0.3228 | $P_3(A_2)=$ 0.3226 | $P_4(A_2)=$ 0.323 | $P_5(A_2)=$ 0.3228 |
| | $P_2(T_2)=$ 0.323 | $P_3(T_2)=$ 0.3228 | $P_4(T_2)=$ 0.3231 | $P_5(T_2)=$ 0.3231 |
| | $P_2(C_2)=$ 0.1773 | $P_3(C_2)=$ 0.1778 | $P_4(C_2)=$ 0.1772 | $P_5(C_2)=$ 0.1775 |
| | $P_2(G_2)=$ 0.177 | $P_3(G_2)=$ 0.1768 | $P_4(G_2)=$ 0.1768 | $P_5(G_2)=$ 0.1765 |
| | | $P_3(A_3)=$ 0.3225 | $P_4(A_3)=$ 0.3226 | $P_5(A_3)=$ 0.3227 |
| | | $P_3(T_3)=$ 0.3235 | $P_4(T_3)=$ 0.3231 | $P_5(T_3)=$ 0.3228 |
| | | $P_3(C_3)=$ 0.1773 | $P_4(C_3)=$ 0.1775 | $P_5(C_3)=$ 0.1777 |
| | | $P_3(G_3)=$ 0.1767 | $P_4(G_3)=$ 0.1769 | $P_5(G_3)=$ 0.1768 |
| | | | $P_4(A_4)=$ 0.3226 | $P_5(A_4)=$ 0.3222 |
| | | | $P_4(T_4)=$ 0.3229 | $P_5(T_4)=$ 0.3231 |
| | | | $P_4(C_4)=$ 0.1774 | $P_5(C_4)=$ 0.1773 |
| | | | $P_4(G_4)=$ 0.1771 | $P_5(G_4)=$ 0.1773 |
| | | | | $P_5(A_5)=$ 0.3227 |
| | | | | $P_5(T_5)=$ 0.3232 |
| | | | | $P_5(C_5)=$ 0.1774 |
| | | | | $P_5(G_5)=$ 0.1767 |

Fig. A3/5. The table of probabilities of subgroups of tetra-groups in the sequence: Caenorhabditis elegans chromosome V,

NCBI Reference Sequence: NC_003283.11

LOCUS NC_003283 20924180 bp DNA linear CON 11-OCT-2017
DEFINITION Caenorhabditis elegans chromosome V. ACCESSION NC_003283
VERSION NC_003283.11
https://www.ncbi.nlm.nih.gov/nuccore/NC_003283.11

| NUCLEOTIDES | DOUBLETS | TRIPLETS | 4-PLETS | 5-PLETS |
|---|---|---|---|---|
| $P_1(A_1)=$ 0.3244 | $P_2(A_1)=$ 0.3244 | $P_3(A_1)=$ 0.3245 | $P_4(A_1)=$ 0.3244 | $P_5(A_1)=$ 0.3241 |
| $P_1(T_1)=$ 0.3236 | $P_2(T_1)=$ 0.3235 | $P_3(T_1)=$ 0.3237 | $P_4(T_1)=$ 0.3237 | $P_5(T_1)=$ 0.3236 |
| $P_1(C_1)=$ 0.1761 | $P_2(C_1)=$ 0.1761 | $P_3(C_1)=$ 0.1757 | $P_4(C_1)=$ 0.1759 | $P_5(C_1)=$ 0.1761 |
| $P_1(G_1)=$ 0.176 | $P_2(G_1)=$ 0.176 | $P_3(G_1)=$ 0.176 | $P_4(G_1)=$ 0.176 | $P_5(G_1)=$ 0.1762 |
| | $P_2(A_2)=$ 0.3243 | $P_3(A_2)=$ 0.324 | $P_4(A_2)=$ 0.3241 | $P_5(A_2)=$ 0.3246 |
| | $P_2(T_2)=$ 0.3237 | $P_3(T_2)=$ 0.3234 | $P_4(T_2)=$ 0.3237 | $P_5(T_2)=$ 0.3231 |
| | $P_2(C_2)=$ 0.1761 | $P_3(C_2)=$ 0.1765 | $P_4(C_2)=$ 0.1762 | $P_5(C_2)=$ 0.1761 |
| | $P_2(G_2)=$ 0.1759 | $P_3(G_2)=$ 0.176 | $P_4(G_2)=$ 0.176 | $P_5(G_2)=$ 0.1762 |
| | | $P_3(A_3)=$ 0.3245 | $P_4(A_3)=$ 0.3245 | $P_5(A_3)=$ 0.3243 |
| | | $P_3(T_3)=$ 0.3237 | $P_4(T_3)=$ 0.3232 | $P_5(T_3)=$ 0.3238 |
| | | $P_3(C_3)=$ 0.1759 | $P_4(C_3)=$ 0.1762 | $P_5(C_3)=$ 0.176 |
| | | $P_3(G_3)=$ 0.1758 | $P_4(G_3)=$ 0.1761 | $P_5(G_3)=$ 0.1759 |
| | | | $P_4(A_4)=$ 0.3244 | $P_5(A_4)=$ 0.3245 |
| | | | $P_4(T_4)=$ 0.3238 | $P_5(T_4)=$ 0.3237 |
| | | | $P_4(C_4)=$ 0.176 | $P_5(C_4)=$ 0.1761 |
| | | | $P_4(G_4)=$ 0.1758 | $P_5(G_4)=$ 0.1757 |
| | | | | $P_5(A_5)=$ 0.3243 |
| | | | | $P_5(T_5)=$ 0.3239 |
| | | | | $P_5(C_5)=$ 0.1761 |
| | | | | $P_5(G_5)=$ 0.1758 |

Fig. A3/6. The table of probabilities of subgroups of tetra-groups in the sequence: Caenorhabditis elegans chromosome X,
NCBI Reference Sequence: NC_003284.9

LOCUS NC_003284 17718942 bp DNA linear CON 11-OCT-2017
DEFINITION Caenorhabditis elegans chromosome X. ACCESSION NC_003284
VERSION NC_003284.9
https://www.ncbi.nlm.nih.gov/nuccore/NC_003284.9

## Appendix 4. Symmetries of tetra-group probabilities in the complete set of chromosomes of *Drosophila melanogaster*

The appendix 4 represents data about the fulfillment of the described tetra-group rules in the complete set of chromosomes of *Drosophila melanogaster*. This fruit fly is used as a model organism in the study of genetics, development and disease for long ago. All initial data are taken from the

CenBank
(https://www.ncbi.nlm.nih.gov/genome/?term=drosophila+melanogaster).

| NUCLEOTIDES | DOUBLETS | TRIPLETS | 4-PLETS | 5-PLETS |
|---|---|---|---|---|
| $P_1(A_1)$= 0.2868 | $P_2(A_1)$= 0.2867 | $P_3(A_1)$= 0.2871 | $P_4(A_1)$= 0.2868 | $P_5(A_1)$= 0.2871 |
| $P_1(T_1)$= 0.2886 | $P_2(T_1)$= 0.2885 | $P_3(T_1)$= 0.2885 | $P_4(T_1)$= 0.2884 | $P_5(T_1)$= 0.2887 |
| $P_1(C_1)$= 0.212 | $P_2(C_1)$= 0.212 | $P_3(C_1)$= 0.2119 | $P_4(C_1)$= 0.212 | $P_5(C_1)$= 0.2119 |
| $P_1(G_1)$= 0.2127 | $P_2(G_1)$= 0.2127 | $P_3(G_1)$= 0.2124 | $P_4(G_1)$= 0.2128 | $P_5(G_1)$= 0.2124 |
| | $P_2(A_2)$= 0.2869 | $P_3(A_2)$= 0.2868 | $P_4(A_2)$= 0.2868 | $P_5(A_2)$= 0.2868 |
| | $P_2(T_2)$= 0.2886 | $P_3(T_2)$= 0.2889 | $P_4(T_2)$= 0.2887 | $P_5(T_2)$= 0.2885 |
| | $P_2(C_2)$= 0.2119 | $P_3(C_2)$= 0.2116 | $P_4(C_2)$= 0.212 | $P_5(C_2)$= 0.2118 |
| | $P_2(G_2)$= 0.2126 | $P_3(G_2)$= 0.2127 | $P_4(G_2)$= 0.2124 | $P_5(G_2)$= 0.213 |
| | | $P_3(A_3)$= 0.2864 | $P_4(A_3)$= 0.2867 | $P_5(A_3)$= 0.2866 |
| | | $P_3(T_3)$= 0.2883 | $P_4(T_3)$= 0.2887 | $P_5(T_3)$= 0.2887 |
| | | $P_3(C_3)$= 0.2123 | $P_4(C_3)$= 0.212 | $P_5(C_3)$= 0.212 |
| | | $P_3(G_3)$= 0.2129 | $P_4(G_3)$= 0.2126 | $P_5(G_3)$= 0.2127 |
| | | | $P_4(A_4)$= 0.2869 | $P_5(A_4)$= 0.2865 |
| | | | $P_4(T_4)$= 0.2885 | $P_5(T_4)$= 0.2886 |
| | | | $P_4(C_4)$= 0.2117 | $P_5(C_4)$= 0.2121 |
| | | | $P_4(G_4)$= 0.2128 | $P_5(G_4)$= 0.2129 |
| | | | | $P_5(A_5)$= 0.287 |
| | | | | $P_5(T_5)$= 0.2885 |
| | | | | $P_5(C_5)$= 0.212 |
| | | | | $P_5(G_5)$= 0.2125 |

Fig. A4/1. The table of probabilities of subgroups of tetra-groups in the sequence: Drosophila melanogaster chromosome X,
NCBI Reference Sequence: NC_004354.4

LOCUS NC_004354 23542271 bp DNA linear CON 14-MAR-2017 DEFINITION Drosophila melanogaster chromosome X. ACCESSION NC_004354 NW_001844882 NW_001845036 NW_001848859 VERSION NC_004354.4
https://www.ncbi.nlm.nih.gov/nuccore/NC_004354.4

| NUCLEOTIDES | DOUBLETS | TRIPLETS | 4-PLETS | 5-PLETS |
|---|---|---|---|---|
| $P_1(A_1)=$ 0.2915 | $P_2(A_1)=$ 0.2914 | $P_3(A_1)=$ 0.2915 | $P_4(A_1)=$ 0.2912 | $P_5(A_1)=$ 0.2915 |
| $P_1(T_1)=$ 0.2907 | $P_2(T_1)=$ 0.2908 | $P_3(T_1)=$ 0.2911 | $P_4(T_1)=$ 0.2909 | $P_5(T_1)=$ 0.2909 |
| $P_1(C_1)=$ 0.2089 | $P_2(C_1)=$ 0.2089 | $P_3(C_1)=$ 0.2089 | $P_4(C_1)=$ 0.209 | $P_5(C_1)=$ 0.2087 |
| $P_1(G_1)=$ 0.2089 | $P_2(G_1)=$ 0.209 | $P_3(G_1)=$ 0.2085 | $P_4(G_1)=$ 0.209 | $P_5(G_1)=$ 0.2089 |
| | $P_2(A_2)=$ 0.2915 | $P_3(A_2)=$ 0.2915 | $P_4(A_2)=$ 0.2915 | $P_5(A_2)=$ 0.2915 |
| | $P_2(T_2)=$ 0.2907 | $P_3(T_2)=$ 0.2912 | $P_4(T_2)=$ 0.2909 | $P_5(T_2)=$ 0.2903 |
| | $P_2(C_2)=$ 0.2089 | $P_3(C_2)=$ 0.2085 | $P_4(C_2)=$ 0.209 | $P_5(C_2)=$ 0.2089 |
| | $P_2(G_2)=$ 0.2089 | $P_3(G_2)=$ 0.2087 | $P_4(G_2)=$ 0.2086 | $P_5(G_2)=$ 0.2093 |
| | | $P_3(A_3)=$ 0.2913 | $P_4(A_3)=$ 0.2916 | $P_5(A_3)=$ 0.2914 |
| | | $P_3(T_3)=$ 0.2899 | $P_4(T_3)=$ 0.2907 | $P_5(T_3)=$ 0.291 |
| | | $P_3(C_3)=$ 0.2092 | $P_4(C_3)=$ 0.2087 | $P_5(C_3)=$ 0.2088 |
| | | $P_3(G_3)=$ 0.2096 | $P_4(G_3)=$ 0.209 | $P_5(G_3)=$ 0.2088 |
| | | | $P_4(A_4)=$ 0.2915 | $P_5(A_4)=$ 0.2913 |
| | | | $P_4(T_4)=$ 0.2905 | $P_5(T_4)=$ 0.2907 |
| | | | $P_4(C_4)=$ 0.2089 | $P_5(C_4)=$ 0.2091 |
| | | | $P_4(G_4)=$ 0.2091 | $P_5(G_4)=$ 0.2089 |
| | | | | $P_5(A_5)=$ 0.2916 |
| | | | | $P_5(T_5)=$ 0.2908 |
| | | | | $P_5(C_5)=$ 0.2090 |
| | | | | $P_5(G_5)=$ 0.2086 |

Fig. A4/2. The table of probabilities of subgroups of tetra-groups in the sequence: Drosophila melanogaster chromosome 2L,

NCBI Reference Sequence: NT_033779.5

LOCUS NT_033779 23513712 bp DNA linear CON 14-MAR-2017 DEFINITION Drosophila melanogaster chromosome 2L. ACCESSION NT_033779 NW_001848855 VERSION NT_033779.5
https://www.ncbi.nlm.nih.gov/nuccore/NT_033779.5

| NUCLEOTIDES | DOUBLETS | TRIPLETS | 4-PLETS | 5-PLETS |
|---|---|---|---|---|
| $P_1(A_1)=$ 0.2877 | $P_2(A_1)=$ 0.2877 | $P_3(A_1)=$ 0.2877 | $P_4(A_1)=$ 0.2874 | $P_5(A_1)=$ 0.2878 |
| $P_1(T_1)=$ 0.2862 | $P_2(T_1)=$ 0.2861 | $P_3(T_1)=$ 0.2859 | $P_4(T_1)=$ 0.2863 | $P_5(T_1)=$ 0.286 |
| $P_1(C_1)=$ 0.2134 | $P_2(C_1)=$ 0.2134 | $P_3(C_1)=$ 0.2136 | $P_4(C_1)=$ 0.2133 | $P_5(C_1)=$ 0.2133 |
| $P_1(G_1)=$ 0.2127 | $P_2(G_1)=$ 0.2129 | $P_3(G_1)=$ 0.2128 | $P_4(G_1)=$ 0.213 | $P_5(G_1)=$ 0.2129 |
| | $P_2(A_2)=$ 0.2877 | $P_3(A_2)=$ 0.2879 | $P_4(A_2)=$ 0.2878 | $P_5(A_2)=$ 0.2878 |
| | $P_2(T_2)=$ 0.2863 | $P_3(T_2)=$ 0.2862 | $P_4(T_2)=$ 0.2864 | $P_5(T_2)=$ 0.2864 |
| | $P_2(C_2)=$ 0.2135 | $P_3(C_2)=$ 0.2134 | $P_4(C_2)=$ 0.2133 | $P_5(C_2)=$ 0.2133 |
| | $P_2(G_2)=$ 0.2125 | $P_3(G_2)=$ 0.2126 | $P_4(G_2)=$ 0.2126 | $P_5(G_2)=$ 0.2125 |
| | | $P_3(A_3)=$ 0.2875 | $P_4(A_3)=$ 0.2879 | $P_5(A_3)=$ 0.2878 |
| | | $P_3(T_3)=$ 0.2865 | $P_4(T_3)=$ 0.2859 | $P_5(T_3)=$ 0.2862 |
| | | $P_3(C_3)=$ 0.2133 | $P_4(C_3)=$ 0.2134 | $P_5(C_3)=$ 0.2136 |
| | | $P_3(G_3)=$ 0.2127 | $P_4(G_3)=$ 0.2128 | $P_5(G_3)=$ 0.2124 |
| | | | $P_4(A_4)=$ 0.2876 | $P_5(A_4)=$ 0.2877 |
| | | | $P_4(T_4)=$ 0.2863 | $P_5(T_4)=$ 0.2863 |
| | | | $P_4(C_4)=$ 0.2137 | $P_5(C_4)=$ 0.2133 |
| | | | $P_4(G_4)=$ 0.2124 | $P_5(G_4)=$ 0.2128 |
| | | | | $P_5(A_5)=$ 0.2874 |
| | | | | $P_5(T_5)=$ 0.2862 |
| | | | | $P_5(C_5)=$ 0.2135 |
| | | | | $P_5(G_5)=$ 0.2128 |

Fig. A4/3. The table of probabilities of subgroups of tetra-groups in the sequence: Drosophila melanogaster chromosome 2R,

NCBI Reference Sequence: NT_033778.4

LOCUS NT_033778 25286936 bp DNA linear CON 14-MAR-2017 DEFINITION Drosophila melanogaster chromosome 2R. ACCESSION NT_033778 NW_001844732 NW_001844738 NW_001848856 VERSION NT_033778.4

https://www.ncbi.nlm.nih.gov/nuccore/NT_033778.4

| NUCLEOTIDES | DOUBLETS | TRIPLETS | 4-PLETS | 5-PLETS |
|---|---|---|---|---|
| $P_1(A_1)$= 0.2909 | $P_2(A_1)$= 0.2909 | $P_3(A_1)$= 0.2909 | $P_4(A_1)$= 0.2908 | $P_5(A_1)$= 0.2909 |
| $P_1(T_1)$= 0.2929 | $P_2(T_1)$= 0.2928 | $P_3(T_1)$= 0.2924 | $P_4(T_1)$= 0.2929 | $P_5(T_1)$= 0.2929 |
| $P_1(C_1)$= 0.2081 | $P_2(C_1)$= 0.2082 | $P_3(C_1)$= 0.2084 | $P_4(C_1)$= 0.2083 | $P_5(C_1)$= 0.2079 |
| $P_1(G_1)$= 0.2081 | $P_2(G_1)$= 0.2080 | $P_3(G_1)$= 0.2083 | $P_4(G_1)$= 0.208 | $P_5(G_1)$= 0.2082 |
| | $P_2(A_2)$= 0.2909 | $P_3(A_2)$= 0.2913 | $P_4(A_2)$= 0.2909 | $P_5(A_2)$= 0.291 |
| | $P_2(T_2)$= 0.2929 | $P_3(T_2)$= 0.2933 | $P_4(T_2)$= 0.2931 | $P_5(T_2)$= 0.2928 |
| | $P_2(C_2)$= 0.208 | $P_3(C_2)$= 0.208 | $P_4(C_2)$= 0.2078 | $P_5(C_2)$= 0.2082 |
| | $P_2(G_2)$= 0.2081 | $P_3(G_2)$= 0.2074 | $P_4(G_2)$= 0.2082 | $P_5(G_2)$= 0.208 |
| | | $P_3(A_3)$= 0.2906 | $P_4(A_3)$= 0.291 | $P_5(A_3)$= 0.291 |
| | | $P_3(T_3)$= 0.2929 | $P_4(T_3)$= 0.2927 | $P_5(T_3)$= 0.2927 |
| | | $P_3(C_3)$= 0.2080 | $P_4(C_3)$= 0.2082 | $P_5(C_3)$= 0.2082 |
| | | $P_3(G_3)$= 0.2085 | $P_4(G_3)$= 0.2081 | $P_5(G_3)$= 0.2081 |
| | | | $P_4(A_4)$= 0.291 | $P_5(A_4)$= 0.2911 |
| | | | $P_4(T_4)$= 0.2927 | $P_5(T_4)$= 0.2928 |
| | | | $P_4(C_4)$= 0.2083 | $P_5(C_4)$= 0.2081 |
| | | | $P_4(G_4)$= 0.208 | $P_5(G_4)$= 0.208 |
| | | | | $P_5(A_5)$= 0.2906 |
| | | | | $P_5(T_5)$= 0.2932 |
| | | | | $P_5(C_5)$= 0.2082 |
| | | | | $P_5(G_5)$= 0.208 |

Fig. A4/4. The table of probabilities of subgroups of tetra-groups in the sequence: Drosophila melanogaster chromosome 3L,

NCBI Reference Sequence: NT_037436.4

LOCUS NT_037436 28110227 bp DNA linear CON 14-MAR-2017 DEFINITION Drosophila melanogaster chromosome 3L. ACCESSION NT_037436 NW_001844849 NW_001848857 VERSION NT_037436.4

https://www.ncbi.nlm.nih.gov/nuccore/NT_037436.4

| NUCLEOTIDES | DOUBLETS | TRIPLETS | 4-PLETS | 5-PLETS |
|---|---|---|---|---|
| $P_1(A_1)$= 0.2872 | $P_2(A_1)$= 0.2873 | $P_3(A_1)$= 0.2874 | $P_4(A_1)$= 0.2872 | $P_5(A_1)$= 0.2872 |
| $P_1(T_1)$= 0.2869 | $P_2(T_1)$= 0.2869 | $P_3(T_1)$= 0.2871 | $P_4(T_1)$= 0.2869 | $P_5(T_1)$= 0.287 |
| $P_1(C_1)$= 0.2132 | $P_2(C_1)$= 0.2131 | $P_3(C_1)$= 0.2131 | $P_4(C_1)$= 0.2132 | $P_5(C_1)$= 0.2133 |
| $P_1(G_1)$= 0.2127 | $P_2(G_1)$= 0.2127 | $P_3(G_1)$= 0.2125 | $P_4(G_1)$= 0.2127 | $P_5(G_1)$= 0.2125 |
| | $P_2(A_2)$= 0.2871 | $P_3(A_2)$= 0.2871 | $P_4(A_2)$= 0.2872 | $P_5(A_2)$= 0.2872 |
| | $P_2(T_2)$= 0.287 | $P_3(T_2)$= 0.287 | $P_4(T_2)$= 0.2868 | $P_5(T_2)$= 0.2868 |
| | $P_2(C_2)$= 0.2132 | $P_3(C_2)$= 0.2131 | $P_4(C_2)$= 0.2133 | $P_5(C_2)$= 0.213 |
| | $P_2(G_2)$= 0.2127 | $P_3(G_2)$= 0.2129 | $P_4(G_2)$= 0.2127 | $P_5(G_2)$= 0.2129 |
| | | $P_3(A_3)$= 0.2871 | $P_4(A_3)$= 0.2873 | $P_5(A_3)$= 0.2874 |
| | | $P_3(T_3)$= 0.2867 | $P_4(T_3)$= 0.2869 | $P_5(T_3)$= 0.287 |
| | | $P_3(C_3)$= 0.2135 | $P_4(C_3)$= 0.2131 | $P_5(C_3)$= 0.2132 |
| | | $P_3(G_3)$= 0.2127 | $P_4(G_3)$= 0.2127 | $P_5(G_3)$= 0.2124 |
| | | | $P_4(A_4)$= 0.287 | $P_5(A_4)$= 0.2869 |
| | | | $P_4(T_4)$= 0.2871 | $P_5(T_4)$= 0.287 |
| | | | $P_4(C_4)$= 0.2132 | $P_5(C_4)$= 0.2133 |
| | | | $P_4(G_4)$= 0.2127 | $P_5(G_4)$= 0.2128 |
| | | | | $P_5(A_5)$= 0.2872 |
| | | | | $P_5(T_5)$= 0.2869 |
| | | | | $P_5(C_5)$= 0.2132 |
| | | | | $P_5(G_5)$= 0.2128 |

Fig. A4/5. The table of probabilities of subgroups of tetra-groups in the sequence: Drosophila melanogaster chromosome 3R,

NCBI Reference Sequence: NT_033777.3

LOCUS NT_033777 32079331 bp DNA linear CON 14-MAR-2017 DEFINITION Drosophila melanogaster chromosome 3R. ACCESSION NT_033777 NW_001844733 NW_001844734 NW_001844736 NW_001844852 NW_001844855 NW_001844895 NW_001848858 VERSION NT_033777.3

https://www.ncbi.nlm.nih.gov/nuccore/NT_033777.3

| NUCLEOTIDES | DOUBLETS | TRIPLETS | 4-PLETS | 5-PLETS |
|---|---|---|---|---|

| $P_1(A_1)=$ 0.3195 | $P_2(A_1)=$ 0.3191 | $P_3(A_1)=$ 0.3199 | $P_4(A_1)=$ 0.3184 | $P_5(A_1)=$ 0.3218 |
|---|---|---|---|---|
| $P_1(T_1)=$ 0.328 | $P_2(T_1)=$ 0.3285 | $P_3(T_1)=$ 0.3290 | $P_4(T_1)=$ 0.3288 | $P_5(T_1)=$ 0.3262 |
| $P_1(C_1)=$ 0.1747 | $P_2(C_1)=$ 0.175 | $P_3(C_1)=$ 0.1734 | $P_4(C_1)=$ 0.1752 | $P_5(C_1)=$ 0.1747 |
| $P_1(G_1)=$ 0.1778 | $P_2(G_1)=$ 0.1774 | $P_3(G_1)=$ 0.1777 | $P_4(G_1)=$ 0.1775 | $P_5(G_1)=$ 0.1773 |
| | $P_2(A_2)=$ 0.3198 | $P_3(A_2)=$ 0.3197 | $P_4(A_2)=$ 0.3199 | $P_5(A_2)=$ 0.3193 |
| | $P_2(T_2)=$ 0.3276 | $P_3(T_2)=$ 0.3273 | $P_4(T_2)=$ 0.3286 | $P_5(T_2)=$ 0.3292 |
| | $P_2(C_2)=$ 0.1744 | $P_3(C_2)=$ 0.1747 | $P_4(C_2)=$ 0.1739 | $P_5(C_2)=$ 0.1744 |
| | $P_2(G_2)=$ 0.1781 | $P_3(G_2)=$ 0.1784 | $P_4(G_2)=$ 0.1776 | $P_5(G_2)=$ 0.1771 |
| | | $P_3(A_3)=$ 0.3188 | $P_4(A_3)=$ 0.3198 | $P_5(A_3)=$ 0.3174 |
| | | $P_3(T_3)=$ 0.3279 | $P_4(T_3)=$ 0.3281 | $P_5(T_3)=$ 0.3296 |
| | | $P_3(C_3)=$ 0.1760 | $P_4(C_3)=$ 0.1747 | $P_5(C_3)=$ 0.1744 |
| | | $P_3(G_3)=$ 0.1773 | $P_4(G_3)=$ 0.1774 | $P_5(G_3)=$ 0.1786 |
| | | | $P_4(A_4)=$ 0.3197 | $P_5(A_4)=$ 0.3194 |
| | | | $P_4(T_4)=$ 0.3266 | $P_5(T_4)=$ 0.3275 |
| | | | $P_4(C_4)=$ 0.175 | $P_5(C_4)=$ 0.1755 |
| | | | $P_4(G_4)=$ 0.1787 | $P_5(G_4)=$ 0.1776 |
| | | | | $P_5(A_5)=$ 0.3193 |
| | | | | $P_5(T_5)=$ 0.3278 |
| | | | | $P_5(C_5)=$ 0.1746 |
| | | | | $P_5(G_5)=$ 0.1783 |

Fig. A4/6. The table of probabilities of subgroups of tetra-groups in the sequence: Drosophila melanogaster chromosome 4,

NCBI Reference Sequence: NC_004353.4

LOCUS NC_004353 1348131 bp DNA linear CON 14-MAR-2017
DEFINITION Drosophila melanogaster chromosome 4. ACCESSION NC_004353 NW_001845041 VERSION NC_004353.4
https://www.ncbi.nlm.nih.gov/nuccore/NC_004353.4

| **NUCLEOTIDES** | **DOUBLETS** | **TRIPLETS** | **4-PLETS** | **5-PLETS** |
|---|---|---|---|---|

| $P_1(A_1)=$ 0.3099 | $P_2(A_1)=$ 0.3098 | $P_3(A_1)=$ 0.3104 | $P_4(A_1)=$ 0.3102 | $P_5(A_1)=$ 0.3105 |
|---|---|---|---|---|
| $P_1(T_1)=$ 0.2958 | $P_2(T_1)=$ 0.2960 | $P_3(T_1)=$ 0.2955 | $P_4(T_1)=$ 0.2959 | $P_5(T_1)=$ 0.2956 |
| $P_1(C_1)=$ 0.2002 | $P_2(C_1)=$ 0.2003 | $P_3(C_1)=$ 0.2002 | $P_4(C_1)=$ 0.2001 | $P_5(C_1)=$ 0.2003 |
| $P_1(G_1)=$ 0.1940 | $P_2(G_1)=$ 0.1939 | $P_3(G_1)=$ 0.1938 | $P_4(G_1)=$ 0.1938 | $P_5(G_1)=$ 0.1936 |
| | $P_2(A_2)=$ 0.3100 | $P_3(A_2)=$ 0.3098 | $P_4(A_2)=$ 0.3095 | $P_5(A_2)=$ 0.31 |
| | $P_2(T_2)=$ 0.2957 | $P_3(T_2)=$ 0.2954 | $P_4(T_2)=$ 0.2961 | $P_5(T_2)=$ 0.2957 |
| | $P_2(C_2)=$ 0.2002 | $P_3(C_2)=$ 0.2003 | $P_4(C_2)=$ 0.2000 | $P_5(C_2)=$ 0.2004 |
| | $P_2(G_2)=$ 0.1941 | $P_3(G_2)=$ 0.1945 | $P_4(G_2)=$ 0.1944 | $P_5(G_2)=$ 0.1939 |
| | | $P_3(A_3)=$ 0.3096 | $P_4(A_3)=$ 0.3094 | $P_5(A_3)=$ 0.3099 |
| | | $P_3(T_3)=$ 0.2965 | $P_4(T_3)=$ 0.2961 | $P_5(T_3)=$ 0.2956 |
| | | $P_3(C_3)=$ 0.2001 | $P_4(C_3)=$ 0.2004 | $P_5(C_3)=$ 0.2006 |
| | | $P_3(G_3)=$ 0.1938 | $P_4(G_3)=$ 0.1941 | $P_5(G_3)=$ 0.194 |
| | | | $P_4(A_4)=$ 0.3105 | $P_5(A_4)=$ 0.31 |
| | | | $P_4(T_4)=$ 0.2953 | $P_5(T_4)=$ 0.2962 |
| | | | $P_4(C_4)=$ 0.2004 | $P_5(C_4)=$ 0.1992 |
| | | | $P_4(G_4)=$ 0.1938 | $P_5(G_4)=$ 0.1946 |
| | | | | $P_5(A_5)=$ 0.3093 |
| | | | | $P_5(T_5)=$ 0.2960 |
| | | | | $P_5(C_5)=$ 0.2007 |
| | | | | $P_5(G_5)=$ 0.1941 |

Fig. A4/7. The table of probabilities of subgroups of tetra-groups in the sequence: Drosophila melanogaster chromosome Y,

NCBI Reference Sequence: NC_024512.1

LOCUS NC_024512 3667352 bp DNA linear CON 14-MAR-2017 DEFINITION Drosophila melanogaster chromosome Y. ACCESSION NC_024512 NW_001844735 NW_001844843 NW_001844868 NW_001844869 NW_001844874 NW_001844917 NW_001845004 NW_001845447 NW_001845711 NW_001845825 NW_001846080 NW_001846110 NW_001847094 NW_001848860 VERSION NC_024512.1
https://www.ncbi.nlm.nih.gov/nuccore/NC_024512.1

## Appendix 5. Symmetries of tetra-group probabilities in the complete set of chromosomes of *Arabidopsis Thaliana*

The appendix 5 represents data about the fulfillment of the described tetra-group rules in the complete set of chromosomes of *Arabidopsis Thaliana*. This small flowering plant has been used for over fifty years to study plant mutations and for classical genetic analysis. It became the first plant genome to be fully sequenced; it has a small genome of ~120 Mb. All initial data about chromosomes are taken from the CenBank (https://www.ncbi.nlm.nih.gov/genome/4 ).

| **NUCLEOTIDES** | **DOUBLETS** | **TRIPLETS** | **4-PLETS** | **5-PLETS** |
|---|---|---|---|---|
| **$P_1(A_1)=$ 0.3208** | **$P_2(A_1)=$ 0.3207** | **$P_3(A_1)=$ 0.3212** | **$P_4(A_1)=$ 0.3206** | **$P_5(A_1)=$ 0.3209** |
| **$P_1(T_1)=$ 0.3204** | **$P_2(T_1)=$ 0.3204** | **$P_3(T_1)=$ 0.3204** | **$P_4(T_1)=$ 0.3207** | **$P_5(T_1)=$ 0.3201** |
| **$P_1(C_1)=$ 0.1796** | **$P_2(C_1)=$ 0.1796** | **$P_3(C_1)=$ 0.1794** | **$P_4(C_1)=$ 0.1797** | **$P_5(C_1)=$ 0.1797** |
| **$P_1(G_1)=$ 0.1791** | **$P_2(G_1)=$ 0.1792** | **$P_3(G_1)=$ 0.1790** | **$P_4(G_1)=$ 0.179** | **$P_5(G_1)=$ 0.1793** |
| | **$P_2(A_2)=$ 0.3209** | **$P_3(A_2)=$ 0.3206** | **$P_4(A_2)=$ 0.3209** | **$P_5(A_2)=$ 0.3209** |
| | **$P_2(T_2)=$ 0.3204** | **$P_3(T_2)=$ 0.3207** | **$P_4(T_2)=$ 0.3203** | **$P_5(T_2)=$ 0.3207** |
| | **$P_2(C_2)=$ 0.1796** | **$P_3(C_2)=$ 0.1798** | **$P_4(C_2)=$ 0.1797** | **$P_5(C_2)=$ 0.1795** |
| | **$P_2(G_2)=$ 0.1791** | **$P_3(G_2)=$ 0.1789** | **$P_4(G_2)=$ 0.1791** | **$P_5(G_2)=$ 0.179** |
| | | **$P_3(A_3)=$ 0.3207** | **$P_4(A_3)=$ 0.3208** | **$P_5(A_3)=$ 0.3207** |
| | | **$P_3(T_3)=$ 0.3202** | **$P_4(T_3)=$ 0.3202** | **$P_5(T_3)=$ 0.3206** |
| | | **$P_3(C_3)=$ 0.1796** | **$P_4(C_3)=$ 0.1796** | **$P_5(C_3)=$ 0.1795** |
| | | **$P_3(G_3)=$ 0.1795** | **$P_4(G_3)=$ 0.1793** | **$P_5(G_3)=$ 0.1792** |
| | | | **$P_4(A_4)=$ 0.321** | **$P_5(A_4)=$ 0.3211** |
| | | | **$P_4(T_4)=$ 0.3205** | **$P_5(T_4)=$ 0.3204** |
| | | | **$P_4(C_4)=$ 0.1794** | **$P_5(C_4)=$ 0.1794** |
| | | | **$P_4(G_4)=$ 0.1791** | **$P_5(G_4)=$ 0.1791** |
| | | | | **$P_5(A_5)=$ 0.3206** |
| | | | | **$P_5(T_5)=$ 0.3204** |
| | | | | **$P_5(C_5)=$ 0.1799** |
| | | | | **$P_5(G_5)=$ 0.1791** |

Fig. A5/1. The table of probabilities of subgroups of tetra-groups in the sequence: Arabidopsis thaliana chromosome 1 sequence,

NCBI Reference Sequence: NC_003070.9

LOCUS NC_003070 30427671 bp DNA linear CON 20-MAR-2017 DEFINITION Arabidopsis thaliana chromosome 1 sequence. ACCESSION NC_003070 VERSION NC_003070.9

https://www.ncbi.nlm.nih.gov/nuccore/NC_003070.9

| NUCLEOTIDES | DOUBLETS | TRIPLETS | 4-PLETS | 5-PLETS |
|---|---|---|---|---|
| $P_1(A_1)=$ 0.3207 | $P_2(A_1)=$ 0.3205 | $P_3(A_1)=$ 0.3205 | $P_4(A_1)=$ 0.3207 | $P_5(A_1)=$ 0.3205 |
| $P_1(T_1)=$ 0.3207 | $P_2(T_1)=$ 0.3208 | $P_3(T_1)=$ 0.321 | $P_4(T_1)=$ 0.3208 | $P_5(T_1)=$ 0.3205 |
| $P_1(C_1)=$ 0.1799 | $P_2(C_1)=$ 0.18 | $P_3(C_1)=$ 0.1797 | $P_4(C_1)=$ 0.1797 | $P_5(C_1)=$ 0.1805 |
| $P_1(G_1)=$ 0.1788 | $P_2(G_1)=$ 0.1787 | $P_3(G_1)=$ 0.1788 | $P_4(G_1)=$ 0.1788 | $P_5(G_1)=$ 0.1785 |
| | $P_2(A_2)=$ 0.3208 | $P_3(A_2)=$ 0.3204 | $P_4(A_2)=$ 0.3207 | $P_5(A_2)=$ 0.3206 |
| | $P_2(T_2)=$ 0.3206 | $P_3(T_2)=$ 0.3208 | $P_4(T_2)=$ 0.3206 | $P_5(T_2)=$ 0.3204 |
| | $P_2(C_2)=$ 0.1798 | $P_3(C_2)=$ 0.1802 | $P_4(C_2)=$ 0.1799 | $P_5(C_2)=$ 0.18 |
| | $P_2(G_2)=$ 0.1788 | $P_3(G_2)=$ 0.1785 | $P_4(G_2)=$ 0.1788 | $P_5(G_2)=$ 0.1789 |
| | | $P_3(A_3)=$ 0.321 | $P_4(A_3)=$ 0.3204 | $P_5(A_3)=$ 0.3209 |
| | | $P_3(T_3)=$ 0.3203 | $P_4(T_3)=$ 0.3207 | $P_5(T_3)=$ 0.3209 |
| | | $P_3(C_3)=$ 0.1797 | $P_4(C_3)=$ 0.1802 | $P_5(C_3)=$ 0.1794 |
| | | $P_3(G_3)=$ 0.1790 | $P_4(G_3)=$ 0.1787 | $P_5(G_3)=$ 0.1788 |
| | | | $P_4(A_4)=$ 0.3209 | $P_5(A_4)=$ 0.3209 |
| | | | $P_4(T_4)=$ 0.3207 | $P_5(T_4)=$ 0.3207 |
| | | | $P_4(C_4)=$ 0.1798 | $P_5(C_4)=$ 0.1796 |
| | | | $P_4(G_4)=$ 0.1787 | $P_5(G_4)=$ 0.1789 |
| | | | | $P_5(A_5)=$ 0.3204 |
| | | | | $P_5(T_5)=$ 0.321 |
| | | | | $P_5(C_5)=$ 0.1799 |
| | | | | $P_5(G_5)=$ 0.1787 |

Fig. A5/2. The table of probabilities of subgroups of tetra-groups in the sequence: Arabidopsis thaliana chromosome 2 sequence,

NCBI Reference Sequence: NC_003071.7

LOCUS NC_003071 19698289 bp DNA linear CON 20-MAR-2017 DEFINITION Arabidopsis thaliana chromosome 2 sequence. ACCESSION NC_003071 VERSION NC_003071.7

https://www.ncbi.nlm.nih.gov/nuccore/NC_003071.7

| NUCLEOTIDES | DOUBLETS | TRIPLETS | 4-PLETS | 5-PLETS |
|---|---|---|---|---|

| $P_1(A_1)=$ 0.3191 | $P_2(A_1)=$ 0.3194 | $P_3(A_1)=$ 0.3195 | $P_4(A_1)=$ 0.3192 | $P_5(A_1)=$ 0.3191 |
|---|---|---|---|---|
| $P_1(T_1)=$ 0.3176 | $P_2(T_1)=$ 0.3175 | $P_3(T_1)=$ 0.3171 | $P_4(T_1)=$ 0.3176 | $P_5(T_1)=$ 0.3175 |
| $P_1(C_1)=$ 0.1816 | $P_2(C_1)=$ 0.1815 | $P_3(C_1)=$ 0.1816 | $P_4(C_1)=$ 0.1815 | $P_5(C_1)=$ 0.1816 |
| $P_1(G_1)=$ 0.1817 | $P_2(G_1)=$ 0.1817 | $P_3(G_1)=$ 0.1818 | $P_4(G_1)=$ 0.1817 | $P_5(G_1)=$ 0.1818 |
| | $P_2(A_2)=$ 0.3189 | $P_3(A_2)=$ 0.3192 | $P_4(A_2)=$ 0.319 | $P_5(A_2)=$ 0.3194 |
| | $P_2(T_2)=$ 0.3177 | $P_3(T_2)=$ 0.3179 | $P_4(T_2)=$ 0.3175 | $P_5(T_2)=$ 0.3172 |
| | $P_2(C_2)=$ 0.1816 | $P_3(C_2)=$ 0.1813 | $P_4(C_2)=$ 0.1816 | $P_5(C_2)=$ 0.1813 |
| | $P_2(G_2)=$ 0.1818 | $P_3(G_2)=$ 0.1816 | $P_4(G_2)=$ 0.1819 | $P_5(G_2)=$ 0.182 |
| | | $P_3(A_3)=$ 0.3187 | $P_4(A_3)=$ 0.3195 | $P_5(A_3)=$ 0.3189 |
| | | $P_3(T_3)=$ 0.3176 | $P_4(T_3)=$ 0.3173 | $P_5(T_3)=$ 0.3178 |
| | | $P_3(C_3)=$ 0.1818 | $P_4(C_3)=$ 0.1816 | $P_5(C_3)=$ 0.1818 |
| | | $P_3(G_3)=$ 0.1818 | $P_4(G_3)=$ 0.1816 | $P_5(G_3)=$ 0.1816 |
| | | | $P_4(A_4)=$ 0.3189 | $P_5(A_4)=$ 0.319 |
| | | | $P_4(T_4)=$ 0.3178 | $P_5(T_4)=$ 0.3178 |
| | | | $P_4(C_4)=$ 0.1816 | $P_5(C_4)=$ 0.1815 |
| | | | $P_4(G_4)=$ 0.1818 | $P_5(G_4)=$ 0.1816 |
| | | | | $P_5(A_5)=$ 0.3192 |
| | | | | $P_5(T_5)=$ 0.3175 |
| | | | | $P_5(C_5)=$ 0.1815 |
| | | | | $P_5(G_5)=$ 0.1818 |

Fig. A5/3. The table of probabilities of subgroups of tetra-groups in the sequence: Arabidopsis thaliana chromosome 3 sequence,

NCBI Reference Sequence: NC_003074.8

LOCUS NC_003074 23459830 bp DNA linear CON 20-MAR-2017 DEFINITION Arabidopsis thaliana chromosome 3 sequence. ACCESSION NC_003074 VERSION NC_003074.8

https://www.ncbi.nlm.nih.gov/nuccore/NC_003074.8?report=fasta

| NUCLEOTIDES | DOUBLETS | TRIPLETS | 4-PLETS | 5-PLETS |
|---|---|---|---|---|
| $P_1(A_1)=$ 0.3197 | $P_2(A_1)=$ 0.3197 | $P_3(A_1)=$ 0.3198 | $P_4(A_1)=$ 0.3195 | $P_5(A_1)=$ 0.3199 |

| $P_1(T_1)=$ 0.3183 | $P_2(T_1)=$ 0.3182 | $P_3(T_1)=$ 0.318 | $P_4(T_1)=$ 0.3183 | $P_5(T_1)=$ 0.3183 |
|---|---|---|---|---|
| $P_1(C_1)=$ 0.1814 | $P_2(C_1)=$ 0.1814 | $P_3(C_1)=$ 0.1812 | $P_4(C_1)=$ 0.1815 | $P_5(C_1)=$ 0.1813 |
| $P_1(G_1)=$ 0.1806 | $P_2(G_1)=$ 0.1807 | $P_3(G_1)=$ 0.1810 | $P_4(G_1)=$ 0.1807 | $P_5(G_1)=$ 0.1806 |
| | $P_2(A_2)=$ 0.3196 | $P_3(A_2)=$ 0.3195 | $P_4(A_2)=$ 0.3197 | $P_5(A_2)=$ 0.3199 |
| | $P_2(T_2)=$ 0.3183 | $P_3(T_2)=$ 0.3183 | $P_4(T_2)=$ 0.3183 | $P_5(T_2)=$ 0.3184 |
| | $P_2(C_2)=$ 0.1815 | $P_3(C_2)=$ 0.1817 | $P_4(C_2)=$ 0.1815 | $P_5(C_2)=$ 0.1813 |
| | $P_2(G_2)=$ 0.1805 | $P_3(G_2)=$ 0.1805 | $P_4(G_2)=$ 0.1805 | $P_5(G_2)=$ 0.1804 |
| | | $P_3(A_3)=$ 0.3197 | $P_4(A_3)=$ 0.32 | $P_5(A_3)=$ 0.3195 |
| | | $P_3(T_3)=$ 0.3185 | $P_4(T_3)=$ 0.3181 | $P_5(T_3)=$ 0.3182 |
| | | $P_3(C_3)=$ 0.1814 | $P_4(C_3)=$ 0.1813 | $P_5(C_3)=$ 0.1817 |
| | | $P_3(G_3)=$ 0.1803 | $P_4(G_3)=$ 0.1807 | $P_5(G_3)=$ 0.1806 |
| | | | $P_4(A_4)=$ 0.3196 | $P_5(A_4)=$ 0.3197 |
| | | | $P_4(T_4)=$ 0.3184 | $P_5(T_4)=$ 0.3182 |
| | | | $P_4(C_4)=$ 0.1815 | $P_5(C_4)=$ 0.1813 |
| | | | $P_4(G_4)=$ 0.1805 | $P_5(G_4)=$ 0.1808 |
| | | | | $P_5(A_5)=$ 0.3196 |
| | | | | $P_5(T_5)=$ 0.3183 |
| | | | | $P_5(C_5)=$ 0.1816 |
| | | | | $P_5(G_5)=$ 0.1806 |

Fig. A5/4. The table of probabilities of subgroups of tetra-groups in the sequence: Arabidopsis thaliana chromosome 4 sequence,
NCBI Reference Sequence: NC_003075.7
LOCUS NC_003075 18585056 bp DNA linear CON 20-MAR-2017
DEFINITION Arabidopsis thaliana chromosome 4 sequence. ACCESSION NC_003075 VERSION NC_003075.7
https://www.ncbi.nlm.nih.gov/nuccore/NC_003075.7

| NUCLEOTIDES | DOUBLETS | TRIPLETS | 4-PLETS | 5-PLETS |
|---|---|---|---|---|
| $P_1(A_1)=$ 0.3197 | $P_2(A_1)=$ 0.3197 | $P_3(A_1)=$ 0.3196 | $P_4(A_1)=$ 0.3195 | $P_5(A_1)=$ 0.3195 |
| $P_1(T_1)=$ 0.3209 | $P_2(T_1)=$ 0.3208 | $P_3(T_1)=$ 0.321 | $P_4(T_1)=$ 0.3209 | $P_5(T_1)=$ 0.3208 |
| $P_1(C_1)=$ 0.1792 | $P_2(C_1)=$ 0.1793 | $P_3(C_1)=$ 0.1796 | $P_4(C_1)=$ 0.1794 | $P_5(C_1)=$ 0.1796 |
| $P_1(G_1)=$ 0.1802 | $P_2(G_1)=$ 0.1802 | $P_3(G_1)=$ 0.1797 | $P_4(G_1)=$ 0.1802 | $P_5(G_1)=$ 0.1802 |
| | $P_2(A_2)=$ 0.3198 | $P_3(A_2)=$ 0.3199 | $P_4(A_2)=$ 0.3198 | $P_5(A_2)=$ 0.3198 |
| | $P_2(T_2)=$ 0.3209 | $P_3(T_2)=$ 0.3206 | $P_4(T_2)=$ 0.3207 | $P_5(T_2)=$ 0.3208 |

| | $P_2(C_2)=$ 0.1791 | $P_3(C_2)=$ 0.179 | $P_4(C_2)=$ 0.1793 | $P_5(C_2)=$ 0.1792 |
|---|---|---|---|---|
| | $P_2(G_2)=$ 0.1802 | $P_3(G_2)=$ 0.1805 | $P_4(G_2)=$ 0.1802 | $P_5(G_2)=$ 0.1802 |
| | | $P_3(A_3)=$ 0.3198 | $P_4(A_3)=$ 0.3199 | $P_5(A_3)=$ 0.3201 |
| | | $P_3(T_3)=$ 0.321 | $P_4(T_3)=$ 0.3208 | $P_5(T_3)=$ 0.3208 |
| | | $P_3(C_3)=$ 0.179 | $P_4(C_3)=$ 0.1792 | $P_5(C_3)=$ 0.179 |
| | | $P_3(G_3)=$ 0.1803 | $P_4(G_3)=$ 0.1802 | $P_5(G_3)=$ 0.1801 |
| | | | $P_4(A_4)=$ 0.3199 | $P_5(A_4)=$ 0.3198 |
| | | | $P_4(T_4)=$ 0.3211 | $P_5(T_4)=$ 0.321 |
| | | | $P_4(C_4)=$ 0.1789 | $P_5(C_4)=$ 0.1791 |
| | | | $P_4(G_4)=$ 0.1801 | $P_5(G_4)=$ 0.1802 |
| | | | | $P_5(A_5)=$ 0.3196 |
| | | | | $P_5(T_5)=$ 0.3211 |
| | | | | $P_5(C_5)=$ 0.1791 |
| | | | | $P_5(G_5)=$ 0.1802 |

Fig. A5/5. The table of probabilities of subgroups of tetra-groups in the sequence: Arabidopsis thaliana chromosome 5 sequence,

NCBI Reference Sequence: NC_003076.8

LOCUS NC_003076 26975502 bp DNA linear CON 20-MAR-2017 DEFINITION Arabidopsis thaliana chromosome 5 sequence. ACCESSION NC_003076 VERSION NC_003076.8

https://www.ncbi.nlm.nih.gov/nuccore/NC_003076.8

## Appendix 6. Symmetries of tetra-group probabilities in the complete set of chromosomes of Mus musculus

The appendix 6 represents data about the fulfillment of the described tetra-group rules in the complete set of chromosomes of Mus musculus (house mouse). The laboratory mouse is a major model organism for basic mammalian biology, human disease, and genome evolution. All initial data are taken from the CenBank https://www.ncbi.nlm.nih.gov/genome?term=mus%20musculus. To show the fulfilment of the tetra-group rules of symmetries in texts at different initial levels of positional convolutions in the fractal genetic trees FGT-2, FGT-3, etc, tables of fluctucation intervals for values of probabilities for the set of all texts at each of levels for all chromosomes are given below (see the Section 11 about fractal genetic trees FGT-2, FGT-3, etc. and also about tables of fluctuation intervals). The fluctution intervals of the probabilities $P_n(A_k)$, $P_n(T_k)$, $P_n(C_k)$ and $P_n(G_k)$ for each initial long DNA-text are shown in the columns «Level 0».

| | Level | Level | Level | Level | Level | Level | Level |
|---|---|---|---|---|---|---|---|

|  | 0 | 1/0 | 1/1 | 2/00 | 2/01 | 2/10 | 2/11 |
|---|---|---|---|---|---|---|---|
| $P_n(A_k)\in$ | 0.2945÷0.2946 | 0.2945÷0.2948 | 0.2944÷0.2945 | 0.2944÷0.2945 | 0.2944÷0.2949 | 0.2943÷0.2948 | 0.2944÷0.2945 |
| $P_n(T_k)\in$ | 0.2939÷0.294 | 0.2939÷0.2939 | 0.2938÷0.294 | 0.2938÷0.2941 | 0.2937÷0.294 | 0.2937÷0.2938 | 0.2939÷0.2942 |
| $P_n(C_k)\in$ | 0.2057÷0.2058 | 0.2057÷0.2057 | 0.2058÷0.2058 | 0.2056÷0.2058 | 0.2056÷0.2056 | 0.2058÷0.2058 | 0.2057÷0.2058 |
| $P_n(G_k)\in$ | 0.2055÷0.2057 | 0.2055÷0.2056 | 0.2055÷0.2057 | 0.2056÷0.2056 | 0.2054÷0.2054 | 0.2056÷0.2057 | 0.2055÷0.2056 |

|  | Level 0 | Level 1/0 | Level 1/1 | Level 1/2 | Level 2/00 | Level 2/01 |
|---|---|---|---|---|---|---|
| $P_n(A_k)\in$ | 0.2945÷0.2946 | 0.2943÷0.2945 | 0.2945÷0.2946 | 0.2944÷0.2944 | 0.2942÷0.2942 | 0.2942÷0.2945 |
| $P_n(T_k)\in$ | 0.2939÷0.294 | 0.2938÷0.294 | 0.2938÷0.2938 | 0.2938÷0.294 | 0.2936÷0.2942 | 0.2937÷0.2942 |
| $P_n(C_k)\in$ | 0.2057÷0.2058 | 0.2057÷0.206 | 0.2057÷0.2058 | 0.2057÷0.2059 | 0.2057÷0.2058 | 0.2052÷0.2059 |
| $P_n(G_k)\in$ | 0.2055÷0.2057 | 0.2055÷0.2055 | 0.2054÷0.2057 | 0.2056÷0.2058 | 0.205÷0.2058 | 0.2054÷0.2054 |

|  | Level 2/02 | Level 2/10 | Level 2/11 | Level 2/12 | Level 2/20 | Level 2/21 | Level 2/22 |
|---|---|---|---|---|---|---|---|
| $P_n(A_k)\in$ | 0.2943÷0.2945 | 0.2944÷0.2949 | 0.2944÷0.2947 | 0.2945÷0.2946 | 0.2943÷0.2946 | 0.2943÷0.2943 | 0.2943÷0.2944 |
| $P_n(T_k)\in$ | 0.2938÷0.2942 | 0.2938÷0.2939 | 0.2936÷0.2941 | 0.2937÷0.2937 | 0.2938÷0.2938 | 0.2936÷0.294 | 0.2938÷0.2942 |
| $P_n(C_k)\in$ | 0.2056÷0.2058 | 0.2056÷0.2057 | 0.2055÷0.2055 | 0.2055÷0.206 | 0.2056÷0.2058 | 0.2055÷0.2058 | 0.2056÷0.2057 |
| $P_n(G_k)\in$ | 0.2052÷0.2055 | 0.2051÷0.2055 | 0.2054÷0.2057 | 0.2052÷0.2058 | 0.2056÷0.2058 | 0.2054÷0.2058 | 0.2054÷0.2057 |

Fig. A6/1. Tables of fluctuation intervals of probabilities $P_n(A_k)$, $P_n(T_k)$, $P_n(C_k)$ and $P_n(G_k)$ for the set of all texts at each of levels of convolutions in the FGT-2 (upper table) and FGT-3 (bottom tables) in the case of the sequence: Mus musculus strain C57BL/6J chromosome 1, GRCm38.p4 C57BL/6J. NCBI Reference Sequence: NC_000067.6. 195471971 bp. https://www.ncbi.nlm.nih.gov/nuccore/NC_000067.6. Here n = 1, 2, 3, 4, 5, k ≤ n. Level 0 corresponds to the initial DNA-text with its different n-letter representations. Other levels in this table correspond to sets of all daughter texts at appropriate levels of positional convolutions in the FGT-2 and the FGT-3.

| | Level 0 | Level 1/0 | Level 1/1 | Level 2/00 | Level 2/01 | Level 2/10 | Level 2/11 |
|---|---|---|---|---|---|---|---|
| $P_n(A_k)\in$ | 0.2892÷ 0.2894 | 0.2893÷ 0.2895 | 0.2892÷ 0.2895 | 0.2891÷ 0.2893 | 0.2893÷ 0.2894 | 0.2892÷ 0.2892 | 0.2892÷ 0.2892 |
| $P_n(T_k)\in$ | 0.2897÷ 0.2898 | 0.2897÷ 0.2898 | 0.2897÷ 0.2899 | 0.2896÷ 0.2896 | 0.2896÷ 0.2897 | 0.2896÷ 0.2898 | 0.2897÷ 0.2899 |
| $P_n(C_k)\in$ | 0.2102÷ 0.2103 | 0.2101÷ 0.2103 | 0.2103÷ 0.2104 | 0.2099÷ 0.2104 | 0.21÷ 0.2105 | 0.2102÷ 0.2103 | 0.2102÷ 0.2105 |
| $P_n(G_k)\in$ | 0.2104÷ 0.2105 | 0.2104÷ 0.2104 | 0.2104÷ 0.2105 | 0.2104÷ 0.2107 | 0.2104÷ 0.2104 | 0.2103÷ 0.2107 | 0.2104÷ 0.2104 |

| | Level 0 | Level 1/0 | Level 1/1 | Level 1/2 | Level 2/00 | Level 2/01 |
|---|---|---|---|---|---|---|
| $P_n(A_k)\in$ | 0.2892÷ 0.2894 | 0.2892÷ 0.2895 | 0.2892÷ 0.2895 | 0.289÷ 0.2895 | 0.2891÷ 0.2893 | 0.2893÷ 0.2897 |
| $P_n(T_k)\in$ | 0.2897÷ 0.2898 | 0.2896÷ 0.2897 | 0.2896÷ 0.2897 | 0.2897÷ 0.2898 | 0.2895÷ 0.2895 | 0.2892÷ 0.2897 |
| $P_n(C_k)\in$ | 0.2102÷ 0.2103 | 0.2101÷ 0.2102 | 0.2101÷ 0.2104 | 0.2102÷ 0.2103 | 0.2099÷ 0.2106 | 0.2101÷ 0.2103 |
| $P_n(G_k)\in$ | 0.2104÷ 0.2105 | 0.2103÷ 0.2106 | 0.2104÷ 0.2104 | 0.2103÷ 0.2104 | 0.2103÷ 0.2106 | 0.2099÷ 0.2103 |

| | Level 2/02 | Level 2/10 | Level 2/11 | Level 2/12 | Level 2/20 | Level 2/21 | Level 2/22 |
|---|---|---|---|---|---|---|---|
| $P_n(A_k)\in$ | 0.289÷ 0.2894 | 0.2892÷ 0.2896 | 0.2889÷ 0.2893 | 0.2889÷ 0.2896 | 0.2889÷ 0.2893 | 0.2891÷ 0.2891 | 0.2889÷ 0.2896 |
| $P_n(T_k)\in$ | 0.2896÷ 0.2898 | 0.2894÷ 0.2894 | 0.2896÷ 0.2899 | 0.2894÷ 0.2896 | 0.2896÷ 0.2898 | 0.2896÷ 0.2902 | 0.2894÷ 0.2898 |
| $P_n(C_k)\in$ | 0.2099÷ 0.2102 | 0.21÷ 0.2105 | 0.21÷ 0.2101 | 0.2101÷ 0.2103 | 0.21÷ 0.2105 | 0.2103÷ 0.2107 | 0.2102÷ 0.2103 |
| $P_n(G_k)\in$ | 0.2104÷ 0.2106 | 0.2102÷ 0.2105 | 0.2105÷ 0.2108 | 0.2105÷ 0.2105 | 0.2101÷ 0.2104 | 0.21÷ 0.21 | 0.2101÷ 0.2103 |

Fig. A6/2. Tables of fluctuation intervals of probabilities $P_n(A_k)$, $P_n(T_k)$, $P_n(C_k)$ and $P_n(G_k)$ for the set of all texts at each of levels of convolutions in the FGT-2 (upper table) and FGT-3 (bottom tables) in the case of the sequence: Mus musculus strain C57BL/6J chromosome 2, GRCm38.p4 C57BL/6J, NCBI Reference Sequence: NC_000068.7. 182113224 bp. https://www.ncbi.nlm.nih.gov/nuccore/NC_000068.7. Here n = 1, 2, 3, 4, 5, k ≤ n. Level 0 corresponds to the initial DNA-text with its different n-letter representations. Other levels in this table correspond to sets of all daughter texts at appropriate levels of positional convolutions in the FGT-2 and the FGT-3.

|  | Level 0 | Level 1/0 | Level 1/1 | Level 2/00 | Level 2/01 | Level 2/10 | Level 2/11 |
|---|---|---|---|---|---|---|---|
| $P_n(A_k)\in$ | 0.2973÷ 0.2974 | 0.2973÷ 0.2973 | 0.2972÷ 0.2973 | 0.2971÷ 0.2972 | 0.2971÷ 0.2975 | 0.2972÷ 0.2975 | 0.2971÷ 0.2971 |
| $P_n(T_k)\in$ | 0.298÷ 0.2982 | 0.298÷ 0.2982 | 0.298÷ 0.2982 | 0.298÷ 0.2982 | 0.2979÷ 0.2982 | 0.2979÷ 0.298 | 0.298÷ 0.2983 |
| $P_n(C_k)\in$ | 0.2019÷ 0.2021 | 0.2019÷ 0.202 | 0.2019÷ 0.2021 | 0.2018÷ 0.2022 | 0.2019÷ 0.2019 | 0.2019÷ 0.2019 | 0.2018÷ 0.2023 |
| $P_n(G_k)\in$ | 0.2024÷ 0.2024 | 0.2023÷ 0.2025 | 0.2022÷ 0.2025 | 0.2022÷ 0.2024 | 0.2021÷ 0.2024 | 0.2024÷ 0.2026 | 0.2021÷ 0.2024 |

|  | Level 0 | Level 1/0 | Level 1/1 | Level 1/2 | Level 2/00 | Level 2/01 |
|---|---|---|---|---|---|---|
| $P_n(A_k)\in$ | 0.2973÷ 0.2974 | 0.2972÷ 0.2972 | 0.2973÷ 0.2976 | 0.2971÷ 0.2974 | 0.2969÷ 0.2972 | 0.2972÷ 0.2978 |
| $P_n(T_k)\in$ | 0.298÷ 0.2982 | 0.2979÷ 0.2981 | 0.2981÷ 0.2984 | 0.2981÷ 0.2981 | 0.2978÷ 0.2978 | 0.2977÷ 0.298 |
| $P_n(C_k)\in$ | 0.2019÷ 0.2021 | 0.202÷ 0.2021 | 0.2017÷ 0.2017 | 0.2019÷ 0.202 | 0.2019÷ 0.2022 | 0.2017÷ 0.2023 |
| $P_n(G_k)\in$ | 0.2024÷ 0.2024 | 0.2021÷ 0.2025 | 0.2022÷ 0.2022 | 0.2023÷ 0.2025 | 0.2022÷ 0.2028 | 0.202÷ 0.202 |

|  | Level 2/02 | Level 2/10 | Level 2/11 | Level 2/12 | Level 2/20 | Level 2/21 | Level 2/22 |
|---|---|---|---|---|---|---|---|
| $P_n(A_k)\in$ | 0.2971÷ 0.2971 | 0.2971÷ 0.2973 | 0.297÷ 0.2973 | 0.2973÷ 0.2977 | 0.2968÷ 0.2972 | 0.2971÷ 0.2976 | 0.2969÷ 0.2973 |
| $P_n(T_k)\in$ | 0.2976÷ 0.2978 | 0.298÷ 0.2985 | 0.2979÷ 0.2983 | 0.2979÷ 0.2987 | 0.2979÷ 0.2982 | 0.298÷ 0.2983 | 0.2979÷ 0.2979 |
| $P_n(C_k)\in$ | 0.202÷ 0.2023 | 0.2016÷ 0.2017 | 0.2018÷ 0.202 | 0.2016÷ 0.2016 | 0.2019÷ 0.2022 | 0.2019÷ 0.2021 | 0.2019÷ 0.2022 |
| $P_n(G_k)\in$ | 0.2021÷ 0.2028 | 0.2022÷ 0.2025 | 0.2023÷ 0.2024 | 0.2019÷ 0.202 | 0.2023÷ 0.2025 | 0.2021÷ 0.2021 | 0.2022÷ 0.2026 |

Fig. A6/3. Tables of fluctuation intervals of probabilities $P_n(A_k)$, $P_n(T_k)$, $P_n(C_k)$ and $P_n(G_k)$ for the set of all texts at each of levels of convolutions in the FGT-2 (upper table) and FGT-3 (bottom tables) in the case of the sequence: Mus musculus strain C57BL/6J chromosome 3, GRCm38.p4 C57BL/6J. NCBI Reference Sequence: NC_000069.6. 160039680 bp. https://www.ncbi.nlm.nih.gov/nuccore/NC_000069.6. Here n = 1, 2, 3, 4, 5, k ≤ n. Level 0 corresponds to the initial DNA-text with its different n-letter representations. Other levels in this table correspond to sets of all daughter texts at appropriate levels of positional convolutions in the FGT-2 and the FGT-3.

|  | Level 0 | Level 1/0 | Level 1/1 | Level 2/00 | Level 2/01 | Level 2/10 | Level 2/11 |
|---|---|---|---|---|---|---|---|
| $P_n(A_k)\in$ | 0.2881÷ 0.2882 | 0.2881÷ 0.2882 | 0.288÷ 0.2882 | 0.2879÷ 0.2884 | 0.288÷ 0.2884 | 0.288÷ 0.2882 | 0.2879÷ 0.2881 |
| $P_n(T_k)\in$ | 0.2887÷ 0.289 | 0.2887÷ 0.2888 | 0.2887÷ 0.2891 | 0.2886÷ 0.2886 | 0.2886÷ 0.2888 | 0.2885÷ 0.289 | 0.2889÷ 0.2891 |
| $P_n(C_k)\in$ | 0.2113÷ 0.2114 | 0.2113÷ 0.2114 | 0.2113÷ 0.2114 | 0.2112÷ 0.2114 | 0.2112÷ 0.2113 | 0.2113÷ 0.2114 | 0.2113÷ 0.2114 |
| $P_n(G_k)\in$ | 0.2114÷ 0.2114 | 0.2114÷ 0.2116 | 0.2114÷ 0.2114 | 0.2114÷ 0.2115 | 0.2114÷ 0.2116 | 0.2113÷ 0.2114 | 0.2114÷ 0.2115 |

|  | Level 0 | Level 1/0 | Level 1/1 | Level 1/2 | Level 2/00 | Level 2/01 |
|---|---|---|---|---|---|---|
| $P_n(A_k)\in$ | 0.2881÷ 0.2882 | 0.2881÷ 0.2882 | 0.2879÷ 0.288 | 0.2881÷ 0.2881 | 0.2879÷ 0.2882 | 0.288÷ 0.2887 |
| $P_n(T_k)\in$ | 0.2887÷ 0.289 | 0.2886÷ 0.2888 | 0.2887÷ 0.2887 | 0.2887÷ 0.2889 | 0.2886÷ 0.2888 | 0.2884÷ 0.2887 |
| $P_n(C_k)\in$ | 0.2113÷ 0.2114 | 0.2113÷ 0.2114 | 0.2112÷ 0.2115 | 0.2113÷ 0.2115 | 0.2114÷ 0.2117 | 0.2111÷ 0.2111 |
| $P_n(G_k)\in$ | 0.2114÷ 0.2114 | 0.2113÷ 0.2116 | 0.2114÷ 0.2117 | 0.2113÷ 0.2115 | 0.2111÷ 0.2113 | 0.2113÷ 0.2115 |

|  | Level 2/02 | Level 2/10 | Level 2/11 | Level 2/12 | Level 2/20 | Level 2/21 | Level 2/22 |
|---|---|---|---|---|---|---|---|
| $P_n(A_k)\in$ | 0.2881÷ 0.2881 | 0.2877÷ 0.288 | 0.2878÷ 0.2878 | 0.2876÷ 0.2879 | 0.2878÷ 0.2883 | 0.2881÷ 0.2882 | 0.2881÷ 0.2881 |
| $P_n(T_k)\in$ | 0.2886÷ 0.2892 | 0.2887÷ 0.2888 | 0.2886÷ 0.2891 | 0.2884÷ 0.2887 | 0.2886÷ 0.2888 | 0.2886÷ 0.2887 | 0.2886÷ 0.2892 |
| $P_n(C_k)\in$ | 0.2112÷ 0.2113 | 0.2111÷ 0.2111 | 0.2111÷ 0.2117 | 0.2113÷ 0.2116 | 0.211÷ 0.211 | 0.211÷ 0.2115 | 0.2112÷ 0.2112 |
| $P_n(G_k)\in$ | 0.2111÷ 0.2114 | 0.2113÷ 0.2122 | 0.2112÷ 0.2115 | 0.2114÷ 0.2118 | 0.2111÷ 0.2119 | 0.2113÷ 0.2115 | 0.2111÷ 0.2115 |

Fig. A6/4. Tables of fluctuation intervals of probabilities $P_n(A_k)$, $P_n(T_k)$, $P_n(C_k)$ and $P_n(G_k)$ for the set of all texts at each of levels of convolutions in the FGT-2 (upper table) and FGT-3 (bottom tables) in the case of the sequence: Mus musculus strain C57BL/6J chromosome 4, GRCm38.p4 C57BL/6J. NCBI Reference Sequence: NC_000070.6. 156508116 bp. https://www.ncbi.nlm.nih.gov/nuccore/NC_000070.6. Here n = 1, 2, 3, 4, 5, k ≤ n. Level 0 corresponds to the initial DNA-text with its different n-letter representations. Other levels in this table correspond to sets of all daughter texts at appropriate levels of positional convolutions in the FGT-2 and the FGT-3.

|  | Level 0 | Level 1/0 | Level 1/1 | Level 2/00 | Level 2/01 | Level 2/10 | Level 2/11 |
|---|---|---|---|---|---|---|---|
| $P_n(A_k)\in$ | 0.2872÷ 0.2872 | 0.2872÷ 0.2873 | 0.2871÷ 0.2872 | 0.2871÷ 0.2873 | 0.2872÷ 0.2873 | 0.2869÷ 0.287 | 0.287÷ 0.2872 |
| $P_n(T_k)\in$ | 0.2873÷ 0.2875 | 0.2872÷ 0.2873 | 0.2873÷ 0.2877 | 0.2871÷ 0.2872 | 0.2872÷ 0.2874 | 0.2872÷ 0.2874 | 0.2874÷ 0.2877 |
| $P_n(C_k)\in$ | 0.2125÷ 0.2126 | 0.2126÷ 0.2126 | 0.2123÷ 0.2125 | 0.2125÷ 0.2128 | 0.2124÷ 0.2127 | 0.2122÷ 0.2129 | 0.2124÷ 0.2125 |
| $P_n(G_k)\in$ | 0.2126÷ 0.2127 | 0.2126÷ 0.2127 | 0.2125÷ 0.2126 | 0.2125÷ 0.2127 | 0.2125÷ 0.2126 | 0.2125÷ 0.2128 | 0.2124÷ 0.2126 |

|  | Level 0 | Level 1/0 | Level 1/1 | Level 1/2 | Level 2/00 | Level 2/01 |
|---|---|---|---|---|---|---|
| $P_n(A_k)\in$ | 0.2872÷ 0.2872 | 0.287÷ 0.2871 | 0.2871÷ 0.2873 | 0.2872÷ 0.2873 | 0.2868÷ 0.2868 | 0.287÷ 0.2872 |
| $P_n(T_k)\in$ | 0.2873÷ 0.2875 | 0.2873÷ 0.2875 | 0.2872÷ 0.2873 | 0.2873÷ 0.2877 | 0.287÷ 0.2876 | 0.2872÷ 0.2873 |
| $P_n(C_k)\in$ | 0.2125÷ 0.2126 | 0.2125÷ 0.2128 | 0.2125÷ 0.2127 | 0.2124÷ 0.2125 | 0.2123÷ 0.2125 | 0.2124÷ 0.2129 |
| $P_n(G_k)\in$ | 0.2126÷ 0.2127 | 0.2126÷ 0.2126 | 0.2126÷ 0.2127 | 0.2124÷ 0.2125 | 0.2126÷ 0.2131 | 0.2124÷ 0.2126 |

|  | Level 2/02 | Level 2/10 | Level 2/11 | Level 2/12 | Level 2/20 | Level 2/21 | Level 2/22 |
|---|---|---|---|---|---|---|---|
| $P_n(A_k)\in$ | 0.2868÷ 0.2871 | 0.287÷ 0.2876 | 0.2867÷ 0.2872 | 0.2872÷ 0.2873 | 0.287÷ 0.2871 | 0.2869÷ 0.2872 | 0.287÷ 0.2871 |
| $P_n(T_k)\in$ | 0.2868÷ 0.2874 | 0.2869÷ 0.2875 | 0.2871÷ 0.2879 | 0.287÷ 0.2872 | 0.2873÷ 0.2876 | 0.2871÷ 0.2872 | 0.2872÷ 0.288 |
| $P_n(C_k)\in$ | 0.2125÷ 0.2131 | 0.2123÷ 0.2123 | 0.2124÷ 0.2124 | 0.2124÷ 0.2128 | 0.2124÷ 0.2127 | 0.2123÷ 0.2129 | 0.2123÷ 0.2126 |
| $P_n(G_k)\in$ | 0.2121÷ 0.2123 | 0.2125÷ 0.2125 | 0.2125÷ 0.2125 | 0.2125÷ 0.2127 | 0.2123÷ 0.2127 | 0.2124÷ 0.2127 | 0.2121÷ 0.2123 |

Fig. A6/5. Tables of fluctuation intervals of probabilities $P_n(A_k)$, $P_n(T_k)$, $P_n(C_k)$ and $P_n(G_k)$ for the set of all texts at each of levels of convolutions in the FGT-2 (upper table) and FGT-3 (bottom tables) in the case of the sequence: Mus musculus strain C57BL/6J chromosome 5, GRCm38.p4 C57BL/6J. NCBI Reference Sequence: NC_000071.6. 151834684 bp. https://www.ncbi.nlm.nih.gov/nuccore/NC_000071.6. Here n = 1, 2, 3, 4, 5, k ≤ n. Level 0 corresponds to the initial DNA-text with its different n-letter representations. Other levels in this table correspond to sets of all daughter texts at appropriate levels of positional convolutions in the FGT-2 and the FGT-3.

| | Level 0 | Level 1/0 | Level 1/1 | Level 2/00 | Level 2/01 | Level 2/10 | Level 2/11 |
|---|---|---|---|---|---|---|---|
| $P_n(A_k)\in$ | 0.2927÷ 0.2928 | 0.2927÷ 0.2927 | 0.2926÷ 0.2929 | 0.2927÷ 0.2928 | 0.2926÷ 0.2927 | 0.2925÷ 0.2928 | 0.2925÷ 0.293 |
| $P_n(T_k)\in$ | 0.293÷ 0.293 | 0.293÷ 0.2932 | 0.2928÷ 0.2928 | 0.2929÷ 0.2929 | 0.2931÷ 0.2932 | 0.2928÷ 0.293 | 0.2927÷ 0.2928 |
| $P_n(C_k)\in$ | 0.2071÷ 0.2072 | 0.207÷ 0.2072 | 0.2071÷ 0.2072 | 0.2069÷ 0.2074 | 0.2066÷ 0.2071 | 0.2071÷ 0.2072 | 0.207÷ 0.2071 |
| $P_n(G_k)\in$ | 0.2069÷ 0.207 | 0.2068÷ 0.207 | 0.2069÷ 0.2071 | 0.2067÷ 0.2069 | 0.2067÷ 0.207 | 0.2069÷ 0.207 | 0.2069÷ 0.2072 |

| | Level 0 | Level 1/0 | Level 1/1 | Level 1/2 | Level 2/00 | Level 2/01 |
|---|---|---|---|---|---|---|
| $P_n(A_k)\in$ | 0.2927÷ 0.2928 | 0.2925÷ 0.2929 | 0.2926÷ 0.2928 | 0.2927÷ 0.293 | 0.2925÷ 0.2925 | 0.2921÷ 0.2929 |
| $P_n(T_k)\in$ | 0.293÷ 0.293 | 0.2931÷ 0.2931 | 0.2927÷ 0.2929 | 0.2929÷ 0.2929 | 0.2928÷ 0.293 | 0.2928÷ 0.2932 |
| $P_n(C_k)\in$ | 0.2071÷ 0.2072 | 0.207÷ 0.207 | 0.2071÷ 0.2072 | 0.207÷ 0.2072 | 0.2068÷ 0.2071 | 0.2069÷ 0.2069 |
| $P_n(G_k)\in$ | 0.2069÷ 0.207 | 0.2068÷ 0.207 | 0.207÷ 0.2071 | 0.2068÷ 0.2068 | 0.2067÷ 0.2073 | 0.2067÷ 0.207 |

| | Level 2/02 | Level 2/10 | Level 2/11 | Level 2/12 | Level 2/20 | Level 2/21 | Level 2/22 |
|---|---|---|---|---|---|---|---|
| $P_n(A_k)\in$ | 0.2925÷ 0.2926 | 0.2925÷ 0.2929 | 0.2925÷ 0.2927 | 0.2925÷ 0.2925 | 0.2927÷ 0.2928 | 0.2925÷ 0.293 | 0.2923÷ 0.2927 |
| $P_n(T_k)\in$ | 0.293÷ 0.293 | 0.2926÷ 0.2929 | 0.2926÷ 0.2927 | 0.2928÷ 0.2929 | 0.2928÷ 0.2933 | 0.2929÷ 0.2933 | 0.2926÷ 0.2929 |
| $P_n(C_k)\in$ | 0.2069÷ 0.2071 | 0.2069÷ 0.2073 | 0.2071÷ 0.2072 | 0.207÷ 0.2073 | 0.207÷ 0.2073 | 0.2069÷ 0.2069 | 0.2068÷ 0.2073 |
| $P_n(G_k)\in$ | 0.2068÷ 0.2074 | 0.2067÷ 0.2069 | 0.2069÷ 0.2074 | 0.2069÷ 0.2073 | 0.2064÷ 0.2067 | 0.2066÷ 0.2068 | 0.2068÷ 0.207 |

Fig. A6/6. Tables of fluctuation intervals of probabilities $P_n(A_k)$, $P_n(T_k)$, $P_n(C_k)$ and $P_n(G_k)$ for the set of all texts at each of levels of convolutions in the FGT-2 (upper table) and FGT-3 (bottom tables) in the case of the sequence: Mus musculus strain C57BL/6J chromosome 6, GRCm38.p4 C57BL/6J. NCBI Reference Sequence: NC_000072.6. 149736546 bp. https://www.ncbi.nlm.nih.gov/nuccore/NC_000072.6. Here n = 1, 2, 3, 4, 5, k ≤ n. Level 0 corresponds to the initial DNA-text with its different n-letter representations. Other levels in this table correspond to sets of all daughter texts at appropriate levels of positional convolutions in the FGT-2 and the FGT-3.

|  | Level 0 | Level 1/0 | Level 1/1 | Level 2/00 | Level 2/01 | Level 2/10 | Level 2/11 |
|---|---|---|---|---|---|---|---|
| $P_n(A_k) \in$ | 0.2837÷0.284 | 0.2839÷0.284 | 0.2836÷0.2839 | 0.2838÷0.2838 | 0.2839÷0.2841 | 0.2835÷0.2836 | 0.2837÷0.2838 |
| $P_n(T_k) \in$ | 0.2855÷0.2855 | 0.2854÷0.2854 | 0.2855÷0.2855 | 0.2853÷0.2855 | 0.2853÷0.2854 | 0.2854÷0.2857 | 0.2854÷0.2857 |
| $P_n(C_k) \in$ | 0.2152÷0.2153 | 0.2152÷0.2155 | 0.2153÷0.2153 | 0.2153÷0.2157 | 0.215÷0.2154 | 0.2153÷0.2155 | 0.2152÷0.2153 |
| $P_n(G_k) \in$ | 0.215÷0.2151 | 0.215÷0.2151 | 0.2151÷0.2153 | 0.2149÷0.215 | 0.215÷0.2152 | 0.2148÷0.2153 | 0.2151÷0.2151 |

|  | Level 0 | Level 1/0 | Level 1/1 | Level 1/2 | Level 2/00 | Level 2/01 |
|---|---|---|---|---|---|---|
| $P_n(A_k) \in$ | 0.2837÷0.284 | 0.2838÷0.2839 | 0.2836÷0.2836 | 0.2836÷0.284 | 0.2837÷0.2837 | 0.2837÷0.2839 |
| $P_n(T_k) \in$ | 0.2855÷0.2855 | 0.2853÷0.2855 | 0.2853÷0.2856 | 0.2855÷0.2856 | 0.2851÷0.2853 | 0.2853÷0.2857 |
| $P_n(C_k) \in$ | 0.2152÷0.2153 | 0.2152÷0.2154 | 0.2153÷0.2156 | 0.215÷0.2152 | 0.215÷0.2156 | 0.215÷0.2152 |
| $P_n(G_k) \in$ | 0.215÷0.2151 | 0.215÷0.2152 | 0.215÷0.2151 | 0.2149÷0.2152 | 0.2149÷0.2154 | 0.2148÷0.2151 |

|  | Level 2/02 | Level 2/10 | Level 2/11 | Level 2/12 | Level 2/20 | Level 2/21 | Level 2/22 |
|---|---|---|---|---|---|---|---|
| $P_n(A_k) \in$ | 0.2834÷0.2834 | 0.2836÷0.2837 | 0.2833÷0.284 | 0.2834÷0.2834 | 0.2834÷0.2839 | 0.2835÷0.2836 | 0.2836÷0.2839 |
| $P_n(T_k) \in$ | 0.2852÷0.2859 | 0.2853÷0.2858 | 0.285÷0.2852 | 0.2853÷0.2858 | 0.2856÷0.2856 | 0.2855÷0.2857 | 0.2853÷0.2856 |
| $P_n(C_k) \in$ | 0.2149÷0.2155 | 0.2151÷0.2154 | 0.2155÷0.2156 | 0.2151÷0.2154 | 0.2149÷0.2157 | 0.2151÷0.2153 | 0.2149÷0.2154 |
| $P_n(G_k) \in$ | 0.2149÷0.2152 | 0.2148÷0.2152 | 0.2151÷0.2151 | 0.2148÷0.2153 | 0.2147÷0.2148 | 0.2148÷0.2154 | 0.2147÷0.2151 |

Fig. A6/7. Tables of fluctuation intervals of probabilities $P_n(A_k)$, $P_n(T_k)$, $P_n(C_k)$ and $P_n(G_k)$ for the set of all texts at each of levels of convolutions in the FGT-2 (upper table) and FGT-3 (bottom tables) in the case of the sequence: Mus musculus strain C57BL/6J chromosome 7, GRCm38.p4 C57BL/6J. NCBI Reference Sequence: NC_000073.6. 145441459 bp . https://www.ncbi.nlm.nih.gov/nuccore/NC_000073.6. Here n = 1, 2, 3, 4, 5, k ≤ n. Level 0 corresponds to the initial DNA-text with its different n-letter representations. Other levels in this table correspond to sets of all daughter texts at appropriate levels of positional convolutions in the FGT-2 and the FGT-3.

| | Level 0 | Level 1/0 | Level 1/1 | Level 2/00 | Level 2/01 | Level 2/10 | Level 2/11 |
|---|---|---|---|---|---|---|---|
| $P_n(A_k)\in$ | 0.2883÷ 0.2884 | 0.2882÷ 0.2884 | 0.2882÷ 0.2885 | 0.2882÷ 0.2883 | 0.288÷ 0.288 | 0.2882÷ 0.2885 | 0.2882÷ 0.2885 |
| $P_n(T_k)\in$ | 0.2878÷ 0.2879 | 0.2877÷ 0.2879 | 0.2878÷ 0.288 | 0.2878÷ 0.2878 | 0.2874÷ 0.288 | 0.2877÷ 0.2881 | 0.2876÷ 0.288 |
| $P_n(C_k)\in$ | 0.2118÷ 0.2119 | 0.2118÷ 0.2119 | 0.2117÷ 0.2117 | 0.2117÷ 0.2122 | 0.2119÷ 0.212 | 0.2116÷ 0.212 | 0.2116÷ 0.2118 |
| $P_n(G_k)\in$ | 0.2117÷ 0.2119 | 0.2117÷ 0.2118 | 0.2115÷ 0.2118 | 0.2115÷ 0.2117 | 0.2116÷ 0.212 | 0.2114÷ 0.2115 | 0.2116÷ 0.2117 |

| | Level 0 | Level 1/0 | Level 1/1 | Level 1/2 | Level 2/00 | Level 2/01 |
|---|---|---|---|---|---|---|
| $P_n(A_k)\in$ | 0.2883÷ 0.2884 | 0.2881÷ 0.2882 | 0.2883÷ 0.2885 | 0.2882÷ 0.2883 | 0.2881÷ 0.2882 | 0.2879÷ 0.2887 |
| $P_n(T_k)\in$ | 0.2878÷ 0.2879 | 0.2878÷ 0.2881 | 0.2877÷ 0.2878 | 0.2878÷ 0.2879 | 0.2875÷ 0.288 | 0.2876÷ 0.2877 |
| $P_n(C_k)\in$ | 0.2118÷ 0.2119 | 0.2117÷ 0.2118 | 0.2116÷ 0.2117 | 0.2118÷ 0.2118 | 0.2116÷ 0.2118 | 0.2115÷ 0.2118 |
| $P_n(G_k)\in$ | 0.2117÷ 0.2119 | 0.2116÷ 0.2119 | 0.2116÷ 0.212 | 0.2114÷ 0.212 | 0.2114÷ 0.2119 | 0.2115÷ 0.2117 |

| | Level 2/02 | Level 2/10 | Level 2/11 | Level 2/12 | Level 2/20 | Level 2/21 | Level 2/22 |
|---|---|---|---|---|---|---|---|
| $P_n(A_k)\in$ | 0.2876÷ 0.2882 | 0.2883÷ 0.2887 | 0.2882÷ 0.2883 | 0.2879÷ 0.2887 | 0.2881÷ 0.2881 | 0.288÷ 0.289 | 0.2881÷ 0.2885 |
| $P_n(T_k)\in$ | 0.2877÷ 0.2881 | 0.2877÷ 0.288 | 0.2876÷ 0.2879 | 0.2873÷ 0.2874 | 0.2878÷ 0.2884 | 0.2878÷ 0.2878 | 0.2876÷ 0.2877 |
| $P_n(C_k)\in$ | 0.2117÷ 0.2117 | 0.2116÷ 0.2117 | 0.2116÷ 0.2118 | 0.2113÷ 0.2119 | 0.2115÷ 0.2118 | 0.2118÷ 0.2118 | 0.2115÷ 0.2118 |
| $P_n(G_k)\in$ | 0.2116÷ 0.2119 | 0.2115÷ 0.2116 | 0.2116÷ 0.212 | 0.2113÷ 0.212 | 0.2115÷ 0.2117 | 0.2114÷ 0.2114 | 0.2111÷ 0.212 |

Fig. A6/8. Tables of fluctuation intervals of probabilities $P_n(A_k)$, $P_n(T_k)$, $P_n(C_k)$ and $P_n(G_k)$ for the set of all texts at each of levels of convolutions in the FGT-2 (upper table) and FGT-3 (bottom tables) in the case of the sequence: Mus musculus strain C57BL/6J, chromosome 8, GRCm38.p4 C57BL/6J. NCBI Reference Sequence: NC_000074.6. 129401213 bp. https://www.ncbi.nlm.nih.gov/nuccore/NC_000074.6 . Here n = 1, 2, 3, 4, 5, k ≤ n. Level 0 corresponds to the initial DNA-text with its different n-letter representations. Other levels in this table correspond to sets of all daughter texts at appropriate levels of positional convolutions in the FGT-2 and the FGT-3.

| | Level 0 | Level 1/0 | Level 1/1 | Level 2/00 | Level 2/01 | Level 2/10 | Level 2/11 |
|---|---|---|---|---|---|---|---|
| $P_n(A_k)\in$ | 0.2865÷0.2866 | 0.2863÷0.2866 | 0.2864÷0.2867 | 0.2863÷0.2866 | 0.2861÷0.2865 | 0.2863÷0.2864 | 0.2865÷0.2867 |
| $P_n(T_k)\in$ | 0.2862÷0.2864 | 0.2863÷0.2865 | 0.2861÷0.2863 | 0.286÷0.2864 | 0.2863÷0.2866 | 0.2861÷0.2865 | 0.2859÷0.2862 |
| $P_n(C_k)\in$ | 0.2135÷0.2135 | 0.2133÷0.2135 | 0.2135÷0.2135 | 0.2132÷0.2134 | 0.2134÷0.2136 | 0.2134÷0.2137 | 0.2133÷0.2136 |
| $P_n(G_k)\in$ | 0.2134÷0.2134 | 0.2133÷0.2133 | 0.2134÷0.2135 | 0.2133÷0.2135 | 0.2132÷0.2133 | 0.2134÷0.2134 | 0.2133÷0.2134 |

| | Level 0 | Level 1/0 | Level 1/1 | Level 1/2 | Level 2/00 | Level 2/01 |
|---|---|---|---|---|---|---|
| $P_n(A_k)\in$ | 0.2865÷0.2866 | 0.2865÷0.2867 | 0.2864÷0.2869 | 0.2862÷0.2864 | 0.2864÷0.2864 | 0.2858÷0.2868 |
| $P_n(T_k)\in$ | 0.2862÷0.2864 | 0.2861÷0.2862 | 0.286÷0.2862 | 0.286÷0.2863 | 0.2859÷0.2865 | 0.2861÷0.2866 |
| $P_n(C_k)\in$ | 0.2135÷0.2135 | 0.2132÷0.2136 | 0.2134÷0.2134 | 0.2135÷0.2138 | 0.2132÷0.2136 | 0.213÷0.213 |
| $P_n(G_k)\in$ | 0.2134÷0.2134 | 0.2133÷0.2135 | 0.2134÷0.2135 | 0.2133÷0.2135 | 0.2129÷0.2135 | 0.2132÷0.2136 |

| | Level 2/02 | Level 2/10 | Level 2/11 | Level 2/12 | Level 2/20 | Level 2/21 | Level 2/22 |
|---|---|---|---|---|---|---|---|
| $P_n(A_k)\in$ | 0.2865÷0.2867 | 0.286÷0.286 | 0.2861÷0.2867 | 0.2863÷0.2867 | 0.286÷0.2863 | 0.2863÷0.2868 | 0.2861÷0.2862 |
| $P_n(T_k)\in$ | 0.286÷0.2862 | 0.286÷0.2866 | 0.2861÷0.2862 | 0.2858÷0.2864 | 0.2861÷0.2868 | 0.2857÷0.2864 | 0.286÷0.2864 |
| $P_n(C_k)\in$ | 0.2133÷0.2137 | 0.2134÷0.2139 | 0.2133÷0.2134 | 0.2132÷0.2132 | 0.2133÷0.2134 | 0.2132÷0.2132 | 0.2133÷0.214 |
| $P_n(G_k)\in$ | 0.2132÷0.2134 | 0.2131÷0.2135 | 0.2131÷0.2137 | 0.2132÷0.2137 | 0.2132÷0.2135 | 0.2129÷0.2136 | 0.213÷0.2134 |

Fig. A6/9. Tables of fluctuation intervals of probabilities $P_n(A_k)$, $P_n(T_k)$, $P_n(C_k)$ and $P_n(G_k)$ for the set of all texts at each of levels of convolutions in the FGT-2 (upper table) and FGT-3 (bottom tables) in the case of the sequence: Mus musculus strain C57BL/6J chromosome 9, GRCm38.p4 C57BL/6J. NCBI Reference Sequence: NC_000075.6. 124595110 bp. https://www.ncbi.nlm.nih.gov/nuccore/NC_000075.6. Here n = 1, 2, 3, 4, 5, k ≤ n. Level 0 corresponds to the initial DNA-text with its different n-letter representations. Other levels in this table correspond to sets of all daughter texts at appropriate levels of positional convolutions in the FGT-2 and the FGT-3.

|  | Level 0 | Level 1/0 | Level 1/1 | Level 2/00 | Level 2/01 | Level 2/10 | Level 2/11 |
|---|---|---|---|---|---|---|---|
| $P_n(A_k)\in$ | 0.2925÷ 0.2926 | 0.2924÷ 0.2924 | 0.2925÷ 0.2925 | 0.2922÷ 0.2926 | 0.2925÷ 0.2926 | 0.2926÷ 0.2926 | 0.2925÷ 0.2925 |
| $P_n(T_k)\in$ | 0.2932÷ 0.2933 | 0.2931÷ 0.2933 | 0.2933÷ 0.2934 | 0.293÷ 0.2931 | 0.293÷ 0.2933 | 0.2932÷ 0.2935 | 0.2932÷ 0.2933 |
| $P_n(C_k)\in$ | 0.2067÷ 0.2069 | 0.2067÷ 0.2069 | 0.2067÷ 0.2069 | 0.2066÷ 0.2071 | 0.2067÷ 0.2068 | 0.2066÷ 0.2068 | 0.2066÷ 0.2069 |
| $P_n(G_k)\in$ | 0.2071÷ 0.2072 | 0.207÷ 0.2074 | 0.2071÷ 0.2072 | 0.2072÷ 0.2073 | 0.2069÷ 0.2073 | 0.207÷ 0.2071 | 0.2071÷ 0.2073 |

|  | Level 0 | Level 1/0 | Level 1/1 | Level 1/2 | Level 2/00 | Level 2/01 |
|---|---|---|---|---|---|---|
| $P_n(A_k)\in$ | 0.2925÷ 0.2926 | 0.2925÷ 0.2927 | 0.2923÷ 0.2924 | 0.2924÷ 0.2925 | 0.2924÷ 0.2926 | 0.2925÷ 0.2928 |
| $P_n(T_k)\in$ | 0.2932÷ 0.2933 | 0.293÷ 0.2933 | 0.2933÷ 0.2937 | 0.293÷ 0.2935 | 0.2928÷ 0.2932 | 0.2929÷ 0.2933 |
| $P_n(C_k)\in$ | 0.2067÷ 0.2069 | 0.2067÷ 0.2067 | 0.2066÷ 0.2068 | 0.2066÷ 0.2068 | 0.2065÷ 0.2069 | 0.2063÷ 0.2065 |
| $P_n(G_k)\in$ | 0.2071÷ 0.2072 | 0.207÷ 0.2072 | 0.2069÷ 0.2071 | 0.2071÷ 0.2072 | 0.2069÷ 0.2072 | 0.2069÷ 0.2074 |

|  | Level 2/02 | Level 2/10 | Level 2/11 | Level 2/12 | Level 2/20 | Level 2/21 | Level 2/22 |
|---|---|---|---|---|---|---|---|
| $P_n(A_k)\in$ | 0.2921÷ 0.2923 | 0.2922÷ 0.2926 | 0.2921÷ 0.2929 | 0.2924÷ 0.2924 | 0.2921÷ 0.293 | 0.2925÷ 0.2929 | 0.2922÷ 0.2926 |
| $P_n(T_k)\in$ | 0.293÷ 0.2936 | 0.2934÷ 0.2938 | 0.293÷ 0.2932 | 0.293÷ 0.2938 | 0.2931÷ 0.2933 | 0.2928÷ 0.2933 | 0.293÷ 0.2936 |
| $P_n(C_k)\in$ | 0.2065÷ 0.2065 | 0.2066÷ 0.2066 | 0.2065÷ 0.207 | 0.2063÷ 0.2069 | 0.2064÷ 0.2064 | 0.2064÷ 0.2065 | 0.2066÷ 0.207 |
| $P_n(G_k)\in$ | 0.2072÷ 0.2075 | 0.2067÷ 0.207 | 0.207÷ 0.207 | 0.2068÷ 0.2068 | 0.207÷ 0.2073 | 0.2072÷ 0.2072 | 0.2068÷ 0.2069 |

Fig. A6/10. Tables of fluctuation intervals of probabilities $P_n(A_k)$, $P_n(T_k)$, $P_n(C_k)$ and $P_n(G_k)$ for the set of all texts at each of levels of convolutions in the FGT-2 (upper table) and FGT-3 (bottom tables) in the case of the sequence: Mus musculus strain C57BL/6J chromosome 10, GRCm38.p4 C57BL/6J. NCBI Reference Sequence: NC_000076.6. 130694993 bp. https://www.ncbi.nlm.nih.gov/nuccore/NC_000076.6 . Here n = 1, 2, 3, 4, 5, k ≤ n. Level 0 corresponds to the initial DNA-text with its different n-letter representations. Other levels in this table correspond to sets of all daughter texts at appropriate levels of positional convolutions in the FGT-2 and the FGT-3.

| | Level 0 | Level 1/0 | Level 1/1 | Level 2/00 | Level 2/01 | Level 2/10 | Level 2/11 |
|---|---|---|---|---|---|---|---|
| $P_n(A_k)\in$ | 0.2812÷0.2814 | 0.2811÷0.2811 | 0.2812÷0.2814 | 0.2809÷0.2811 | 0.2811÷0.2813 | 0.2812÷0.2813 | 0.2811÷0.2811 |
| $P_n(T_k)\in$ | 0.2805÷0.2806 | 0.2805÷0.2806 | 0.2802÷0.2808 | 0.2804÷0.2807 | 0.2804÷0.2804 | 0.2802÷0.2805 | 0.2803÷0.2811 |
| $P_n(C_k)\in$ | 0.2191÷0.2191 | 0.219÷0.2192 | 0.2189÷0.2189 | 0.2188÷0.219 | 0.219÷0.2192 | 0.2189÷0.219 | 0.2188÷0.2188 |
| $P_n(G_k)\in$ | 0.2189÷0.2189 | 0.2189÷0.2191 | 0.2188÷0.2189 | 0.2188÷0.2192 | 0.2188÷0.219 | 0.2188÷0.2191 | 0.2188÷0.219 |

| | Level 0 | Level 1/0 | Level 1/1 | Level 1/2 | Level 2/00 | Level 2/01 |
|---|---|---|---|---|---|---|
| $P_n(A_k)\in$ | 0.2812÷0.2814 | 0.2811÷0.2811 | 0.2811÷0.2812 | 0.2811÷0.2818 | 0.2811÷0.2811 | 0.2809÷0.2813 |
| $P_n(T_k)\in$ | 0.2805÷0.2806 | 0.2804÷0.2808 | 0.2803÷0.2803 | 0.2804÷0.2804 | 0.2803÷0.2806 | 0.2803÷0.2807 |
| $P_n(C_k)\in$ | 0.2191÷0.2191 | 0.219÷0.2192 | 0.2189÷0.2193 | 0.2189÷0.219 | 0.2187÷0.2194 | 0.2189÷0.219 |
| $P_n(G_k)\in$ | 0.2189÷0.2189 | 0.2188÷0.2189 | 0.2189÷0.2192 | 0.2188÷0.2188 | 0.2187÷0.219 | 0.2187÷0.219 |

| | Level 2/02 | Level 2/10 | Level 2/11 | Level 2/12 | Level 2/20 | Level 2/21 | Level 2/22 |
|---|---|---|---|---|---|---|---|
| $P_n(A_k)\in$ | 0.2809÷0.2809 | 0.2808÷0.2814 | 0.2811÷0.2813 | 0.2808÷0.2808 | 0.2811÷0.2811 | 0.2811÷0.2813 | 0.281÷0.2817 |
| $P_n(T_k)\in$ | 0.2803÷0.2808 | 0.2801÷0.2807 | 0.2803÷0.2808 | 0.28÷0.2802 | 0.2803÷0.2812 | 0.2805÷0.2807 | 0.2803÷0.2804 |
| $P_n(C_k)\in$ | 0.2192÷0.2192 | 0.2187÷0.219 | 0.2187÷0.2191 | 0.2187÷0.2193 | 0.2186÷0.2187 | 0.219÷0.2192 | 0.2189÷0.2189 |
| $P_n(G_k)\in$ | 0.2185÷0.2191 | 0.219÷0.219 | 0.2186÷0.2188 | 0.219÷0.2196 | 0.2186÷0.219 | 0.2186÷0.2188 | 0.2188÷0.219 |

Fig. A6/11. Tables of fluctuation intervals of probabilities $P_n(A_k)$, $P_n(T_k)$, $P_n(C_k)$ and $P_n(G_k)$ for the set of all texts at each of levels of convolutions in the FGT-2 (upper table) and FGT-3 (bottom tables) in the case of the sequence: Mus musculus strain C57BL/6J, chromosome 11, GRCm38.p4 C57BL/6J. NCBI Reference Sequence: NC_000077.6 122082543 bp. https://www.ncbi.nlm.nih.gov/nuccore/NC_000077.6 . Here n = 1, 2, 3, 4, 5, k ≤ n. Level 0 corresponds to the initial DNA-text with its different n-letter representations. Other levels in this table correspond to sets of all daughter texts at appropriate levels of positional convolutions in the FGT-2 and the FGT-3.

| | Level 0 | Level 1/0 | Level 1/1 | Level 2/00 | Level 2/01 | Level 2/10 | Level 2/11 |
|---|---|---|---|---|---|---|---|

| $P_n(A_k)\in$ | 0.2898÷ 0.2899 | 0.2898÷ 0.2901 | 0.2897÷ 0.2898 | 0.2896÷ 0.29 | 0.2896÷ 0.2902 | 0.2896÷ 0.2899 | 0.2897÷ 0.2898 |
|---|---|---|---|---|---|---|---|
| $P_n(T_k)\in$ | 0.2926÷ 0.2928 | 0.2926÷ 0.2928 | 0.2924÷ 0.2927 | 0.2926÷ 0.2927 | 0.2926÷ 0.2929 | 0.2925÷ 0.2926 | 0.2922÷ 0.2927 |
| $P_n(C_k)\in$ | 0.2083÷ 0.2085 | 0.2083÷ 0.2084 | 0.2085÷ 0.2085 | 0.2083÷ 0.2086 | 0.2082÷ 0.2083 | 0.2084÷ 0.2087 | 0.2084÷ 0.2085 |
| $P_n(G_k)\in$ | 0.2088÷ 0.2088 | 0.2087÷ 0.2087 | 0.2089÷ 0.209 | 0.2086÷ 0.2088 | 0.2086÷ 0.2086 | 0.2088÷ 0.2089 | 0.2088÷ 0.209 |

| | Level 0 | Level 1/0 | Level 1/1 | Level 1/2 | Level 2/00 | Level 2/01 |
|---|---|---|---|---|---|---|
| $P_n(A_k)\in$ | 0.2898÷ 0.2899 | 0.2895÷ 0.29 | 0.2898÷ 0.29 | 0.2898÷ 0.2899 | 0.2892÷ 0.2892 | 0.2892÷ 0.2897 |
| $P_n(T_k)\in$ | 0.2926÷ 0.2928 | 0.2925÷ 0.2927 | 0.2925÷ 0.2928 | 0.2926÷ 0.2929 | 0.2923÷ 0.2927 | 0.2927÷ 0.2927 |
| $P_n(C_k)\in$ | 0.2083÷ 0.2085 | 0.2084÷ 0.2084 | 0.2082÷ 0.2084 | 0.2083÷ 0.2084 | 0.2083÷ 0.2088 | 0.2082÷ 0.2086 |
| $P_n(G_k)\in$ | 0.2088÷ 0.2088 | 0.2089÷ 0.2089 | 0.2087÷ 0.2088 | 0.2086÷ 0.2088 | 0.2085÷ 0.2093 | 0.2088÷ 0.2091 |

| | Level 2/02 | Level 2/10 | Level 2/11 | Level 2/12 | Level 2/20 | Level 2/21 | Level 2/22 |
|---|---|---|---|---|---|---|---|
| $P_n(A_k)\in$ | 0.2895÷ 0.2899 | 0.2897÷ 0.29 | 0.2895÷ 0.29 | 0.29÷ 0.29 | 0.2897÷ 0.2899 | 0.2898÷ 0.2901 | 0.2897÷ 0.2898 |
| $P_n(T_k)\in$ | 0.2922÷ 0.2926 | 0.2923÷ 0.2927 | 0.2925÷ 0.2928 | 0.2923÷ 0.2924 | 0.2926÷ 0.2932 | 0.2924÷ 0.2927 | 0.2924÷ 0.2929 |
| $P_n(C_k)\in$ | 0.2084÷ 0.2084 | 0.208÷ 0.2086 | 0.208÷ 0.2085 | 0.208÷ 0.2087 | 0.208÷ 0.2084 | 0.2082÷ 0.2085 | 0.2082÷ 0.2085 |
| $P_n(G_k)\in$ | 0.2085÷ 0.209 | 0.2087÷ 0.2087 | 0.2087÷ 0.2087 | 0.2085÷ 0.2088 | 0.2086÷ 0.2086 | 0.2086÷ 0.2087 | 0.2085÷ 0.2088 |

Fig. A6/12. Tables of fluctuation intervals of probabilities $P_n(A_k)$, $P_n(T_k)$, $P_n(C_k)$ and $P_n(G_k)$ for the set of all texts at each of levels of convolutions in the FGT-2 (upper table) and FGT-3 (bottom tables) in the case of the sequence: Mus musculus strain C57BL/6J chromosome 12, GRCm38.p4 C57BL/6J. NCBI Reference Sequence: NC_000078.6. 120129022 bp. https://www.ncbi.nlm.nih.gov/nuccore/NC_000078.6 . Here n = 1, 2, 3, 4, 5, k ≤ n. Level 0 corresponds to the initial DNA-text with its different n-letter representations. Other levels in this table correspond to sets of all daughter texts at appropriate levels of positional convolutions in the FGT-2 and the FGT-3.

| | Level 0 | Level 1/0 | Level 1/1 | Level 2/00 | Level 2/01 | Level 2/10 | Level 2/11 |
|---|---|---|---|---|---|---|---|
| $P_n(A_k)\in$ | 0.2924÷ 0.2925 | 0.2923÷ 0.2924 | 0.2923÷ 0.2926 | 0.2922÷ 0.2926 | 0.2921÷ 0.2921 | 0.2922÷ 0.2923 | 0.2921÷ 0.2926 |
| $P_n(T_k)\in$ | 0.2911÷ 0.2913 | 0.2911÷ 0.2913 | 0.2911÷ 0.2914 | 0.291÷ 0.2913 | 0.2912÷ 0.2914 | 0.291÷ 0.2913 | 0.2911÷ 0.2917 |
| $P_n(C_k)\in$ | 0.208÷ 0.2081 | 0.2079÷ 0.208 | 0.2079÷ 0.2079 | 0.2079÷ 0.2082 | 0.208÷ 0.2081 | 0.2081÷ 0.2081 | 0.2077÷ 0.2077 |
| $P_n(G_k)\in$ | 0.2079÷ 0.2081 | 0.208÷ 0.2083 | 0.2078÷ 0.2081 | 0.2079÷ 0.2079 | 0.2079÷ 0.2084 | 0.2078÷ 0.2083 | 0.2079÷ 0.2081 |

| | Level 0 | Level 1/0 | Level 1/1 | Level 1/2 | Level 2/00 | Level 2/01 |
|---|---|---|---|---|---|---|
| $P_n(A_k)\in$ | 0.2924÷ 0.2925 | 0.2921÷ 0.2921 | 0.2923÷ 0.2924 | 0.2924÷ 0.2927 | 0.292÷ 0.2922 | 0.292÷ 0.2924 |
| $P_n(T_k)\in$ | 0.2911÷ 0.2913 | 0.2912÷ 0.2914 | 0.2911÷ 0.2914 | 0.291÷ 0.2913 | 0.291÷ 0.2913 | 0.2909÷ 0.2911 |
| $P_n(C_k)\in$ | 0.208÷ 0.2081 | 0.2078÷ 0.2083 | 0.208÷ 0.2082 | 0.208÷ 0.208 | 0.2077÷ 0.2086 | 0.2076÷ 0.208 |
| $P_n(G_k)\in$ | 0.2079÷ 0.2081 | 0.2079÷ 0.2082 | 0.2079÷ 0.208 | 0.208÷ 0.208 | 0.2079÷ 0.2079 | 0.2077÷ 0.2084 |

| | Level 2/02 | Level 2/10 | Level 2/11 | Level 2/12 | Level 2/20 | Level 2/21 | Level 2/22 |
|---|---|---|---|---|---|---|---|
| $P_n(A_k)\in$ | 0.2921÷ 0.2923 | 0.2919÷ 0.2925 | 0.2923÷ 0.2927 | 0.2921÷ 0.2921 | 0.2922÷ 0.2931 | 0.2922÷ 0.2926 | 0.2924÷ 0.2931 |
| $P_n(T_k)\in$ | 0.2912÷ 0.2914 | 0.2909÷ 0.2914 | 0.291÷ 0.2911 | 0.2911÷ 0.2917 | 0.2906÷ 0.2906 | 0.2908÷ 0.2912 | 0.2909÷ 0.2913 |
| $P_n(C_k)\in$ | 0.2074÷ 0.2078 | 0.208÷ 0.208 | 0.2078÷ 0.2082 | 0.208÷ 0.208 | 0.2078÷ 0.2083 | 0.2077÷ 0.2079 | 0.2077÷ 0.2078 |
| $P_n(G_k)\in$ | 0.2079÷ 0.2085 | 0.2077÷ 0.2081 | 0.2079÷ 0.208 | 0.2076÷ 0.2082 | 0.2078÷ 0.2079 | 0.2079÷ 0.2083 | 0.2078÷ 0.2078 |

Fig. A6/13. Tables of fluctuation intervals of probabilities $P_n(A_k)$, $P_n(T_k)$, $P_n(C_k)$ and $P_n(G_k)$ for the set of all texts at each of levels of convolutions in the FGT-2 (upper table) and FGT-3 (bottom tables) in the case of the sequence: Mus musculus strain C57BL/6J chromosome 13, GRCm38.p4 C57BL/6J. NCBI Reference Sequence: NC_000079.6. 120421639 bp. https://www.ncbi.nlm.nih.gov/nuccore/NC_000079.6 . Here n = 1, 2, 3, 4, 5, k ≤ n. Level 0 corresponds to the initial DNA-text with its different n-letter representations. Other levels in this table correspond to sets of all daughter texts at appropriate levels of positional convolutions in the FGT-2 and the FGT-3.

| | Level 0 | Level 1/0 | Level 1/1 | Level 2/00 | Level 2/01 | Level 2/10 | Level 2/11 |
|---|---|---|---|---|---|---|---|
| $P_n(A_k)\in$ | 0.2939÷ 0.2939 | 0.2938÷ 0.294 | 0.2939÷ 0.2939 | 0.2937÷ 0.2938 | 0.2937÷ 0.294 | 0.2937÷ 0.2937 | 0.2938÷ 0.2938 |
| $P_n(T_k)\in$ | 0.2943÷ 0.2944 | 0.2943÷ 0.2944 | 0.2943÷ 0.2943 | 0.2944÷ 0.2945 | 0.2942÷ 0.2944 | 0.2941÷ 0.2944 | 0.2942÷ 0.2942 |
| $P_n(C_k)\in$ | 0.2055÷ 0.2057 | 0.2055÷ 0.2057 | 0.2055÷ 0.2058 | 0.2054÷ 0.2058 | 0.2055÷ 0.2058 | 0.2054÷ 0.2059 | 0.2053÷ 0.2058 |
| $P_n(G_k)\in$ | 0.2058÷ 0.206 | 0.2058÷ 0.2059 | 0.2058÷ 0.206 | 0.2059÷ 0.2059 | 0.2057÷ 0.2058 | 0.2058÷ 0.206 | 0.2056÷ 0.2062 |

| | Level 0 | Level 1/0 | Level 1/1 | Level 1/2 | Level 2/00 | Level 2/01 |
|---|---|---|---|---|---|---|
| $P_n(A_k)\in$ | 0.2939÷ 0.2939 | 0.2939÷ 0.2939 | 0.2936÷ 0.2937 | 0.2937÷ 0.2937 | 0.2937÷ 0.2942 | 0.2938÷ 0.2941 |
| $P_n(T_k)\in$ | 0.2943÷ 0.2944 | 0.2941÷ 0.2946 | 0.2943÷ 0.2945 | 0.2943÷ 0.2946 | 0.2939÷ 0.2939 | 0.2942÷ 0.2943 |
| $P_n(C_k)\in$ | 0.2055÷ 0.2057 | 0.2053÷ 0.2053 | 0.2056÷ 0.2057 | 0.2056÷ 0.2057 | 0.2053÷ 0.2056 | 0.2051÷ 0.2057 |
| $P_n(G_k)\in$ | 0.2058÷ 0.206 | 0.2059÷ 0.2061 | 0.2057÷ 0.2061 | 0.2057÷ 0.206 | 0.2057÷ 0.2062 | 0.2058÷ 0.2059 |

| | Level 2/02 | Level 2/10 | Level 2/11 | Level 2/12 | Level 2/20 | Level 2/21 | Level 2/22 |
|---|---|---|---|---|---|---|---|
| $P_n(A_k)\in$ | 0.2935÷ 0.2939 | 0.2936÷ 0.2939 | 0.2931÷ 0.2936 | 0.2934÷ 0.2934 | 0.2936÷ 0.2937 | 0.2933÷ 0.2941 | 0.2937÷ 0.2939 |
| $P_n(T_k)\in$ | 0.294÷ 0.2949 | 0.294÷ 0.2946 | 0.2943÷ 0.2945 | 0.2941÷ 0.2944 | 0.294÷ 0.2947 | 0.2939÷ 0.2939 | 0.2941÷ 0.2942 |
| $P_n(C_k)\in$ | 0.2053÷ 0.2053 | 0.2056÷ 0.2058 | 0.2054÷ 0.2057 | 0.2055÷ 0.206 | 0.2055÷ 0.2059 | 0.2055÷ 0.2059 | 0.2057÷ 0.2058 |
| $P_n(G_k)\in$ | 0.2057÷ 0.2059 | 0.2056÷ 0.2056 | 0.2057÷ 0.2062 | 0.2054÷ 0.2062 | 0.2054÷ 0.2058 | 0.2056÷ 0.206 | 0.2055÷ 0.206 |

Fig. A6/14. Tables of fluctuation intervals of probabilities $P_n(A_k)$, $P_n(T_k)$, $P_n(C_k)$ and $P_n(G_k)$ for the set of all texts at each of levels of convolutions in the FGT-2 (upper table) and FGT-3 (bottom tables) in the case of the sequence: Mus musculus strain C57BL/6J chromosome 14, GRCm38.p4 C57BL/6J. NCBI Reference Sequence: NC_000080.6. 124902244 bp. https://www.ncbi.nlm.nih.gov/nuccore/NC_000080.6 . Here n = 1, 2, 3, 4, 5, k ≤ n. Level 0 corresponds to the initial DNA-text with its different n-letter representations. Other levels in this table correspond to sets of all daughter texts at appropriate levels of positional convolutions in the FGT-2 and the FGT-3.

|  | Level 0 | Level 1/0 | Level 1/1 | Level 2/00 | Level 2/01 | Level 2/10 | Level 2/11 |
|---|---|---|---|---|---|---|---|
| $P_n(A_k)\in$ | 0.2898÷ 0.2898 | 0.2897÷ 0.29 | 0.2895÷ 0.2895 | 0.2896÷ 0.2902 | 0.2897÷ 0.2901 | 0.2896÷ 0.2899 | 0.2894÷ 0.2894 |
| $P_n(T_k)\in$ | 0.2904÷ 0.2907 | 0.2903÷ 0.2904 | 0.2905÷ 0.2909 | 0.2902÷ 0.2903 | 0.2902÷ 0.2906 | 0.2904÷ 0.2904 | 0.2904÷ 0.2911 |
| $P_n(C_k)\in$ | 0.2098÷ 0.2098 | 0.2096÷ 0.2097 | 0.2098÷ 0.2099 | 0.2096÷ 0.2099 | 0.2097÷ 0.2099 | 0.2096÷ 0.2101 | 0.2097÷ 0.2099 |
| $P_n(G_k)\in$ | 0.2096÷ 0.2096 | 0.2096÷ 0.2098 | 0.2095÷ 0.2097 | 0.2096÷ 0.2096 | 0.2095÷ 0.2095 | 0.2095÷ 0.2096 | 0.2094÷ 0.2096 |

|  | Level 0 | Level 1/0 | Level 1/1 | Level 1/2 | Level 2/00 | Level 2/01 |
|---|---|---|---|---|---|---|
| $P_n(A_k)\in$ | 0.2898÷ 0.2898 | 0.2898÷ 0.2899 | 0.2897÷ 0.29 | 0.2897÷ 0.2897 | 0.2895÷ 0.2899 | 0.2893÷ 0.2893 |
| $P_n(T_k)\in$ | 0.2904÷ 0.2907 | 0.2903÷ 0.2906 | 0.2903÷ 0.2906 | 0.2904÷ 0.2908 | 0.2904÷ 0.2909 | 0.2901÷ 0.2908 |
| $P_n(C_k)\in$ | 0.2098÷ 0.2098 | 0.2096÷ 0.2097 | 0.2096÷ 0.2096 | 0.2097÷ 0.2098 | 0.2094÷ 0.2098 | 0.2098÷ 0.2099 |
| $P_n(G_k)\in$ | 0.2096÷ 0.2096 | 0.2094÷ 0.2097 | 0.2094÷ 0.2098 | 0.2096÷ 0.2096 | 0.2094÷ 0.2095 | 0.2093÷ 0.21 |

|  | Level 2/02 | Level 2/10 | Level 2/11 | Level 2/12 | Level 2/20 | Level 2/21 | Level 2/22 |
|---|---|---|---|---|---|---|---|
| $P_n(A_k)\in$ | 0.29÷ 0.2901 | 0.2895÷ 0.2898 | 0.2895÷ 0.2895 | 0.2894÷ 0.2899 | 0.2894÷ 0.2896 | 0.2897÷ 0.2897 | 0.2893÷ 0.2897 |
| $P_n(T_k)\in$ | 0.2901÷ 0.2913 | 0.2903÷ 0.2912 | 0.29÷ 0.2905 | 0.2899÷ 0.2908 | 0.2903÷ 0.2905 | 0.2901÷ 0.2907 | 0.2903÷ 0.2905 |
| $P_n(C_k)\in$ | 0.2092÷ 0.2092 | 0.2094÷ 0.21 | 0.2094÷ 0.2102 | 0.2096÷ 0.2097 | 0.2096÷ 0.2101 | 0.2096÷ 0.21 | 0.2094÷ 0.2097 |
| $P_n(G_k)\in$ | 0.2092÷ 0.2094 | 0.209÷ 0.209 | 0.2089÷ 0.2098 | 0.2093÷ 0.2096 | 0.2092÷ 0.2098 | 0.2095÷ 0.2095 | 0.2096÷ 0.2101 |

Fig. A6/15. Tables of fluctuation intervals of probabilities $P_n(A_k)$, $P_n(T_k)$, $P_n(C_k)$ and $P_n(G_k)$ for the set of all texts at each of levels of convolutions in the FGT-2 (upper table) and FGT-3 (bottom tables) in the case of the sequence: Mus musculus strain C57BL/6J chromosome 15, GRCm38.p4 C57BL/6JNCBI Reference Sequence: NC_000081.6, LOCUS NC_000081 104043685 bp DNA linear CON 22-JUN-2016 DEFINITION Mus musculus strain C57BL/6J chromosome 15, GRCm38.p4 C57BL/6J. ACCESSION NC_000081 GPC_000000788 VERSION NC_000081.6 https://www.ncbi.nlm.nih.gov/nuccore/NC_000081.6 . Here n = 1, 2, 3, 4, 5, k ≤ n. Level 0 corresponds to the initial DNA-text with its different n-letter representations. Other levels in this table correspond to sets of all daughter texts

at appropriate levels of positional convolutions in the FGT-2 and the FGT-3.

| | Level 0 | Level 1/0 | Level 1/1 | Level 2/00 | Level 2/01 | Level 2/10 | Level 2/11 |
|---|---|---|---|---|---|---|---|
| $P_n(A_k)\in$ | 0.2949÷ 0.2951 | 0.2948÷ 0.2951 | 0.2949÷ 0.2952 | 0.2948÷ 0.2952 | 0.2947÷ 0.2952 | 0.2949÷ 0.2951 | 0.2947÷ 0.2953 |
| $P_n(T_k)\in$ | 0.2954÷ 0.2956 | 0.2954÷ 0.2954 | 0.2954÷ 0.2957 | 0.2953÷ 0.2956 | 0.2953÷ 0.2954 | 0.2954÷ 0.2954 | 0.2953÷ 0.2956 |
| $P_n(C_k)\in$ | 0.2045÷ 0.2045 | 0.2044÷ 0.2049 | 0.2042÷ 0.2044 | 0.2042÷ 0.2047 | 0.2042÷ 0.2048 | 0.2041÷ 0.2047 | 0.2044÷ 0.2044 |
| $P_n(G_k)\in$ | 0.2047÷ 0.2047 | 0.2047÷ 0.2047 | 0.2047÷ 0.2047 | 0.2045÷ 0.2045 | 0.2046÷ 0.2046 | 0.2047÷ 0.2048 | 0.2045÷ 0.2047 |

| | Level 0 | Level 1/0 | Level 1/1 | Level 1/2 | Level 2/00 | Level 2/01 |
|---|---|---|---|---|---|---|
| $P_n(A_k)\in$ | 0.2949÷ 0.2951 | 0.2947÷ 0.2947 | 0.2948÷ 0.295 | 0.295÷ 0.2954 | 0.2945÷ 0.2951 | 0.2947÷ 0.295 |
| $P_n(T_k)\in$ | 0.2954÷ 0.2956 | 0.2953÷ 0.2957 | 0.2955÷ 0.2958 | 0.2953÷ 0.2953 | 0.2953÷ 0.2956 | 0.2952÷ 0.2956 |
| $P_n(C_k)\in$ | 0.2045÷ 0.2045 | 0.2044÷ 0.2046 | 0.2044÷ 0.2046 | 0.2043÷ 0.2044 | 0.2042÷ 0.2044 | 0.2041÷ 0.2047 |
| $P_n(G_k)\in$ | 0.2047÷ 0.2047 | 0.2046÷ 0.2049 | 0.2046÷ 0.2047 | 0.2047÷ 0.2049 | 0.2045÷ 0.2049 | 0.2047÷ 0.2047 |

| | Level 2/02 | Level 2/10 | Level 2/11 | Level 2/12 | Level 2/20 | Level 2/21 | Level 2/22 |
|---|---|---|---|---|---|---|---|
| $P_n(A_k)\in$ | 0.2947÷ 0.2948 | 0.2944÷ 0.2948 | 0.2946÷ 0.2948 | 0.2947÷ 0.2956 | 0.2944÷ 0.2944 | 0.2948÷ 0.2953 | 0.2949÷ 0.2951 |
| $P_n(T_k)\in$ | 0.2949÷ 0.2958 | 0.2954÷ 0.2959 | 0.2954÷ 0.2963 | 0.2951÷ 0.2958 | 0.2948÷ 0.296 | 0.2947÷ 0.2957 | 0.2951÷ 0.2953 |
| $P_n(C_k)\in$ | 0.2043÷ 0.2047 | 0.2039÷ 0.2043 | 0.2043÷ 0.2045 | 0.2043÷ 0.2044 | 0.2045÷ 0.2046 | 0.2039÷ 0.2044 | 0.2045÷ 0.2047 |
| $P_n(G_k)\in$ | 0.2046÷ 0.2047 | 0.2043÷ 0.2049 | 0.2041÷ 0.2045 | 0.2042÷ 0.2042 | 0.2045÷ 0.205 | 0.2046÷ 0.2046 | 0.2044÷ 0.2049 |

Fig. A6/16. Tables of fluctuation intervals of probabilities $P_n(A_k)$, $P_n(T_k)$, $P_n(C_k)$ and $P_n(G_k)$ for the set of all texts at each of levels of convolutions in the FGT-2

(upper table) and FGT-3 (bottom tables) in the case of the sequence: Mus musculus strain C57BL/6J chromosome 16, GRCm38.p4 C57BL/6J, NCBI Reference Sequence: NC_000082.6, LOCUS NC_000082 98207768 bp DNA linear CON 22-JUN-2016 DEFINITION Mus musculus strain C57BL/6J chromosome 16, GRCm38.p4 C57BL/6J. ACCESSION NC_000082 GPC_000000789 VERSION NC_000082.6, https://www.ncbi.nlm.nih.gov/nuccore/NC_000082.6 . Here n = 1, 2, 3, 4, 5, k ≤ n. Level 0 corresponds to the initial DNA-text with its different n-letter representations. Other levels in this table correspond to sets of all daughter texts at appropriate levels of positional convolutions in the FGT-2 and the FGT-3.

| | Level 0 | Level 1/0 | Level 1/1 | Level 2/00 | Level 2/01 | Level 2/10 | Level 2/11 |
|---|---|---|---|---|---|---|---|
| $P_n(A_k)\in$ | 0.286÷ 0.2864 | 0.2862÷ 0.2862 | 0.2859÷ 0.2863 | 0.2861÷ 0.2864 | 0.2863÷ 0.2863 | 0.2859÷ 0.2864 | 0.2858÷ 0.2862 |
| $P_n(T_k)\in$ | 0.2867÷ 0.2867 | 0.2866÷ 0.2869 | 0.2867÷ 0.2868 | 0.2866÷ 0.2869 | 0.2864÷ 0.287 | 0.2866÷ 0.2869 | 0.2866÷ 0.2869 |
| $P_n(C_k)\in$ | 0.2134÷ 0.2136 | 0.2134÷ 0.2138 | 0.2133÷ 0.2136 | 0.2133÷ 0.2133 | 0.2134÷ 0.2136 | 0.2131÷ 0.2134 | 0.2134÷ 0.2135 |
| $P_n(G_k)\in$ | 0.2133÷ 0.2133 | 0.2131÷ 0.2131 | 0.2133÷ 0.2133 | 0.213÷ 0.2134 | 0.2132÷ 0.2132 | 0.2132÷ 0.2134 | 0.2132÷ 0.2134 |

| | Level 0 | Level 1/0 | Level 1/1 | Level 1/2 | Level 2/00 | Level 2/01 |
|---|---|---|---|---|---|---|
| $P_n(A_k)\in$ | 0.286÷ 0.2864 | 0.2859÷ 0.286 | 0.2861÷ 0.2864 | 0.2859÷ 0.2863 | 0.2857÷ 0.2864 | 0.2858÷ 0.2862 |
| $P_n(T_k)\in$ | 0.2867÷ 0.2867 | 0.2868÷ 0.2868 | 0.2863÷ 0.2867 | 0.2866÷ 0.2869 | 0.2867÷ 0.2869 | 0.2866÷ 0.287 |
| $P_n(C_k)\in$ | 0.2134÷ 0.2136 | 0.2133÷ 0.2139 | 0.2134÷ 0.2138 | 0.2134÷ 0.2137 | 0.2129÷ 0.2132 | 0.2133÷ 0.2135 |
| $P_n(G_k)\in$ | 0.2133÷ 0.2133 | 0.2132÷ 0.2132 | 0.2131÷ 0.2131 | 0.213÷ 0.2131 | 0.2132÷ 0.2135 | 0.2131÷ 0.2132 |

| | Level 2/02 | Level 2/10 | Level 2/11 | Level 2/12 | Level 2/20 | Level 2/21 | Level 2/22 |
|---|---|---|---|---|---|---|---|
| $P_n(A_k)\in$ | 0.2858÷ 0.2859 | 0.286÷ 0.2865 | 0.2859÷ 0.2863 | 0.2861÷ 0.2863 | 0.2859÷ 0.2863 | 0.2861÷ 0.2867 | 0.2851÷ 0.2862 |
| $P_n(T_k)\in$ | 0.2864÷ 0.2867 | 0.2863÷ 0.2869 | 0.2862÷ 0.2863 | 0.2861÷ 0.287 | 0.2863÷ 0.2867 | 0.2864÷ 0.2867 | 0.2867÷ 0.2869 |
| $P_n(C_k)\in$ | 0.2132÷ 0.2142 | 0.2133÷ 0.2135 | 0.2132÷ 0.2134 | 0.2131÷ 0.2135 | 0.2133÷ 0.2133 | 0.2133÷ 0.2133 | 0.2132÷ 0.214 |
| $P_n(G_k)\in$ | 0.2131÷ 0.2132 | 0.2126÷ 0.2131 | 0.2134÷ 0.214 | 0.213÷ 0.2131 | 0.213÷ 0.2137 | 0.213÷ 0.2133 | 0.2129÷ 0.2129 |

Fig. A6/17. Tables of fluctuation intervals of probabilities $P_n(A_k)$, $P_n(T_k)$, $P_n(C_k)$ and $P_n(G_k)$ for the set of all texts at each of levels of convolutions in the FGT-2 (upper table) and FGT-3 (bottom tables) in the case of the sequence: Mus musculus strain C57BL/6J chromosome 17, GRCm38.p4 C57BL/6J , CBI Reference Sequence: NC_000083.6LOCUS NC_000083 94987271 bp DNA linear CON 22-JUN-2016 DEFINITION Mus musculus strain C57BL/6J chromosome 17, GRCm38.p4 C57BL/6J. ACCESSION NC_000083 GPC_000000790 VERSION NC_000083.6, https://www.ncbi.nlm.nih.gov/nuccore/NC_000083.6 . Here n = 1, 2, 3, 4, 5, k ≤ n. Level 0 corresponds to the initial DNA-text with its different n-letter representations. Other levels in this table correspond to sets of all daughter texts at appropriate levels of positional convolutions in the FGT-2 and the FGT-3.

| | Level 0 | Level 1/0 | Level 1/1 | Level 2/00 | Level 2/01 | Level 2/10 | Level 2/11 |
|---|---|---|---|---|---|---|---|
| $P_n(A_k)\in$ | 0.2928÷ 0.2928 | 0.2928÷ 0.2931 | 0.2926÷ 0.2926 | 0.2924÷ 0.293 | 0.2928÷ 0.2933 | 0.2924÷ 0.293 | 0.2927÷ 0.2927 |
| $P_n(T_k)\in$ | 0.2925÷ 0.2926 | 0.2926÷ 0.2929 | 0.2924÷ 0.2926 | 0.2927÷ 0.293 | 0.2924÷ 0.293 | 0.2925÷ 0.2927 | 0.2923÷ 0.2924 |
| $P_n(C_k)\in$ | 0.2068÷ 0.207 | 0.2067÷ 0.2068 | 0.2069÷ 0.2071 | 0.2066÷ 0.2066 | 0.2066÷ 0.2066 | 0.2069÷ 0.2069 | 0.2068÷ 0.2071 |
| $P_n(G_k)\in$ | 0.2074÷ 0.2076 | 0.2072÷ 0.2072 | 0.2073÷ 0.2077 | 0.2072÷ 0.2073 | 0.2071÷ 0.2071 | 0.2072÷ 0.2074 | 0.2075÷ 0.2078 |

| | Level 0 | Level 1/0 | Level 1/1 | Level 1/2 | Level 2/00 | Level 2/01 |
|---|---|---|---|---|---|---|
| $P_n(A_k)\in$ | 0.2928÷ 0.2928 | 0.2928÷ 0.2932 | 0.2925÷ 0.293 | 0.2928÷ 0.2928 | 0.2926÷ 0.2928 | 0.2927÷ 0.2927 |
| $P_n(T_k)\in$ | 0.2925÷ 0.2926 | 0.2923÷ 0.2924 | 0.2923÷ 0.2925 | 0.2925÷ 0.2927 | 0.2923÷ 0.2929 | 0.2922÷ 0.2923 |
| $P_n(C_k)\in$ | 0.2068÷ 0.207 | 0.2067÷ 0.2069 | 0.2068÷ 0.207 | 0.2068÷ 0.2069 | 0.2065÷ 0.2069 | 0.2066÷ 0.2071 |
| $P_n(G_k)\in$ | 0.2074÷ 0.2076 | 0.2072÷ 0.2076 | 0.2074÷ 0.2074 | 0.2073÷ 0.2076 | 0.2073÷ 0.2073 | 0.207÷ 0.2079 |

| | Level 2/02 | Level 2/10 | Level 2/11 | Level 2/12 | Level 2/20 | Level 2/21 | Level 2/22 |
|---|---|---|---|---|---|---|---|
| $P_n(A_k)\in$ | 0.2928÷ 0.293 | 0.2924÷ 0.2924 | 0.2924÷ 0.293 | 0.2924÷ 0.2931 | 0.2924÷ 0.2929 | 0.2924÷ 0.293 | 0.2924÷ 0.2927 |
| $P_n(T_k)\in$ | 0.2923÷ 0.2925 | 0.2922÷ 0.2931 | 0.292÷ 0.2929 | 0.2924÷ 0.2928 | 0.2921÷ 0.2931 | 0.2924÷ 0.2924 | 0.2922÷ 0.2926 |
| $P_n(C_k)\in$ | 0.2064÷ | 0.2066÷ | 0.2067÷ | 0.2066÷ | 0.2068÷ | 0.2065÷ | 0.2068÷ |

| | 0.207 | 0.207 | 0.2067 | 0.2067 | 0.2069 | 0.207 | 0.2069 |
|---|---|---|---|---|---|---|---|
| $P_n(G_k)\in$ | 0.207÷ 0.2075 | 0.2074÷ 0.2075 | 0.2072÷ 0.2075 | 0.2072÷ 0.2074 | 0.2071÷ 0.2071 | 0.2071÷ 0.2076 | 0.2068÷ 0.2078 |

Fig. A6/18. Tables of fluctuation intervals of probabilities $P_n(A_k)$, $P_n(T_k)$, $P_n(C_k)$ and $P_n(G_k)$ for the set of all texts at each of levels of convolutions in the FGT-2 (upper table) and FGT-3 (bottom tables) in the case of the sequence: Mus musculus strain C57BL/6J chromosome 18, GRCm38.p4 C57BL/6J, NCBI Reference Sequence: NC_000084.6, LOCUS NC_000084 90702639 bp DNA linear CON 22-JUN-2016 DEFINITION Mus musculus strain C57BL/6J chromosome 18, GRCm38.p4 C57BL/6J. ACCESSION NC_000084 GPC_000000791 VERSION NC_000084.6, https://www.ncbi.nlm.nih.gov/nuccore/NC_000084.6 . Here n = 1, 2, 3, 4, 5, k ≤ n. Level 0 corresponds to the initial DNA-text with its different n-letter representations. Other levels in this table correspond to sets of all daughter texts at appropriate levels of positional convolutions in the FGT-2 and the FGT-3.

| | Level 0 | Level 1/0 | Level 1/1 | Level 2/00 | Level 2/01 | Level 2/10 | Level 2/11 |
|---|---|---|---|---|---|---|---|
| $P_n(A_k)\in$ | 0.2873÷ 0.2876 | 0.2871÷ 0.2871 | 0.2873÷ 0.2878 | 0.2867÷ 0.2879 | 0.2871÷ 0.2876 | 0.2873÷ 0.2873 | 0.2872÷ 0.2877 |
| $P_n(T_k)\in$ | 0.285÷ 0.2852 | 0.2849÷ 0.2853 | 0.2849÷ 0.2853 | 0.2849÷ 0.2849 | 0.2848÷ 0.2848 | 0.2847÷ 0.2855 | 0.2851÷ 0.2854 |
| $P_n(C_k)\in$ | 0.2137÷ 0.2139 | 0.2137÷ 0.214 | 0.2137÷ 0.2138 | 0.2135÷ 0.2138 | 0.2138÷ 0.2141 | 0.2138÷ 0.2138 | 0.2134÷ 0.2138 |
| $P_n(G_k)\in$ | 0.2132÷ 0.2133 | 0.2132÷ 0.2136 | 0.2131÷ 0.2131 | 0.2132÷ 0.2134 | 0.2131÷ 0.2136 | 0.2131÷ 0.2134 | 0.2131÷ 0.2131 |

| | Level 0 | Level 1/0 | Level 1/1 | Level 1/2 | Level 2/00 | Level 2/01 |
|---|---|---|---|---|---|---|
| $P_n(A_k)\in$ | 0.2873÷ 0.2876 | 0.2872÷ 0.2875 | 0.2872÷ 0.2872 | 0.2871÷ 0.2874 | 0.2867÷ 0.2871 | 0.2869÷ 0.288 |
| $P_n(T_k)\in$ | 0.285÷ 0.2852 | 0.2851÷ 0.2854 | 0.2849÷ 0.2853 | 0.2848÷ 0.2852 | 0.2849÷ 0.2853 | 0.2851÷ 0.2852 |
| $P_n(C_k)\in$ | 0.2137÷ 0.2139 | 0.2136÷ 0.214 | 0.2137÷ 0.2138 | 0.2137÷ 0.2141 | 0.2139÷ 0.214 | 0.213÷ 0.2135 |
| $P_n(G_k)\in$ | 0.2132÷ 0.2133 | 0.213÷ 0.213 | 0.2133÷ 0.2137 | 0.2131÷ 0.2133 | 0.2126÷ 0.2135 | 0.2131÷ 0.2132 |

| | Level 2/02 | Level 2/10 | Level 2/11 | Level 2/12 | Level 2/20 | Level 2/21 | Level 2/22 |
|---|---|---|---|---|---|---|---|
| $P_n(A_k)\in$ | 0.2873÷ 0.2874 | 0.2873÷ 0.2874 | 0.287÷ 0.2873 | 0.2869÷ 0.2869 | 0.287÷ 0.2877 | 0.2869÷ 0.2869 | 0.2869÷ 0.2874 |
| $P_n(T_k)\in$ | 0.2846÷ 0.2851 | 0.2847÷ 0.2853 | 0.2844÷ 0.2857 | 0.2846÷ 0.2854 | 0.2847÷ 0.2847 | 0.2849÷ 0.2856 | 0.2843÷ 0.2854 |
| $P_n(C_k)\in$ | 0.2135÷ 0.2144 | 0.2134÷ 0.2134 | 0.2134÷ 0.2135 | 0.2134÷ 0.2141 | 0.2132÷ 0.214 | 0.2133÷ 0.2141 | 0.2135÷ 0.2137 |
| $P_n(G_k)\in$ | 0.2127÷ 0.2131 | 0.2128÷ 0.2139 | 0.2132÷ 0.2134 | 0.2129÷ 0.2136 | 0.2125÷ 0.2136 | 0.2134÷ 0.2134 | 0.2128÷ 0.2135 |

Fig. A6/19. Tables of fluctuation intervals of probabilities $P_n(A_k)$, $P_n(T_k)$, $P_n(C_k)$ and $P_n(G_k)$ for the set of all texts at each of levels of convolutions in the FGT-2 (upper table) and FGT-3 (bottom tables) in the case of the sequence: Mus musculus strain C57BL/6J chromosome 19, GRCm38.p4 C57BL/6J, NCBI Reference Sequence: NC_000085.6, LOCUS NC_000085 61431566 bp DNA linear CON 22-JUN-2016 DEFINITION Mus musculus strain C57BL/6J chromosome 19, GRCm38.p4 C57BL/6J. ACCESSION NC_000085 GPC_000000792 VERSION NC_000085.6, https://www.ncbi.nlm.nih.gov/nuccore/NC_000085.6 . Here n = 1, 2, 3, 4, 5, k ≤ n. Level 0 corresponds to the initial DNA-text with its different n-letter representations. Other levels in this table correspond to sets of all daughter texts at appropriate levels of positional convolutions in the FGT-2 and the FGT-3.

| | Level 0 | Level 1/0 | Level 1/1 | Level 2/00 | Level 2/01 | Level 2/10 | Level 2/11 |
|---|---|---|---|---|---|---|---|
| $P_n(A_k)\in$ | 0.3037÷ 0.3039 | 0.3037÷ 0.3037 | 0.3036÷ 0.3039 | 0.3035÷ 0.3037 | 0.3036÷ 0.3038 | 0.3034÷ 0.3038 | 0.3036÷ 0.3038 |
| $P_n(T_k)\in$ | 0.3036÷ 0.3036 | 0.3035÷ 0.3035 | 0.3035÷ 0.3035 | 0.3035÷ 0.3037 | 0.3035÷ 0.3035 | 0.3033÷ 0.3037 | 0.3035÷ 0.3038 |
| $P_n(C_k)\in$ | 0.1961÷ 0.1963 | 0.196÷ 0.1962 | 0.1961÷ 0.1962 | 0.1962÷ 0.1965 | 0.1959÷ 0.1961 | 0.196÷ 0.1962 | 0.196÷ 0.196 |
| $P_n(G_k)\in$ | 0.1962÷ 0.1963 | 0.1961÷ 0.1966 | 0.1962÷ 0.1964 | 0.196÷ 0.1961 | 0.1961÷ 0.1966 | 0.1963÷ 0.1963 | 0.1962÷ 0.1963 |

| | Level 0 | Level 1/0 | Level 1/1 | Level 1/2 | Level 2/00 | Level 2/01 |
|---|---|---|---|---|---|---|
| $P_n(A_k)\in$ | 0.3037÷ 0.3039 | 0.3037÷ 0.3039 | 0.3034÷ 0.3034 | 0.3035÷ 0.3038 | 0.3036÷ 0.3038 | 0.3036÷ 0.3039 |
| $P_n(T_k)\in$ | 0.3036÷ 0.3036 | 0.3035÷ 0.3039 | 0.3035÷ 0.3039 | 0.3034÷ 0.3034 | 0.3034÷ 0.3038 | 0.3033÷ 0.3036 |
| $P_n(C_k)\in$ | 0.1961÷ | 0.1959÷ | 0.196÷ | 0.1962÷ | 0.1958÷ | 0.1961÷ |

| | 0.1963 | 0.1961 | 0.1963 | 0.1964 | 0.1961 | 0.1963 |
|---|---|---|---|---|---|---|
| $P_n(G_k)\in$ | 0.1962÷ 0.1963 | 0.196÷ 0.1961 | 0.1962÷ 0.1964 | 0.1962÷ 0.1965 | 0.1961÷ 0.1963 | 0.1961÷ 0.1962 |

| | Level 2/02 | Level 2/10 | Level 2/11 | Level 2/12 | Level 2/20 | Level 2/21 | Level 2/22 |
|---|---|---|---|---|---|---|---|
| $P_n(A_k)\in$ | 0.3034÷ 0.3041 | 0.3033÷ 0.3038 | 0.3035÷ 0.3041 | 0.3031÷ 0.3031 | 0.3033÷ 0.3036 | 0.3034÷ 0.3037 | 0.3034÷ 0.3042 |
| $P_n(T_k)\in$ | 0.3037÷ 0.3037 | 0.3034÷ 0.3041 | 0.3034÷ 0.3035 | 0.3035÷ 0.3043 | 0.3034÷ 0.3034 | 0.3033÷ 0.3036 | 0.3032÷ 0.3032 |
| $P_n(C_k)\in$ | 0.1959÷ 0.1962 | 0.1958÷ 0.1958 | 0.196÷ 0.1961 | 0.196÷ 0.1962 | 0.1961÷ 0.1963 | 0.1962÷ 0.1963 | 0.196÷ 0.1961 |
| $P_n(G_k)\in$ | 0.1959÷ 0.1959 | 0.1962÷ 0.1963 | 0.1958÷ 0.1963 | 0.196÷ 0.1963 | 0.1961÷ 0.1966 | 0.1961÷ 0.1964 | 0.1959÷ 0.1965 |

Fig. A6/20. Tables of fluctuation intervals of probabilities $P_n(A_k)$, $P_n(T_k)$, $P_n(C_k)$ and $P_n(G_k)$ for the set of all texts at each of levels of convolutions in the FGT-2 (upper table) and FGT-3 (bottom tables) in the case of the sequence: Mus musculus strain C57BL/6J chromosome X, GRCm38.p4 C57BL/6J, NCBI Reference Sequence: NC_000086.7, LOCUS NC_000086 171031299 bp DNA linear CON 22-JUN-2016 DEFINITION Mus musculus strain C57BL/6J chromosome X, GRCm38.p4 C57BL/6J. ACCESSION NC_000086 GPC_000000793 VERSION NC_000086.7, https://www.ncbi.nlm.nih.gov/nuccore/NC_000086.7 . Here n = 1, 2, 3, 4, 5, $k \le n$. Level 0 corresponds to the initial DNA-text with its different n-letter representations. Other levels in this table correspond to sets of all daughter texts at appropriate levels of positional convolutions in the FGT-2 and the FGT-3.

| | Level 0 | Level 1/0 | Level 1/1 | Level 2/00 | Level 2/01 | Level 2/10 | Level 2/11 |
|---|---|---|---|---|---|---|---|
| $P_n(A_k)\in$ | 0.3045÷ 0.3045 | 0.3043÷ 0.3046 | 0.3043÷ 0.3044 | 0.3039÷ 0.3043 | 0.3044÷ 0.3046 | 0.3042÷ 0.305 | 0.3044÷ 0.3044 |
| $P_n(T_k)\in$ | 0.3064÷ 0.3066 | 0.3063÷ 0.3067 | 0.3064÷ 0.3066 | 0.3062÷ 0.3066 | 0.3063÷ 0.3066 | 0.3062÷ 0.3065 | 0.3065÷ 0.3068 |
| $P_n(C_k)\in$ | 0.1948÷ 0.1948 | 0.1947÷ 0.1948 | 0.1948÷ 0.1948 | 0.1947÷ 0.1951 | 0.1946÷ 0.1948 | 0.1946÷ 0.1946 | 0.1947÷ 0.1947 |
| $P_n(G_k)\in$ | 0.1937÷ 0.1942 | 0.1937÷ 0.1939 | 0.1938÷ 0.1942 | 0.1936÷ 0.1941 | 0.1936÷ 0.194 | 0.1938÷ 0.1938 | 0.1937÷ 0.194 |

| | Level 0 | Level 1/0 | Level 1/1 | Level 1/2 | Level 2/00 | Level 2/01 |
|---|---|---|---|---|---|---|
| $P_n(A_k)\in$ | 0.3045÷ 0.3045 | 0.3041÷ 0.3046 | 0.3043÷ 0.3046 | 0.3045÷ 0.3045 | 0.3039÷ 0.3046 | 0.3043÷ 0.3049 |
| $P_n(T_k)\in$ | 0.3064÷ 0.3066 | 0.3062÷ 0.3067 | 0.3065÷ 0.3065 | 0.3064÷ 0.3068 | 0.3061÷ 0.3061 | 0.306÷ 0.306 |
| $P_n(C_k)\in$ | 0.1948÷ 0.1948 | 0.1947÷ 0.1947 | 0.1946÷ 0.1948 | 0.1946÷ 0.1946 | 0.1945÷ 0.1953 | 0.1946÷ 0.1948 |
| $P_n(G_k)\in$ | 0.1937÷ 0.1942 | 0.194÷ 0.194 | 0.1934÷ 0.1942 | 0.1937÷ 0.194 | 0.1937÷ 0.194 | 0.1937÷ 0.1943 |

| | Level 2/02 | Level 2/10 | Level 2/11 | Level 2/12 | Level 2/20 | Level 2/21 | Level 2/22 |
|---|---|---|---|---|---|---|---|
| $P_n(A_k)\in$ | 0.304÷ 0.3042 | 0.3042÷ 0.3047 | 0.3044÷ 0.305 | 0.3042÷ 0.3046 | 0.3041÷ 0.3048 | 0.3045÷ 0.3052 | 0.3042÷ 0.3045 |
| $P_n(T_k)\in$ | 0.3061÷ 0.3064 | 0.3062÷ 0.3064 | 0.306÷ 0.3063 | 0.3065÷ 0.3068 | 0.3064÷ 0.3064 | 0.3059÷ 0.3061 | 0.3064÷ 0.3065 |
| $P_n(C_k)\in$ | 0.1945÷ 0.1951 | 0.1947÷ 0.1953 | 0.1942÷ 0.1948 | 0.1944÷ 0.1948 | 0.1946÷ 0.1953 | 0.1946÷ 0.195 | 0.1945÷ 0.1945 |
| $P_n(G_k)\in$ | 0.1939÷ 0.1944 | 0.1936÷ 0.1936 | 0.1932÷ 0.1939 | 0.1934÷ 0.1939 | 0.1935÷ 0.1935 | 0.1935÷ 0.1937 | 0.1937÷ 0.1945 |

Fig. A6/21. Tables of fluctuation intervals of probabilities $P_n(A_k)$, $P_n(T_k)$, $P_n(C_k)$ and $P_n(G_k)$ for the set of all texts at each of levels of convolutions in the FGT-2 (upper table) and FGT-3 (bottom tables) in the case of the sequence: Mus musculus strain C57BL/6J chromosome Y, GRCm38.p4 C57BL/6J

NCBI Reference Sequence: NC_000087.7, LOCUS NC_000087 91744698 bp DNA linear CON 22-JUN-2016 DEFINITION Mus musculus strain C57BL/6J chromosome Y, GRCm38.p4 C57BL/6J. ACCESSION NC_000087 GPC_000000794 VERSION NC_000087.7, https://www.ncbi.nlm.nih.gov/nuccore/NC_000087.7 . Here n = 1, 2, 3, 4, 5, k ≤ n. Level 0 corresponds to the initial DNA-text with its different n-letter representations. Other levels in this table correspond to sets of all daughter texts at appropriate levels of positional convolutions in the FGT-2 and the FGT-3.

| Chromo-somes | Fluctuations $P_n(A_k)$ | Fluctuations $P_n(T_k)$ | Fluctuations $P_n(C_k)$ | Fluctuations $P_n(G_k)$ |
|---|---|---|---|---|
| 1 | 0.2945÷0.2946 | 0.2939÷0.2940 | 0.2057÷0.2058 | 0.2055÷0.2057 |
| 2 | 0.2892÷0.2894 | 0.2897÷0.2898 | 0.2102÷0.2103 | 0.2104÷0.2105 |
| 3 | 0.2973÷0.2974 | 0.2980÷0.2982 | 0.2019÷0.2021 | 0.2024÷0.2024 |
| 4 | 0.2881÷0.2882 | 0.2887÷0.2890 | 0.2113÷0.2114 | 0.2114÷0.2114 |
| 5 | 0.2872÷0.2872 | 0.2873÷0.2875 | 0.2125÷0.2126 | 0.2126÷0.2127 |
| 6 | 0.2927÷0.2928 | 0.2930÷0.2930 | 0.2071÷0.2072 | 0.2069÷0.2070 |
| 7 | 0.2837÷0.2840 | 0.2855÷0.2855 | 0.2152÷0.2153 | 0.2150÷0.2151 |
| 8 | 0.2883÷0.2884 | 0.2878÷0.2879 | 0.2118÷0.2119 | 0.2117÷0.2119 |
| 9 | 0.2865÷0.2866 | 0.2862÷0.2864 | 0.2135÷0.2135 | 0.2134÷0.2134 |
| 10 | 0.2925÷0.2926 | 0.2932÷0.2933 | 0.2067÷0.2069 | 0.2071÷0.2072 |
| 11 | 0.2812÷0.2814 | 0.2805÷0.2806 | 0.2191÷0.2191 | 0.2189÷0.2189 |
| 12 | 0.2898÷0.2899 | 0.2926÷0.2928 | 0.2083÷0.2085 | 0.2088÷0.2088 |

| 13 | 0.2924÷0.2925 | 0.2911÷0.2913 | 0.2080÷0.2081 | 0.2079÷0.2081 |
|---|---|---|---|---|
| 14 | 0.2939÷0.2939 | 0.2943÷0.2944 | 0.2055÷0.2057 | 0.2058÷0.206 |
| 15 | 0.2898÷0.2898 | 0.2904÷0.2907 | 0.2098÷0.2098 | 0.2096÷0.2096 |
| 16 | 0.2949÷0.2951 | 0.2954÷0.2956 | 0.2045÷0.2045 | 0.2047÷0.2047 |
| 17 | 0.2860÷0.2864 | 0.2867÷0.2867 | 0.2134÷0.2136 | 0.2133÷0.2133 |
| 18 | 0.2928÷0.2928 | 0.2925÷0.2926 | 0.2068÷0.2070 | 0.2074÷0.2076 |
| 19 | 0.2873÷0.2876 | 0.2850÷0.2852 | 0.2137÷0.2139 | 0.2132÷0.2133 |
| X | 0.3037÷0.3039 | 0.3036÷0.3036 | 0.1961÷0.1963 | 0.1962÷0.1963 |
| Y | 0.3045÷0.3045 | 0.3064÷0.3066 | 0.1948÷0.1948 | 0.1937÷0.1942 |

Fig. A6/22. The table of fluctuations of collective probabilities $P_n(A_k)$, $P_n(T_k)$, $P_n(C_k)$ and $P_n(G_k)$ for initial DNA-texts of all chromosomes of Mus musculus. These small fluctuations exist under changes of values n and k (n = 1, 2, 3, 4, 5; k ≤ n). The probabilities are shown in fractions of a unit.

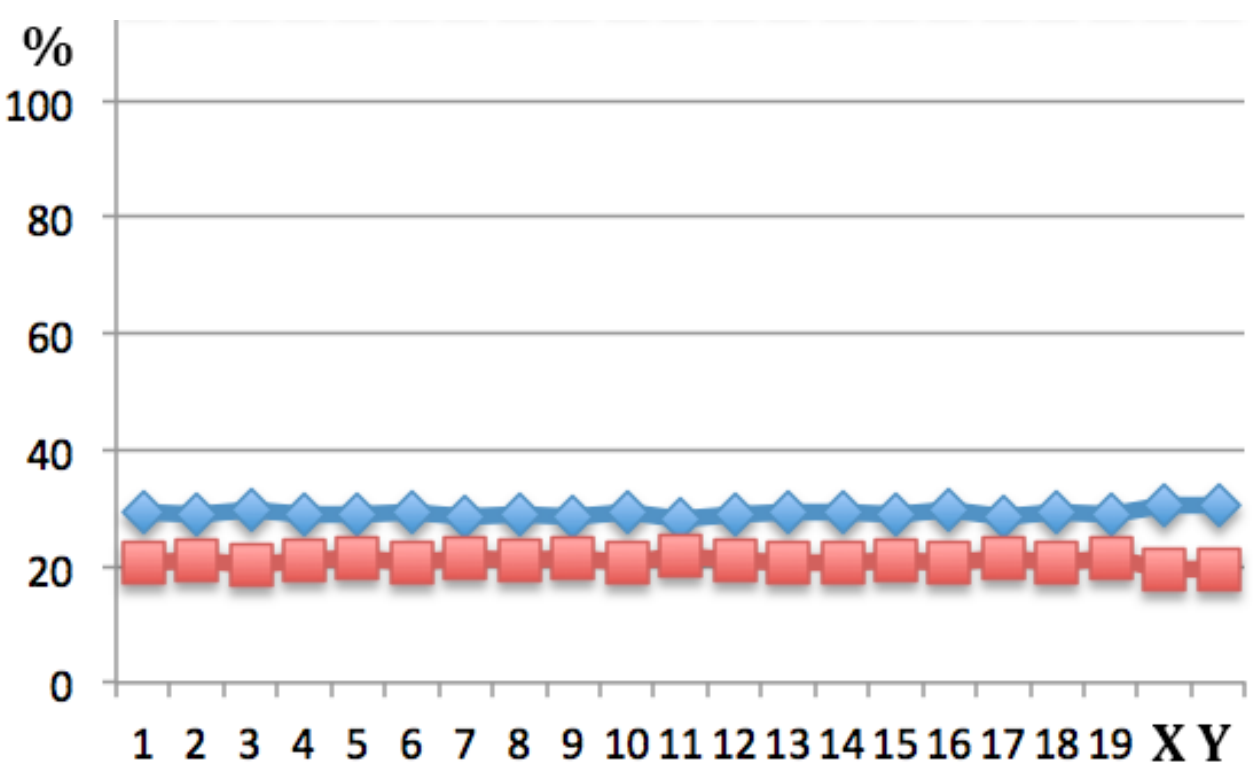

Fig. A6/23. The graphic representation of average values of probabilities $P_n(A_k)$ , $P_n(T_k)$, $P_n(C_k)$ and $P_n(G_k)$ in percentage for DNA-texts of chromosomes of *Mus Musculus*. The abscissa axis contains denotations of chromosomes, and the ordinate axis contains average values of these probabilities in percent. The symbol of a blue diamond corresponds the average values of $P_n(A_k) \approx P_n(T_k)$, and the symbol of a red square corresponds the average values of $P_n(C_k) \approx P_n(G_k)$.

## Appendix 7. Confirmations of model predictions on symmetries of collective probabilities for $4^s$-groups of n-plets in long DNA-texts

This Appendix represents a few initial confirmations of our model predictions made above in the end of the Section 5. These confirmations were received in the result of analysis of the single DNA-sequence with one million letters in it (this sequence was already analysed above for the table in Fig. 4). For this sequence, the table in Fig. A7/1 shows collective probabilities of $4^2$-groups, each of which contains subgroups defined by one of 16 doublets. Denotations in the table are similar for denotations used above. For example, $P_2(CC_1)$ means the individual probability of the doublet CC in this sequence; $P_3(CC_1)$ means the collective probability of the CC-subgroup of 4 triplets CCC, CCT, CCG and CCA with the doublet CC in their beginnings; $P_4(CC_3)$ means the collective probability of the CC-subgroup of 16 tetraplets (like AACC, ATCC, TGCC, etc.) with the same doublet CC in their ends.

One can see from the table in Fig. A7/1 that, for example, $P_3(AA_1) \approx P_3(AA_2) \approx P_4(AA_1) \approx P_4(AA_2) \approx P_4(AA_3)$. These approximate equalities correspond to the first and the second rules of tetra-group symmetries, which were described in the Section 3 for another case, that is for "4-groups of the first degree": C-subgroups, T-subgroups, G-subgroups and A-subgroups (see Figs. 1, 2). Data of the table in Fig. A7/1 testify in favor that – in line with our model prediction - similar rules of symmetries of collective probabilities are also fulfiled for $4^2$-groups («tetra-groups of the second degree»), subgroups of which are defined by 16 doublets: CC-subgroups, TG-subgroups, etc.

| $P_2(AA)$=0,10501 | $P_3(AA_1)$=0,10541 | $P_4(AA_1)$=0,10462 |
|---|---|---|
| | $P_3(AA_2)$=0,10442 | $P_4(AA_2)$=0,10458 |
| | | $P_4(AA_3)$=0,10495 |
| $P_2(AC)$=0,04687 | $P_3(AC_1)$=0,04675 | $P_4(AC_1)$=0,0469 |
| | $P_3(AC_2)$=0,04714 | $P_4(AC_2)$=0,04667 |
| | | $P_4(AC_3)$=0,04676 |
| $P_2(AG)$=0,06300 | $P_3(AG_1)$=0,06348 | $P_4(AG_1)$=0,06327 |
| | $P_3(AG_2)$=0,06382 | $P_4(AG_2)$=0,06417 |
| | | $P_4(AG_3)$=0,06273 |
| $P_2(AT)$=0,09242 | $P_3(AT_1)$=0,09233 | $P_4(AT_1)$=0,09264 |
| | $P_3(AT_2)$=0,09241 | $P_4(AT_2)$=0,09344 |
| | | $P_4(AT_3)$=0,09220 |
| $P_2(CA)$=0,065442 | $P_3(CA_1)$=0,06527 | $P_4(CA_1)$=0,06568 |
| | $P_3(CA_2)$=0,06551 | $P_4(CA_2)$=0,06572 |
| | | $P_4(CA_3)$=0,06520 |
| $P_2(CC)$= 0,03821 | $P_3(CC_1)$=0,03852 | $P_4(CC_1)$=0,03814 |
| | $P_3(CC_2)$=0,03852 | $P_4(CC_2)$=0,03834 |
| | | $P_4(CC_3)$=0,03827 |
| $P_2(CG_1)$=0,00560 | $P_3(CG_1)$=0,00584 | $P_4(CG_1)$=0,00569 |
| | $P_3(CG_2)$=0,00542 | $P_4(CG_2)$=0,00536 |
| | | $P_4(CG_3)$=0,00551 |
| $P_2(CT)$=0,067066 | $P_3(CT_1)$=0,067041 | $P_4(CT_1)$=0,06721 |
| | $P_3(CT_2)$=0,06762 | $P_4(CT_2)$=0,067244 |
| | | $P_4(CT_3)$=0,06692 |

| $P_2(GA)=0,05575$ | $P_3(GA_1)=0,05562$ | $P_4(GA_1)=0,05600$ |
|---|---|---|
| | $P_3(GA_2)=0,05548$ | $P_4(GA_2)=0,05490$ |
| | | $P_4(GA_3)=0,05550$ |
| $P_2(GC)=0,03341$ | $P_3(GC_1)=0,03282$ | $P_4(GC_1)=0,03342$ |
| | $P_3(GC_2)=0,03356$ | $P_4(GC_2)=0,03279$ |
| | | $P_4(GC_3)=0,03340$ |
| $P_2(GG)=0,03962$ | $P_3(GG_1)=0,03932$ | $P_4(GG_1)=0,03890$ |
| | $P_3(GG_2)=0,03951$ | $P_4(GG_2)=0,03924$ |
| | | $P_4(GG_3)=0,04034$ |
| $P_2(GT)=0,05256$ | $P_3(GT_1)=0,05277$ | $P_4(GT_1)=0,05263$ |
| | $P_3(GT_2)=0,05224$ | $P_4(GT_2)=0,05184$ |
| | | $P_4(GT_3)=0,05250$ |
| $P_2(TA)=0,08153$ | $P_3(TA_1)=0,08149$ | $P_4(TA_1)=0,08211$ |
| | $P_3(TA_2)=0,08140$ | $P_4(TA_2)=0,08144$ |
| | | $P_4(TA_3)=0,08094$ |
| $P_2(TC)=0,05858$ | $P_3(TC_1)=0,05834$ | $P_4(TC_1)=0,05814$ |
| | $P_3(TC_2)=0,05836$ | $P_4(TC_2)=0,05810$ |
| | | $P_4(TC_3)=0,05902$ |
| $P_2(TG)=0,07196$ | $P_3(TG_1)=0,07215$ | $P_4(TG_1)=0,07157$ |
| | $P_3(TG_2)=0,07222$ | $P_4(TG_2)=0,07298$ |
| | | $P_4(TG_3)=0,07234$ |
| $P_2(TT)=0,12297$ | $P_3(TT_1)=0,12284$ | $P_4(TT_1)=0,12254$ |
| | $P_3(TT_2)=0,12299$ | $P_4(TT_2)=0,12250$ |
| | | $P_4(TT_3)=0,12339$ |

Fig. A7/1. Analysis of the collective probabilities of $4^2$-groups of n-plets in the Homo sapiens chromosome 7 sequence, 1000000 bp, encode region ENm012, accession NT_086368, version NT_086368.3, https://www.ncbi.nlm.nih.gov/nuccore/NT_086368.3. Each of $4^2$-groups contains subgroups defined by one of 16 doublets. There are shown the individual probabilities of each of 16 doublets (like $P_2(AA)$, etc.) and also collective probabilities for those sets of triplets and 4-plets (like $P_3(AA_1)$, $P_4(AA_3)$, etc.), where each of 16 doublets plays the decisive atributive role, taking its position in the beginning or in the middle or in the end of these n-plets (n=3, 4). See additional explanations in the text.

The table in Fig. A7/2 shows collective probabilities for the case of $4^3$-groups, each of which contains subgroups defined by one of 64 triplets. In this table the denotation $P_3(CCC)$ means the individual probability of the triplet CCC in the considered sequence; $P_4(CCC_1)$ means the collective probability of the CCC-subgroup of 4 tetraplets CCCA, CCCT, CCCG and CCCC with the triplet CCC in their beginnings; $P_4(CCC_3)$ means the collective probability of the CCC-subgroup of 16 tetraplets with the same triplet CCC in their ends (like ACCC, TCCC, GCCC, CCCC), etc. One can see from the table in Fig. A7/2 that, for example, $P_3(AAA) \approx P_4(AAA_1)$. Data of the table in Fig. A7/1 testify in favor that – in line with the prediction of our model in the Section 5 - similar rules of symmetries of collective probabilities are also fulfiled for $4^3$-groups («tetra-groups of the third

degree»), subgroups of which are defined by 64 triplets: CCC-subgroups, TGA-subgroups, etc.

| TRIPLETS | 4-PLETS | TRIPLETS | 4-PLETS |
| --- | --- | --- | --- |
| $P_3$(AAA)=0,04086 | $P_4$(AAA$_1$)=0,04049 | $P_3$(TAA)=0,02588 | $P_4$(TAA$_1$)=0,02604 |
| $P_3$(AAC)=0,01459 | $P_4$(AAC$_1$)=0,01439 | $P_3$(TAC)=0,01277 | $P_4$(TAC$_1$)=0,01276 |
| $P_3$(AAG)=0,01961 | $P_4$(AAG$_1$)=0,01958 | $P_3$(TAG)=0,01507 | $P_4$(TAG$_1$)=0,01489 |
| $P_3$(AAT)=0,03036 | $P_4$(AAT$_1$)=0,03062 | $P_3$(TAT)=0,02777 | $P_4$(TAT$_1$)=0,02841 |
| $P_3$(ACA)=0,01949 | $P_4$(ACA$_1$)=0,01958 | $P_3$(TCA)=0,01999 | $P_4$(TCA$_1$)=0,01970 |
| $P_3$(ACC)=0,00886 | $P_4$(ACC$_1$)=0,00896 | $P_3$(TCC)=0,01286 | $P_4$(TCC$_1$)=0,01295 |
| $P_3$(ACG)=0,00157 | $P_4$(ACG$_1$)=0,00155 | $P_3$(TCG)=0,00145 | $P_4$(TCG$_1$)=0,00148 |
| $P_3$(ACT)=0,01683 | $P_4$(ACT$_1$)=0,01689 | $P_3$(TCT)=0,02404 | $P_4$(TCT$_1$)=0,02400 |
| $P_3$(AGA)=0,02103 | $P_4$(AGA$_1$)=0,02077 | $P_3$(TGA)=0,02031 | $P_4$(TGA$_1$)=0,02016 |
| $P_3$(AGC)=0,01136 | $P_4$(AGC$_1$)=0,01114 | $P_3$(TGC)=0,01327 | $P_4$(TGC$_1$)=0,01296 |
| $P_3$(AGG)=0,01373 | $P_4$(AGG$_1$)=0,01384 | $P_3$(TGG)=0,01590 | $P_4$(TGG$_1$)=0,01560 |
| $P_3$(AGT)=0,01737 | $P_4$(AGT$_1$)=0,01752 | $P_3$(TGT)=0,02268 | $P_4$(TGT$_1$)=0,02286 |
| $P_3$(ATA)=0,02617 | $P_4$(ATA$_1$)=0,02595 | $P_3$(TTA)=0,02783 | $P_4$(TTA$_1$)=0,02801 |
| $P_3$(ATC)=0,01365 | $P_4$(ATC$_1$)=0,01370 | $P_3$(TTC)=0,02259 | $P_4$(TTC$_1$)=0,02235 |
| $P_3$(ATG)=0,01988 | $P_4$(ATG$_1$)=0,02024 | $P_3$(TTG)=0,02225 | $P_4$(TTG$_1$)=0,02220 |
| $P_3$(ATT)=0,03264 | $P_4$(ATT$_1$)=0,03275 | $P_3$(TTT)=0,05017 | $P_4$(TTT$_1$)=0,05000 |
| | | | |
| $P_3$(CAA)=0,01751 | $P_4$(CAA$_1$)=0,01786 | $P_3$(GAA)=0,02018 | $P_4$(GAA$_1$)=0,02019 |
| $P_3$(CAC)=0,01183 | $P_4$(CAC$_1$)=0,01174 | $P_3$(GAC)=0,00796 | $P_4$(GAC$_1$)=0,00778 |
| $P_3$(CAG)=0,01580 | $P_4$(CAG$_1$)=0,01610 | $P_3$(GAG)=0,01334 | $P_4$(GAG$_1$)=0,01360 |
| $P_3$(CAT)=0,02014 | $P_4$(CAT$_1$)=0,01998 | $P_3$(GAT)=0,01415 | $P_4$(GAT$_1$)=0,01444 |
| $P_3$(CCA)=0,01402 | $P_4$(CCA$_1$)=0,01427 | $P_3$(GCA)=0,01201 | $P_4$(GCA$_1$)=0,01218 |
| $P_3$(CCC)=0,00861 | $P_4$(CCC$_1$)=0,00850 | $P_3$(GCC)=0,00756 | $P_4$(GCC$_1$)=0,00792 |
| $P_3$(CCG)=0,00129 | $P_4$(CCG$_1$)=0,00117 | $P_3$(GCG)=0,00111 | $P_4$(GCG$_1$)=0,00116 |
| $P_3$(CCT)=0,01460 | $P_4$(CCT$_1$)=0,01420 | $P_3$(GCT)=0,01214 | $P_4$(GCT$_1$)=0,01216 |
| $P_3$(CGA)=0,00152 | $P_4$(CGA$_1$)=0,00147 | $P_3$(GGA)=0,01262 | $P_4$(GGA$_1$)=0,01250 |
| $P_3$(CGC)=0,00120 | $P_4$(CGC$_1$)=0,00108 | $P_3$(GGC)=0,00775 | $P_4$(GGC$_1$)=0,00762 |
| $P_3$(CGG)=0,00126 | $P_4$(CGG$_1$)=0,00120 | $P_3$(GGG)=0,00863 | $P_4$(GGG$_1$)=0,00860 |
| $P_3$(CGT)=0,00187 | $P_4$(CGT$_1$)=0,00193 | $P_3$(GGT)=0,01033 | $P_4$(GGT$_1$)=0,01019 |
| $P_3$(CTA)=0,01412 | $P_4$(CTA$_1$)=0,01425 | $P_3$(GTA)=0,01328 | $P_4$(GTA$_1$)=0,01323 |
| $P_3$(CTC)=0,01356 | $P_4$(CTC$_1$)=0,01353 | $P_3$(GTC)=0,00856 | $P_4$(GTC$_1$)=0,00852 |
| $P_3$(CTG)=0,01682 | $P_4$(CTG$_1$)=0,01704 | $P_3$(GTG)=0,01328 | $P_4$(GTG$_1$)=0,01350 |
| $P_3$(CTT)=0,02254 | $P_4$(CTT$_1$)=0,02239 | $P_3$(GTT)=0,01766 | $P_4$(GTT$_1$)=0,01738 |

Fig. A7/2. Analysis of the collective probabilities of $4^3$-groups of n-plets in the Homo sapiens chromosome 7 sequence, 1000000 bp, encode region ENm012, accession NT_086368, version NT_086368.3, [https://www.ncbi.nlm.nih.gov/nuccore/NT_086368.3](https://www.ncbi.nlm.nih.gov/nuccore/NT_086368.3). Each of $4^3$-groups contains subgroups defined by one of 64 triplets. There are shown the individual probabilities of each of 64 triplets (like $P_3$(AAA), etc.) and also collective probabilities for those sets of 4-plets (like $P_4$(AAA$_1$), etc.), where each of 64 triplets plays the decisive atributive role, taking its position in the beginning of these 4-plets. See additional explanations in the text.

## Appendix 8. Symmetries of tetra-group probabilities in genomes of microorganisms living in extreme environments

The https://en.wikipedia.org/wiki/Extremophile website contains a table of microorganisms living under extreme conditions of high and low temperatures, radiation, acidic and alkaline environments, and drying. For the test, the author used 1-2 organisms from each category of the table. The following data were received on the collective probabilities $P_n(A_k)$, $P_n(T_k)$, $P_n(C_k)$ and $P_n(G_k)$ of all members of the corresponding tetra-groups of n-plets (or tetra-alphabets of n-plets) (n = 1, 2, 3, 4, 5; k ≤n ) in the genomes of these organisms (the initial data on the genomes were taken from GenBank).

| Ranges of fluctuations of collective probabilities for different positions in n-plets | | | | | | | |
|---|---|---|---|---|---|---|---|
| $P_n(A_k)$ | 0.2239÷ 0.227 | $P_n(T_k)$ | 0.2242÷ 0.2267 | $P_n(C_k)$ | 0.2725÷ 0.2764 | $P_n(G_k)$ | 0.2739÷ 0.2761 |

Detailed data on collective probabilities $P_n(A_k)$, $P_n(T_k)$, $P_n(C_k)$ and $P_n(G_k)$ (n = 1, 2, 3, 4, 5; k ≤n ) for each position in n-plets:

| **NUCLEOTIDES** | | **DOUBLETS** | | **TRIPLETS** | | **4-PLETS** | | **5-PLETS** | |
|---|---|---|---|---|---|---|---|---|---|
| $P_1(A_1)=$ | 0.2256 | $P_2(A_1)=$ | 0.2258 | $P_3(A_1)=$ | 0.226 | $P_4(A_1)=$ | 0.2256 | $P_5(A_1)=$ | 0.2259 |
| $P_1(T_1)=$ | 0.2254 | $P_2(T_1)=$ | 0.2255 | $P_3(T_1)=$ | 0.2252 | $P_4(T_1)=$ | 0.2256 | $P_5(T_1)=$ | 0.2257 |
| $P_1(C_1)=$ | 0.274 | $P_2(C_1)=$ | 0.274 | $P_3(C_1)=$ | 0.273 | $P_4(C_1)=$ | 0.2745 | $P_5(C_1)=$ | 0.2739 |
| $P_1(G_1)=$ | 0.2751 | $P_2(G_1)=$ | 0.2747 | $P_3(G_1)=$ | 0.2759 | $P_4(G_1)=$ | 0.2743 | $P_5(G_1)=$ | 0.2745 |
| | | $P_2(A_2)=$ | 0.2254 | $P_3(A_2)=$ | 0.2269 | $P_4(A_2)=$ | 0.2258 | $P_5(A_2)=$ | 0.2255 |
| | | $P_2(T_2)=$ | 0.2252 | $P_3(T_2)=$ | 0.2267 | $P_4(T_2)=$ | 0.226 | $P_5(T_2)=$ | 0.2247 |
| | | $P_2(C_2)=$ | 0.274 | $P_3(C_2)=$ | 0.2725 | $P_4(C_2)=$ | 0.2736 | $P_5(C_2)=$ | 0.2746 |
| | | $P_2(G_2)=$ | 0.2754 | $P_3(G_2)=$ | 0.2739 | $P_4(G_2)=$ | 0.2747 | $P_5(G_2)=$ | 0.2752 |
| | | | | $P_3(A_3)=$ | 0.2239 | $P_4(A_3)=$ | 0.226 | $P_5(A_3)=$ | 0.227 |
| | | | | $P_3(T_3)=$ | 0.2242 | $P_4(T_3)=$ | 0.2254 | $P_5(T_3)=$ | 0.2254 |
| | | | | $P_3(C_3)=$ | 0.2764 | $P_4(C_3)=$ | 0.2735 | $P_5(C_3)=$ | 0.2729 |
| | | | | $P_3(G_3)=$ | 0.2754 | $P_4(G_3)=$ | 0.2752 | $P_5(G_3)=$ | 0.2747 |
| | | | | | | $P_4(A_4)=$ | 0.2251 | $P_5(A_4)=$ | 0.2242 |
| | | | | | | $P_4(T_4)=$ | 0.2245 | $P_5(T_4)=$ | 0.2267 |
| | | | | | | $P_4(C_4)=$ | 0.2743 | $P_5(C_4)=$ | 0.274 |
| | | | | | | $P_4(G_4)=$ | 0.2761 | $P_5(G_4)=$ | 0.2751 |
| | | | | | | | | $P_5(A_5)=$ | 0.2253 |
| | | | | | | | | $P_5(T_5)=$ | 0.2244 |
| | | | | | | | | $P_5(C_5)=$ | 0.2745 |
| | | | | | | | | $P_5(G_5)=$ | 0.2759 |

Fig. 8/1. Pyrolobus fumarii 1A, complete genome, 1843267 bp (Submarine hydrothermal vents), *https://www.ncbi.nlm.nih.gov/nuccore/NC_015931.1*

| Ranges of fluctuations of collective probabilities for different positions in n-plets | | | | | | | |
|---|---|---|---|---|---|---|---|
| $P_n(A_k)$ | $P_n(A_k)$ | $P_n(T_k)$ | 0.2947÷ 0.297 | $P_n(C_k)$ | 0.2008÷ 0.2081 | $P_n(G_k)$ | 0.2024÷ 0.206 |

Detailed data on collective probabilities $P_n(A_k)$, $P_n(T_k)$, $P_n(C_k)$ and $P_n(G_k)$ ($n = 1, 2, 3, 4, 5$; $k \leq n$ ) for each position in n-plets:

| **NUCLEOTIDES** | | **DOUBLETS** | | **TRIPLETS** | | **4-PLETS** | | **5-PLETS** | |
|---|---|---|---|---|---|---|---|---|---|
| $P_1(A_1)=$ | 0.2962 | $P_2(A_1)=$ | 0.2969 | $P_3(A_1)=$ | 0.2988 | $P_4(A_1)=$ | 0.296 | $P_5(A_1)=$ | 0.2964 |
| $P_1(T_1)=$ | 0.2961 | $P_2(T_1)=$ | 0.2955 | $P_3(T_1)=$ | 0.297 | $P_4(T_1)=$ | 0.2956 | $P_5(T_1)=$ | 0.2963 |
| $P_1(C_1)=$ | 0.2037 | $P_2(C_1)=$ | 0.2035 | $P_3(C_1)=$ | 0.2008 | $P_4(C_1)=$ | 0.2041 | $P_5(C_1)=$ | 0.2037 |
| $P_1(G_1)=$ | 0.204 | $P_2(G_1)=$ | 0.2042 | $P_3(G_1)=$ | 0.2034 | $P_4(G_1)=$ | 0.2042 | $P_5(G_1)=$ | 0.2035 |
| | | $P_2(A_2)=$ | 0.2954 | $P_3(A_2)=$ | 0.2968 | $P_4(A_2)=$ | 0.2951 | $P_5(A_2)=$ | 0.2962 |
| | | $P_2(T_2)=$ | 0.2968 | $P_3(T_2)=$ | 0.2947 | $P_4(T_2)=$ | 0.297 | $P_5(T_2)=$ | 0.296 |
| | | $P_2(C_2)=$ | 0.2038 | $P_3(C_2)=$ | 0.2021 | $P_4(C_2)=$ | 0.2042 | $P_5(C_2)=$ | 0.2041 |
| | | $P_2(G_2)=$ | 0.2039 | $P_3(G_2)=$ | 0.2064 | $P_4(G_2)=$ | 0.2038 | $P_5(G_2)=$ | 0.2037 |
| | | | | $P_3(A_3)=$ | 0.2929 | $P_4(A_3)=$ | 0.2978 | $P_5(A_3)=$ | 0.2965 |
| | | | | $P_3(T_3)=$ | 0.2967 | $P_4(T_3)=$ | 0.2953 | $P_5(T_3)=$ | 0.2957 |
| | | | | $P_3(C_3)=$ | 0.2081 | $P_4(C_3)=$ | 0.2028 | $P_5(C_3)=$ | 0.2038 |
| | | | | $P_3(G_3)=$ | 0.2024 | $P_4(G_3)=$ | 0.2041 | $P_5(G_3)=$ | 0.204 |
| | | | | | | $P_4(A_4)=$ | 0.2958 | $P_5(A_4)=$ | 0.2962 |
| | | | | | | $P_4(T_4)=$ | 0.2967 | $P_5(T_4)=$ | 0.2958 |
| | | | | | | $P_4(C_4)=$ | 0.2035 | $P_5(C_4)=$ | 0.2034 |
| | | | | | | $P_4(G_4)=$ | 0.2041 | $P_5(G_4)=$ | 0.2046 |
| | | | | | | | | $P_5(A_5)=$ | 0.2955 |
| | | | | | | | | $P_5(T_5)=$ | 0.2969 |
| | | | | | | | | $P_5(C_5)=$ | 0.2031 |
| | | | | | | | | $P_5(G_5)=$ | 0.204 |

Fig. 8/2. Pyrococcus furiosus DSM 3638, complete genome, 1908256 bp (Submarine hydrothermal vents), *https://www.ncbi.nlm.nih.gov/nuccore/NC_003413.1*

| Ranges of fluctuations of collective probabilities for different positions in n-plets | | | | | | | |
|---|---|---|---|---|---|---|---|
| $P_n(A_k)$ | 0.2315÷ 0.2343 | $P_n(T_k)$ | 0.2311÷ 0.2329 | $P_n(C_k)$ | 0.2648÷ 0.2694 | $P_n(G_k)$ | 0.2662÷ 0.2684 |

Detailed data on collective probabilities $P_n(A_k)$, $P_n(T_k)$, $P_n(C_k)$ and $P_n(G_k)$ (n = 1, 2, 3, 4, 5; $k \leq n$ ) for each position in n-plets:

| **NUCLEOTIDES** | | **DOUBLETS** | | **TRIPLETS** | | **4-PLETS** | | **5-PLETS** | |
|---|---|---|---|---|---|---|---|---|---|
| $P_1(A_1)$= | 0.2326 | $P_2(A_1)$= | 0.2325 | $P_3(A_1)$= | 0.232 | $P_4(A_1)$= | 0.2329 | $P_5(A_1)$= | 0.232 |
| $P_1(T_1)$= | 0.2323 | $P_2(T_1)$= | 0.2328 | $P_3(T_1)$= | 0.2311 | $P_4(T_1)$= | 0.2327 | $P_5(T_1)$= | 0.2319 |
| $P_1(C_1)$= | 0.2678 | $P_2(C_1)$= | 0.2677 | $P_3(C_1)$= | 0.2692 | $P_4(C_1)$= | 0.2675 | $P_5(C_1)$= | 0.2681 |
| $P_1(G_1)$= | 0.2673 | $P_2(G_1)$= | 0.2671 | $P_3(G_1)$= | 0.2677 | $P_4(G_1)$= | 0.2669 | $P_5(G_1)$= | 0.268 |
| | | $P_2(A_2)$= | 0.2327 | $P_3(A_2)$= | 0.2343 | $P_4(A_2)$= | 0.2334 | $P_5(A_2)$= | 0.2331 |
| | | $P_2(T_2)$= | 0.2317 | $P_3(T_2)$= | 0.2329 | $P_4(T_2)$= | 0.2322 | $P_5(T_2)$= | 0.2317 |
| | | $P_2(C_2)$= | 0.2679 | $P_3(C_2)$= | 0.2648 | $P_4(C_2)$= | 0.2672 | $P_5(C_2)$= | 0.2682 |
| | | $P_2(G_2)$= | 0.2676 | $P_3(G_2)$= | 0.268 | $P_4(G_2)$= | 0.2671 | $P_5(G_2)$= | 0.267 |
| | | | | $P_3(A_3)$= | 0.2315 | $P_4(A_3)$= | 0.232 | $P_5(A_3)$= | 0.2325 |
| | | | | $P_3(T_3)$= | 0.2327 | $P_4(T_3)$= | 0.2329 | $P_5(T_3)$= | 0.2325 |
| | | | | $P_3(C_3)$= | 0.2694 | $P_4(C_3)$= | 0.2679 | $P_5(C_3)$= | 0.2679 |
| | | | | $P_3(G_3)$= | 0.2663 | $P_4(G_3)$= | 0.2672 | $P_5(G_3)$= | 0.2671 |
| | | | | | | $P_4(A_4)$= | 0.232 | $P_5(A_4)$= | 0.232 |
| | | | | | | $P_4(T_4)$= | 0.2312 | $P_5(T_4)$= | 0.2323 |
| | | | | | | $P_4(C_4)$= | 0.2686 | $P_5(C_4)$= | 0.2673 |
| | | | | | | $P_4(G_4)$= | 0.2681 | $P_5(G_4)$= | 0.2684 |
| | | | | | | | | $P_5(A_5)$= | 0.2334 |
| | | | | | | | | $P_5(T_5)$= | 0.2329 |
| | | | | | | | | $P_5(C_5)$= | 0.2675 |
| | | | | | | | | $P_5(G_5)$= | 0.2662 |

Fig. 8/3. Synechococcus lividus PCC 6715 chromosome, complete genome, 2659739 bp (low temperature conditions), https://www.ncbi.nlm.nih.gov/nuccore/NZ_CP018092.1 )

| Ranges of fluctuations of collective probabilities for different positions in n-plets | | | | | | | |
|---|---|---|---|---|---|---|---|
| $P_n(A_k)$ | 0.2845÷ 0.2869 | $P_n(T_k)$ | 0.2846÷ 0.2866 | $P_n(C_k)$ | 0.2129÷ 0.2159 | $P_n(G_k)$ | 0.213÷ 0.2157 |

Detailed data on collective probabilities $P_n(A_k)$, $P_n(T_k)$, $P_n(C_k)$ and $P_n(G_k)$ (n = 1, 2, 3, 4, 5; $k \leq n$ ) for each position in n-plets:

| **NUCLEOTIDES** | | **DOUBLETS** | | **TRIPLETS** | | **4-PLETS** | | **5-PLETS** | |
|---|---|---|---|---|---|---|---|---|---|
| $P_1(A_1)$= | 0.2855 | $P_2(A_1)$= | 0.2854 | $P_3(A_1)$= | 0.2845 | $P_4(A_1)$= | 0.2857 | $P_5(A_1)$= | 0.285 |
| $P_1(T_1)$= | 0.2858 | $P_2(T_1)$= | 0.2858 | $P_3(T_1)$= | 0.2866 | $P_4(T_1)$= | 0.2854 | $P_5(T_1)$= | 0.2849 |
| $P_1(C_1)$= | 0.2146 | $P_2(C_1)$= | 0.2148 | $P_3(C_1)$= | 0.2159 | $P_4(C_1)$= | 0.2151 | $P_5(C_1)$= | 0.2156 |
| $P_1(G_1)$= | 0.2142 | $P_2(G_1)$= | 0.214 | $P_3(G_1)$= | 0.213 | $P_4(G_1)$= | 0.2138 | $P_5(G_1)$= | 0.2145 |
| | | $P_2(A_2)$= | 0.2856 | $P_3(A_2)$= | 0.2851 | $P_4(A_2)$= | 0.286 | $P_5(A_2)$= | 0.2854 |
| | | $P_2(T_2)$= | 0.2857 | $P_3(T_2)$= | 0.2861 | $P_4(T_2)$= | 0.2859 | $P_5(T_2)$= | 0.2859 |
| | | $P_2(C_2)$= | 0.2144 | $P_3(C_2)$= | 0.215 | $P_4(C_2)$= | 0.2141 | $P_5(C_2)$= | 0.2145 |
| | | $P_2(G_2)$= | 0.2144 | $P_3(G_2)$= | 0.2139 | $P_4(G_2)$= | 0.214 | $P_5(G_2)$= | 0.2143 |
| | | | | $P_3(A_3)$= | 0.2869 | $P_4(A_3)$= | 0.2851 | $P_5(A_3)$= | 0.286 |
| | | | | $P_3(T_3)$= | 0.2846 | $P_4(T_3)$= | 0.2862 | $P_5(T_3)$= | 0.2857 |
| | | | | $P_3(C_3)$= | 0.2129 | $P_4(C_3)$= | 0.2146 | $P_5(C_3)$= | 0.2144 |
| | | | | $P_3(G_3)$= | 0.2157 | $P_4(G_3)$= | 0.2141 | $P_5(G_3)$= | 0.2139 |
| | | | | | | $P_4(A_4)$= | 0.2851 | $P_5(A_4)$= | 0.2859 |
| | | | | | | $P_4(T_4)$= | 0.2856 | $P_5(T_4)$= | 0.286 |
| | | | | | | $P_4(C_4)$= | 0.2147 | $P_5(C_4)$= | 0.2141 |
| | | | | | | $P_4(G_4)$= | 0.2147 | $P_5(G_4)$= | 0.214 |
| | | | | | | | | $P_5(A_5)$= | 0.2851 |
| | | | | | | | | $P_5(T_5)$= | 0.2864 |
| | | | | | | | | $P_5(C_5)$= | 0.2144 |
| | | | | | | | | $P_5(G_5)$= | 0.2141 |

Fig. 8/4. Psychrobacter alimentarius strain PAMC 27889 chromosome, complete genome, 3332539 bp (Soda lakes), *https://www.ncbi.nlm.nih.gov/nuccore/NZ_CP014945.1*

| Ranges of fluctuations of collective probabilities for different positions in n-plets | | | | | | | |
|---|---|---|---|---|---|---|---|
| $P_n(A_k)$ | 0.3485÷ 0.3565 | $P_n(T_k)$ | 0.3455÷ 0.348 | $P_n(C_k)$ | 0.1434÷ 0.1495 | $P_n(G_k)$ | 0.1522÷ 0.1564 |

Detailed data on collective probabilities $P_n(A_k)$, $P_n(T_k)$, $P_n(C_k)$ and $P_n(G_k)$ ($n$ = 1, 2, 3, 4, 5; $k \leq n$ ) for each position in n-plets:

| **NUCLEOTIDES** | | **DOUBLETS** | | **TRIPLETS** | | **4-PLETS** | | **5-PLETS** | |
|---|---|---|---|---|---|---|---|---|---|
| $P_1(A_1)$= | 0.3521 | $P_2(A_1)$= | 0.3522 | $P_3(A_1)$= | 0.3515 | $P_4(A_1)$= | 0.3525 | $P_5(A_1)$= | 0.3525 |
| $P_1(T_1)$= | 0.347 | $P_2(T_1)$= | 0.3466 | $P_3(T_1)$= | 0.3475 | $P_4(T_1)$= | 0.3458 | $P_5(T_1)$= | 0.3469 |
| $P_1(C_1)$= | 0.1473 | $P_2(C_1)$= | 0.1477 | $P_3(C_1)$= | 0.1489 | $P_4(C_1)$= | 0.1481 | $P_5(C_1)$= | 0.148 |
| $P_1(G_1)$= | 0.1536 | $P_2(G_1)$= | 0.1535 | $P_3(G_1)$= | 0.1522 | $P_4(G_1)$= | 0.1537 | $P_5(G_1)$= | 0.1525 |
| | | $P_2(A_2)$= | 0.3521 | $P_3(A_2)$= | 0.3565 | $P_4(A_2)$= | 0.3524 | $P_5(A_2)$= | 0.3525 |
| | | $P_2(T_2)$= | 0.3474 | $P_3(T_2)$= | 0.348 | $P_4(T_2)$= | 0.3471 | $P_5(T_2)$= | 0.3458 |
| | | $P_2(C_2)$= | 0.1468 | $P_3(C_2)$= | 0.1434 | $P_4(C_2)$= | 0.147 | $P_5(C_2)$= | 0.1471 |
| | | $P_2(G_2)$= | 0.1537 | $P_3(G_2)$= | 0.1522 | $P_4(G_2)$= | 0.1535 | $P_5(G_2)$= | 0.1546 |
| | | | | $P_3(A_3)$= | 0.3485 | $P_4(A_3)$= | 0.3519 | $P_5(A_3)$= | 0.353 |
| | | | | $P_3(T_3)$= | 0.3455 | $P_4(T_3)$= | 0.3475 | $P_5(T_3)$= | 0.3475 |
| | | | | $P_3(C_3)$= | 0.1495 | $P_4(C_3)$= | 0.1473 | $P_5(C_3)$= | 0.1466 |
| | | | | $P_3(G_3)$= | 0.1564 | $P_4(G_3)$= | 0.1534 | $P_5(G_3)$= | 0.1529 |
| | | | | | | $P_4(A_4)$= | 0.3518 | $P_5(A_4)$= | 0.3516 |
| | | | | | | $P_4(T_4)$= | 0.3476 | $P_5(T_4)$= | 0.3471 |
| | | | | | | $P_4(C_4)$= | 0.1467 | $P_5(C_4)$= | 0.1472 |
| | | | | | | $P_4(G_4)$= | 0.1539 | $P_5(G_4)$= | 0.1541 |
| | | | | | | | | $P_5(A_5)$= | 0.3511 |
| | | | | | | | | $P_5(T_5)$= | 0.3477 |
| | | | | | | | | $P_5(C_5)$= | 0.1473 |
| | | | | | | | | $P_5(G_5)$= | 0.1539 |

Fig. 8/5. Clostridium paradoxum JW-YL-7 = DSM 7308 strain JW-YL-7 ctg1, whole genome shotgun sequence, 1855173 bp (Volcanic springs, acid mine drainage), https://www.ncbi.nlm.nih.gov/nuccore/LSFY01000001.1

| Ranges of fluctuations of collective probabilities for different positions in n-plets | | | | | | | |
|---|---|---|---|---|---|---|---|
| $P_n(A_k)$ | 0.1624÷ 0.166 | $P_n(T_k)$ | 0.1622÷ 0.1674 | $P_n(C_k)$ | 0.3341÷ 0.3379 | $P_n(G_k)$ | 0.3325÷ 0.3375 |

Detailed data on collective probabilities $P_n(A_k)$, $P_n(T_k)$, $P_n(C_k)$ and $P_n(G_k)$ (n = 1, 2, 3, 4, 5; k ≤n ) for each position in n-plets:

| **NUCLEOTIDES** | | **DOUBLETS** | | **TRIPLETS** | | **4-PLETS** | | **5-PLETS** | |
|---|---|---|---|---|---|---|---|---|---|
| $P_1(A_1)$= | 0.1648 | $P_2(A_1)$= | 0.1644 | $P_3(A_1)$= | 0.1624 | $P_4(A_1)$= | 0.1638 | $P_5(A_1)$= | 0.1653 |
| $P_1(T_1)$= | 0.1651 | $P_2(T_1)$= | 0.165 | $P_3(T_1)$= | 0.1622 | $P_4(T_1)$= | 0.165 | $P_5(T_1)$= | 0.1652 |
| $P_1(C_1)$= | 0.3354 | $P_2(C_1)$= | 0.3357 | $P_3(C_1)$= | 0.3379 | $P_4(C_1)$= | 0.3365 | $P_5(C_1)$= | 0.335 |
| $P_1(G_1)$= | 0.3347 | $P_2(G_1)$= | 0.335 | $P_3(G_1)$= | 0.3375 | $P_4(G_1)$= | 0.3347 | $P_5(G_1)$= | 0.3344 |
| | | $P_2(A_2)$= | 0.1653 | $P_3(A_2)$= | 0.166 | $P_4(A_2)$= | 0.1654 | $P_5(A_2)$= | 0.1647 |
| | | $P_2(T_2)$= | 0.1652 | $P_3(T_2)$= | 0.1656 | $P_4(T_2)$= | 0.1656 | $P_5(T_2)$= | 0.1646 |
| | | $P_2(C_2)$= | 0.3351 | $P_3(C_2)$= | 0.3342 | $P_4(C_2)$= | 0.3348 | $P_5(C_2)$= | 0.336 |
| | | $P_2(G_2)$= | 0.3344 | $P_3(G_2)$= | 0.3341 | $P_4(G_2)$= | 0.3343 | $P_5(G_2)$= | 0.3347 |
| | | | | $P_3(A_3)$= | 0.166 | $P_4(A_3)$= | 0.1649 | $P_5(A_3)$= | 0.1648 |
| | | | | $P_3(T_3)$= | 0.1674 | $P_4(T_3)$= | 0.1649 | $P_5(T_3)$= | 0.1649 |
| | | | | $P_3(C_3)$= | 0.3341 | $P_4(C_3)$= | 0.3349 | $P_5(C_3)$= | 0.3354 |
| | | | | $P_3(G_3)$= | 0.3325 | $P_4(G_3)$= | 0.3352 | $P_5(G_3)$= | 0.3349 |
| | | | | | | $P_4(A_4)$= | 0.1651 | $P_5(A_4)$= | 0.1648 |
| | | | | | | $P_4(T_4)$= | 0.1648 | $P_5(T_4)$= | 0.1656 |
| | | | | | | $P_4(C_4)$= | 0.3355 | $P_5(C_4)$= | 0.335 |
| | | | | | | $P_4(G_4)$= | 0.3346 | $P_5(G_4)$= | 0.3346 |
| | | | | | | | | $P_5(A_5)$= | 0.1645 |
| | | | | | | | | $P_5(T_5)$= | 0.165 |
| | | | | | | | | $P_5(C_5)$= | 0.3356 |
| | | | | | | | | $P_5(G_5)$= | 0.3349 |

Fig. 8/6. Deinococcus radiodurans R1 chromosome 1, complete sequence, 2648638 bp, (Cosmic rays, X-rays, radioactive decay), https://www.ncbi.nlm.nih.gov/nuccore/NC_001263.1

| Ranges of fluctuations of collective probabilities for different positions in n-plets | | | | | | | |
|---|---|---|---|---|---|---|---|
| $P_n(A_k)$ | 0.158÷ 0.1623 | $P_n(T_k)$ | 0.1583÷ 0.1636 | $P_n(C_k)$ | 0.338÷ 0.3414 | $P_n(G_k)$ | 0.3361÷ 0.3426 |

Detailed data on collective probabilities $P_n(A_k)$, $P_n(T_k)$, $P_n(C_k)$ and $P_n(G_k)$ ($n = 1, 2, 3, 4, 5;\ k \leq n$ ) for each position in n-plets:

| **NUCLEOTIDES** | | **DOUBLETS** | | **TRIPLETS** | | **4-PLETS** | | **5-PLETS** | |
|---|---|---|---|---|---|---|---|---|---|
| $P_1(A_1)=$ | 0.1605 | $P_2(A_1)=$ | 0.1609 | $P_3(A_1)=$ | 0.1623 | $P_4(A_1)=$ | 0.1614 | $P_5(A_1)=$ | 0.1605 |
| $P_1(T_1)=$ | 0.1603 | $P_2(T_1)=$ | 0.1599 | $P_3(T_1)=$ | 0.1636 | $P_4(T_1)=$ | 0.1597 | $P_5(T_1)=$ | 0.1601 |
| $P_1(C_1)=$ | 0.3401 | $P_2(C_1)=$ | 0.3405 | $P_3(C_1)=$ | 0.338 | $P_4(C_1)=$ | 0.341 | $P_5(C_1)=$ | 0.3395 |
| $P_1(G_1)=$ | 0.3391 | $P_2(G_1)=$ | 0.3387 | $P_3(G_1)=$ | 0.3361 | $P_4(G_1)=$ | 0.3379 | $P_5(G_1)=$ | 0.34 |
| | | $P_2(A_2)=$ | 0.1602 | $P_3(A_2)=$ | 0.1612 | $P_4(A_2)=$ | 0.1603 | $P_5(A_2)=$ | 0.1608 |
| | | $P_2(T_2)=$ | 0.1607 | $P_3(T_2)=$ | 0.1592 | $P_4(T_2)=$ | 0.1605 | $P_5(T_2)=$ | 0.1607 |
| | | $P_2(C_2)=$ | 0.3397 | $P_3(C_2)=$ | 0.3411 | $P_4(C_2)=$ | 0.3396 | $P_5(C_2)=$ | 0.3402 |
| | | $P_2(G_2)=$ | 0.3394 | $P_3(G_2)=$ | 0.3385 | $P_4(G_2)=$ | 0.3396 | $P_5(G_2)=$ | 0.3383 |
| | | | | $P_3(A_3)=$ | 0.158 | $P_4(A_3)=$ | 0.1604 | $P_5(A_3)=$ | 0.16 |
| | | | | $P_3(T_3)=$ | 0.1583 | $P_4(T_3)=$ | 0.1602 | $P_5(T_3)=$ | 0.161 |
| | | | | $P_3(C_3)=$ | 0.3411 | $P_4(C_3)=$ | 0.3399 | $P_5(C_3)=$ | 0.3398 |
| | | | | $P_3(G_3)=$ | 0.3426 | $P_4(G_3)=$ | 0.3395 | $P_5(G_3)=$ | 0.3393 |
| | | | | | | $P_4(A_4)=$ | 0.16 | $P_5(A_4)=$ | 0.1604 |
| | | | | | | $P_4(T_4)=$ | 0.161 | $P_5(T_4)=$ | 0.1603 |
| | | | | | | $P_4(C_4)=$ | 0.3398 | $P_5(C_4)=$ | 0.3414 |
| | | | | | | $P_4(G_4)=$ | 0.3392 | $P_5(G_4)=$ | 0.3379 |
| | | | | | | | | $P_5(A_5)=$ | 0.1609 |
| | | | | | | | | $P_5(T_5)=$ | 0.1597 |
| | | | | | | | | $P_5(C_5)=$ | 0.3396 |
| | | | | | | | | $P_5(G_5)=$ | 0.3398 |

Fig. 8/7. Halobacterium sp. NRC-1, complete genome, 2014239 bp (High salt concentration), https://www.ncbi.nlm.nih.gov/nuccore/NC_002607.1

| Ranges of fluctuations of collective probabilities for different positions in n-plets | | | | | | | |
|---|---|---|---|---|---|---|---|
| $P_n(A_k)$ | 0.2767÷ 0.2783 | $P_n(T_k)$ | 0.2772÷ 0.279 | $P_n(C_k)$ | 0.222÷ 0.223 | $P_n(G_k)$ | 0.2205÷ 0.223 |

Detailed data on collective probabilities $P_n(A_k)$, $P_n(T_k)$, $P_n(C_k)$ and $P_n(G_k)$ (n = 1, 2, 3, 4, 5; $k \leq n$ ) for each position in n-plets:

| **NUCLEOTIDES** | | **DOUBLETS** | | **TRIPLETS** | | **4-PLETS** | | **5-PLETS** | |
|---|---|---|---|---|---|---|---|---|---|
| $P_1(A_1)$= | 0.2774 | $P_2(A_1)$= | 0.277 | $P_3(A_1)$= | 0.2773 | $P_4(A_1)$= | 0.277 | $P_5(A_1)$= | 0.2776 |
| $P_1(T_1)$= | 0.2782 | $P_2(T_1)$= | 0.2782 | $P_3(T_1)$= | 0.279 | $P_4(T_1)$= | 0.2781 | $P_5(T_1)$= | 0.2781 |
| $P_1(C_1)$= | 0.2226 | $P_2(C_1)$= | 0.2226 | $P_3(C_1)$= | 0.222 | $P_4(C_1)$= | 0.2224 | $P_5(C_1)$= | 0.2225 |
| $P_1(G_1)$= | 0.2218 | $P_2(G_1)$= | 0.2222 | $P_3(G_1)$= | 0.2217 | $P_4(G_1)$= | 0.2225 | $P_5(G_1)$= | 0.2217 |
| | | $P_2(A_2)$= | 0.2779 | $P_3(A_2)$= | 0.2767 | $P_4(A_2)$= | 0.2777 | $P_5(A_2)$= | 0.2772 |
| | | $P_2(T_2)$= | 0.2781 | $P_3(T_2)$= | 0.2772 | $P_4(T_2)$= | 0.2777 | $P_5(T_2)$= | 0.2784 |
| | | $P_2(C_2)$= | 0.2225 | $P_3(C_2)$= | 0.2228 | $P_4(C_2)$= | 0.223 | $P_5(C_2)$= | 0.2226 |
| | | $P_2(G_2)$= | 0.2215 | $P_3(G_2)$= | 0.2233 | $P_4(G_2)$= | 0.2216 | $P_5(G_2)$= | 0.2218 |
| | | | | $P_3(A_3)$= | 0.2783 | $P_4(A_3)$= | 0.277 | $P_5(A_3)$= | 0.2774 |
| | | | | $P_3(T_3)$= | 0.2783 | $P_4(T_3)$= | 0.2783 | $P_5(T_3)$= | 0.2781 |
| | | | | $P_3(C_3)$= | 0.2229 | $P_4(C_3)$= | 0.2229 | $P_5(C_3)$= | 0.2223 |
| | | | | $P_3(G_3)$= | 0.2205 | $P_4(G_3)$= | 0.2218 | $P_5(G_3)$= | 0.2222 |
| | | | | | | $P_4(A_4)$= | 0.278 | $P_5(A_4)$= | 0.2776 |
| | | | | | | $P_4(T_4)$= | 0.2785 | $P_5(T_4)$= | 0.2781 |
| | | | | | | $P_4(C_4)$= | 0.222 | $P_5(C_4)$= | 0.2227 |
| | | | | | | $P_4(G_4)$= | 0.2214 | $P_5(G_4)$= | 0.2216 |
| | | | | | | | | $P_5(A_5)$= | 0.2773 |
| | | | | | | | | $P_5(T_5)$= | 0.2781 |
| | | | | | | | | $P_5(C_5)$= | 0.2228 |
| | | | | | | | | $P_5(G_5)$= | 0.2218 |

Fig. 8/8. Chroococcidiopsis thermalis PCC 7203, complete genome, 6315792 bp, (Desiccation ), https://www.ncbi.nlm.nih.gov/nuccore/NC_019695.1

## Appendix 9. Symmetries of tetra-group probabilities in genomes of ferns

This Appendix consists of data about 2 species of ferns analysed in our researches.

| Ranges of fluctuations of collective probabilities for different positions in n-plets | | | | | | | |
|---|---|---|---|---|---|---|---|
| $P_n(A_k)$ | 0.2407÷ 0.2441 | $P_n(T_k)$ | 0.242÷ 0.2433 | $P_n(C_k)$ | 0.2527÷ 0.2591 | $P_n(G_k)$ | 0.2566÷ 0.2621 |

Detailed data on collective probabilities $P_n(A_k)$, $P_n(T_k)$, $P_n(C_k)$ and $P_n(G_k)$ ($n = 1, 2, 3, 4, 5;\ k \leq n$) for each position in n-plets:

| **NUCLEOTIDES** | | **DOUBLETS** | | **TRIPLETS** | | **4-PLETS** | | **5-PLETS** | |
|---|---|---|---|---|---|---|---|---|---|
| $P_1(A_1)=$ | 0.242 | $P_2(A_1)=$ | 0.2417 | $P_3(A_1)=$ | 0.241 | $P_4(A_1)=$ | 0.2417 | $P_5(A_1)=$ | 0.2422 |
| $P_1(T_1)=$ | 0.2427 | $P_2(T_1)=$ | 0.2431 | $P_3(T_1)=$ | 0.2432 | $P_4(T_1)=$ | 0.2429 | $P_5(T_1)=$ | 0.2426 |
| $P_1(C_1)=$ | 0.2556 | $P_2(C_1)=$ | 0.2553 | $P_3(C_1)=$ | 0.2591 | $P_4(C_1)=$ | 0.2551 | $P_5(C_1)=$ | 0.2558 |
| $P_1(G_1)=$ | 0.2597 | $P_2(G_1)=$ | 0.26 | $P_3(G_1)=$ | 0.2566 | $P_4(G_1)=$ | 0.2603 | $P_5(G_1)=$ | 0.2594 |
| | | $P_2(A_2)=$ | 0.2424 | $P_3(A_2)=$ | 0.241 | $P_4(A_2)=$ | 0.2425 | $P_5(A_2)=$ | 0.2429 |
| | | $P_2(T_2)=$ | 0.2423 | $P_3(T_2)=$ | 0.242 | $P_4(T_2)=$ | 0.2422 | $P_5(T_2)=$ | 0.2423 |
| | | $P_2(C_2)=$ | 0.2559 | $P_3(C_2)=$ | 0.2548 | $P_4(C_2)=$ | 0.2555 | $P_5(C_2)=$ | 0.2553 |
| | | $P_2(G_2)=$ | 0.2594 | $P_3(G_2)=$ | 0.2621 | $P_4(G_2)=$ | 0.2598 | $P_5(G_2)=$ | 0.2594 |
| | | | | $P_3(A_3)=$ | 0.2441 | $P_4(A_3)=$ | 0.2416 | $P_5(A_3)=$ | 0.242 |
| | | | | $P_3(T_3)=$ | 0.2428 | $P_4(T_3)=$ | 0.2433 | $P_5(T_3)=$ | 0.2425 |
| | | | | $P_3(C_3)=$ | 0.2527 | $P_4(C_3)=$ | 0.2554 | $P_5(C_3)=$ | 0.255 |
| | | | | $P_3(G_3)=$ | 0.2604 | $P_4(G_3)=$ | 0.2597 | $P_5(G_3)=$ | 0.2605 |
| | | | | | | $P_4(A_4)=$ | 0.2424 | $P_5(A_4)=$ | 0.2407 |
| | | | | | | $P_4(T_4)=$ | 0.2424 | $P_5(T_4)=$ | 0.2431 |
| | | | | | | $P_4(C_4)=$ | 0.2562 | $P_5(C_4)=$ | 0.256 |
| | | | | | | $P_4(G_4)=$ | 0.259 | $P_5(G_4)=$ | 0.2602 |
| | | | | | | | | $P_5(A_5)=$ | 0.2424 |
| | | | | | | | | $P_5(T_5)=$ | 0.2429 |
| | | | | | | | | $P_5(C_5)=$ | 0.2557 |
| | | | | | | | | $P_5(G_5)=$ | 0.259 |

Fig. 9/1. Neisseria meningitidis MC58 chromosome, complete genome, 2272360 bp, https://www.ncbi.nlm.nih.gov/nuccore/NC_003112.2 .

| Ranges of fluctuations of collective probabilities for different positions in n-plets | | | | | | | |
|---|---|---|---|---|---|---|---|
| $P_n(A_k)$ | 0.3071÷ 0.3089 | $P_n(T_k)$ | 0.3067÷ 0.3083 | $P_n(C_k)$ | 0.1912÷ 0.1928 | $P_n(G_k)$ | 0.1918÷ 0.1929 |

Detailed data on collective probabilities $P_n(A_k)$, $P_n(T_k)$, $P_n(C_k)$ and $P_n(G_k)$ (n = 1, 2, 3, 4, 5; $k \leq n$ ) for each position in n-plets:

| **NUCLEOTIDES** | | **DOUBLETS** | | **TRIPLETS** | | **4-PLETS** | | **5-PLETS** | |
|---|---|---|---|---|---|---|---|---|---|
| $P_1(A_1)$= | 0.308 | $P_2(A_1)$= | 0.3077 | $P_3(A_1)$= | 0.3071 | $P_4(A_1)$= | 0.3078 | $P_5(A_1)$= | 0.3079 |
| $P_1(T_1)$= | 0.3075 | $P_2(T_1)$= | 0.3076 | $P_3(T_1)$= | 0.3072 | $P_4(T_1)$= | 0.3075 | $P_5(T_1)$= | 0.3073 |
| $P_1(C_1)$= | 0.1921 | $P_2(C_1)$= | 0.1924 | $P_3(C_1)$= | 0.1928 | $P_4(C_1)$= | 0.1923 | $P_5(C_1)$= | 0.1923 |
| $P_1(G_1)$= | 0.1924 | $P_2(G_1)$= | 0.1923 | $P_3(G_1)$= | 0.1929 | $P_4(G_1)$= | 0.1925 | $P_5(G_1)$= | 0.1925 |
| | | $P_2(A_2)$= | 0.3083 | $P_3(A_2)$= | 0.308 | $P_4(A_2)$= | 0.3081 | $P_5(A_2)$= | 0.3078 |
| | | $P_2(T_2)$= | 0.3074 | $P_3(T_2)$= | 0.3072 | $P_4(T_2)$= | 0.3071 | $P_5(T_2)$= | 0.3078 |
| | | $P_2(C_2)$= | 0.1918 | $P_3(C_2)$= | 0.1922 | $P_4(C_2)$= | 0.192 | $P_5(C_2)$= | 0.1921 |
| | | $P_2(G_2)$= | 0.1925 | $P_3(G_2)$= | 0.1925 | $P_4(G_2)$= | 0.1929 | $P_5(G_2)$= | 0.1923 |
| | | | | $P_3(A_3)$= | 0.3089 | $P_4(A_3)$= | 0.3077 | $P_5(A_3)$= | 0.3083 |
| | | | | $P_3(T_3)$= | 0.3081 | $P_4(T_3)$= | 0.3077 | $P_5(T_3)$= | 0.3067 |
| | | | | $P_3(C_3)$= | 0.1912 | $P_4(C_3)$= | 0.1924 | $P_5(C_3)$= | 0.1925 |
| | | | | $P_3(G_3)$= | 0.1918 | $P_4(G_3)$= | 0.1921 | $P_5(G_3)$= | 0.1924 |
| | | | | | | $P_4(A_4)$= | 0.3085 | $P_5(A_4)$= | 0.3084 |
| | | | | | | $P_4(T_4)$= | 0.3077 | $P_5(T_4)$= | 0.3075 |
| | | | | | | $P_4(C_4)$= | 0.1916 | $P_5(C_4)$= | 0.1914 |
| | | | | | | $P_4(G_4)$= | 0.1922 | $P_5(G_4)$= | 0.1927 |
| | | | | | | | | $P_5(A_5)$= | 0.3076 |
| | | | | | | | | $P_5(T_5)$= | 0.3083 |
| | | | | | | | | $P_5(C_5)$= | 0.1921 |
| | | | | | | | | $P_5(G_5)$= | 0.192 |

Fig. 9/2. 'Nostoc azollae' 0708, complete genome, 5354700 bp, https://www.ncbi.nlm.nih.gov/nuccore/CP002059.1

## Appendix 10. Symmetries of tetra-group probabilities in genomes of a moss

This Appendix shows ranges of fluctuation of all 27 chromosomes of a moss Physcomitrella patens ecotype Gransden 2004, Phypa V3, whole genome shotgun sequence This moss is used as a model organism for studies on plant evolution, development, and physiology.

| Chromosome | $P_n(A_k)$ | $P_n(T_k)$ | $P_n(C_k)$ | $P_n(G_k)$ |
|---|---|---|---|---|
| 1 | 0.331÷ 0.3315 | 0.3311÷ 0.3317 | 0.1686÷ 0.1693 | 0.1683÷ 0.1688 |
| 2 | 0.3306÷ 0.3315 | 0.3321÷ 0.3333 | 0.1677÷ 0.1691 | 0.1675÷ 0.168 |
| 3 | 0.3306÷ 0.3313 | 0.3296÷ 0.3304 | 0.169÷ 0.1695 | 0.1695÷ 0.17 |
| 4 | 0.3315÷ 0.3322 | 0.3319÷ 0.3325 | 0.1683÷ 0.1687 | 0.1671÷ 0.1676 |
| 5 | 0.3322÷ 0.3327 | 0.3331÷ 0.3336 | 0.1668÷ 0.1672 | 0.1669÷ 0.1675 |
| 6 | 0.332÷ 0.3326 | 0.331÷ 0.3316 | 0.1683÷ 0.1689 | 0.1676÷ 0.1681 |
| 7 | 0.3309÷ 0.3318 | 0.3306÷ 0.3313 | 0.1683÷ 0.169 | 0.1689÷ 0.1696 |
| 8 | 0.3323÷ 0.3331 | 0.3324÷ 0.3329 | 0.1673÷ 0.1678 | 0.1667÷ 0.1674 |
| 9 | 0.3314÷ 0.3319 | 0.3309÷ 0.332 | 0.1678÷ 0.1685 | 0.1685÷ 0.1689 |
| 10 | 0.3322÷ 0.3331 | 0.3322÷ 0.333 | 0.1663÷ 0.1671 | 0.1675÷ 0.1684 |
| 11 | 0.3286÷ 0.329 | 0.33÷ 0.3307 | 0.171÷ 0.1714 | 0.1694÷ 0.1702 |
| 12 | 0.3306÷ 0.3314 | 0.3313÷ 0.3319 | 0.1681÷ 0.1688 | 0.1685÷ 0.1692 |
| 13 | 0.3319÷ 0.333 | 0.3328÷ 0.3334 | 0.1668÷ 0.1675 | 0.167÷ 0.1675 |
| 14 | 0.3319÷ 0.3326 | 0.3309÷ 0.3316 | 0.1673÷ 0.1678 | 0.1687÷ 0.1692 |
| 15 | 0.3308÷ 0.332 | 0.3323÷ 0.3331 | 0.1683÷ 0.1687 | 0.1672÷ 0.1678 |
| 16 | 0.3301÷ 0.3303 | 0.3297÷ 0.3306 | 0.1694÷ 0.1701 | 0.1696÷ 0.1703 |
| 17 | 0.3303÷ 0.3309 | 0.3302÷ 0.3307 | 0.1698÷ 0.1709 | 0.1686÷ 0.1694 |
| 18 | 0.3309÷ 0.3321 | 0.3319÷ 0.333 | 0.1677÷ 0.1683 | 0.1678÷ 0.1685 |
| 19 | 0.332÷ 0.333 | 0.3323÷ 0.333 | 0.1672÷ 0.1681 | 0.1669÷ 0.1674 |
| 20 | 0.3314÷ 0.332 | 0.3303÷ 0.3307 | 0.1693÷ 0.1698 | 0.1678÷ 0.1684 |
| 21 | 0.3308÷ 0.3319 | 0.331÷ 0.332 | 0.168÷ 0.1692 | 0.1679÷ 0.1694 |
| 22 | 0.33÷ 0.3314 | 0.3312÷ 0.3328 | 0.1685÷ 0.1696 | 0.1677÷ 0.1687 |
| 23 | 0.3307÷ 0.3312 | 0.3313÷ 0.3323 | 0.1691÷ 0.1698 | 0.1677÷ 0.1684 |
| 24 | 0.3304÷ 0.3314 | 0.3306÷ 0.3316 | 0.1683÷ 0.169 | 0.169÷ 0.1695 |
| 25 | 0.3335÷ 0.3347 | 0.333÷ 0.334 | 0.1653÷ 0.1666 | 0.1662÷ 0.1673 |
| 26 | 0.3303÷ 0.3313 | 0.3317÷ 0.3322 | 0.1685÷ 0.1692 | 0.1682÷ 0.1687 |
| 27 | 0.3305÷ 0.3321 | 0.3294÷ 0.3306 | 0.1687÷ 0.1708 | 0.1685÷ 0.1699 |

Fig. 10/1. Ranges of fluctuations of collective probabilities $P_n(A_k)$, $P_n(T_k)$, $P_n(C_k)$ and $P_n(G_k)$ ($n = 1, 2, 3, 4, 5$; $k \le n$ ) for different positions in n-plets for all 27 chromosomes Physcomitrella patens ecotype Gransden 2004 (initial data - from https://www.ncbi.nlm.nih.gov/genome/?term=Physcomitrella+patens).

## Appendix 11. Symmetries of tetra-group probabilities in genomes of monocots

This Appendix shows ranges of fluctuations of the collective probabilities in all 11 chromosomes of a monocot Musa acuminata.

| Chromosome | $P_n(A_k)$ | $P_n(A_k)$ | $P_n(A_k)$ | $P_n(A_k)$ |
|---|---|---|---|---|
| 1 | 0.3065÷ 0.3074 | 0.3063÷ 0.3071 | 0.1929÷ 0.1933 | 0.193÷ 0.1936 |
| 2 | 0.3048÷ 0.3056 | 0.3029÷ 0.3034 | 0.1955÷ 0.1962 | 0.1954÷ 0.1962 |
| 3 | 0.3046÷ 0.3052 | 0.3035÷ 0.304 | 0.1955÷ 0.1962 | 0.195÷ 0.1956 |
| 4 | 0.3054÷ 0.3061 | 0.3081÷ 0.3086 | 0.1927÷ 0.1933 | 0.1926÷ 0.1931 |
| 5 | 0.3069÷ 0.3075 | 0.3055÷ 0.3063 | 0.1934÷ 0.1941 | 0.193÷ 0.1933 |

| 6 | 0.305÷ 0.3058 | 0.3051÷ 0.3056 | 0.1944÷ 0.195 | 0.1941÷ 0.1949 |
|---|---|---|---|---|
| 7 | 0.305÷ 0.3056 | 0.3045÷ 0.3053 | 0.1944÷ 0.195 | 0.1948÷ 0.1956 |
| 8 | 0.3058÷ 0.3065 | 0.3056÷ 0.3063 | 0.1937÷ 0.1941 | 0.1938÷ 0.1941 |
| 9 | 0.3073÷ 0.3076 | 0.3064÷ 0.3069 | 0.1929÷ 0.1931 | 0.1925÷ 0.1932 |
| 10 | 0.3045÷ 0.3049 | 0.3054÷ 0.3058 | 0.1947÷ 0.1953 | 0.1943÷ 0.1947 |
| 11 | 0.306÷ 0.3065 | 0.3046÷ 0.3049 | 0.1945÷ 0.1951 | 0.1939÷ 0.1944 |

Fig. 11/1. Ranges of fluctuations of collective probabilities $P_n(A_k)$, $P_n(T_k)$, $P_n(C_k)$ and $P_n(G_k)$ (n = 1, 2, 3, 4, 5; $k \leq n$) for different positions in n-plets for all 11 chromosomes of Musa acuminata (initial data were taken from https://www.ncbi.nlm.nih.gov/genome/10976).

## Appendix 12. Symmetries of tetra-group probabilities in genomes of shrubs

This Appendix shows ranges of fluctuations of collective probabilities in all 13 chromosomes of a shrub Gossypium arboretum.

| **Chromosome** | **$P_n(A_k)$** | **$P_n(A_k)$** | **$P_n(A_k)$** | **$P_n(A_k)$** |
|---|---|---|---|---|
| 1 | 0.2661÷ 0.2677 | 0.2634÷ 0.2655 | 0.2336÷ 0.2351 | 0.2339÷ 0.2355 |
| 2 | 0.265÷ 0.2666 | 0.2644÷ 0.2658 | 0.2341÷ 0.2358 | 0.233÷ 0.2342 |
| 3 | 0.2654÷ 0.2666 | 0.2658÷ 0.2675 | 0.234÷ 0.2356 | 0.2318÷ 0.2332 |
| 4 | 0.2638÷ 0.2652 | 0.2652÷ 0.2675 | 0.2345÷ 0.236 | 0.2331÷ 0.235 |
| 5 | 0.2647÷ 0.2667 | 0.2674÷ 0.2689 | 0.2341÷ 0.2354 | 0.2307÷ 0.2327 |
| 6 | 0.2658÷ 0.2676 | 0.2677÷ 0.2697 | 0.2311÷ 0.2331 | 0.231÷ 0.2331 |
| 7 | 0.2677÷ 0.2692 | 0.2666÷ 0.2689 | 0.2318÷ 0.2339 | 0.2299÷ 0.2322 |
| 8 | 0.268÷ 0.2703 | 0.2668÷ 0.2697 | 0.2298÷ 0.2324 | 0.2304÷ 0.2327 |
| 9 | 0.2665÷ 0.2687 | 0.2687÷ 0.2711 | 0.229÷ 0.2332 | 0.2294÷ 0.2324 |
| 10 | 0.2669÷ 0.2693 | 0.2671÷ 0.2722 | 0.2288÷ 0.2333 | 0.2294÷ 0.232 |
| 11 | 0.2681÷ 0.2709 | 0.2692÷ 0.2716 | 0.2284÷ 0.2305 | 0.2303÷ 0.2318 |
| 12 | 0.2697÷ 0.2738 | 0.2657÷ 0.2689 | 0.2287÷ 0.2332 | 0.2297÷ 0.2318 |
| 13 | 0.2685÷ 0.271 | 0.2675÷ 0.2706 | 0.229÷ 0.2314 | 0.2294÷ 0.2312 |

Fig. 12/1. Ranges of fluctuations of collective probabilities $P_n(A_k)$, $P_n(T_k)$, $P_n(C_k)$ and $P_n(G_k)$ (n = 1, 2, 3, 4, 5; $k \leq n$) for different positions in n-plets for all 13 chromosomes of Gossypium arboretum (initial data were taken from https://www.ncbi.nlm.nih.gov/genome/10948).

## Appendix 13. Symmetries of tetra-group probabilities in genomes of trees

This Appendix shows ranges of fluctuations of collective probailities in all 10 chromosomes of a tree Theobroma cacao. Initial data about these chromosomes are given on the following 10 consecutive sites: https://www.ncbi.nlm.nih.gov/nuccore/CM001879.1, https://www.ncbi.nlm.nih.gov/nuccore/CM001880.1, https://www.ncbi.nlm.nih.gov/nuccore/CM001881.1 , https://www.ncbi.nlm.nih.gov/nuccore/CM001882.1 , https://www.ncbi.nlm.nih.gov/nuccore/CM001883.1 , https://www.ncbi.nlm.nih.gov/nuccore/CM001884.1 , https://www.ncbi.nlm.nih.gov/nuccore/CM001885.1 , https://www.ncbi.nlm.nih.gov/nuccore/CM001886.1 , https://www.ncbi.nlm.nih.gov/nuccore/CM001887.1 , https://www.ncbi.nlm.nih.gov/nuccore/CM001888.1 .

| **Chromosome** | **$P_n(A_k)$** | **$P_n(A_k)$** | **$P_n(A_k)$** | **$P_n(A_k)$** |
|---|---|---|---|---|
| 1 | 0.3295÷ 0.3299 | 0.3282÷ 0.3286 | 0.1708÷ 0.1712 | 0.1708÷ 0.1711 |
| 2 | 0.3288÷ 0.3291 | 0.3292÷ 0.3297 | 0.171÷ 0.1714 | 0.1702÷ 0.1706 |
| 3 | 0.3285÷ 0.3289 | 0.3291÷ 0.3295 | 0.1706÷ 0.171 | 0.1711÷ 0.1713 |
| 4 | 0.3289÷ 0.3292 | 0.3292÷ 0.3297 | 0.1703÷ 0.1706 | 0.1708÷ 0.1713 |
| 5 | 0.3305÷ 0.3312 | 0.3297÷ 0.3303 | 0.1695÷ 0.1702 | 0.1692÷ 0.1695 |
| 6 | 0.3296÷ 0.33 | 0.3291÷ 0.3294 | 0.1698÷ 0.1703 | 0.1707÷ 0.1711 |
| 7 | 0.3292÷ 0.3299 | 0.3295÷ 0.3301 | 0.1696÷ 0.1701 | 0.1706÷ 0.171 |
| 8 | 0.3278÷ 0.3286 | 0.3285÷ 0.3291 | 0.1715÷ 0.1719 | 0.1711÷ 0.1716 |
| 9 | 0.3293÷ 0.3299 | 0.331÷ 0.3313 | 0.1695÷ 0.1698 | 0.1693÷ 0.1698 |
| 10 | 0.3307÷ 0.3313 | 0.3309÷ 0.3314 | 0.1686÷ 0.1694 | 0.1687÷ 0.1693 |

Fig. 13/1. Ranges of fluctuations of collective probabilities $P_n(A_k)$, $P_n(T_k)$, $P_n(C_k)$ and $P_n(G_k)$ (n = 1, 2, 3, 4, 5; $k \leq n$) for different positions in n-plets for all 10 chromosomes of a tree Theobroma cacao.

**Acknowledgments:** Some results of this paper have been possible due to a long-term cooperation between Russian and Hungarian Academies of Sciences on the topic "Non-linear models and symmetrologic analysis in biomechanics, bioinformatics, and the theory of self-organizing systems", where S.V. Petoukhov was a scientific chief from the Russian Academy of Sciences. The author is grateful to G. Darvas, I. Stepanyan and G. Tolokonnikov for their collaboration. A special thanks is to V. Svirin for his computer programs, which were made on the basis of the author's technical tasks and descriptions, to study automatically long

DNA-sequences (initial author's results were obtained by a half-manually manner). The author notes with gratitude the support from the side of Z.B. Hu in carrying out this study. The author is very grateful to E. Fimmel, M. Gumbel, L. Strüngmann and A. Karpuzoglu for their selective verification and the confirmation of some of described results about tetra-groups symmetries by means of their independently created computer program and also for their fruitful discussion of the described results; it was done during the author's internship in autumn 2017 at the Institute of Mathematical Biology of the Mannheim University of Applied Sciences (Germany) on the basis of a scholarship provided by the German Academic Exchange Service (DAAD).